\documentclass{article}

\usepackage[lmargin=1.25in,rmargin=1.25in,tmargin=1.45in,bmargin=1.45in]{geometry}

\begin{document}

\title{{\footnotesize $~~~~~~~~~~~~~~$ $~~~~~~~~~~~~~~$ $~~~~~~~~~~~~~~$ $~~~~~~$ http://relativity.livingreviews.org/Articles/lrr-2013-5/}\\
{\footnotesize $~~~~~~~~~~~~~~$ $~~~~~~~~~~~~~~$ $~~~~~~~~~~~~~~$ $~~~~~~~~~~~~~~$ $~~~~~~~~~~~~~~$ Living Reviews in Relativity 16 (2013) 5}\\$~~~~~~~~~~~~~~$\\Quantum Spacetime Phenomenology}

\author{Giovanni Amelino-Camelia\\
{\small Dipart.~Fisica Univ.~La Sapienza and Sez.~Roma1 INFN}\\
{\small P.le Moro 2, I-00185 Roma, Italy}}

\maketitle

\begin{abstract}
I review the current status of phenomenological programs inspired
by quantum-spacetime research.
I stress in particular
the significance of results establishing that certain
data analyses provide sensitivity to effects introduced
genuinely at the Planck scale. And my main focus is on phenomenological
programs that
managed to affect the directions taken by studies of quantum-spacetime theories.
\end{abstract}

\newpage

\tableofcontents

\newpage


\section{Introduction and Preliminaries}
\label{intro}

\subsection{The ``Quantum-Gravity problem'' as seen by a
  phenomenologist}

Our present description of the fundamental laws of Nature is based on
two disconnected pieces: ``quantum mechanics'' and ``general relativity''.
On the quantum-mechanics side our most significant successes were obtained
applying relativistic quantum field theory, which turns out to be the appropriate
formalization of
(special-)relativistic quantum mechanics. This
theory neglects gravitational effects and is formulated in a flat/Minkowskian spacetime
background. Interesting results (but so far with little experimental support)
can be obtained by reformulating this theory in certain curved spacetime backgrounds,
but there is no rigorous generalization allowing for dynamics of gravitational fields.
The only known way for having a manageable formulation of some gravitational effects
within quantum field theory is to adopt the perspective of effective
field theory~\cite{burgessLRR,hanwillen},  which allows Lagrangians that are not renormalizable.
At leading order, this effective theory just gives us back EinsteinÕs general relativity,
but beyond leading order it predicts corrections proportional to powers of $E^2/E_p^2$, where $E$ is characteristic
energy scale of the process under consideration (typically the center of mass energy for a scattering experiment)
and $E_p$ is the Planck scale ($E_p \sim 10^{28}$~eV). The effective-field-theory description
evidently breaks down at energies $E$ of order the Planck scale, leaving unanswered~\cite{burgessLRR,hanwillen}
most of the core issues concerning the interplay between gravity and quantum mechanics.
And most importantly the
experiments which have formed our trust in quantum
mechanics are nearly exclusively experiments in which gravitational effects
are negligible at the presently-achievable levels of
experimental sensitivity (some of the rare instances where the outcome of a quantum-mechanical
measurement is affected by gravitational effects, such as the one reported in Ref.~\cite{natureNeutron},
will be discussed later in this review).

On the gravity side our present description is based on general
relativity. This a classical-mechanics theory which neglects all quantum properties of
particles.  Our
trust in general relativity has emerged in experimental studies
and observations in which gravitational interactions cannot be
neglected, such as the motion of planets around the Sun. Planets
are ``composed'' of a huge number of fundamental particles, and the
additive nature of energy (playing in such contexts roughly the role of ``gravitational charge")
is such that the energy of a planet is
very large, in spite of the fact that each composing fundamental
particle carries only small energy. As a result for planets
gravitational interactions dominate over other interactions.
Moreover, a planet satisfies the conditions under which
quantum theory is in the classical limit: in the description of
the orbits of the planets the quantum properties of the composing
particles can be safely neglected.

General relativity and relativistic quantum mechanics do have
some ``shared tools'', such as the notion of spacetime, but they handle these entities in
profoundly different manner. The differences are indeed so profound that
it might be natural to expect only one or the other language to be successful,
but instead they have both
been extremely successful. This is possible
 because of the type of experiments in which they have
been tested so far, with two sharply separated classes of
experiments, allowing  complementary approximations.

Of course, while somewhat puzzling from a philosopher's perspective,
all this would not on its own amount to a scientific problem. In the experiments
we are presently able to perform and at the level of sensitivities
we are presently able to achieve there is no problem. But a scientific
problem, which may well deserve to be called ``quantum-gravity problem'',
is found if we consider, for example, the structure of the scattering experiments
done in particle-physics laboratories. There are no surprises in
the analysis of processes with, say, an ``in'' state with two
particles each with an energy of $10^{12}$~eV.
Relativistic quantum mechanics makes definite
predictions for the (distributions/probabilities of)
results of this type of measurement procedures, and our
experiments fully confirm the validity of these predictions. We
are presently unable to redo the same experiments having as ``in
state'' two particles with energy of, say, $10^{30}$~eV (i.e., energy
higher than the Planck scale), but, nonetheless, if one factors out gravity,
relativistic quantum mechanics makes a definite prediction for
these conceivable (but presently undoable) experiments. However,
for collisions of particles of $10^{30}$~eV energy
 the gravitational interactions predicted by
general relativity are very strong and gravity should not be
negligible. On the other hand also the quantum properties
predicted for the particles by relativistic quantum mechanics (for
example the fuzziness of their trajectories) cannot be neglected,
contrary to the ``desires'' of the classical mechanics of our present description
of gravity. One could
naively attempt to apply both theories simultaneously, but
it is well established that such attempts do not produce anything
meaningful (for example by encountering uncontrollable divergences).
As mentioned above, a framework where these issues can be raised in
very precise manner is the one of effective quantum field theory, and the break down of the
effective quantum field theory of gravitation
at the Planck scale signals the challenges that are here concerning me.

This ``trans-Planckian collisions'' picture is one
(not necessarily the best, but a sufficiently clear) way to introduce a
quantum-gravity problem. But is the conceivable measurement procedure I just
discussed truly sufficient to introduce a scientific problem? One ingredient
appears to be missing: the measurement procedure is conceivable but presently
we are unable to perform it. Moreover, one could argue that
mankind might never be able to perform the measurement
procedure I just discussed. There appears to be no need to elaborate
predictions for the outcomes of that measurement procedure.
However, it is easy to see that the measurement procedure I just discussed
contains the elements of a true scientific problem.
One relevant point can be made considering the
experimental/observational evidence we are gathering about the ``early Universe''.
This evidence strongly supports the idea that in the early
Universe particles with energies comparable to the Planck energy
scale $E_p$ were abundant, and that these particles played a key
role in those early stages of evolution of the Universe. This does
not provide us with opportunities for ``good experiments''
(controlled repeatable experiments), but it does represent a
context in which proposals for the quantum-gravity/Planck-scale realm could be
tested. Different scenarios for the physical theory that applies
in the quantum-gravity realm could be compared on the basis of
their description of the early Universe. The detailed analysis of
a given physical theory for the quantum-gravity realm could allow
us to establish some characteristic predictions for the early
Universe and for some manifestations in our observations (cosmology)
of those early stages of evolution
of the Universe. The theory would be testable on the basis of
those predictions for our present observations. These
early-Universe considerations therefore provide an opportunity
for comparison between the predictions of a
quantum-gravity theory and measurement results.
And it might not be necessary to resort to cosmology: the fact
that (in setting up the quantum-gravity problem)
we have established some objective limitations of our present
theories implies that some qualitatively new
effects will be predicted by the theory that applies to the
quantum-gravity realm. These effects should dominate in that
realm (in particular, they will affect profoundly the results of
measurements done on particles with Planck-scale energy), but
 they should always be present. For processes involving
particles with energy $E$ much smaller than $E_p$ the implications
of a typical quantum-gravity theory will be rather marginal but
not altogether absent. The magnitude of the associated effects
should be suppressed by some small overall coefficients,
probably given by powers of the ratio $E/E_p$; small but different from zero.

We therefore do have a genuine ``quantum-gravity problem'', and this
problem has been studied for more than
70~years~\cite{stachelearly}. Unfortunately most of this research has
been conducted assuming that no guidance could be
obtained from experiments. But of course, if there is to be
a ``science'' of the quantum-gravity problem, this problem must be treated just
like any other scientific problem, seeking desperately the
guidance of experimental facts, and letting those facts take the
lead in the development of new concepts. Clearly physicists
must hope this works
also for the quantum-gravity problem, or else abandon it to the
appetites of philosophers.

It is unfortunately true that there is a certain level of risk that
experiments might never give us any clear
lead toward quantum gravity, especially if we are correct in expecting
that the magnitude of the characteristic effects of the new
theory should be set by
the tiny Planck length ($\ell_p \equiv 1/E_p \sim 10^{-35}$~m, the inverse
of the huge Planck scale in natural units). But even if the new effects were
really so small we could still try to uncover experimentally some
manifestations of quantum gravity. This is hard, and there is no
guarantee of success, but we must try.
As I shall stress again in later parts of this first section,
let me notice here that some degree of optimism could be inspired by considering, for example,
the prediction of proton decay within certain modern grandunified
theories of particle physics. The decay probability for a proton
in those theories is really very small, suppressed by the fourth
power of the ratio between the mass of the proton and the
grandunification scale (a scale which is only some three orders of magnitude
smaller than the Planck scale), but meaningful tests of scenarios for proton decay in
grandunified theories have been devised.

While the possibility of a ``quantum gravity phenomenology''~\cite{polonpap}
could be considered, on the basis of these arguments, even in the early days
of quantum-gravity research, a
sizable effort finally matured only since the second half of the 1990s.
In particular, only over this recent period
 we have the first cases of phenomenological
programs that truly affect the directions taken by more formal work in quantum gravity.
And, especially in relation to this healthy two-way cross-influence between
formal theory and phenomenology, a prominent role has been played
by proposals testing features that could be manifestations of
 spacetime quantization.
The expectation that the fundamental description of spacetime should
not be given by a classical geometry is shared by a large majority of quantum-gravity
researchers. And as a result the phenomenology inspired by this expectation has had influence
on a sizable part of the recent quantum-gravity literature.
My goal here is primarily the one of reviewing
the main results and proposals produced by this emerging area
of phenomenology
centered on the possibility of spacetime quantization.


\subsection{Quantum spacetime vs quantum black hole and graviton exchange}
The notion of ``quantum-gravity research'' can have a different meaning for different
researchers. This is due both to the many sides of the quantum-gravity problem
and the fact that most  researchers arrive to the study
of quantum gravity from earlier interests in other areas of physics research.
Because of its nature the quantum-gravity problem has a different appearance,
for example, to a particle physicist and to a relativist.

In particular, this affects the perception of
the implications of the  ``double role'' of gravitational fields:
unlike all other fields studied in fundamental physics
the gravitational field is not just used to
describe  ``gravitational interactions'' but also characterizes
the structure of spacetime itself.
The structure of Einstein's theory of gravitational phenomena
tells us both of the geometry of spacetime, which should be described
in terms of smooth Reimannian manifolds, and of the implications
of Einstein's equations for dynamics.
But in most approaches these two sides of gravity are not handled
on the same footing. Particularly from the perspective of a particle physicist
it makes sense to focus on contexts amenable to
treatment assuming some given Reimannian-manifold spacetime background
and ``gravitons'' as mediators of ``perturbative gravitational interactions''. Other researchers,
typically not coming from a particle-physics background,  are
instead primarily interested in speculations for how to  replace
Reimannian manifolds in the description of the structure
of spacetime, and contemplating a regime describing perturbative gravitational interactions
is not one of their main concerns.

Because of my objectives,  it is appropriate for me to locate
early on in this review the quantum-spacetime issues within the broader
spectrum of quantum-gravity research.

\subsubsection{The quantum-black-hole regime}

We do expect that there is a regime of physics where quantum-gravity
does not simply amount to small corrections to our currently adopted theories,
but rather our current theories should be there completely inapplicable.
An example of this is the class of hypothetical situations discussed in my
opening remarks: if we consider a collision with impact parameter
of order the Planck length between two particles which exchange in the collision
an energy of order the Planck scale then our current theories do not even
give us a reliable first approximation of the outcome.

Such collisions would create a concentration of energy comparable to the Planck scale
in a region of Planck-length size. And we have no previous experience with systems
concentrated in a Planck-length region with rest energy\footnote{For simplicity I am assuming
the description of the collision is being given in the center-of-mass frame. For example
we might have two identical particles which in that frame have the same gigantic energy
but propagate in opposite directions.} of order the Planck scale.
In such cases the pillars of our current description of the laws of physics
come in very explicit conflict. On one side we have quantum mechanics, with its
characteristic property that a rest energy $M$ can only be localized within
a region of size the Compton wavelength
$$r_{C} \sim \frac{\hbar}{M}~.$$
On the other side general relativity assigns to any localized (point-like)
amount $M$ of rest energy a region of size its Schwarzschild radius
$$r_{S} \sim G_N M \sim \frac{\ell_p^2 M}{\hbar}~,$$
where I denoted Newton's constant by $G_N$  and I denoted again by $\ell_p$
the Planck length ($\ell_p \equiv \sqrt{\hbar G}\sim 10^{-35}\mathrm{\ m}$, in units with speed-of-light
scale set to unity, $c=1$).

If $M \sim \hbar/\ell_p$ (rest energy of order the Planck scale)
the Compton and the Schwarzschild radii
are of the same order of magnitude and quantum mechanics cannot ignore
gravitation but at the same time gravitation cannot ignore quantum mechanics.
Evidently we can get nowhere attempting to investigate this issue by just
combining\footnote{Note however that this type of Heisenberg-microscope-type issues
can be studied in the some of the frameworks under consideration for the quantum-gravity problem.
In particular, this has attracted some interest by proponents
 of asymptotic safety, as illustrated by the study reported in Ref.~\cite{percacciHeisenberg}
 (for an alternative perspective see Ref.~\cite{dvaliHeisenberg}).}
 somehow the Standard Model of particle physics and the general-relativistic
classical description of gravitational phenomena.

Another context of similar conceptual content can be imagined
if we take for granted (which of course we can do only as working assumptions)
the existence of Hawking radiation. We could then start with an isolated
macroscopic black hole and attempt to describe its whole future evolution.
As long as the black hole remains macroscopic, but looses weight through
Hawking radiation, we can imagine to be able to devise a reliable first
approximation. But when then the black hole reaches Planck-length size
(and Planck-scale rest energy)
we are again left without any even approximate answers.

The description of these types of ``quantum-black-hole regimes''
(description which I shall use rather generically, including for example the regime
characteristic of the very early Universe)
is evidently an example of cases such that we could only have a satisfactory
picture by understanding how both of the two roles of the gravitational fields
need to be revised (how spacetime structure should be then described and how
the gravitational-interaction aspects of gravitation should be then described).

Providing a description of such a quantum-black-hole regime is probably
the most fascinating
challenge for quantum-gravity research, but evidently it is not a promising
avenue for actually discovering quantum-gravity effects experimentally.
As I shall mention, somewhat incidentally in a few points of this review,
this expectation would of course change if surprisingly gravitation turns out to be
much stronger than we presently expect, so that at least in some contexts
its strength is not characterized by the Planck scale. But
this review adopts the conservative view that quantum-gravity effects are
at least roughly as small as we expect, and therefore characterized roughly
by the Planck scale. And if that is the case it is hard to even imagine a future
when we gain access to a quantum-black-hole regime.

A key assumption of this review is that quantum gravity will manifest itself
experimentally in the shape of small corrections to contexts which we are able
to describe in first approximation within our current theories.

\subsubsection{The graviton-exchange regime}
For particle physicists (and therefore for at least part of the legitimate
overall perspective on the quantum-gravity problem)
the most natural opportunities in which  quantum gravity
could introduce small corrections are in contexts involving the
gravitational-interaction aspects of quantum gravity.

Rather than attempting to give general definitions let me offer
a clear example. These are studies of long-range corrections to the Newtonian
limit of gravitation, where gravity does look like a Newton-force interaction.
By focusing on long-range features one stays far from the trouble zone
mentioned in the previous subsection. But there are still issues of considerable
interest, at least conceptually, that a quantum gravity should address
in that regime. It is natural to expect that the description of gravity in
terms of a Newton-force interaction would also show traces of the new
laws that quantum gravity will bring about.

This possibility can be investigated coherently (but without any guarantee of a reliable answer)
with effective-field-theory
techniques applied to the nonrenormalizable theory of quantum gravity
obtained by linearizing the Einstein-Hilbert theory before quantization.
It essentially turns into an exercise of exploring the properties that
such an effective theory attributes
 to
gravitons.
And one does derive a correction to Newton's potential with
behaviour~\cite{donoghue1993,bellucci1996,kirilin2002,donoghue2003,kirilin2004,woodard2010}
$$\Delta V_{Newton} \sim \frac{\hbar G^2 M}{r^3}\sim \frac{\ell_p^2}{r^2} V_{Newton}~,$$
where $M$ is the mass of the source of the gravitational potential and on the
right-hand side I highlighted the fact that this correction would come in suppressed with
respect to the standard leading Newtonian term by a factor given by the square
of the ratio of the Planck length versus the distance scale at which the potential is probed.

This illustrates the sort of effects one may look for within schemes centered
on a background Minkowski spacetime and properties of the graviton.
In this specific case the effect is unmanageably small\footnote{Our familiarity with
the Newtonian regime of gravity extends down to distance scales no shorter than $\sim 10^{-6}\mathrm{\ m}$
so the factor $\ell_p^2/r^2$ would invite us to determine a Newtonian potential
with accuracy of at least parts in $10^{58}$.},
but in principle one could look for other effects
of this sort which might be observably large in some applications.

\subsubsection{The quantum-spacetime regime}
Having given some example of the way in which a quantum-gravity might change
our description of gravitational interactions, let me now turn
to the complementary type of issues that are in focus when one
studies the idea of spacetime quantization.

The nature of the quantum-gravity problem tells us in many ways that
the ultimate description of spacetime structure is not going to be
in terms of a smooth classical geometry.
We do not have at present enough information to deduce how our formalization
of spacetime should change, but it must change.
The collection of arguments in support of this expectation
(see, e.g.,
Refs.~\cite{mead,venekonmen1,venekonmen2,venekonmen3,venekonmen4,padma,dopliPLB,ahlu1994,ng1994,gacmpla,garay,casadio})
is impressive and relies both on aspects of the quantum-gravity problem
and on analyses of proposed approaches to the solution of the quantum-gravity
problem.

Surely some very dramatic
manifestations of
 spacetime quantization should be expected in what I labeled as
 the quantum-black-hole regime.
 But, as already stressed above,
 it is hard to even imagine managing to derive evidence of
 spacetime quantization from experimental access to that regime.
It is easy to see that our best chance for uncovering non-classical properties
of spacetime is to focus on the implications
of spacetime quantization for the ``Minkowski limit'' (or perhaps the ``de Sitter limit'')
of quantum gravity.
Our data on contexts we presently describe as involving particle propagation
in a background Minkowski spacetime is abundant and of high quality.
If the fundamental formalization of spacetime is not in terms of a smooth classical
geometry then we should find some traces of spacetime quantization
also in those well-studied contexts.
The effects are likely to
be very small, but the quality of the data available to us in this
quantum-spacetime regime is very high, occasionally high enough
to probe spacetime structure with Planck-scale sensitivity.

This is the main theme of my review. I do not elaborate further on it here
since it will of course take full shape in the following.

\subsubsection{Aside on the classical-gravity regime}
It is an interesting aspect of how the quantum-gravity community
is fragmented to observe that it is sometimes difficult to explain to
a relativist how graviton-exchange studies could be seen
as part of quantum-gravity research and it is difficult to explain to
particle physicists how studies of particles not interacting
gravitationally in a quantum-spacetime can play a role in quantum-gravity
research.
I hope this subsection proves useful in this respect.

Let me also discuss one more aspect of the interplay between quantum mechanics
and gravitation which can be of interest from a quantum-gravity perspective, even though
at first sight it does not look like quantum gravity at all. These are studies of quantum mechanics
in a curved background spacetime, without assuming spacetime is quantized
and without including any graviton-like contribution to the interactions.
No aspect of gravity is quantized in such studies  but they concern
a regime which must be present as a limiting case of quantum
gravity, and therefore by studying this regime we are establishing
constraints on how quantum-gravity might look like.

On the conceptual side perhaps the most significant example of how quantum
mechanics in curved spacetime backgrounds can provide important hints
toward quantum gravity is provided by studies of black-hole thermodynamics.
And it is a regime of physics where we do have some experimental access
mainly through studies of the quantum properties
of particles in cases when the geometry of spacetime
near the surface of the Earth (essentially gravity of Earth, the acceleration $g$)
does matter. I shall mention a couple of these experimental studies in the next subsection.

While my focus here is on quantum-spacetime studies, it will be occasionally
useful for me to adopt the perspective of quantum mechanics in curved
classical background spacetimes.

\subsection{20th century quantum-gravity phenomenology}
In order to fully expose the change of perspective which matured over the last
decade it is useful to first discuss briefly some
 earlier analyses that  made contact with experiments/observations and are relevant
for the understanding of the interplay between general relativity
and quantum mechanics.

Some of the works produced by Chandrasekhar  in the 1930s already
fit this criterion.
In particular, the renowned Chandrasekhar limit~\cite{chandralimit1,chandralimit2},
which describes the maximum mass of a white-dwarf star,
was obtained introducing
some quantum-mechanical properties of particles
(essentially Pauli's exclusion principle)
within an analysis of gravitational phenomena.

A fully rigorous derivation of the
Chandrasekhar limit would of course require ``quantum gravity'', but not all of it:
it would suffice to master one special
limit of quantum gravity, the ``classical-gravity limit'',
in which one takes into account the quantum
properties of matter fields (particles) in presence of rather
strong spacetime curvature (treated however classically).
By testing experimentally the
Chandrasekhar-limit formula one is therefore to some extent
probing (the classical-gravity limit of) quantum gravity.

Also relevant for the
classical-gravity limit of quantum gravity
are the relatively more recent  studies of the implications of the Earth's gravitational field
in matter-interferometry experiments.
Experiments investigating these effects
have been conducted since the mid 1970s
and are often called ``COW experiments'' from the initials of
Colella, Overhauser and Werner who performed the
first such experiment~\cite{cow}.
The main target of these studies is the form of the
Schr\"{o}dinger equation in presence of the Earth gravitational field,
which could be naturally conjectured to be of the form\footnote{Consistently
with what is done in the relevant literature I write Eq.(\ref{coweq}) without
assuming exact validity of the equivalence principle, and therefore introducing
different symbols $M_I$ and $M_G$ for the inertial
and gravitational mass respectively.}
\begin{eqnarray}
\left[ - \left( {1 \over 2 \, M_I} \right) \vec{\nabla}^2
+ M_G \, \phi(\vec{r}) \right] \psi(t,\vec{r}) = i  \,
{\partial \, \psi(t,\vec{r}) \over \partial t}
\label{coweq}
\end{eqnarray}
for the description of the dynamics of matter (with wave
function $\psi(t,\vec{r})$) in presence of the Earth's
gravitational potential $\phi(\vec{r})$.

The COW experiments exploit the fact that
the Earth's gravitational potential puts together the contributions
of a very large number of particles (all the
particles composing the Earth)
and, as a result, in spite of its per-particle weakness,
the overall gravitational field is large enough to introduce observable
effects.

Valuable
reading material relevant for these COW experiments can be
found in Refs.~\cite{sakurai,gasperiniEP,dharamEP}.
While the basic message is that a gravity-improved
Schr\"{o}dinger  equation of the form~(\ref{coweq})
is indeed essentially applicable, some interesting discussions
have been generated by these COW experiments, particularly as a result
of the data reported by one such experiment~\cite{cowEPviol}
(data whose reliability is still being debated),
which some authors have interpreted a a possible manifestation
of a violation of the Equivalence Principle.

In the same category of studies relevant for the classical-gravity
limit of quantum gravity I should mention some proposals put forward
mainly by Anandan (see, e.g., Ref.~\cite{anan1,anan2}),
already in the mid 1980s,
and some very recent remarkable studies
which test how the gravitational field
affects the structure of quantum states, such as the study reported in
Ref.~\cite{natureNeutron} which I shall discuss in some detail later in this review.

Evidently the study of the classical limit provides only a limited window
on quantum gravity, and surely cannot provide any insight on the possibility
of short-distance spacetime quantization, on which I shall here focus.

A list of early examples of studies raising at least the issue that spacetime structure
might one day be probed with Planck-scale sensitivity should start
with the arguments reported by Mead already in 1965~\cite{meadPHEN}.
There Mead was contemplating broadening of spectral lines possibly resulting
from adopting the
Planck length as the value of the minimum possible uncertainty in position measurements.
Then in works published in the 1980s and early 1990s there were a few
phenomenological studies, adopting the Planck scale as target
and focusing essentially on the possibility that
quantum-mechanical coherence might be spoiled by quantum-gravity
effects.
One example is provided by the studies of Planck-scale-induced
CPT-symmetry violation and violations
of ordinary quantum mechanics reported in Refs.~\cite{ehns,elmn} and references
therein (also see, for aspects concerning mainly the CPT-symmetry aspects,
Refs.~\cite{huetpesk,floreacpt}),
which are particularly
relevant for the analysis of data~\cite{cplear} on the neutral-kaon system.
A quantization of spacetime is encoded in the non-critical-string-theory
formalism adopted in these Refs.~\cite{ehns,elmn}, but only to the extent that
one can view as such the novel description of time
there adopted.
A similar characterization  applies to the
studies reported in Refs.~\cite{perci1,perci0,perci2}, which
considered violations
of ordinary quantum mechanics of a type describable in terms of
the ``primary-state-diffusion'' formalism, with results that could be
relevant for atom interferometry. Also in Refs.~\cite{perci1,perci0,perci2}
the main quantum-spacetime feature is found in the description of time.

From a broader quantum-gravity-problem perspective I should also mention
the possibility of violations of CPT and Lorentz symmetry within string theory
analyzed in Refs.~\cite{kostesamuel,kostcpt}. These studies, like most phenomelogy-relevant
studies inspired by string theory (see related comments later in this review),
do not involve any spacetime quantization and do not necessarily imply that
the magnitude of the effects is set by a Planckian scale. But they should nonetheless
be prominently listed among the early proposals assuming that some of the
theories used in quantum-gravity research might be testable with currently-available
experimental techniques.


\subsection{Genuine Planck-scale sensitivity and the dawn of
  quantum-spacetime phenomenology}

The rather isolated proposals that composed ``20th-century quantum-gravity phenomenology'' were already rather significant. In particular, some
of these studies,
perhaps most notably the ones in Ref.~\cite{elmn} and Ref.~\cite{perci2} ,
were providing first preliminary evidence of the fact that
it might be possible to investigate experimentally the structure
of spacetime at the Planck scale, which is expected to be the main
key for the understanding of the quantum-gravity realm, and should involve
spacetime quantization.
But, in spite of their objective significance, these studies did not manage to
have an impact on the overall development of quantum-gravity
research. For example, all mainstream quantum-gravity reviews
up to the mid 1990s still only mentioned the ``experiments issue''
in the form of some brief folkloristic
statements, such as ``the only way to test Planck scale effects
is to build a particle accelerator
all around our galaxy''.

This fact that up to the mid 1990s the possibility of a quantum-spacetime phenomenology
was mostly ignored resulted in large part from
a common phenomenon of ``human inertia''
that affects some scientific communities, but some role
 was also played by a meaningful technical
observation:
the studies available up to that point
relied on models with the magnitude of the effect set
by a free dimensionless parameter,
and at best  the sensitivity of the experiment was at a level
such that one could argue setting the value of
the dimensionless parameter as a ratio between the Planck length
and one of the characteristic length scales of the relevant physical context.
It is of course true
that this kind of dimensional-analysis
reasoning does not amount to really establishing that  the relevant
candidate quantum-gravity effect is being probed with Planck-scale sensitivity,
and this resulted in a perception that such studies, while deserving some interest,
could not be described objectively as probes of the quantum-gravity realm.
For some theorists a certain level of uneasiness also originated from the fact
that the formalisms adopted in studies such as the ones
in Ref.~\cite{elmn} and Ref.~\cite{perci2}
involved rather virulent departures from quantum mechanics.

Still it did turn out that those earlier attempts to investigate the quantum-gravity
problem experimentally were setting the stage for
a wider acceptance of
quantum-spacetime phenomenology.
The situation
started to evolve rather rapidly when in the span of just a few years,
between 1997 and 2000, several analyses were produced describing
different physical contexts in which effects introduced
genuinely at the Planck sale could be tested.
It started with some analyses of observations of gamma-ray bursts
at sub-MeV energies~\cite{grbgac,gampul,schaefer},
then came some analyses of large laser-light
interferometers~\cite{gacgwi,gacgwiB,bignapap,nggwi},
quickly followed by the first discussions of Planck scale effects
relevant for the analysis of ultra-high-energy cosmic rays~\cite{kifune,ita,gactp}
and the first analyses relevant for observations of TeV gamma rays
from Blazars~\cite{ita,gactp,aus} (also see Ref.~\cite{kluz,billetal}).

In particular, the fact that some of these analyses (as I discuss in detail
later) considered Planck-scale effects amounting to departures from classical
Lorentz symmetry played a key role in their ability to have an impact
on a significant portion of the overall quantum-gravity-research effort.
Classical Lorentz symmetry is a manifestation of the smooth (classical) light-cone
structure of Minkowski spacetime, and it has long been understood that
by introducing new ``quantum features'' (e.g.,
discreteness or noncommutativity of the
spacetime coordinates) in spacetime structure, as some aspects of the ``quantum-gravity
problem'' might invite us to do, Lorentz symmetry may be affected.
And the idea of having some departure from Lorentz symmetry does not necessarily
require violations of ordinary quantum mechanics.
Moreover, by offering an opportunity to test quantum-gravity theories at
 a pure kinematical level, these ``Lorentz-symmetry-test proposals''
 provided a path toward testability that appeared to be accessible even
 to the most ambitious theories that are being considered as candidates
 for the solution of the quantum gravity problem. Some of these theories
 are so complex that one cannot expect (at least not through the work
 of only a few generations of physicists) to extract all of their physical
 predictions, but the kinematics of the ``Minkowski limit'' may well be within
 our reach. An example of this type is provided
 by Loop Quantum Gravity~\cite{crLIVING,ashtelewandREVIEW,leeLQGrev,thieREV,ashtekarLQGreview2012},
  where one is presently unable to
 even formulate many desirable physics questions, but at least some
 (however tentative) progress
 has been made~\cite{gampul,urrutia,thiemLS,kodadsr,bojoLQGphen}
 in the exploration of the kinematics of the Minkowski limit.

From a pure-phenomenology perspective the late-1990s transition
is particularly significant, as I shall discuss in greater detail later,
in as much as it marks a sharp transition toward
falsifiability. Some of the late-1990s phenomenology
proposals concern effects that one can imagine honestly deriving
in a given quantum-gravity theory. Instead the effects described for example
in studies such as the ones reported in
Ref.~\cite{elmn} and Ref.~\cite{perci2} were not really derived
from proposed models but rather they were inspired
by some paths toward the solution of the quantum-gravity problem (the relevant formalisms
were not really manageable to the point of allowing
a rigorous derivation of the nature and size of the effects under study,
but some intuition for the nature and size of the effects was developed
 combining our limited understanding of the formalisms and some heuristics).
 Such a  line of reasoning is certainly valuable,
 and can inspire some meaningful ``new physics'' experimental searches,
 but if the results of the experiments are negative the theoretical
 ideas that motivated them are not falsified: when the link from theory
 to experiments is weak (contaminated by heuristic arguments) it is not
 possible to follow the link in the opposite direction (use negative
 experimental results to falsify the theory).
 Through further developments
 of the work that started in the late 1990s we are now getting close to taking
 quantum-spacetime phenomenology from the mere realm of searches of
 quantum-spacetime effects (which are striking if they are successful but have
 limited impact if they fail) to the one of ``falsification tests'' of some
 theoretical ideas.
 This is a point that I am planning to convey strongly with some key parts
 of this review, together with another sign a maturity of this phenomenology:
 the ability to discriminate between different (but similar) Planck-scale
 physics scenarios.
 Of course, in order for a phenomenology to even get started one must find some
 instances in which the new-physics effects can be distinguished from the
 effects predicted by current theories, but a more mature phenomenology should
 also be able to discriminate between similar (but somewhat different)
  new-physics scenarios.

Together with some (however slow) progress toward establishing the
ability to falsify models and discriminate between models, the phenomenology
work of this
past decade has also shown that
the handful of examples of ``Planck-scale sensitivities'' that
generated excitement between 1997 and 2000 were not a ``one-time
lucky streak'': the list of examples of experimental/observational
contexts in which sensitivity to some effects introduced genuinely
at the Planck scale is established (or found to be realistically within reach)
has continued to grow at a
steady pace, as the content of this review will indicate, and the
number of research groups joining the quantum-spacetime-phenomenology
effort is also growing rapidly.
And it
is not uncommon for recent quantum-gravity
reviews~\cite{ashtereview,crHISTO,leePW,carlip},
even when
the primary focus is on developments on the mathematics side,
to discuss in some detail (and acknowledge the significance of)
the work done in quantum-gravity phenomenology.


\subsection{A simple example of genuine Planck-scale
  sensitivity}
\label{simple-example}

So far my preliminary description of quantum-spacetime phenomenology has
a rather abstract character. It may be useful to now provide a
simple example of analysis that illustrates some of the concepts
I have discussed and renders more explicit the fact that
some of the sensitivity levels now available experimentally
do correspond to
effects introduced genuinely at the Planck scale.

These objectives motivate me to
invite the reader to contemplate the possibility of a
discretization of spacetime on a lattice with $E^{-1}_{p}$ lattice
spacing and a free particle propagating on such a spacetime.
It is well established that in these hypotheses there
are $E^{-2}_{p}$ corrections
to the energy-momentum on-shell relation, which in general
are of the type\footnote{For a detailed discussion of the implications
for the energy-momentum
relation of a given scheme of spacetime discretization see, e.g., Ref.~\cite{thooftDiscret}.}
\begin{equation}
 m^2 \simeq E^2 - \vec{p}^2
 + \sum_{\{m_\mu\}} \eta_{m_0,m_1,m_2,m_3} \left({E^{m_0} p_1^{m_1} p_2^{m_2} p_3^{m_3} \over E^2_{p}}\right)
 +O \left(\frac{E^6}{E_p^4}\right)
~, \label{displeadDISCR}
\end{equation}
where the non-negative integers ${\{m_\mu\}}$ are such that $m_0+m_1+m_2+m_3=4$,
and the parameters $\eta_{m_0,m_1,m_2,m_3}$, which for $E^{-1}_{p}$ lattice spacing
typically turn out to be of order 1 (when non-zero),
reflect the specifics of the chosen discretization.

I should stress that the idea of a rigid-lattice
description of spacetime is not really one of the most advanced
for quantum-gravity research (but see the recent related study in Ref.~\cite{rigidfoam}).
Moreover, while it is easy to describe a free particle on such a lattice,
the more realistic case of interacting fields is very different, and its implications
for the form of the on-shell relation are expected to be significantly more complex
than assumed in (\ref{displeadDISCR}).
In particular, if described within effective field theory the implications for interacting theories
of such a lattice description of spacetime include
departures from special-relativistic on-shellness
for which there is no Planck-scale suppression, and are therefore unacceptable.
This is due to loop corrections,
through a mechanism of the type discussed in Refs.~\cite{suda1,suda2,suda3,tuning}
(on which I shall return later), and assumes one is naturally unwilling to contemplate
extreme fine-tuning.
I feel it is nonetheless very significant that
the however unrealistic case of a free particle propagating in a lattice with Planck-scale lattice spacing
leads to features of the type shown in (\ref{displeadDISCR}).
It shows that features of the type shown in (\ref{displeadDISCR}) have magnitude set
by nothing else but a feature of Planck-scale magnitude introduced in spacetime structure.
So, in spite of the idealizations involved, the smallness of the effects discussed in this subsection
 is plausibly representative of the type
 of magnitude that quantum-spacetime effects could have,
 even though any realistic model of, say, the Standard Model of particle physics in
 a  quantum spacetime should evidently remove those idealizations.

And of course one finds that
in most contexts corrections to the energy-momentum relation
of the type in Eq.~(\ref{displeadDISCR}) are completely
negligible. For example, for the analysis of center-of-mass collisions
between particles of energy $\sim$~1~TeV (such as the ones studied
at the LHC) these correction terms
affect the analysis at the level of 1 part in $10^{32}$.
However (at least if such a modified dispersion relation is part of a framework
with standard laws of energy-momentum conservation)  one
easily finds~\cite{kifune,ita,aus,gactp} significant implications for the
cosmic-ray spectrum. In particular one can consider the ``GZK cutoff'',
which is a key expected
feature of the cosmic-ray spectrum, and is essentially given by the
threshold energy for cosmic-ray protons to produce pions in
collisions with CMBR photons. In the evaluation of the threshold
energy for $p+\gamma_{\mathrm{CMBR}} \rightarrow p+\pi$ the $1/E^2_{p}$ correction
terms of~(\ref{displeadDISCR}) can be very
significant. As I shall discuss in greater detail in Section~\ref{photopion-production},
whereas the classical-spacetime prediction for the
GZK cutoff is around $5 \cdot 10^{19}$~eV, a much higher value of the cutoff
is naturally obtained~\cite{kifune,ita,aus,gactp}
in frameworks with the structure of Eq.~(\ref{displeadDISCR}).
The Planck-scale correction terms in Eq.~(\ref{displeadDISCR}) turn into
corresponding correction terms for the threshold-energy formula, and the
significance of these corrections can be  roughly
estimated with  $\eta E^4 / (\epsilon E^2_{p})$,
where $E$ is the energy of the cosmic-ray proton and $\epsilon$ is the energy
of the CMBR photon, to be compared to $m^2/16 \epsilon$, where $m$ here is the proton mass,
which roughly gives the GZK scale.
Adopting the ``typical quantum-gravitist
estimate''\footnote{$\eta$ here represent one of (or a linear combination of)
the coefficients $\eta_{m_0,m_1,m_2,m_3}$.
Of course the
quantum-gravity intuition for $\eta$ is $|\eta| \sim 1$.
For example, in my simple-minded ``spacetime-lattice picture'' $|\eta| \simeq 1$
is obtained when the lattice spacing is exactly $E^{-1}_{p}$. Sensitivity
to values of $|\eta|$
even smaller than 1 would reflect the ability to probe spacetime structure
down to distance scales even smaller than the Planck length (in
the ``spacetime-lattice picture'' this would correspond to a
lattice spacing of $\sqrt{\eta} E^{-1}_{p}$).}  $|\eta| \sim 1$
it turns out that in the GZK regime the ratio $E/m$
is large enough to compensate for the smallness of the ratio $E/E_p$,
so that a term of the type $E^4 / (\epsilon E^2_{p})$ is not negligible
with respect to $m^2/\epsilon$.
This observation is one of the core ingredients of the quantum-spacetime
phenomenology that has been
done~\cite{kifune,ita,aus,gactp} analyzing GZK-scale cosmic rays.
Of course another key ingredient of those analyses is the quality of
cosmic-ray data, which has improved very significantly over these last few years,
especially as a result of observations performed
at the Pierre Auger Observatory.

Let me here use this cosmic-ray context also as an opportunity
to discuss explicitly a first example of the type of ''amplifier''
which is inevitably needed in quantum-gravity phenomenology.
It is easy to figure out~\cite{polonpap,gactp} that
the large ordinary-physics number that acts as amplifier of the Planck-scale
effect in this case is provided by the ratio between a
cosmic-ray proton ultra-high energy, which can be of order $10^{20}$~eV,
and the mass (rest energy) of the proton.
This is clearly shown by the comparison I made between an estimate of Planck-scale
corrections of order $E^4 / (\epsilon E^2_{p})$ and an estimate of the uncorrected
result of order $m^2/\epsilon$. Evidently $E/m$ is the amplifier of the Planck-scale
corrections, which also implies that these Planck-scale modifications
of the photopion-production threshold formula go very quickly from being significant to being
completely negligible, as the proton energy is decreased.
A cosmic-ray proton with energy $E$ of the
order of $10^{20}$~eV is so highly boosted that $E/m_p \sim 10^{11}$,
and this leads to $E^4 / (\epsilon E^2_{p}) \sim m^2/\epsilon$ in my estimates,
but at accelerator-accessible proton energies (and proton boosts with respect
to its rest frame) the correction is completely negligible.
According to traditional quantum-gravity arguments,
which focus only on the role played by the ratio  $E/E_p$,
one should assume that this analysis could be successful only when $E/E_p \sim 1$;
clearly instead this analysis is successful already at energies
of order $10^{20}$~eV (i.e., some 8 orders of magnitude below the Planck scale).
And this is not surprising since the relevant
Planck-scale effect is an effect of Lorentz symmetry violation, so that of course
large boosts (i.e., in this context, large values of $E/m_p$) can act
as powerful  amplifiers of the effect, even when the energies are not Planckian.


\subsection{Focusing on a neighborhood of the Planck scale}

There is a strikingly large number of arguments pointing to the Planck scale as the
characteristic scale of quantum-gravity effects. Although clearly these arguments are not
all independent, their overall weight must be certainly judged as substantial. I shall not
review them here since they can easily be found in several quantum-gravity reviews,
and there are even some dedicated review papers (see, e.g., Ref.~\cite{garay}).
Faithful to the perspective of this review, I do want to stress one argument in favour
of the Planck scale as the quantum-gravity/quantum-spacetime
scale which is often overlooked, but is in my opinion
particularly significant, especially since it is based (however indirectly) on experimental facts.
These are the well-known experimental facts pointing to a unification of
the coupling ``constants'' of the electroweak forces
and of the strong force.
While gravity usually is not involved in arguments that provide support
for unification of the nongravitational couplings,
it is striking from a quantum-gravity perspective that,
even just using the little information we presently have (mostly at scales
below the TeV scale), our present best extrapolation of the available data on the running
of these coupling constants rather robustly indicates that there will indeed be a unification
and that this unification will occur at a scale which is not very far from the Planck scale.
In spite of the fact that we are not in a position to exclude that it be just a
quantitative accident,
this correspondence between (otherwise completely unrelated) scales
must presently be treated as the clearest hint of new physics
that is available to us.

The present (admittedly preliminary)
status of our understanding of this ``unification puzzle'' might even suggest that there
could be a single stage of full unification of all forces, including gravity. However,
according to the arguments that are presently fashionable among theoretical physicists,
it would seem that the unification of nongravitational coupling constants should occur sizably
above the scale of $(10^{27}eV)^{-1}$ (presently preferred is a value close
to $(2 \cdot 10^{26}eV)^{-1}$) and at such relatively large distance scales gravity
should still be too weak to matter, since it is indeed naively expected that gravity
should be able to compete with the other forces only starting at scales
as short as the Planck length, of $\sim (10^{28}eV)^{-1}$.

Even setting aside this coupling-unification argument, there are other compelling reasons
for attributing to the Planck scale the role of characteristic sale of quantum-gravity effects.
In particular, if one adopts the perspective of the effective-quantum-field-theory description
of gravitational phenomena the case for the Planck scale can be made rather precisely.
A particularly compelling argument in this respect is found in Ref.~\cite{hanwillen}
which focuses on the loss of unitarity within the effective-quantum-field-theory description
of gravitational phenomena.
Unitarity has been a successful criterion for determining the scale
at which other effective quantum field theories break down, such as the Fermi theory of weak interactions.
And it does turn out that the scale at which unitarity is violated for
the effective-quantum-field-theory description
of gravitational phenomena is within an order of magnitude of the Planck scale~\cite{hanwillen}.

But it appears legitimate to consider alternatives to such estimates.
For example, some authors (see, {\it e.g.}, Ref.~\cite{calmethsuPRL})
consider it to be likely  that the ``effective Newton constant''
is also affected by some sort of renormalization-group running,
and if this is the case than the outlook of all these arguments would change significantly.
For the length scale of spacetime quantization, $\ell_{\mathrm{QST}}$,
naively assumed to be given by $\sqrt{G_N(\infty)}$,
where $G_N(\infty)$ is the measured value of the Newton constant (characteristic of
gravity at large distances), any running of gravity would imply an estimate\footnote{Evidently
this might mean that the length scale of spacetime quantization might be somewhat lower or somewhat
higher than the Planck scale. But notice that when reasoning
in terms of $\ell_{\mathrm{QST}} \sim \sqrt{G_N(\ell_{\mathrm{QST}})}$
one should then allow for the possibility that there might not be any spacetime quantization after all
(the self-consistent solution of $\ell_{\mathrm{QST}} \sim \sqrt{G_N(\ell_{\mathrm{QST}})}$
might be $\ell_{\mathrm{QST}} =0$).}
of the type $\ell_{\mathrm{QST}} \sim \sqrt{G_N(\ell_{\mathrm{QST}})}$.

In relation to estimates of the scale of spacetime quantization
these considerations should invite us to consider the Planck length, $\sim 10^{-35}\mathrm{\ m}$
 only
as a crude, very preliminary estimate.
Throughout this review I shall tentatively take into account this issue
by assuming that the scale where nonclassical properties of spacetime emerge
should be somewhere between $\sim 10^{-32}\mathrm{\ m}$ and $\sim 10^{-38}\mathrm{\ m}$, hoping
that 3 orders of magnitude of prudence from above and from below could suffice.

It is striking that these considerations also allow one to be be more optimistic with respect
to the (already intrinsically appealing~\cite{wilczekUNI})
hypothesis of a single stage of unification of all forces, possibly
even at distance scales as ``large''
as $(10^{26}eV)^{-1} \simeq 10^{-33}\mathrm{\ m}$.
And I find that, in relation to this issue, the recent (mini-)burst of interest
in the role of gravity in unification is particularly exciting.
A convincing case is being built concerning the
possibility that gravity might affect the running
of the Standard-Model coupling constants, and this to could
have significant effects for the estimate of the unification scale
(see, e.g., Refs.~\cite{wilczekUNI,tomsNature} and references therein).
And in turn there is a rather robust argument
(see, e.g., Refs.~\cite{calmethsuPRL,calmethsuPRD} and references therein)
suggesting the other fields might affect significantly the
 strength of gravity.

 My personal perspective on the overall balance of this limited insight
 that is available to us is summarized by the attitude I adopted for this
 review in relation to the expectations for the value of the quantum-spacetime scale.
Unsurprisingly I give top priority for this to the only (and however faint) indication
we have from experiments: the values measured for coupling constants
at presently accessible ``ultra-large'' distance scales appear to be arranged in
such a way to produce a unification of nongravitational forces at a much smaller
length scale, which happens to be not distant from where we would naively expect
gravity to come into the picture. This in some sense tells us that our naive estimate
of where gravity becomes ``strong'' (and spacetime turns nonclassical)
cannot be too far off the mark. But at the same time imposes upon us at least
a certain level of prudence: we cannot assume that the quantum-spacetime scale
is exactly the Planck length but we have some encouragement for assuming that
it is within a few orders of magnitude of the Planck length.

In closing this long aside on the quantum-gravity/quantum-spacetime scale
let me stress that even prudently assuming a few orders of uncertainty above
and below the Planck length is not necessarily safe. It is in my opinion
the most natural working assumption in light of the information presently
available to us, but we should be fully aware of the fact that our naive
estimates might be off by more than a few orders of magnitude.
Following the line of reasoning I here adopted this would take the shape
of a solution for $\ell_{\mathrm{QST}} \sim \sqrt{G_N(\ell_{\mathrm{QST}})}$
that unexpectedly turned out to be wildly different form the Planck length.
The outlook of the analysis of the unification of forces appears to
discourage such speculations, but we must keep an open-minded perspective
(more on this toward the end of this review, when I briefly
consider the ``large extra dimensions'' scenario).

\subsection{Characteristics of the experiments}

Having commented on the first ``ingredient''
for the search of experiments
relevant for quantum-spacetime and quantum-gravity,
which is the estimate of the characteristic
scale of this new physics,
let me next comment on a few other ingredients,
starting with  some
intuition for the
type of quantum-spacetime effects that
one might plausibly look for,
and what that requires.

As stressed earlier in this section,
we cannot place much hope of experimental breakthroughs in the
full quantum-black-hole regime.
Our best chances are for studies
of contexts amenable to a description in terms of the properties of particles
in a background quantum spacetime.
And, as also already stressed,
these effects will be minute,
with magnitude governed by some power of the ratio
between the Planck length and the wavelength of the
particles involved.

The presence of these suppression factors
on the one hand reduces sharply our chances of actually discovering
quantum-spacetime effects, but on the other hand simplifies
the problem of figuring out what are the most promising experimental contexts,
since these experimental contexts must enjoy very
special properties which would not go easily unnoticed.
For laboratory experiments
even an optimistic estimate of these suppression factors
leads to a suppression of order $10^{-16}$, which one obtains
by assuming (probably already using some optimism)
that at least some quantum-gravity effects are only linearly
suppressed by the Planck length and taking as particle wavelength
the shorter wavelengths we are able to produce ($\sim 10^{-19}$~m).
In astrophysics (which however limits one to ``observations''
rather than ``experiments'') particles of shorter wavelength
are being studied, but even for the highest energy cosmic rays,
with energy of $\sim 10^{20}$~eV and therefore wavelengths
of $\sim 10^{-27}$~m, a suppression of the type $L_p/\lambda$
would take values of order $10^{-8}$.
It is mostly as a result of this type of considerations
that traditional quantum-gravity reviews considered
the possibility of experimental studies with unmitigated
pessimism.
However, the presence of these large suppression factors
surely cannot suffice for drawing any conclusions.
Even just looking within the subject of particle physics
we know that certain types of small effects can be studied,
as illustrated by the example of the remarkable limits
obtained on proton instability.
The prediction of proton decay within certain grandunified theories
of particle physics is really a small effect, suppressed by the fourth
power of the ratio between the mass of the proton and
grandunification scale, which is only
three orders of magnitude smaller than the Planck scale.
In spite of this horrifying suppression,
of order $[m_{\mathrm{proton}}/E_{gut}]^4 \sim 10^{-64}$,
with a simple idea we have managed to acquire full sensitivity
to the new effect: the proton lifetime predicted by grandunified
theories is of order $10^{39}$~s and quite a few generations
of physicists should invest their entire lifetimes staring at a single
proton before its decay, but by managing to keep under observation
a large number of protons (think for example of a situation in which
$10^{33}$ protons are monitored)
our sensitivity to proton decay is dramatically increased.
In that context the number of protons is the (ordinary-physics)
dimensionless quantity that works as ``amplifier'' of the new-physics
effect.

Outside of particle physics more
success stories of this type are easily found: think for example
of the brownian-motion
studies conducted already a century ago.
Within the 1905 Einstein description one uses
Brownian-motion measurements on macroscopic scales
as evidence for the atomic structure of matter. For the Brownian-motion case the needed
amplifier is provided by the fact that a very large number of microscopic processes
intervenes in each single macroscopic effect that is being measured.

It is hard but clearly not impossible to find
experimental contexts in which there is effectively a large
amplification of some small effects of interest.
And this is the strategy that is adopted~\cite{polonpap}
in the attempts to gain access
to the Planck-scale realm.


\subsection{Paradigm change and test theories of not everything}

Something else that characterizes the work attitude of the community
whose results I am here reviewing is the expectation that the
solution of the quantum gravity problem will require a significant
change of theory paradigm. Members of this community find in the structure
of the quantum-gravity problem sufficient elements for expecting that
the transition form our current theories to a successful theory of quantum gravity
should be no less (probably more) significant then the transition from
classical mechanics to quantum mechanics, the prototypical example
of a change of theory paradigm.

This marks a strong difference in intuition and methodology with respect
to other areas of quantum-gravity research, which do not assume the
need of a paradigm change. If for example the string theory program turned
out to be successful then quantum gravity should take the shape of
 just one more (particularly complex but nonetheless consequential)
step in the exploitation of the current theory paradigm, the one that took
us all the way from QED to the Standard Model of particle physics.

This difference of intuitions even affects the nature of the sort of questions
the different communities ask. The expectation of those not preparing
for a change of theory
paradigm is that one day some brilliant mind will wake up with the correct full
quantum gravity theory, with a single big conceptual jump.
What is expected is  single big conceptual step leading to
 a theory that describes
potentially everything we know so far\footnote{Some colleagues even use the
expression ``theory of everything'' without adding ``we kow so far''. }
Something of the sort of the discovery of QCD: a full theory even though some
of its answers to our questions are not immediately manifest once the theory
is written out (see, e.g., confinement).

The expectation of those who are instead
preparing for a change of theory
paradigm is that
we will get to a mature formulation of quantum gravity
only at the end of a multi-step journey, with each step being
of rather humble nature.
The model here of course is the phase of the ``old quantum theory''.
The change of theory paradigm in going from classical mechanics to
quantum mechanics was of such magnitude that of course we could
not possibly get it right in one single jump. Imagine someone, however brilliant,
looking at black-body radiation and proposing a solution based on
observables described as self-adjoint operators on Hilbert
spaces and all that.
Planck's description of black-body radiation was very far from being a full
formalization of quantum mechanics, and was even internally unsatisfactory,
with a very limited class of contexts and regimes where it could be applied.
It was a theory of very few things,
but it was a necessary step toward quantum mechanics. A similar role in the gradual
emergence of quantum mechanics was played by other theories of limited scope,
such as Einstein's description of the photoelectric effect, Bohr's description
of atoms, and the successful
proposal by de Broglie that wave-particle duality should be applied also to matter.

So while those not preparing for a change of paradigm look for
theories of everything, we are looking for theories of very few things,
like Planck, Einstein, Bohr, de Broglie and other great contributors
to the ultimate advent of quantum mechanics.
Let me here add that even when exploiting a successful theory paradigm
often the next level of exploitation still requires us taking some clumsy steps
based on theories of few things.
Consider for example Fermi's description of weak interactions
in terms of four-fermion-vertex processes. Fermi's theory can be applied to
a limited class of phenomena and only in a relatively narrow regime, and it is
even a theory that is not satisfactory from the perspective of internal
logical consistency. Yet Fermi's theory was an important and necessary step
toward richer and more satisfactory descriptions of weak interactions.

The difference in methodologies is also connected with some practical
considerations, connected with the fact that
the formalisms presently being considered as solutions for the
quantum-gravity problem
are so complex that very little is understood of their truly physical
implications. Some theories of few things can even be inspired by
a given theory of everything:
since it is {\it de facto} impossible to compare to data
present full candidates for quantum gravity one ends
up comparing to data
the predictions of an associated ``test theory'', a model
that is inspired by some
features we do understand
(usually not more than qualitatively or semi-quantitatively)
of the original theory but casts them within
a simple framework that is well suited for comparison to experiments
(but for which there is no actual guarantee of full compatibility with
the original theory).

So in the eyes of some workers these test theories
of few things are needed to bridge the gap
between the experimental data and our present understanding
of the relevant formalisms.
In the eyes of others the
test theories
of few things are just attempts to bridge the gap
between the experimental data available to us and our
limited understanding the quantum-gravity problem.

Essentially in working in quantum-spacetime phenomenology one
must first develop some intuition for some candidate quantum-spacetime effects.
And this can come either from analyzing the
 structure of the formalisms that are being considered
in the search of a solution to the quantum-gravity problem
or  from analyzing
the structure of the quantum-gravity problem.
Once a class of effects is deemed of interest
 some test theories of these candidate effects must be developed
 so that they can
be used as guidance for experimental searches.

From the perspective of a phenomenologist
 some carefully tailored test theories
can also be valuable as some sort of
common language to be used in assessing
the progresses made in improving the sensitivity of experiments,
a language that must be suitable for access both from the side of
experimentalists and from the side of
those working at the development of quantum-gravity
theories.

The possibility to contemplate such ``quantum-gravity theories of not
everything'' is facilitated by the fact
that the ``quantum-gravity problem'' can be described in terms of
several ``subproblems'', each challenging us perhaps as much as some
full open problems of other areas of physics. To mention just a few of
these ``subproblems'' let me notice that: (i)~it appears likely that
the solution of this problem requires a nonclassical description of
spacetime geometry, (ii)~quantum gravity might have to be profoundly
different (from an ``information-theory perspective'') from previous
fundamental-physics theories, as suggested by certain analyses of the
evolution of pure states in a black-hole background, (iii)~the
perturbative expansions that are often needed for the analysis of
experimental data might require the development of new techniques,
since it appears that the ones that rely on perturbative
renormalizability might be unavailable, and (iv)~we must find some way
to reconcile general-relativistic background independence with the
apparent need of quantum mechanics to be formulated in a given
background spacetime.

For each of these aspects of the quantum-gravity problem we can in principle
attempt to devise formalisms, intended
as descriptions of those regimes of the quantum-gravity
realm that are dominantly characterized by the corresponding features.


\subsection{Sensitivities rather than limits}

In providing my description of the present
status of quantum-spacetime phenomenology
I shall here adopt as my ``default mode" the one of characterizing
the sensitivities
that are within reach for certain classes of experiments/observations, with
only a few cases where I discuss both sensitivities and
available experimental limits.
The analysis of sensitivities was the traditional exercise a decade ago,
in the early days of modern quantum-spacetime phenomenology,
since the key objective then was to establish that
sensitivity to effects introduced genuinely at the Planck
scale is achievable. In light of the observation I already reported
in Section~\ref{simple-example}
(and several other observations reported later in this review),
the ``case for existence'' of quantum-spacetime phenomenology is
at this point well settled.

We are now entering a more mature phase in which
we start having the first examples of
 candidate quantum-spacetime effects for which the development of
suitable test theories is approaching a level of maturity
such that  placing experimental bounds (``limits'')
on the parameters of these test theories does deserve intrinsic interest.
However, at the time of writing this review
of quantum-spacetime phenomenology, the transition ``from sensitivities
to limits'' is not yet complete. The cases where I will offer comments on available
experimental limits are cases for which (in my opinion)
this transition has been made satisfactorily. But in several area of quantum-spacetime
phenomenology it is still common practice to discuss
experimental bounds on the basis of a single
little-understood experimental
result (often a single observation in astrophysics) and most of the
test theories are not yet developed to the point that we can attach much significance
to placing limits on their parameters.
This is of course a key issue, and
throughout this review I will find opportunities to
discuss in more detail my concerns  and
offer some remarks that are relevant for completing the needed
transition ``from sensitivities
to limits''.
I do plan to update regularly this review, and with each update
readers should find the emphasis gradually going more and more
from sensitivities to experimental limits.

\subsection{Other limitations of the scopes of this review}
\label{limitations}
After having clarified
that the ``default mode'' of this
 review provides descriptions of sensitivities
 (with occasional characterizations of experimental bounds),
 I should also comment on the types of theory works and especially
 of phenomenology work which are the main focus of this review.
I have prepared other reviews on these and related topics~\cite{polonpap,newjourn}
with a broader perspective but much more limited depth.
Here my main focus is to analyze and review in some detail
the healthy interface between pure theory and phenomenology
of quantum spacetime.
I shall mostly describe the phenomenology proposals but the selection
of which proposals should be included is primarily based
on their proven ability to motivate developments on the pure-theory side
and to react to (take into account adaptively of)  the indications that then emerge from these
pure-theory studies.
This will be the  ``default mode'' of my selection of topics, with some exceptions
allowed in cases where I find that there are promising opportunities for such a healthy
interface to mature over the next few years.

The net result of these goals of the review produces a certain bias toward
proposals for quantum spacetime which originated  from
(or where inspired by) the study of Loop Quantum Gravity
and/or the study of Planck-scale spacetime noncommutativity.
These are the two areas of pure-theory research in which, so far,
the desirable  two-way interface has most concretely materialized: pure-theory specialists
have redirected part of their work toward the topics that phenomenologists
have highlighted as most promising for phenomenology;
and the work of quantum-spacetime phenomenologists has been in turn influenced
by the results then obtained on the pure-theory side.

In addition to a relatively long list of proposals inspired by Loop Quantum Gravity
and/or by Planck-scale spacetime noncommutativity
I shall also comment on a few proposals inspired by other approaches
to spacetime quantization (e.g.
Causal Sets and Noncritical String Theory).
From a broader quantum-gravity-problem perspective
 one should of course also consider
Critical String Theory, which actually remains the most studied candidate
for quantum gravity.
However, I focus here on quantum-spacetime effects and effects
whose natural characteristic scale is the Planck scale, whereas
the phenomenology proposals so far inspired by the
Critical-String-Theory research program
do not revolve around quantum properties of spacetime
and often the characteristic scale of the effects is not naturally the Planck scale.

I shall observe in the next section that
the analysis of Critical String Theory actually
has provided encouragement for the idea that it could be also a model
of spacetime quantization, but the relevant aspects of  Critical String Theory
are still poorly understood and have not produced phenomenological
proposals of the sort I am here reviewing.
I do believe that it is likely that in a not-so-distant future some new opportunities
for quantum-spacetime phenomenology will arise from this avenue.


\subsection{Schematic outline of this review}

The main objective of the next section is to motivate
a list of candidate quantum-spacetime effects,
on the basis of the structure of the quantum-gravity problem and/or
of results obtained in certain theories that are being considered as
relevant to the understanding of the quantum-gravity problem.
The rest of this review attempts to describe the status
of searches of these candidate quantum-spacetime effects.

Choosing which structure to give to
Sections~\ref{symmetry-tests},~\ref{other-areas},~\ref{uvir}
and~\ref{QGC}
was the main challenge faced by my work on this review.
The option which finally prevailed attempts to assign each
phenomenological proposal to a certain area of quantum-spacetime
phenomenology. These of course should be viewed only as tentative assignments,
or at least assignments based on a perception of what could be the primary
targets of a given phenomenological proposal. And there are some visible limitations:
some readers could legitimately argue that a certain subsection which I placed in
one of the sections should instead find a more fitting setting in another section.
Indeed as I was working on this review there were a few subsections
which kept switching from one
section to another. If used wisely, I feel that the structure I gave is still
preferable to some of the alternatives that could have been considered.
For example even such a tentative structure of organization is probably
going to be more easy to use than a long unstructured list of all the many
phenomenological proposals I am considering.
And the option of organizing phenomenological proposals on the basis
of the theories that motivate them, rather than roughly on the basis
of their primary area of relevance in phenomenology, would have been against
the whole spirit of this review.

Section~\ref{symmetry-tests} focuses on effects that amount to Planck-scale departures
from Lorentz/Poincar\'e symmetry, which is the type of effects
on which the most energetic quantum-spacetime phenomenology effort has been so far
directed.
The content of Section~\ref{symmetry-tests} has some overlap with another
\textit{Living Review}~\cite{mattinLRR} that described the status
of modern tests of Lorentz symmetry, and therefore was in part also devoted
to cases in which such tests are motivated by quantum-spacetime research.
My perspective will however be
rather different, focused on the quantum-spacetime-motivated searches and
also using the example of Lorentz/Poincar\'e-symmetry tests
to comment on the level of maturity reached by quantum-spacetime phenomenology
in relation to the falsification of (test) theories and to the discrimination
between different but similar theories. And whereas from the broader viewpoint
of probing the robustness of Lorentz symmetry one should consider
as significant any proposal capable of improving the bounds established within
a given parametrization of departures from Lorentz symmetry,
I shall focus on the demands of Planck-scale sensitivity, as required
by the objectives of research on Planck-scale quantization of spacetime,
which is here my main focus.

Then, in Section~\ref{other-areas}, I describe the status of other areas
of  quantum-spacetime phenomenology in which the Planck-scale also
characterizes the onset of ultraviolet effects, but not of the types that
require departures from Lorentz/Poincar\'e symmetry.

While  the primary objectives of this review
concern ultraviolet effects linked with the Planck-scale structure of spacetime,
in Section~\ref{uvir} I briefly consider the possibility of UV/IR (ultraviolet/infrared)
mixing. In such UV/IR-mixing scenarios
the role of the Planck scale would be in governing the UV side,
and possibly then combining with other scales when IR effects
are considered.

Sections~\ref{symmetry-tests},~\ref{other-areas}, and~\ref{uvir}
concern proposals of (only a few) controlled laboratory experiments and
(several) observations in astrophysics. These are the contexts in which
presently one finds more mature proposals, particularly concerning
the robustness of claims of Planck-scale
sensitivity of some relevant data analyses. However, observations in
cosmology should also provide some very valuable opportunities, and
there are some  ``quantum-spacetime-cosmology'' proposals, to which I
devote Section~\ref{QGC}, that can already be used to expose the great
potential reach of this type of analyses.

While different proposals of quantum-spacetime phenomenology
often involve different formalizations and completely different
experimental techniques, there is a common setup of all proposals
described in  Sections~\ref{symmetry-tests},~\ref{other-areas},~\ref{uvir},
and~\ref{QGC}. This main strategy of quantum-spacetime phenomenology
is summarized in Section~\ref{beyond-standard} , also pondering
what might be some of its  limitations.

 Section~\ref{closing-remarks} offers some closing remarks.

\newpage


\section{Quantum-Gravity Theories, Quantum Spacetime, and Candidate Effects}
\label{candidate-effects}

Before getting to the main task of this review, which concerns
phenomenology proposals,  it is useful to
summarize briefly the motivation for studying
certain candidate quantum-spacetime effects.
The possible sources of motivation come either form analyses of the
structure of the quantum-gravity problem
or from what is emerging
in the development of some theories that
have been proposed as candidate solutions of the quantum-gravity problem.
As already stressed, my main focus here is on effects
which could be linked to {\emph{spacetime quantization}}
at (about) the {\emph{Planck scale}}, and particularly the ones
that were involved in the two-way interface that materialized over this last decade
between phenomenologists and theorists working on the loop-quantum-gravity
approach and spacetime noncommutativity.

In the first part of this section I offer a few comments on some of the approaches
being pursued in the study of the quantum-gravity problem,
mostly focusing on whether or not they support a quantum-spacetime picture
and the role played by the Planck scale.
This part focuses primarily on Loop Quantum Gravity
and spacetime noncommutativity, but I also comment briefly on Critical String Theory
and other approaches.

Then in the second part of this section I list some key candidates as phenomena that
could characterize the quantum-spacetime realm.
 This list is only very tentative but it seems to me we cannot do any better
 than this at the present time.
Indeed, compiling a list of candidate quantum-spacetime effects
is not straightforward. Analogous situations in other
areas of physics are usually such that there are a few new
theories which have started to earn our trust by successfully
describing some otherwise unexplained data, and then often we let
those theories guide us toward new effects that should be looked
for. The theories that are under consideration for the solution
of the quantum-gravity problem, and for a ``quantum'' (non-classical)
description of spacetime,
 cannot yet claim any success in the experimental realm.
Moreover, even if nonetheless
we wanted to use them as guidance for experiments the
complexity of these theories proves to be a formidable obstruction.
In most cases, especially  concerning testable predictions,
the best we can presently do with these theories is analyze their general structure
and use this as a source of intuition for the proposal of a few candidate
effects.
Similarly when we motivate the search of certain quantum-spacetime features
on the basis of our present understanding of the quantum-gravity
 problem we are of course in no way assured that they should still
 find support in
 future better insight on the
 nature of this problem, but it is the best we can do at the present time.

\subsection{Quantum-Gravity Theories and Quantum Spacetime}

\subsubsection{Critical String Theory}
The most studied approach
to the quantum-gravity problem is a version of
String Theory which adopts supersymmetry and works
 in a ``critical'' number of spacetime dimensions.
If this main-stream perspective turned out to be correct it would be bad news
for quantum-spacetime phenomenologists, since the theory is formulated in classical
Minkowski background spacetime.
It would be bad news for phenomenology in general because (critical, supersymmetric)
String Theory is a particularly soft modification of current theories, and
the new effects that can be accommodated by the theory are untestably small,
if all the new features
are indeed introduced (as traditionally assumed)
at a string scale roughly given by the Planck scale.

String Theory is a natural attempt form
a particle-physics perspective, but other perspectives on the quantum-gravity
problem remain unimpressed, particularly considering that most results
on String Theory still only apply in a fixed background Minkowski spacetime.
And it is interesting to notice how the most careful analyses performed
even adopting a String-Theory perspective end up finding that the case
for applicability to the quantum-gravity problem is still rather weak
(see, e.g., Ref.~\cite{giddings2011qgstrings}).

This not withstanding there has been in recent years a more
vigorous effort of development
of a string-inspired phenomenology, with inspiration found in mechanisms that
are however outside the traditional formulation of String Theory.
This string-inspired phenomenology does not involve spacetime quantization
and often does not refer explicitly to the Planck scale, so I shall not
discuss it in detail in this review (although there will be scattered opportunities,
at points of this review, where it becomes indirectly relevant).
The possibility which received most attention in recent years is the one of ``large''
extra dimensions~\cite{anto1,anto2,anto3,addlarge,ledMARCH,rubakovLED}.
The existence of extra dimensions can of course be conceived even outside
String Theory, but it is  noteworthy that in String Theory the criticality
criterion actually requires extra dimensions. If the extra dimensions,
as traditionally assumed, have finite size of the order of the Planck length
then one ends up having associated Planck-scale effects for the low-energy realm
where our experiments and observations take place. This would be a classic
exercise for quantum-gravity phenomenology but it appears that the Planck-scale
suppression of these extra-dimension effects is so strong that they really could
not ever be seen/tested.
The recent interest in the ``large extra dimensions'' scenario originates
from the observation that dimensions of size much larger than the Planck length
(but still microscopic), while not particularly natural from a string-theory perspective,
may well be allowed in string theory~\cite{anto1,anto2,anto3,addlarge,ledMARCH,rubakovLED}.
And for some choices of number and sizes of extra dimensions a rich phenomenology
is produced.

Most other phenomenological proposals inspired by String Theory
essentially make use of the fact that, at least as seen by a traditional particle
physicist, String Theory makes room for several new fields.
The new effects are indeed of types that are naturally described by introducing
new fields in a classical spacetime background, rather than quantum-spacetime features,
and the magnitude of these effects is not naturally governed by the
Planck scale\footnote{One may then
argue that there are some indirect reasons why the Planck scale should appear
in formulas setting the significance of the effects, but the connection with the Planck
scale remains relatively weak.}.

In spite of these profound differences there are some points of contact
between
the Planck-scale quantum-spacetime
phenomenology, which I am here concerned with, and this string phenomenology.
In a quantum spacetime it is necessary to reexamine the issue
of spacetime symmetries, and certain specific scenarios for the fate
of Lorent symmetry come into focus. From a different perspective and in a technically
different way one also finds reasons to scrutinize Lorentz symmetry in string
phenomenology: it is  plausible~\cite{kostesamuel}
that some string-theory tensor fields (most likely some of
the new fields introduced by the theory)
could acquire a nonzero vacuum expectation value, in which case evidently one would
have a ``spontaneous breakdown'' of Lorentz symmetry.
I shall also comment on the possibility that spacetime quantization might affect
the Equivalence Principle. Again, from a different perspective and in a technically
different way, one also finds reasons to scrutinize the Equivalence Principle
in string phenomenology.  And again it is typically due to the extra fields introduced
in string theory: most notably some scenarios involving the dilaton,
a scalar partner to the graviton predicted by string theory,
produce violations of the Equivalence Principle (see, e.g., Ref.~\cite{damourEP2}).

I should here stress, because of the scopes of this review, that
the idea of a quantum spacetime is not completely foreign to String Theory.
It is presently appearing at an undigested and/or indirect level of analysis,
but it is plausible that future evolutions of the string-theory program
might have a primitive/fundamental role for spacetime quantization.
So far the most studied connection with quantum-spacetime ideas
comes from
a mechanism analogous to the emergence of noncommutativity
of position coordinates in the Landau model (see, e.g.
Ref.~\cite{banerLANDAUgood})
which is found to be applicable
to the description of strings in presence of a constant Neveu-Schwarz two-form
(``$B_{\mu \nu}$'') field~\cite{dougnekr,szaboREVIEW}.
It should be stressed that these cases of ``emerging noncommutativity''
(effective descriptions applicable only in certain specific regimes)
do not amount to genuine nonclassicality of spacetime.
Still these recent string-theory results do create a point of contact between
research (and particularly phenomenology)
on fundamental spacetime noncommutativity and string theory, with the peculiarity that
from the string-theory perspective one would not necessarily focus (and typically there
is no focus) on the case of noncommutativity introduced at about the Planck scale,
since it is instead given in terms of the free specification of the field $B_{\mu \nu}$.

For the hope of a possible future reformulation of String Theory in
some way that would accommodate a primitive role for spacetime nonclassicality
my impression is that the key opportunities should be seen in results
suggesting that there are
fundamental limitations for the localization of a spacetime event
in String Theory~\cite{venekonmen1,venekonmen2,venekonmen3,venekonmen4}.
The significance of these results on limitations of localizability in String Theory probably
has not
been appreciated sufficiently. Only a few authors have emphasized the possible
significance of these results~\cite{wittenPT}, but
I would argue that finding such  limitations in a theory originally formulated
in a classical spacetime background may well provide the starting point
for reformulating the theory completely, perhaps codifying spacetime quantization
at a primitive level.

\subsubsection{Loop Quantum Gravity}
The most studied theory framework providing a quantum description of
spacetime is Loop Quantum
Gravity~\cite{crLIVING,ashtelewandREVIEW,leeLQGrev,thieREV,ashtekarLQGreview2012}.
The intuition of many phenomenologists who have looked at (or actually worked on)
Loop Quantum Gravity is that this theory should predict quite a few testable
effects, some of which may well be
testable with existing technologies. However, the complexity of the formalism has proven so far to
be unmanageable from the point of view of obtaining crisp physical predictions.
Among the many challenges I should at least mention
the much debated ``classical-limit problem'', which for example obstructs the way
toward a definite set of predictions for the quasi-Minkowski (or quasi-deSitter, or quasi-FRW)
regime,
which is were of course most of the opportunities for phenomenology
can be found.

One may attempt however
to infer from the general structure of the theory
motivation for the study of some candidate Loop-Quantum-Gravity effects.
And, as I shall stress in several parts of this review, this type of attitude
has generated a healthy interface between phenomenologists
and Loop-Quantum-Gravity theorists.
Most of the relevant proposals are ignited indeed by the quantum properties
of spacetime in Loop-Quantum-Gravity, which appear to be primarily codified
in a discretization of the area and volume
observables~\cite{carloleeAREA,ashtAREA,crLIVING}
In particular, several studies (see later in this review) have argued that the type of discretization
of spacetime observables usually attributed to Loop Quantum Gravity could be
responsible\footnote{While the type of quantum-spacetime effects
considered in the Loop-Quantum-Gravity literature makes it natural to question
the fate of Lorentz symmetry in the quasi-Minkowski limit, I should stress
that at present no fully robust result on this is available, and some authors
(notably Refs.~\cite{rovellispeziale2003,rovellispeziale2011}) have observed that
there could be ways to reconcile what is presently known about Loop Quantum
Gravity with the presence of exact Lorentz symmetry in the quasi-Minkowski
limit.} for Planck-scale departures from Lorentz symmetry.

In addition to a large effort focused on the fate of Lorentz symmetry,
there has been also a rather large effort focused on early-Universe cosmology
inspired by Loop Quantum Gravity.
Among the appealing features of this cosmology work I should
at least mention ``singularity avoidance''.
For
the loop-quantum-gravity approach actually there might be no alternative to avoiding
the big-bang singularity, since ideed, at least as presently understood, Loop Quantum Gravity
describes spacetime has a fundamentally
discrete structure governed by difference (rather than
differential) equations. This discreteness is expected to become
a dominant characteristic of the
framework for processes involving comparably small (Planckian) length scales,
and in particular it should inevitably give rise to a totally unconventional picture
 of the earliest
stages of evolution of the Universe.
Attempts developing a setup for
a quantitative description of these early-universe features
have been put forward for example in
Refs.~\cite{bojoFIRST,bojoFIRSTb,bojoLRR,ashtekarLQC}
and references therein,
but one  must inevitably  resort to rather drastic approximations,
since a full loop-quantum-gravity analysis is not possible at present.

For other areas of phenomenology discussed in this review the influence
of Loop Quantum Gravity has ben less direct, but it appears safe to assume
that it will inevitably grow in the coming years. To give a particularly striking
example let me mention the many proposals here discussed which concern
spacetime fuzziness. It is evident that Loop Quantum Gravity gives
a fuzzy picture of spacetime (in the sense discussed more precisely in later
parts of this review), and it would
be of important guidance for the phenomenologists to have definite
predictions for these features. Even just a semiheuristic derivation
of such effects is beyond the reach of our present
understanding of Loop Quantum Gravity, but it will come.

\subsubsection{Approaches based on spacetime noncommutativity}
Of course, the idea of having a nonclassical fundamental description of spacetime
is central to the study of spacetime noncommutativity.
The formalization that is most applied in the study of the quantum-gravity/quantum-spacetime
problem is mainly based on the formalism of  ``quantum-groups'' and essentially
assumes that the quantum properties of spacetime should be at least to some extent
analogous to the quantum properties of phase space in ordinary quantum mechanics.
Ordinary quantum mechanics introduces some limitations for procedures
intending to obtain a combined determination of both position and momentum,
and this is formalized in terms of noncommutativity of the position and momentum observables.
With spacetime noncommutativity one essentially assumes that spacetime coordinates
should not commute~\cite{dopliPLB,majrue,kpoinap,wessNCYM,gacmaj,balaSPINstat}
among themselves, producing some limitations for the combined
determination of more than one coordinate of a spacetime point/event.
This has been the formalization of spacetime
noncommutativity for which the two-way interface between theory and
phenomenology, which is at center stage in this review, has been most significant.

Looking ahead at the future of quantum-spacetime phenomenology
it appears legitimate to hope that another, perhaps even more compelling,
candidate concept of noncommutative geometry, the one championed
by Connes~\cite{connes1995,connesBOOK}, may provide guidance.
At present the most studied applications of this notion of noncommutative geometry
are focused on giving a fully geometric description of the standard model
of particle physics, with the noncommutativity of geometry used to codify
known properties of particle physics in geometric fashion, while keeping
spacetime as a classical geometry.

Going back to the quantum-group-based description of spacetime noncommutativity
I should stress
that so far the most significant developments
have concerned attempts to describe the Minkowski limit of the quantum-gravity problem,
i.e. noncommutative version of Minkowski spacetime (spacetimes which reproduce
classical Minkowski spacetime in the limit
in which the noncommutativity parameters are taken to 0).
Some related work has also been directed toward quantum versions
of deSitter spacetime, but very little about spacetime dynamics
and only at barely an exploratory level.
This should of course change in the future.
But at the present time this situation could be described
by stating that most work on spacetime noncommutatvity is
considering only one half of the quantum-gravity problem,
the quantum-spacetime aspects (neglecting the gravity aspects).
Because of the double role of the gravitational field,
which in some ways is just like another (e.g. electromagnatic) field given in
spacetime but it is also the field that describes the structure of spacetime,
in quantum-gravity research the idea that this classical field be replaced by a nonclassical
one ends up amounting to two concepts: some sort of quantization of gravitational interactions
(which for example might be mediated by a graviton) and some sort of quantization
of spacetime structure. At present one might say that
only within the Loop Quantum Gravity approach we are truly exploring both aspects of the problem.
String Theory, as long as it is formulated in a classical (background) spacetime,
focuses in a sense on the quantization of the gravitational interaction, and sets aside
(or will address in the future)
the possible ``quantization'' of spacetime~\cite{wittenPT}.
Spacetime noncommutativity is an avenue for exploring the implications of the other
side, the quantization of spacetime geometry.

The  description of (Minkowski-limit) spacetime in terms
of (quantum-group-based) spacetime noncommutativity has proven particularly valuable
in providing intuition for the fate of (Minkowski-limit/Poincar\'e)
spacetime symmetries
at the Planck scale.
Also parity transformations appear to be affected by at least some schemes of spacetime
noncommutativity and this in turn provides motivation for testing CPT symmetry.

Unfortunately spacetime fuzziness, which is the primary intuition that leads most researchers to
noncommutativity, frustratingly remains only vaguely characterized in current research
on noncommutative spacetimes; certainly not characterized with the sharpness needed for
phenomenology.

\subsubsection{Other proposals}
I shall not here attempt to review the overall status of quantum-gravity research.
The challenge of reviewing and offering a perspective on quantum-spacetime
phenomenology is already overwhelming.
And according to the perspective of this phenomenological approach
the central challenge of quantum-gravity research is to find the first
experimental manifestations of the quantum-gravity realm. If not find them, perhaps
just stumble upon them, but get them somehow.
The different formalisms proposed for the study for the quantum-gravity problem
can be very valuable for this objective but only in as much as they provide
intuition for the type of new effects that might characterize the quantum-gravity realm.
In practice, at least for the next few decades, what will be compared to data will
be simple test theories inspired by our understanding of the quantum-gravity problem
or by the intuition obtained in the study of formal theories of quantum gravity.
The possibility of comparing directly to experiments a full quantum-gravity theory
appears to be for a still distant future, as a result of the complexity of these theories
(which prevents us from deriving testable predictions).

I have invested a few pages on string theory, loop quantum gravity and spacetime noncommutativity
for different reasons. Providing
some reasonably detailed comments on string theory was encouraged, in spite of the lack
of a fundamental role for spacetime quantization, by its prominent role in
the quantum-gravity literature.
And, as stressed above, loop quantum gravity and spacetime noncommutativity
are particularly relevant for this review because the scenarios of spacetime quantization
these approaches consider/derive have been a particularly influential source of intuition for proposals
in quantum-spacetime phenomenology. Moreover, it is within the
loop-quantum-gravity and spacetime-noncommutativity communities that
we have so far witnessed the most significant examples of the
 healthy two-way cross-influence between
formal theory and phenomenology.

I shall not offer comparably detailed comments on any other
quantum-gravity formalism, but there are a few that I should mention
because of the significance of their role in quantum-spacetime phenomenology.
First of all let me mention the noncritical ``Liouville String Theory'' approach championed
by Ellis, Mavromatos and Nanopoulos~\cite{emn,emnreview,aemn1,nickREVIEW2010}.
This is a variant of the string-theory approach which (unlike the main-stream critical-string-theory
approach)  adopts the choice of
working in ``noncritical'' number of spacetime dimensions, and describes
time in a novel way.
As it will be evident in several points of this review,
Ellis, Mavromatos, Nanopoulos and collaborators have developed
noncritical Liouville String Theory from a perspective that admirably keeps phenomenology
always at center stage, and this has been a key influence on several quantum-spacetime-phenomenology
research lines.

Another approach for which
there is by now a rather sizable research program aimed at phenomenological consequences
is the one based on ``discrete causal sets''~\cite{sorkinPRLcausalsets,sorkinRideoutCONTINUUM}.
This is an approach of spacetime discretization which exploits the fact
that a Lorentzian metric determines both a geometry and a causal structure
and also determines the metric up to a conformal factor.
One can then take the causal structure as primary, and start with a finite set
of points with a causal ordering, recovering the conformal factor by
counting points. Several opportunities for phenomenology are then produced by the
discretization of spacetime.

Still on the subject of approaches in which a role is played by spacetime discretization
I should also bring to the attention of my readers the recent developments in the
study of Causal Dynamical
Triangulations~\cite{ambjorn1993,lollLRR,ambjornlollPRL,ambjornlollPRD,lollDESKTOP,ambjornloll2010}.
Through causal dynamical triangulations one gives an explicit, nonperturbative
and background-independent, realization of the formal gravitational path integral
on a given differential manifold. And some of the results obtained within this approach
 already provide
elements of valuable intuition for quantum-spacetime phenomenology, as exemplified
by the results providing~\cite{lollFIRSTdimreduction} first  evidence for
a scale-dependent spectral dimension of space-time, varying from four at large
scales to two at scales of the order of the Planck length.
These ``running spectral dimensions" could have very significant applications
in phenomenology, and early signs that this might indeed be the case
can be found in the debate reported in Refs.~\cite{runningSDexp1,runningSDexp2,runningSDexp3}
concerning the implication for primordial gravity waves.

Also particularly important for quantum-spacetime phenomenology is the program
of asymptotically safe quantum gravity.
This is an attempt at the nonperturbative construction of a predictive quantum field theory of the metric tensor centered on the availability of a non-Gaussian renormalization-group fixed
point~\cite{weinbergSAFETY,reuterSAFETY,percacciSAFETY}.
There are a few perspectives from which this asymptotic-safety program
is influencing part of the research on quantum-spacetime phenomenology.
As an example of phenomenology work that was directly inspired by asymptotic safety
I should mention
the expectation that quantum-gravity effects might also be important
in a large-distance regime~\cite{reuter0702051}, with possible relevance
for phenomenology. I shall comment on this later in this review, also in relation
to the idea of ``ultraviolet/infrared mixing'' as a possibility which appears
to be plausible even within other perspectives on quantum gravity and quantum spacetime.
And there are significant indications (see, e.g., Ref.~\cite{reuterMinLength})
that ultimately the description of spacetime in a quantum-gravity with asymptotic safety
will be a quantum-spacetime description. Also significant for quantum-spacetime
phenomenology is the whole idea of running gravitational couplings, which
is of course central to asymptotic safety. As mentioned we tentatively assume
that quantum-spacetime effects originate at the Planck scale, but the Planck scale
is computed in terms of (the infrared value of) Newton's constant and might give
us a misleading intuition for the characteristic scales of spacetime quantization.

There are also some perspectives on the quantum-gravity problem
which at present I do not see as direct opportunities for quantum-spacetime phenomenology,
but certainly are playing the role of ``intuition builders" for the phenomenologists,
affecting the perception of the quantum-gravity problem that guides some
of the relevant research.
Among these I should mention the rather large literature on the ``emergent gravity paradigm"
(see, {\it e.g.},
Refs.~\cite{emergentLIBER,emergentVOLOVI,emergentPADMA,emergentNCST1,emergentNCST2,emergentSINDO,emergentHU}).
This literature actually contains a variety of possible way through which gravity could be described not
as a fundamental aspect of the laws of nature, but rather as an emergent feature.
A simple analogy here is with pion-mediated strong interactions,
which emerge from the quantum chromodynamics of quarks and gluons  at low energies.

And I should mention as another potential ``intuition builder" for the phenomenologists a class
of studies which in various ways place dissipation in connection with aspects of the quantum-gravity
problem (see, {\it e.g.}, Refs.~\cite{thooftDISSIPA,huDISSIPA}).

\subsection{Candidate effects}
From the viewpoint of phenomenologists
the theory proposals I briefly considered in the previous subsection
(all still lacking any experimental success)
can only serve the purpose of inspiring some test theories suitable
for comparison to data.

In this subsection I will briefly motivate a partial list of possible classes of effects
that could characterize the quantum-gravity/quantum-spacetime realm.
And indeed in compiling such a list one ends up using both intuition based
on the general structure of the quantum-gravity problem
and intuition based on what has been so far understood
of theories that predict or assume spacetime quantization.

Both the analysis of the general structure of the quantum-gravity problem and
the analysis of proposed approaches to the solution of the quantum-gravity
problem provide a rather broad collection of intuitions for what might
be the correct ``quantization'' of spacetime (see, e.g.,
Refs.~\cite{mead,venekonmen1,venekonmen2,venekonmen3,venekonmen4,padma,dopliPLB,ahlu1994,ng1994,gacmpla,garay,casadio}),
and in turn this variety of scenarios produces a rather broad collection
of hypothesis concerning possible experimental manifestations
of spacetime quantization.

\subsubsection{Planck-scale departures from classical-spacetime
  symmetries}
From a quantum-spacetime perspective it is natural to expect
that some opportunities for phenomenology might come from tests of spacetime symmetries.
It is relatively easy to test spacetime symmetries very sensitively, and it is natural to expect
that introducing new (``quantum'') features in spacetime structure the symmetries
would also be affected.

Let us consider in particular the Minkowski limit, the one
described by the classical Minkowski spacetime in current
theories: there is a duality one-to-one relation between the
classical Minkowski spacetime and the classical (Lie-) algebra of
Poincar\'e symmetrie. Poincar\'e transformations are smooth
arbitrary-magnitude classical transformations and it is therefore natural
to subject them to scrutiny~\footnote{Of course,
a space with some elements of quantization/discreteness may have
classical continuous symmetries, but only if things are arranged
in an \textit{ad hoc} manner. Typically
quantization/discretization of spacetime observables does lead to
departures from classical spacetime symmetries. So
clearly spacetime-symmetry tests should be a core area of
quantum-spacetime phenomenology, but to be pursued with the awareness
that spacetime quantization does not automatically affect
spacetime symmetries (it typically does but not automatically).}
if the classical Minkowski spacetime is replaced by a
quantized/discretized version.

The most active quantum-spacetime-phenomenology research area is indeed
the one considering possible Planck-scale departures from
Poincar\'e/Lorentz symmetries.
One possibility that of course has been considered in detail is the one
of some symmetry-breaking mechanism affecting Poincar\'e/Lorentz symmetry.
An alternative, which I advocated a few years ago~\cite{gacdsr,gacdsrB}, is the
one of a ``spacetime quantization'' which deforms but does not break some spacetime symmetries.

Besides the analysis of the general structure of the quantum-gravity problem,
encouragement for these Poincar\'e/Lorentz-symmetry studies is found also
within some of the most popular proposals for spacetime quantization.
As mentioned, according to the present understanding of
Loop Quantum Gravity the fundamental description of spacetime involves
some intrinsic discretization~\cite{crLIVING,leeLQGrev}, and,
although very little of robust is presently known about the Minkowski limit
of the theory, several indirect arguments suggest that
this discretization should induce departures
from classical Poincar\'e symmetry.
While most of the Loop-Quantum-Gravity literature on the fate of Poincar\'e symmetries
argues for symmetry violation (see, e.g., Refs.~\cite{gampul,urrutia}),
there are some candidate mechanisms
(see, e.g., Refs.~\cite{kodadsr,jurekkodadsr,smolinDSR2008})
that appear to provide opportunities for a deformation of symmetries
in Loop Quantum Gravity.

And as mentioned a growing number of quantum-gravity researchers are also studying
noncommutative versions of Minkowski spacetime,
which are promising candidates as ``quantum-gravity theories of not everything'',
i.e., opportunities to get insight on some, but definitely not all,
aspects of the quantum-gravity problem.
For the most studied examples,
canonical noncommutativity,
\begin{equation}
[x_\mu,x_\nu] = i \theta_{\mu \nu}~,
\label{canonicalNCSTb}
\end{equation}
 and $\kappa$-Minkowski
noncommutativity,
\begin{equation}
[x_m,t] = \frac{i}{\kappa} x_m ~,~~~[x_m,x_l] = 0~,
\label{kappaminkB}
\end{equation}
the issues relevant for the fate of Poincar\'e symmetry are very much
in focus, and departures from Poincar\'e symmetry appear to be inevitable\footnote{In the
case of canonical noncommutativity evidence of some departures from Poincar\'e symmetry
are found both
if $\theta_{\mu \nu}$ is a fixed tensor~\cite{dougnekr,szaboREVIEW}
or a fixed observer-independent matrix~\cite{chaichan,wessfiore,balaIRUV}. The possibility to
preserve classical Poincar\'e symmetry is instead still not excluded in what was
actually the earliest
approach~\cite{dopliPLB} based on $[x_\mu,x_\nu] = i \theta_{\mu \nu}$, where for $\theta_{\mu \nu}$
one seeks a formulation with richer algebraic properties.}.


\subsubsection{Planck-scale departures from CPT symmetry}\label{cptgeneral}
Arguments suggesting
that CPT violation might arise in the quantum-gravity realm
have a long
tradition~\cite{hawkingCPT,pageCPT,waldCPT,pagegaume1,pagegaume2,emnWcpt,huetpesk,kostcpt,bertocpt}
(and also see, e.g., the more recent Refs.~\cite{ahlucpt,muracpt,klinkCPT}).
And, in light of the scopes of this review I should stress that specifically
the idea of spacetime quantization invites one to place CPT symmetry under scrutiny.
Indeed locality (in addition to unitarity and Lorentz invariance)
is a crucial ingredient for ensuring CPT invariance, and a common feature of
all the proposals for spacetime quantization
is the presence of limitations to locality, at least intended
as limitations to the localizability of a spacetime event.

Unfortunately a proper analysis of CPT symmetry requires a level of
understanding of the formalism that is often beyond our present
reach in the study of formalizations of the concept of quantum spacetime.
 In Loop Quantum Gravity
one should have a good control of the Minkowski (classical-) limit,
and of the description of charged particles in that limit,
and this is still beyond what can be presently done within
Loop Quantum Gravity.

Similar remarks apply to spacetime noncommutativity, although
in that case some indirect arguments relevant for CPT
symmetry can be meaningfully structured. For example, in Ref.~\cite{gacmaj}
it is observed that certain spacetime noncommutativity scenarios
appear to require a deformation of $P$ (parity) transformations, which would of course
result in a corresponding deformation of CPT transformations.

In the mentioned quantum-spacetime picture based on
noncritical Liouville String Theory~\cite{emn,emnchaos},
evidence of violations of CPT symmetry has been reported~\cite{elmn},
and later in this review
I shall comment on the exciting phenomenology which was inspired
by these results.


\subsubsection{Decoherence and modifications of the Heisenberg principle}
It is well established that the availability of a classical spacetime background
has been instrumental to the successful tests of quantum mechanics so far performed.
The applicability of quantum mechanics to a broader class of contexts remains
an open experimental question.
If indeed spacetime is quantized there might be some associated departures
from quantum mechanics.
And this quantum-spacetime intuition fits well with a rather popular
intuition for the broader context of quantum-gravity research,
as discussed for example in Refs.~\cite{hawkingDECO,rubakovDECO}.

Some of the test theories used to model spacetime quantization
have been found to provide motivation for departures from quantum mechanics
in the form of ``decoherence'', loss of quantum coherence~\cite{ng1994,gacmpla,pullinDECOprl}
A description of decoherence
has been inspired by the mentioned noncritical Liouville String Theory~\cite{emn,emnchaos},
and is essentially the core feature of the formalism advocated by
Percival and collaborators~\cite{perci1,perci0,perci2}.

The possibility of modifications of the Heisenberg principle and of the de Broglie relation
has also been much studied in accordance with the intuition that some aspects
of quantum mechanics might need to be adapted to spacetime quantization.
Although the details of the mechanism that produces such modifications
vary significantly from one picture
of spacetime quantization to another~\cite{kempf,ahluDB,kappaHP},
one can develop an intuition of rather general applicability
 by noticing that the form of the
de Broglie relation in ordinary quantum mechanics reflects the properties of the
classical geometry of spacetime that is there assumed. More precisely the
de Broglie relation reflects the properties of the differential calculus
on the spacetime manifold, since ordinary quantum mechanics describes the momentum
observable in terms of a derivative operator (assuming the Heisenberg principle holds), which acting on
wave functions with wavelength $\lambda$ leads to
the de Broglie relation $p= h/\lambda$.
In a nonclassical (``quantum'') spacetime one must adopt new forms of differential
calculus~\cite{diffcalc1,diffcalc2}, and as a result the description of the
momentum observable and its relation to the wavelength of a wave must be
reformulated~\cite{kempf,ahluDB,kappaHP,gacdsr2010}.

While the possibiilty of spacetime quantization provides a particularly direct logical line
toward modifications of laws of quantum mechanics,
one should consider such modifications as natural for the
whole quantum-gravity problem
(even when studied without assuming spacetime quantization).
For example, in String Theory, assuming the availability
of a classical spacetime background, one finds some evidence of modification
of the Heisenberg principle (the ``Generalized Uncertainty Principle''
discussed, e.g., in Refs.~\cite{venekonmen1,venekonmen2,venekonmen3,venekonmen4,wittenPT}).


\subsubsection{Distance fuzziness and spacetime foam}
A description that is often used to give some
 intuition for the effects induced by spacetime
quantization is Wheeler's ``spacetime foam'', even though it does not amount to an operative
definition.
Most authors see it as motivation to look for formalizations of spacetime
in which the distance between two events cannot be sharply determined,
and the metric is correspondingly fuzzy.
As I shall discuss in parts of Section~\ref{other-areas}, a few attempts
to characterize operatively the concept of spacetime foam
and to introduce corresponding test theories have been recently developed.
And a rather rich phenomenology is maturing from these proposals,
often centered both on spacetime fuzziness {\it per se}
and associated decoherence.

Unfortunately,
very little guidance can be obtained from the most studied quantum-spacetime
pictures.
In Loop Quantum Gravity this type of experimentally tangible
characterization of spacetime foam
is not presently available.
And remarkably even with spacetime noncommutativity, an idea that was mainly motivated
by the spacetime-foam intuition of a nonclassical spacetime,
we are presently unable to describe, for example,  the fuzziness
that would intervene in operating an interferometer with the type of crisp
physical characterization needed for phenomenology.

\subsubsection{Planck-scale departures from the Equivalence Principle}
The possibility of violations of the Equivalence Principle has not been extensively studied
from a quantum-spacetime perspective, in spite of the fact that
spacetime quantization does provide some motivation for placing under scrutiny
at least some implications of the Equivalence Principle.
This is at least suggested by the observation that
locality is a key ingredient of the present formulation of the
Equivalence Principle: the Equivalence Principle ensures
that (under appropriate conditions) two point particles would go on the same geodesic
independently of their mass. But it is well established that this is not applicable
to extended bodies, and presumably also not applicable
to ``delocalized point particles'' (point particles whose position is
affected by uncontrolled uncertainties). Presumably also the description
of particles
in a spacetime that is nonclassical (``quantized''), and therefore sets absolute limitations
on the identification of a spacetime point, would require departures from some aspects
of the Equivalence Principle.

Only relatively few studies have been devoted to violations of the Equivalence Principle
from a quantum-spacetime perspective.
Examples are
 the study reported in
Ref.~\cite{camachoEPdeco}, which obtained violations of the Equivalence Principle
from quantum-spacetime-induced decoherence,
the study based on noncritical Liouville String Theory reported
in Ref.~\cite{ellisNONUNI}, and
the study based on metric flluctuations reported in
 Ref.~\cite{clausEP}.

 Also the broader quantum-gravity literature (even without spacetime quantization)
 provides motivation for scrutinizing the Equivalence Principle.
 In particular, a strong phenomenology centered on
violations of the Equivalence Principle was proposed
in the string-theory-inspired studies reported in
 Refs.~\cite{stringsEP,repDamourPOLY1,repDamourPOLY2,damour,damourEP2,repDamourVARYING}
and references
 therein, which actually provide a description of violations of the Equivalence
 Principle\footnote{While I shall not discuss it here, because it has so far attracted
 little interest from a quantum-spacetime perspective, I should encourage
 readers to also become acquainted with the fact that studies
 such as the ones in Refs.~\cite{repDamourPOLY2,repDamourVARYING},
 through the mechanism for violation of the Equivalence Principle,
 also provide motivation for ``varying coupling constants''~\cite{repUzan,repMartins}.}
 at a level which might soon be within our experimental reach.

 Also relevant for the topics of this review is the possibility that violations of the Equivalence Principle
 might be a by-product of violations of Lorentz symmetry.
This is in particular suggested by the analysis reported in Ref.~\cite{kostegrav} where
the gravitational couplings of matter are studied in the presence of Lorentz violation.

\newpage


\section{Quantum-Spacetime Phenomenology of UV corrections to Lorentz Symmetry}
\label{symmetry-tests}

The largest area of quantum-spacetime-phenomenology research concerns
the fate of Lorentz (/Poincar\'e) symmetry at the Planck scale,
focusing on the idea that the conjectured new effects might become
manifest
at low energies (the particle energies accessible to us, which are much
below the Planck scale) in the form of ``UV corrections'',
correction terms
with powers of energy in the numerator and powers of the Planck scale
in the denominator.

Among the possible effects that might signal departures
from Lorentz/Poincar\'e symmetry
the interest has been predominantly directed toward
the study of the form of energy-momentum (dispersion) relation.
This was due both to the (relative) robustness of associated theory
results in quantum-spacetime research and to the availability of very valuable
opportunities of related data analyses. Indeed, as several examples in this section will show,
 over the last decade there were very significant improvements
 of the sensitivity of Lorentz- and Poincar\'e-symmetry tests.

Before discussing some actual phenomenological analyses I find appropriate
to start this section with some preparatory work.
This will include some comments on the ``Minkowski
limit of Quantum Gravity'', which I have already referred to but should
be discussed a bit more carefully in preparation for this section.
And I shall also give a rather broad perspective
on the quantum-spacetime implications for the set up of test theories
suitable for the study of the fate
of Lorentz/Poincar\'e symmetry
at the Planck scale.


\subsection{Some relevant concepts}

\subsubsection{The Minkowski limit}
\label{minkowski-limit}

In our current conceptual framework
Poincar\'e symmetry emerges
in situations that allow
the adoption of a Minkowski metric throughout.
These situations could be described as the ``classical Minkowski limit''.

It is not inconceivable that quantum gravity might
admit a limit in which one can
assume throughout a (expectation value of the) metric of
Minkowski type, but some Planck-scale features
of the fundamental description of spacetime
(such as spacetime discreteness and/or spacetime noncommutativity)
are still not completely negligible.
This ``nontrivial Minkowski limit'' would be such that
 essentially the role of the Planck scale
in the description of gravitational phenomena
can be ignored (so that indeed one can make reference to a fixed Minkowski metric),
but the possible role of the Planck scale in
spacetime structure/kinematics is still significant.
This intuition inspires the work on
quantum-Minkowski spacetimes,
and the analysis of the symmetries of these quantum spacetimes.

It is of course not obvious that the correct quantum gravity should
admit such a nontrivial Minkowski limit. With the little we presently
know about the quantum-gravity problem we must be open to the possibility
that the Minkowski limit would actually be trivial, i.e.,
that whenever the role of the Planck scale
in the description of gravitational phenomena can be neglected
(and the metric is Minkowskian at least on average)
one should also
neglect the role of the Planck scale in spacetime structure.
But the hypothesis of a nontrivial Minkowski limit is worth exploring: it is a plausible
hypothesis and
it would be extremely valuable for us if quantum gravity did admit such a limit,
since it might open a wide range of opportunities for
accessible experimental
verification, as I shall stress in what follows.

When I mention a result on the theory side concerning
the fate of Poincar\'e symmetry at the Planck
scale clearly it must be the case that the authors
have considered (or attempted to consider) the Minkowski limit of
their preferred formalism.

\subsubsection{Three perspectives on the fate of Lorentz symmetry
  at the Planck scale}

It is probably fair to state that each quantum-gravity research line
can be connected with one of
three perspectives on the problem: the particle-physics perspective,
the General-Relativity perspective and the condensed-matter perspective.

From a particle-physics perspective it is natural to attempt to
reproduce as much as possible the successes of the Standard Model
of particle physics.
One is tempted to see gravity simply as one more gauge interaction.
From this particle-physics perspective a
natural solution of the quantum-gravity problem should have
 its core features
described in terms of graviton-like exchange
in a background classical spacetime.
Indeed this structure is found in String Theory, the most developed
among the quantum-gravity approaches that originate from
a particle-physics perspective.

The particle-physics perspective provides no \textit{a priori} reasons
to renounce to exact Poincar\'e symmetry, since
Minkowski classical spacetime is an admissible background spacetime,
and in classical Minkowski there cannot be any a priori obstruction
for classical Poincar\'e symmetry.
Still, a breakdown of Lorentz symmetry,
in the sense of spontaneous symmetry breaking,
is of course possible, and this
possibility has been studied extensively
over the last few years, especially
in String Theory
(see, e.g.,
Ref.~\cite{kostesamuel,dougnekr} and references therein).

Complementary to the particle-physics perspective
is the General-Relativity perspective, whose
core characteristic is the intuition that one should
firmly reject the the possibility to rely on
a background spacetime~\cite{crLIVING,leeLQGrev}.
According to General Relativity the evolution of particles
and the structure of spacetime are selfconsistently connected:
rather than specify a spacetime arena (a spacetime background) beforehand,
the dynamical equations determine at once both the spacetime structure
and the evolution of particles.
Although less publicized, there is also growing awareness
of the fact that, in addition to the concept of background independence,
the development of General Relativity relied heavily
on the careful consideration of the in-principle limitations that
measurement procedures can encounter\footnote{Think for example of the
limitations that the speed-of-light limit imposes on certain setups
for clock synchronization and of the contexts in which it is
impossible to distinguish between a constant acceleration and
the presence of a gravitational field.}.
In light of the various arguments
suggesting that, whenever both quantum mechanics and General Relativity
are taken into account, there should be an in-principle Planck-scale
limitation to the localization of a spacetime
point (an event),
the general-relativity perspective
invites one to renounce to any direct reference
to a classical spacetime~\cite{dopliPLB,ahlu1994,ng1994,gacmpla,garay}.
Indeed this requirement that spacetime be described as fundamentally
nonclassical (``fundamentally quantum''),
so  that the measurability limitations
be reflected by  a corresponding
measurability-limited formalization of spacetime,
is another element of intuition which is guiding quantum-gravity
research from the general-relativity perspective.
This naturally leads one to consider
discretized spacetimes, as in the Loop
Quantum Gravity approach, or noncommutative spacetimes.

Results obtained over the last few years
indicate that this general-relativity perspective
naturally leads, through the emergence of spacetime discreteness and/or
noncommutativity, to some departures from classical Poincar\'e symmetry.
Loop Quantum Gravity and some other discretized-spacetime quantum-gravity
approaches appear to require a description of the familiar
(classical, continuous) Poincar\'e  symmetry
as an approximate symmetry, with departures governed by the Planck scale.
And in the study of noncommutative spacetimes some Planck-scale
departures from Poincar\'e symmetry appear to be inevitable.

The third possibility is a condensed-matter perspective
on the quantum-gravity problem
(see, e.g., Refs.~\cite{volovik,laugh,laughchap}),
in which spacetime itself is seen as a sort of emerging critical-point entity.
Condensed-matter theorists are used to describe the degrees
of freedom that are measured in the laboratory
as collective excitations within
a theoretical framework whose primary description is given
in terms of much different, and often practically inaccessible,
fundamental degrees of freedom.
Close to a critical point some symmetries arise for the
collective-excitation theory, which do not carry the
significance of fundamental symmetries,
and are in fact lost as soon as the theory is probed somewhat
away from the critical point. Notably,
some familiar systems are known to exhibit special-relativistic invariance
in certain limits, even though, at a more fundamental level,
they are described in terms of a nonrelativistic theory.
So clearly from the condensed-matter perspective
on the quantum-gravity problem
it is natural to see the familiar
classical continuous Poincar\'e symmetry only as an approximate symmetry.

Further encouragement for the idea of an emerging
 spacetime  (though not necessarily invoking the condensed-matter perspective)
 comes from the realization~\cite{jacoTHERM,verliTHERM,padmaTHERM}
that the Einstein equations can be viewed as an equation of state,
so in some sense thermodynamics implies general relativity
and the associated microscopic theory might not look much like gravity.

\subsubsection{Aside on broken versus deformed spacetime symmetries}

If the fate of Poincar\'e symmetry at the Planck scale is nontrivial
the simplest  possibility is the one of broken
Poincar\'e symmetry, in the same sense that other symmetries are broken
in physics.
As mentioned, an example of suitable mechanism is provided by the
possibility that a tensor field might
have a vacuum expectation value~\cite{kostesamuel}.

An alternative possibility,
which in recent years has attracted the interest
of a growing number of researchers within
the quantum-spacetime and the
quantum-gravity communities, is the one of
 deformed (rather than broken) spacetime symmetries,
in the sense of the ``doubly-special-relativity'' (DSR) proposal
I put forward a few years ago~\cite{gacdsr}.
I have elsewhere~\cite{gacdsr2010} attempted to expose the compellingness
of this possibility. Still, because of the purposes of this review,
I must take into account that the development of phenomenologically-viable
DSR models is still in its infancy.
In particular,
several authors (see, {\it e.g.}, Refs.~\cite{dsrREVijmp2002,unruhNONLOCAL,dedeoNONLOCAL,sabinePRL})
have highlighted the challenges for the description
of spacetime and in particular spacetime locality
that inevitably arise when contemplating a DSR scenario.
I am confident that some of the most recent DSR studies, particularly those
centered on the analysis of  the
so-called ``relative locality"~\cite{bob,leeLIMITEDinertial,michJUREoct2010,principleRL},
contain the core ideas that in due time will allow us to fully establish a robust DSR picture
of spacetime, but I nonetheless feel that we are still far from the possibility of
developing a robust DSR phenomenology.

Interested readers have available a rather sizable DSR literature
(see, e.g., Ref.~\cite{gacdsr,gacdsrB,kowadsr,gacroxkowa,leedsrPRL,leedsrPRD,jurekDSRnew,rainbowDSR,repJUREK1,repJUREK2,repJUREK3,repAHLUWA,repDaszk,unruhNONLOCAL,repRembie,hossePRD,judesvisser,stefanoDSRinterpr} and references therein),
but for the purposes of this review I shall limit my consideration of DSR ideas for phenomenology to
a single one of the (many) relevant
issues, which is an observation that concerns the
compatibility between modifications of the energy-momentum
dispersion relation and modifications of the law of conservation of energy-momentum.
My main task in this subsection is to illustrate the differences
(in relation to this compatibility issue) between the broken-symmetry
hypothesis and the DSR-deformed-symmetry hypothesis.

The doubly-special-relativity scenario was proposed~\cite{gacdsr}
as a sort of alternative perspective
on the results on Planck-scale departures from Lorentz symmetry which
had been reported in numerous articles~\cite{grbgac,gampul,kifune,ita,gactp,aus,urrutia}
between 1997 and 2000. These studies were advocating a Planck-scale modification
of the energy-momentum dispersion relation, usually of the form
$E^2=p^2+m^2+\eta L_p^n p^2 E^n +O(L_p^{n+1}E^{n+3})$, on the basis
of preliminary findings in the analysis
of several formalisms in use for Planck-scale physics.
The complexity of the formalisms is such that very
little else was known about their physical consequences, but the
evidence of a modification of the dispersion relation was becoming
robust. In all of the relevant papers it was assumed that such
modifications of the dispersion relation would amount to a breakdown
of Lorentz symmetry, with associated emergence of a preferred
class of inertial observers (usually identified with the natural
observer of the cosmic microwave background radiation).

However, it then turned out to be possible~\cite{gacdsr}
to avoid this preferred-frame expectation, following a line of analysis
in many ways analogous to the one familiar from
the developments which led to the emergence of Special
Relativity, now more than a century ago. In Galileian Relativity
there is no observer-independent scale, and in fact
the energy-momentum relation is written as $E=p^2/(2m)$. As experimental
evidence in favour of Maxwell equations started to grow, the fact
that those equations involve a fundamental velocity scale appeared
to require the introduction of a preferred class of inertial
observers. But in the end we figured out that the
situation was not demanding the introduction of a preferred frame,
but rather a modification of the laws of transformation between
inertial observers. Einstein's Special Relativity introduced the
first observer-independent relativistic scale (the velocity scale
$c$), its dispersion relation takes the form $E^2 = c^2 p^2 + c^4
m^2$ (in which $c$ plays a crucial role in relation to
dimensional analysis), and the presence of $c$ in Maxwell's
equations is now understood as a manifestation of the necessity to
deform the Galilei transformations.

It is plausible that
we might be presently confronted with an analogous scenario.
Research in quantum gravity is increasingly providing reasons of
interest in Planck-scale modifications of the dispersion relation,
and, while it was customary to assume
that this would amount to the introduction of a preferred class of
inertial frames (a ``quantum-gravity ether''), the proper
description of these new structures might require yet again a
modification of the laws of transformation between inertial
observers. The new transformation laws would have to be
characterized by two scales ($c$ and $\lambda$) rather than the single
one ($c$) of ordinary Special Relativity.

While the DSR idea came to be proposed in the context
of studies of modifications of the dispersion relation,
one could have other uses for
the second relativistic scale, as stressed in parts of the DSR literature~\cite{gacdsr,gacdsrB,kowadsr,gacroxkowa,leedsrPRL,leedsrPRD,jurekDSRnew,rainbowDSR,repJUREK1,repJUREK2,repJUREK3,repAHLUWA,repDaszk,unruhNONLOCAL,repRembie,hossePRD,judesvisser,stefanoDSRinterpr}.
Instead of promoting to the status of relativistic invariant a modified dispersion
relation, one can have DSR scenarios with undeformed dispersion relation
but, for example, with an observer-independent
an observer-independent bound on the
accuracy achievable in the measurement of distances~\cite{gacdsr2010}.
However, as announced, within the confines of
this quantum-spacetime-phenomenology
review I shall only make use of one DSR argument,
which applies to cases in which
indeed the dispersion relation is modified.
This concerns the fact that in presence of
observer-independent modifications of the dispersion relation
(DSR-)relativistic invariance imposes the presence of associated
modifications of the law of energy-momentum conservation.
More general discussions of this issue are offered in Refs.~\cite{gacdsr,gacdsr2010},
but it is here  suffiient to illustrate it in a specific example.
Let us then consider a dispersion relation whose leading-order deformation
(by a length scale $\lambda$) is given by
\begin{equation}
 E^2 \simeq  \vec{p}^2 + m^2 + \lambda \vec{p}^2 E
 ~.
\label{dispKpoinnew}
\end{equation}
This dispersion relation is clearly an invariant of classical space
rotations,  and of
deformed boost transformations generated by~\cite{gacdsr,gacdsr2010}
\begin{equation}
{\cal B}_j \simeq i p_j \frac{\partial}{\partial E}+ i
\left(E + \frac{\lambda}{2} {\vec p}^2 + \lambda E^2 \right) \frac{\partial}{\partial p_j}
-i \lambda p_j  \left(p_k
\frac{\partial}{\partial p_k} \right )~. \label{dsr1boosts}
\end{equation}

The issue concerning energy-momentum conservation arises
because  both the dispersion
relation and the law of energy-momentum conservation
must be (DSR-)relativistic.
And the boosts (\ref{dsr1boosts}), which enforce relativistically the
modification of the dispersion relation, are incompatible with
the standard form of energy-momentum conservation.
For example, for processes  with
two incoming particles, $a$ and $b$, and two outgoing
particles, $c$ and $d$,
the requirements $E_a + E_b - E_c -E_d=0$ and $p_a + p_b - p_c -p_d=0$
are not observer-independent laws
according to~(\ref{dsr1boosts}).
An example of modification of energy-momentum conservation
which is compatible with (\ref{dsr1boosts}) is~\cite{gacdsr}
\begin{equation}
E_a + E_b + \lambda p_a p_b \simeq
 E_c +E_d +\lambda p_c p_d
~,
\label{conservnewe}
\end{equation}
\begin{equation}
p_a + p_b + \lambda (E_a p_b + E_b p_a)\simeq
 p_c +p_d + \lambda (E_c p_d + E_d p_c)
~.
\label{conservnewp}
\end{equation}
And analogous formulas can be given
for any process with $n$
 incoming particles and $m$ outgoing
particles.
In particular, in the case of a
two-body particle decay $a \rightarrow b+c$
the laws
\begin{equation}
E_a \simeq E_b + E_c + \lambda {p}_b  {p}_c
~,
\label{consDecayA}
\end{equation}
\begin{equation}
p_a \simeq p_b + p_c + \lambda (E_b p_c + E_c p_b)
~.
\label{consDecayB}
\end{equation}
provide an acceptable (observer-independent, covariant according to~(\ref{dsr1boosts}))
possibility.

This observation
provides a general motivation for contemplating
modifications of the law of energy-momentum conservation
in frameworks with modified dispersion relations.
And I shall often test the potential impact on the phenomenology
of introducing such modifications of the conservation
of energy-momentum by using as examples
DSR-inspired laws of the
type (\ref{conservnewe}),(\ref{conservnewp}),(\ref{consDecayA}),(\ref{consDecayB}).
I shall do this without necessarily advocating a DSR interpretation: knowing whether or not
the outcome of tests of modifications of the dispersion relation depends on the possibility of
having also a modification of the momentum-conservation laws is of intrinsic interest, with or without
the DSR intuition.
But I must stress that when the relativistic symmetries are broken
(rather than deformed in the DSR sense) there is no a priori reason
to modify the law of energy-momentum conservation, even when
the dispersion relation is modified. Indeed most authors
adopting modified dispersion relations within a broken-symmetry scenario
keep the law of energy-momentum conservation undeformed.

On the other hand the DSR research program has still not reached the maturity
for providing a fully satisfactory interpretation of the nonlinearities in the conservation laws.
For some time the main challenge came (in addition to the mentioned interpretational challenges
connected with spacetime locality) from arguments suggesting that one might
well replace a given nonlinear setup for a DSR model with one obtained by
redefining nonlinearly the coordinatization of
momentum space (see, {\it e.g.}, Ref.~\cite{repAHLUWA}).
When contemplating such changes of coordinatization of momentum space many interpretational
challenges appeared to arise. In my opinion, also in this direction the recent DSR literature has
made significant progress, by casting the nonlinearities for momentum-space properties
in terms of geometric entities, such as the metric and the affine connection on momentum space
(see, {\it e.g.}, Ref.~\cite{principleRL}).
This novel geometric interpretation is offering several opportunities for addressing the
interpretational challenges, but the process is still far from complete.


\subsection{Preliminaries on test theories with modified dispersion
  relation}

So far the main focus of  Poincar\'e-symmetry tests
planned from a quantum-spacetime-phenomenology perspective
has been on the form of the energy-momentum dispersion relation.
Indeed, certain analyses of formalisms
provide encouragement for the possibility that the Minkowski limit of
quantum gravity might indeed be characterized by modified dispersion relations.
However, the complexity of the formalisms that motivate the study of Planck-scale modifications
of the dispersion relation is such that one has only
partial information on the form of the correction terms and actually
one does not even establish robustly the
presence of modifications of the dispersion relation. Still, in some cases,
most notably within some Loop-Quantum-Gravity studies
and some studies of noncommutative spacetimes, the ``theoretical
evidence'' in favour of modifications of the dispersion relations appears to
be rather robust.

This is exactly the type of situation that I mentioned earlier in this review
as part of a preliminary characterization of the peculiar type of test theories
that must at present be used in quantum-spacetime phenomenology.
It is not possible to compare to data the predictions for departures from Poincar\'e symmetry
 of, say, Loop Quantum Gravity
and/or noncommutative geometry because these theories do not yet provide a sufficiently
rich description of the structures needed for actually doing phenomenology
with modified dispersion relations. What we can compare to data
are some simple models inspired by the little we believe we understand
of the relevant issues within the theories that provide motivation for this
phenomenology.

And the development of such models requires a delicate balancing act.
If we only provide them with the structures we do understand of the original
theories they will be as sterile as the original theories.
So we must add some structure, make some assumptions, but do so
with prudence, limiting as much as possible
the risk of assuming properties that could turn out not
to be verified once we understand the relevant formalisms better.

As this description should suggest, there has been of course a proliferation
of models adopted
by different authors, each reflecting a different intuition on what could
or could not be assumed. Correspondingly, in order to make a serious overall
assessment of the experimental
limits so far established with quantum-spacetime phenomenology
of modified dispersion relations,
one should consider a huge zoo of parameters. Of course,
even the parameters of the same parametrization of modifications
of the dispersion relation when analyzed using different assumptions about
other aspects of the model should really be treated as different/independent
sets of parameters.

I shall here be satisfied with considering some illustrative examples of models,
chosen in such a way to represent possibilities that are qualitatively very different,
and representative of the breadth of possibilities that are under consideration.
These examples of
models will be then used in some relevant parts of this review as ``language'' for the description
of the sensitivity to Planck-scale effects
that is within the reach of certain experimental
analyses.

\subsubsection{With or without standard quantum field theory?}

Before describing actual  test theories I should
discuss at least the most significant among the issues that must be
considered in setting up any such test theory with modified dispersion relation.
This concerns the choice
of whether or not to assume that the test theory should be a
standard low-energy effective quantum field theory.

A significant portion of the quantum-gravity and quantum-spacetime community
is rather skeptical about the results obtained
using low-energy effective field theory in
analyses relevant for the Planck-scale regime.
One of the key reasons for this skepticism is the description given by effective
field theory of the cosmological constant.
The cosmological constant is the most significant experimental fact
 of evident gravitational relevance which could be within the reach of effective field theory.
And current approaches to deriving the cosmological constant within
effective field theory produce results which
are some 120 orders of magnitude greater
than allowed by observations\footnote{And this assessment
does not improve much upon observing
that exact supersymmetry could protect from the emergence of any energy density of the sort
relevant for such cosmological-constant studies.
In fact, Nature clearly does not have supersymmetry at least
up to the TeV scale, and this would still lead to a natural prediction
of the cosmological constant which is some 60 orders of magnitude too high.}.

However, just like
there are several
researchers who are skeptical about any results obtained
using low-energy effective field theory in
analyses relevant for the quantum-gravity/quantum-spacetime regime,
there are
also quite a few researchers
who feel that it should be ok to assume
a description in terms of effective field theory for
all low-energy (sub-Planckian) manifestations
of the quantum-gravity/quantum-spacetime regime.

Adopting a strict phenomenologist viewpoint perhaps the most important
observation is that for several of the effects discussed in this section on
UV corrections to Lorentz Symmetry, and for some of the effects
discussed in later sections,
studies based on effective quantum field theory can only be performed
with a rather strongly ``pragmatic'' attitude.
One would like to confine the new effects to unexplored high-energy regimes,
by adjusting bare parameters accordingly,
but, as I shall stress again later,
quantum corrections produce~\cite{suda1,suda2,suda3,tuning} effects that are nonetheless significant
at accessible low energies, unless one allows for rather severe fine-tuning.
On the other hand we do not have enough clues
concerning setups alternative to quantum-field theory that could be used.
For example, as I
shall discuss in some detail later, some attempts are
centered on density-matrix formalisms that go beyond quantum mechanics,
but those are (however legitimate)
mere speculations at the present time.
Nonetheless several of the phenomenologists involved, myself included,
feel that in such a situation phenomenology cannot be
stopped by the theory impasse, even at the risk of later discovering
that the whole (or a sizable part of)
phenomenological effort was not on sound conceptual bases.

But I stress that even when contemplating the possibility of physics outside the domain of
effective quantum field theory one inevitably must at least come
to terms with the success of effective field theory in reproducing a vast class
of experimental data.
In this respect, at least for studies of Planck-scale departures from classical-spacetime
relativistic symmetries I find particularly intriguing a potential  ``order-of-limits issue''.
The effective-field-theory description
 description might be applicable
only in reference frames in which the process of interest
is essentially occurring in its center of mass
(no ``Planck-large boost''~\cite{dsrgzk}
with respect to center-of-mass frame).
The field theoretic description could emerge in
a sort of ``low-boost limit'', rather than the expected
low-energy limit.
The regime of low boosts with respect to the center-of-mass frame
is often indistinguishable from the low-energy limit.
For example, from a Planck-scale perspective, our laboratory
experiments (even the ones conducted at, e.g., CERN, DESY, SLAC, \dots)
are both low-boost (with respect to the center of mass frame)
and low-energy.
However, some contexts that are of interest in quantum-gravity phenomenology,
such as the collisions between ultra-high-energy cosmic-ray protons and
CMBR photons, are situations where all the energies of the particles
are still tiny with respect to the Planck energy scale,
but the boost with respect to the center-of-mass frame
 could be considered to be ``large''
from a Planck-scale perspective: the
Lorentz factor $\gamma$ with respect
to the proton rest frame is much greater than the ratio between
the Planck scale and
the proton mass
\begin{equation}
\gamma =E/m_{\mathrm{proton}} \gg E_p/E~.
\label{boostingUHECRphotons}
\end{equation}

Another interesting scenario concerning the nature of the limit through which
quantum-spacetime physics should reproduce ordinary physics
is suggested by results on
field theories
in noncommutative spacetimes.
One can observe that a spacetime characterized by
an uncertainty relation
of the type
\begin{equation}
\delta x \, \delta y \ge \theta (x,y)
\label{gupNCST}
\end{equation}
never really behaves as a classical spacetime, not even at very low energies.
In fact, according to this type of uncertainty relation,
a low-energy process involving soft momentum exchange
in the $x$ direction (large $\delta x$) should somehow be connected
to the exchange of a hard momentum in the $y$ direction
($\delta y \ge \theta/\delta x$), and this feature cannot be faithfully
captured by our ordinary field-theory formalisms.
For the so-called ``canonical noncommutative spacetimes''
one does obtain a plausible-looking field theory~\cite{dougnekr},
but the results actually show that it is not possible to rely
on an ordinary effective low-energy
quantum-field-theory description because of the presence
of ``UV/IR mixing''\cite{dougnekr,susskind} (a mechanism
such that the high-energy sector of the theory does not decouple
from the low-energy sector, which in turn affects
very severely
the outlook of analyses based on
an ordinary effective low-energy quantum-field-theory description).
For other (non-canonical) noncommutative spacetimes
we are still struggling in the search of a satisfactory formulation of
a quantum field theory~\cite{lukieFT,gacmich}, and it is at this point
legitimate to worry that such a formulation of dynamics in those
spacetimes does not exist.

And the assumption of availability of
an ordinary effective low-energy quantum-field-theory description
has also been challenged
by some perspectives on the Loop-Quantum-Gravity approach.
For example the arguments presented in Ref.~\cite{gampulDENSI}
suggest that in several contexts in which one
would naively expect a low-energy field theory description
Loop Quantum Gravity might instead require a density-matrix
description with features going beyond the reach of effective quantum field theory.

\subsubsection{Other key features of test theories with modified
  dispersion relation}

In order to be applicable to a significant ensemble of experimental
contexts a test theory should of course specify much more than
the form of the dispersion relation.
In light of the type of data that we expect to have access to
(see later) besides the choice of working within or without
low-energy effective quantum field theory there are at least three other
issues that the formulation of such a test theory should clearly
address:
\begin{list}{}{}

\item (i) is the modification of the dispersion relation ``universal''?
or should one instead allow different modification parameters for different
particles?

\smallskip

\item (ii) in presence of a modified dispersion relation between the energy $E$
and the momentum $p$ of a particle should
we still assume the validity of the relation $v = dE/dp$ between
the speed of a particle and its dispersion relation?

\smallskip

\item (iii) in presence of a modified dispersion relation should
we still assume the validity of the standard law of
energy-momentum conservation?

\end{list}

Unfortunately on these three key points the quantum-spacetime
pictures which are providing motivation for the study
of Planck-scale modifications of the dispersion relation
are not giving us much guidance yet.

For example, in Loop Quantum Gravity,
while we do have some (however fragile and indirect)
evidence that the dispersion relation should
be modified, we do not yet have a clear indication concerning
whether the law of energy-momentum conservation should also
be modified and we also cannot yet establish whether
the relation $v=dE/dp$ should be preserved.

Similarly in the analysis of noncommutative
spacetimes we are close to establishing
rather robustly the presence of modifications of the dispersion relation,
but other aspects of the relevant theories have not yet been clarified.
While most of the literature for canonical noncommutative spacetimes
assumes~\cite{dougnekr,susskind}
that the law of energy-momentum conservation should not be modified,
most of the literature for $\kappa$-Minkowski spacetime
argues in favor of a modification
of the law of energy-momentum conservation.
There is also still no consensus
on the relation between speed and dispersion relation,
and particularly in the $\kappa$-Minkowski literature
some departures from the $v=dE/dp$ relation are actively
considered~\cite{kosinski:2002gu,mignemi:2003ab,daszkiewicz:2003yr,jurekREV}.
And at least for canonical noncommutative spacetimes the possibility
of a nonuniversal dispersion relation is considered
extensively~\cite{dougnekr,susskind}.

Concerning the relation $v = dE/dp$ it may be useful to stress that
it can be obtained assuming that a Hamiltonian
description is still available, $v = dx/dt \sim [x,H(p)]$,
and that the Heisenberg uncertainty principle
still holds exactly ($ [x,p] = 1 \rightarrow x \sim \partial / \partial p$).
The possibility of modifications of the Hamiltonian description is of course
an aspect of the debate on ``Planck-scale dynamics'' that was in part discussed
in Subsection.~3.2.1.
And concerning the
Heisenberg uncertainty principle I have already mentioned some arguments
that invite us to contemplate modifications.

\subsubsection{A test theory for pure kinematics}
With so many possible alternative ingredients to mix
one can of course produce a large variety of test theories.
As mentioned, I intend to focus on some illustrative examples of
test theories for my characterization
of achievable experimental sensitivities.

My first example is a test theory of very limited scope,
since it is conceived to only describe pure-kinematics effects.
This will restrict strongly the class of experiments that can be
analyzed in terms of this test theory, but the
advantage is that the limits obtained on the parameters of this
test theory will have rather wide applicability (they will apply
to any quantum-spacetime theory with that form of kinematics, independently
of the description of dynamics).

The first element of this test theory, introduced from
a quantum-spacetime-phenomenology perspective in Refs.~\cite{grbgac,aemn1},
is a ``universal'' (same for all particles)
dispersion relation of the form
\begin{equation}
 m^2 \simeq E^2 - \vec{p}^2
+  \eta \vec{p}^2 \left({E^n \over E^n_{p}}\right)
~,
\label{displeadbisNEW}
\end{equation}
with real $\eta$ of order 1 and integer $n$ ($>0$).
This formula is compatible with some
of the results obtained in the Loop-Quantum-Gravity approach
and reflects some results obtained for theories in $\kappa$-Minkowski
noncommutative spacetime.

Already in the first studies~\cite{grbgac} that proposed a phenomenology
based on~(\ref{displeadbisNEW}) it was assumed
that  even at the Planck scale the familiar
description of ``group velocity'', obtained from the dispersion relation
according to $v=dE/dp$, would hold.

And in other early
phenomenology works~\cite{kifune,ita,gactp,aus}
based on~(\ref{displeadbisNEW})
it was assumed that the law of energy-momentum conservation
should not be modified at the Planck scale, so that, for example,
in a $a + b \rightarrow c + d$ particle-physics process one would have
\begin{equation}
E_a + E_b = E_c + E_d
~,
\label{econs}
\end{equation}
\begin{equation}
\vec{p}_a + \vec{p}_b = \vec{p}_c + \vec{p}_d
~.
\label{pcons}
\end{equation}

In the following I will refer to this test theory as the ``PKV0 test
theory'', where ``PK'' reflects its ``Pure-Kinematics'' nature, ``V''
reflects its ``Lorentz-symmetry Violation'' content, and ``0'' reflects the fact
that it combines the dispersion relation ({\ref{displeadbisNEW}) with what appears
to be the most elementary set of assumptions concerning other key aspects of
the physics: universality of the dispersion relation, $v=dE/dp$,
and unmodified law of energy-momentum conservation.

This rudimentary framework is a good starting point for exploring
the relevant phenomenology. But one should also consider
some of the possible variants.
For example, as mentioned, the undeformed conservation of energy-momentum
is relativistically incompatible with the deformation of the dispersion relation
(so, in particular, the PKV0 test theory requires a preferred frame).
Modifications of the law of energy-momentum conservation would
be required in a DSR picture, and may be naturally considered even in other
scenarios\footnote{As stressed earlier
in this section, one can restore a relativistic formulation by
appropriately {\emph{matching}} the modification of the dispersion relation
and a modification of energy-momentum conservation.
When the modifications of the dispersion relation and of energy-momentum conservation
(even when both present) do not match one has a framework
which requires a preferred frame.}.

Evidently also the universality of the effect can and should be challenged.
And there are indeed (as I shall stress again later in this review)
several proposals of test theories with different
magnitude of the effects for different particles~\cite{mattinLRR,tedAP}.
Let me just mention, in closing this subsection, a case which
is particularly challenging for phenomenology,
which is the case of the variant of the PKV0 test theory allowing for
 nonuniversality such that the effects
are restricted only to photons~\cite{ellisNONUNI,unoEdue},
thereby limiting significantly the class of observations/experiments
that could test the scenario (see however Ref.~\cite{stefaSIGLemn}).

\subsubsection{A test theory based on low-energy effective field
  theory}

The restriction to pure kinematics has the merit to allow us
to establish constraints that are applicable to a relatively
large class of quantum-spacetime scenarios (different formulations
of dynamics would still be subject to the relevant constraints),
but it also restricts severely the type of experimental
contexts that can be considered, since it is only in rare instances
(and only to some extent) that one can qualify an analysis as purely
kinematical.
The desire to be able to analyze a wider class of experimental contexts
is therefore providing motivation for the development of
test theories more ambitious than the PKV0 test theory,
with at least some elements of dynamics.
This is of course rather reasonable, as long as one proceeds
with awareness of the fact that, in light of the situation on the theory
side, for test theories adopting a given description of dynamics
there is a risk that we may eventually find out that
none of the quantum-gravity approaches which are being pursued
is reflected in the test theory.

When planning to devise a test theory that includes the possibility
to describe dynamics of course the first natural candidate
(not withstanding the concerns reviewed in Subsection~3.2.1) is  the framework
of low-energy effective quantum field theory.
In this subsection I want to discuss a test theory
which is indeed based on low-energy effective field theory,
and has emerged primarily\footnote{While Myers and Pospelov
have the merit of alerting the community to several opportunities and issues within
a simple model, we now understand that many of these aspects uncovered
through their simple model
(such as birefringence) are common aspects of a more general
class of field-theory models with rotationally invariant operators of odd dimension
(see, {\it e.g.}, Ref.~\cite{kosteEM}).}
from the analysis reported by Myers and Pospelov
in Ref.~\cite{rob}. Motivated mainly by the perspective of Loop Quantum Gravity
advocated in Ref.~\cite{gampul},
this test theory explores the possibility of a linear-in-$L_p$
modification of the dispersion relation
\begin{equation}
 m^2 \simeq E^2 - \vec{p}^2
+  \eta \vec{p}^2 L_p E
~,
\label{dispROB}
\end{equation}
 i.e., the case $n=1$ of Eq.~(\ref{displeadbisNEW}).
Perhaps the most notable outcome of the exercise of introducing
such a dispersion relation within an effective low-energy field-theory
setup is the observation~\cite{rob} that for the case of electromagnetic radiation,
assuming essentially only that the effects are characterized
mainly by an external four-vector, one arrives
at a single possible correction term for the Lagrangian density:
\begin{equation}
 \mathcal L=-\frac{1}{4}F_{\mu\nu}F^{\mu\nu}
 +\frac{1}{2E_p} n^\alpha F_{\alpha\delta}n^\sigma
  \partial_\sigma(n_\beta\varepsilon^{\beta\delta\gamma\lambda}F_{\gamma\lambda})
 \label{eq:D5lagrangian}
\end{equation}
where the four-vector $n_\alpha$ parameterizes the effect.

This is also a framework for broken Lorentz symmetry, since
the (dimensionless)
components of $n_\alpha$ take different values in different reference frames,
transforming indeed as the components of a four-vector.
And a full-scope phenomenology for this proposal should
explore~\cite{gacBIREFgiulia} the four-dimensional parameter space, $n_\alpha$,
taking into account the characteristic frame dependence of the parameters $n_\alpha$.
As I shall discuss in later parts of this section,
there is already a rather sizeable literature on this phenomenology,
but still mainly focused on what turns out to be the simplest possibility
for the Myers-Pospelov framework, which relies on the assumption that one is
in a reference frame where $n_\alpha$
only has a time component, $n_\alpha = (n_0,0,0,0)$.
Then, upon introducing the convenient notation  $\xi \equiv (n_0)^3$,
one can  rewrite~(\ref{eq:D5lagrangian}) as
\begin{equation}
 \mathcal L=-\frac{1}{4}F_{\mu\nu}F^{\mu\nu}+\frac{\xi}{2E_{p} }
  \varepsilon^{jkl} F_{0 j} \partial_0F_{k l}\, ,
 \label{eq:MP}
\end{equation}
and in particular one can exploit the simplifications provided
by spatial isotropy.
And a key feature that arises is birefringence:
within this setup it turns
out that when right-circular polarized photons satisfy the
dispersion relation $E^2 \simeq p^2 + \eta_\gamma p^3$ then necessarily
left-circular polarized photons satisfy the ``opposite sign''
dispersion relation $E^2 \simeq p^2 - \eta_\gamma p^3$.

In the same spirit one can add spin-$1/2$ particles to the model,
but for them the structure of the framework does not introduce
constraints on the parameters, and in particular there can be
 two independent parameters $\eta_+$ and $\eta_-$
to characterize the
modification of the dispersion relation for fermions
of different helicity:
\begin{equation}
 m^2 \simeq E^2 - \vec{p}^2
+  \eta_+ \vec{p}^2 \left({E \over E_{p}}\right)
~,
\label{displeadNONUNIgpmpmin3}
\end{equation}
in the positive-helicity case, and
\begin{equation}
 m^2 \simeq E^2 - \vec{p}^2
+  \eta_- \vec{p}^2 \left({E \over E_{p}}\right)
~,
\label{displeadNONUNIgpmpmin4}
\end{equation}
in the negative-helicity case.
The formalism is compatible with the possibility
to introduce further independent parameters for each additional
fermion in the theory (so that, e.g., protons would have different
values of $\eta_+$ and $\eta_-$ with respect to the ones of electrons).
And there is no constraint on the relation between $\eta_+$ and $\eta_-$, but
the consistency of the framework requires~\cite{tedAP} that for particle-antiparticle pairs
the deformation should have opposite sign on opposite helicities,
so that, for example, $\eta_+^{\mathrm{(electron)}} = - \eta_-^{\mathrm{(positron)}}$
and $\eta_-^{\mathrm{(electron)}} = - \eta_+^{\mathrm{(positron)}}$.

In some investigations one might prefer to look at particularly
meaningful portions of this large parameter space.
For example, one might consider~\cite{newjourn} the possibility that
the deformation for all spin-$1/2$ particles be characterized by only
two parameters, the same two parameters for all particle-antiparticle pairs
(leaving open however some possible sign ambiguities to accommodate the possibility
to choose between, for example, $\eta_+^{\mathrm{(muon)}} = \eta_+^{\mathrm{(electron)}} = - \eta_-^{\mathrm{(positron)}}$
and $\eta_+^{\mathrm{(muon)}} = \eta_+^{\mathrm{(positron)}}
= -\eta_-^{\mathrm{(electron)}} $).
In the following I will refer to this test theory as the ``FTV0 test
theory'', where ``FT'' reflects its adoption of a ``low-energy effective Field Theory''
description, ``V''
reflects its ``Lorentz-symmetry Violation'' content, and ``0'' reflects the ``minimalistic''
assumption of ``universality for spin-$1/2$ particles''.

\subsubsection{More on ``pure-kinematics'' and ``field-theory-based''
  phenomenology}

 Before starting my characterization
of experimental sensitivities in terms of the parameters of some test theories
I find appropriate to add a few remarks warning about some difficulties that
are inevitably encountered.

For the pure-kinematics test theories some key difficulties originate
from the fact that sometimes an effect due to modification
of dynamics can take a form that is not easily
distinguished from a pure-kinematics effect. And other times
one deals with an analysis of effects that appear to be exclusively
sensitive to kinematics but then at the stage of converting experimental results
into bounds on parameters some level of dependence
on dynamics arises. An example of this latter possibility will be provided
by my description of particle-decay thresholds in test theories that
violate Lorentz symmetry. The derivation of the equations that
characterize the threshold requires only the knowledge of the laws
of kinematics. And if according to the kinematics of a given test theory
a certain particle at a certain energy cannot decay then observation of the
decay allows to set robust pure-kinematics limits on the parameters.
But if the test theory predicts that a certain particle at a certain energy
can decay then by not finding such decays we are not in a position to
truly establish pure-kinematics limits on the parameters of the test theory.
If the decay is kinematically allowed but not seen it is possible that
the laws of dynamics prevent it from occurring (small decay amplitude).

Of course, by adopting low-energy quantum field theory this type
of limitations are removed, but other issues must be taken into
account, particularly in association with the fact that the
FTV0 quantum field theory is not renormalizable.
Quantum-field-theory-based  descriptions of Planck-scale departures from
Lorentz symmetry can only be developed
with a rather strongly ``pragmatic'' attitude.
In particular, for the FTV0 test theory, with its
Planck-scale suppressed effects
at tree level,
some authors (notably Refs.~\cite{suda1,suda2,suda3,tuning}) have argued
that the loop expansion could effectively generate additional terms
of modification of the dispersion relation
that are unsuppressed by the cut-off scale of the (nonrenormalizable)
field theory.
Of course, the parameters of the field theory can be fine-tuned to eliminate
the unwanted large effects, but the needed level of fine tuning
is usually rather unpleasant.
While certainly undesirable, this severe fine-tuning
problem should not discourage us from considering the FTV0 test theory,
at least not at this early stage of the development of the relevant phenomenology.
Actually some of the most successful theories used in fundamental
physics are affected by severe fine tuning. It is not uncommon to eventually
discover
that the fine tuning is only apparent, and for example some hidden symmetry is
actually ``naturally'' setting up the hierarchy of parameters.

In particular, it is already established that supersymmetry
can tame the fine-tuning issue~\cite{pospeSYMM,pospeSYMM2}.
If one extends supersymmetric quantum electrodynamics by
adding interactions
with external vector and tensor backgrounds that violate Lorentz symmetry
at the Planck scale then
exact supersymmetry requires that such interactions
correspond to operators of dimension five or higher, so that no fine-tuning
is needed in order to suppress the unwanted operators of dimension lower than five.
Of course, supersymmetry can only be an approximate symmetry
of the physical world, and the effects of the scale of soft-supersymmetry-breaking
masses contorls the renormalization-group evolution of dimension five Lorentz-violating
 operators and
their mixing with dimension three Lorentz-violating operators~\cite{pospeSYMM,pospeSYMM2}.

And it has also been established~\cite{pospeHL}
that if Lorentz-violation occurs in the gravitational sector,
then the violations of Lorentz symmetry induced on
the matter sector do not require severe fine-tuning.
In particular, this has been investigated by  coupling the Standard Model
of particle physics
to a Horava-Lifshitz description of gravitational phenomena.

Actually the study of Planck-scale departures from Lorentz symmetry
may even find some encouragement in perspectives based on renormalization theory,
at least in as much as it has been shown~\cite{anselmi1,anselmi2,horava,visser}
that some field theories modified
by Lorentz-violating terms are actually rather well behaved in the ultraviolet.


\subsection{Photon stability}
\label{photon-stability}

\subsubsection{Photon stability and modified dispersion relations}

The first example of Planck-scale sensitivity that I discuss
is the case of a process which is
kinematically forbidden in presence of exact
Lorentz symmetry but becomes kinematically allowed
in presence of certain
departures from Lorentz symmetry.
It has been established (see, e.g.,
Refs.~\cite{tedOLDgood,gacpion,seth,orfeupion}) that when Lorentz
symmetry is broken at the Planck scale there can be significant
implications for certain decay processes. At the qualitative level
the most significant novelty would be the possibility for massless
particles to decay. And
certain observations in astrophysics,
which allow us to establish
that photons of energies up to $\sim 10^{14}$~eV
are stable,
can be then used~\cite{tedOLDgood,gacpion,seth,orfeupion}
to set limits on schemes for departures
from Lorentz symmetry.

For my purposes it suffices here
to consider the process $\gamma \rightarrow e^+ e^-$.
Let us start from the perspective of the PKV0 test theory, and therefore
adopt the dispersion relation~(\ref{displeadbisNEW})
and unmodified energy-momentum conservation.
One easily finds a relation between
the energy $E_\gamma$ of the incoming photon, the opening angle $\theta$
between the outgoing electron-positron pair, and the energy $E_+$ of
the outgoing positron (of course the energy of the outgoing electron
is simply given by $E_\gamma - E_+$).
Setting $n=1$ in~(\ref{displeadbisNEW}) one finds that,
for the region of phase space with $m_e \ll E_\gamma \ll E_p$,
this relation takes the form
\begin{eqnarray}
\cos(\theta)    \! \simeq \!& \frac{E_+ (E_\gamma -E_+) + m_e^2
- \eta  E_\gamma E_+ (E_\gamma -E_+)/E_p}{ E_+ (E_\gamma -E_+)} ~,
\label{gammathresh}
\end{eqnarray}
where $m_e$ is the electron mass.

The fact that for $\eta = 0$ Eq.~(\ref{gammathresh}) would
require $\cos(\theta) > 1$ reflects the fact that, if Lorentz symmetry
is preserved, the process $\gamma \rightarrow e^+ e^-$ is kinematically
forbidden. For $\eta < 0$ the process is still forbidden, but for
positive $\eta$ high-energy photons can decay
into an electron-positron pair. In fact,
for $E_\gamma \gg (m_e^2 E_p/|\eta |)^{1/3}$
one finds that
there is a region of
phase space where $\cos(\theta) < 1$, i.e., there is a physical
phase space available for the decay.

The energy scale $(m_e^2 E_p)^{1/3} \sim 10^{13}$~eV is not
too high for testing, since, as mentioned, in
astrophysics we see photons of energies up to $\sim 10^{14}$~eV
that are stable (they clearly travel safely some large astrophysical
distances).
The level of sensitivity that is within the reach of these studies
therefore goes at least down to values of (positive) $\eta$ of order 1
and somewhat smaller than 1. This is what one describes as ``Planck-scale
sensitivity'' in the quantum-spacetime phenomenology literature: having set the
dimensionful deformation parameter to the Planck-scale value the coefficient
of the term that can be tested is of order 1 or smaller.
Specifically for the case of the photon-stability analysis
it is however rather challenging to transform
this Planck-scale sensitivity into actual experimental limits.

Within PKV0 kinematics, for $n=1$ and
positive $\eta$ of order 1, it would have been natural
to expect that photons with $\sim 10^{14}$~eV energy
are unstable.
But the fact that the decay of $10^{14}$~eV photons is allowed
by PKV0 kinematics  of course does not guarantee
that these photons should rapidly decay. It depends on the relevant
probability amplitude, whose evaluation goes beyond the reach
of kinematics.
Still it is likely that these observations are very significant for theories
that are compatible with PKV0 kinematics. For a theory that
is compatible with PKV0 kinematics (with positive $\eta$)
this evidence of stability of photons
imposes the identification of a dynamical mechanism that essentially
prevents photon decay. If one finds no such mechanism the theory
is ``ruled out'' (or at least its parameters are severely constrained),
but in principle one could look endlessy for such a mechanism.
A balanced approach to this issue must take into account that
quantum-spacetime physics may well modify both kinematics and
the strength (and nature) of interactions  at a certain scale, and it might in principle
do this in ways that cannot be accommodated within the confines of effective
quantum field theory, but one should take notice of the fact that,
even in some new (to-be-discovered) framework outside  effective
quantum field theory, it is unlikely that there will be very large ``conspiracies" between
the modifications of kinematics and the modifications of
the strength of interaction. In principle models based on pure kinematics are
immune from certain bounds on parameters which are derived
also using descriptions of the interactions, and it is conceivable that
in the correct theory the actual bound would be somewhat shifted from the
value derived within effective quantum field theory, but in order
to contemplate large
differences in the bounds one would need to advocate
very large and {\it ad hoc}
modifications of the strength of interactions, large enough
 to compensate for the often dramatic implications of the modifications of kinematics.
 The challenge then is to find satisfactory criteria for confining speculations
 about variations
 of the strengths of interaction only within a certain plausible range.
 To my knowledge this has not yet been attempted, but it would deserve high priority.

Of course a completely analogous calculation can be done
within the FTV0 test theory, and there one can easily
arrive at the conclusion\cite{maccione:2008iw}
 that the FTV0 description of dynamics
should not suppress significantly the photon-decay process.
However, as mentioned, consistency with
the effective-field-theory setup
requires that the two polarizations of the photon acquire
opposite-sign modifications of the dispersion relation.
We observe in astrophysics some photons
of energies up to $\sim 10^{14}$~eV
that are stable over large distances,
but as far as we know those photons could be all,
say, right-circular polarized (or all left-circular polarized).
This evidence of stability of photons therefore is only applicable
to the portion of the FTV0 parameter space in which both polarizations
should be unstable (a subset of the region
with $|\eta_+|>|\eta_\gamma|$ and $|\eta_-|>|\eta_\gamma|$).

\subsubsection{Photon stability and modified energy-momentum conservation}

So far I have discussed photon stability assuming that only the
dispersion relation is modified. If the modification of the dispersion relation
is instead combined with a modification of the law of energy-momentum
conservation the results can change very significantly.
In order to expose these changes in rather striking fashion
let me consider the example of DSR-inspired
 laws of
energy-momentum conservation
for the case of $\gamma \rightarrow e^+ e^-$:
\begin{eqnarray}
E_\gamma \simeq E_+ + E_- - \eta \vec{p}_+ {\cdot} \vec{p}_- ~,
\label{econsgammastab}
\end{eqnarray}
\begin{eqnarray}
 \vec{p}_\gamma \simeq \vec{p}_+ + \vec{p}_- - \eta E_+ \vec{p}_-
 - \eta E_- \vec{p}_+ ~.
\label{pconsgammastab}
\end{eqnarray}
Using these in place of ordinary conservation of energy-momentum one ends
up with a result for $\cos(\theta)$ which is still of the form $(A+B)/A$
but now with $A = 2 E_+ (E_\gamma -E_+) +
\lambda  E_\gamma E_+ (E_\gamma -E_+)$ and $B=2 m_e^2$:
\begin{eqnarray}
\cos(\theta)    \! \simeq \!& \frac{
2 E_+ (E_\gamma -E_+) +
\eta  E_\gamma E_+ (E_\gamma -E_+)
 + 2 m_e^2}{2 E_+ (E_\gamma -E_+) +
\eta  E_\gamma E_+ (E_\gamma -E_+)} ~,
\label{gammathreshDSR}
\end{eqnarray}
Evidently this formula always
gives $\cos(\theta) > 1$, so there are combinations
of modifications of the dispersion relation and modifications
of energy-momentum conservation such that $\gamma \rightarrow e^+ e^-$
is still forbidden.

If the modification of the dispersion relation and the modification
of the law of energy-momentum conservation are not matched exactly
to get this result then one can have the possibility of photon decay,
but in some cases it can be further suppressed (in addition to the Planck-scale
suppression) by the partial compensation between the two modifications.

The fact that the matching between
modification of the dispersion relation and modification
of the law of energy-momentum conservation
that produces a stable photon is obtained using a DSR-inspired setup
is not surprising~\cite{gacdsr2010}.
The relativistic properties of the framework are clearly at stake in this
derivation.
A threshold-energy requirement for particle decay
(such as the $E_\gamma \gg (m_e^2 E_p/|\eta |)^{1/3}$ mentioned above)
cannot be introduced as an observer-independent law, and is therefore
incompatible with any relativistic (even DSR-relativistic) formulation
of the laws of physics.
In fact, different observers assign different values to the energy of a particle
and therefore in presence of a threshold-energy requirement for particle decay
a given particle should be allowed to decay according to some observers
while being totally stable for others.


\subsection{Pair-production threshold anomalies and gamma-ray
  observations}
\label{pair-production}

Another opportunity to investigate quantum-spacetime-inspired
Planck-scale departures from
Lorentz symmetry is provided by certain types of energy thresholds
for particle-production processes that are relevant in
astrophysics. This is a very powerful tool for quantum-spacetime
phenomenology~\cite{kifune,ita,gactp,aus,steckGlashow,lehnertTHRESH,repLiberati,maREVIEW},
and in fact at the beginning of
this review I chose the evaluation of the threshold energy
for photopion production, $p+\gamma_{CMBR} \rightarrow p+\pi$,
 as the basis for illustrating how the sensitivity levels that
 are within our reach can be placed in rather natural connection
 with effects introduced at the Planck scale.

I shall discuss the photopion production threshold analysis in more detail
in the next subsection. Here I consider instead
the electron-positron pair production process,  $\gamma \gamma \rightarrow e^+ e^-$.

\subsubsection{Modified dispersion relations and $\gamma \gamma \rightarrow e^+ e^-$}
\label{photons}

The threshold for $\gamma \gamma \rightarrow e^+ e^-$
is  relevant for studies of
 the opacity
of our Universe to photons.
In particular, according to the conventional (classical-spacetime)
description,
the infrared diffuse extragalactic background
should give rise to strong absorption of ``TeV photons''
(here understood as photons with energy $1\mathrm{\ TeV} < E < 30\mathrm{\ TeV}$),
but this prediction must of course
be reassessed in presence of violations of
 Lorentz symmetry.

To show that this is the case, let me start once again from the
perspective of the PKV0 test theory,
and analyze a collision between
a soft photon of energy $\epsilon$
and a high-energy photon of energy $E$ which might produce an
electron-positron pair.
Using the dispersion relation~(\ref{displeadbisNEW}) (for $n=1$)
and the (unmodified) law of energy-momentum conservation,
one finds that for given soft-photon energy $\epsilon$,
the process $\gamma \gamma \rightarrow e^+ e^-$
is allowed only if $E$ is greater than a certain
threshold energy $E_{th}$ which depends on $\epsilon$ and $m_e^2$,
as implicitly codified in the formula
(valid for $\epsilon \ll m_e \ll E_{th} \ll E_p$)
\begin{equation}
E_{th} \epsilon + \eta \frac{E_{th}^3}{8 E_p} \simeq m_e^2
~.
\label{thrTRE}
\end{equation}
The special-relativistic result $E_{th} = m_e^2 /\epsilon$
corresponds of course to the $\eta \rightarrow 0$ limit
of~(\ref{thrTRE}).
For $|\eta | \sim 1$ the Planck-scale correction can be
safely neglected as long as $\epsilon \gg (m_e^4/E_p)^{1/3}$.
But eventually, for sufficiently small values of $\epsilon$
(and correspondingly large values of $E_{th}$) the
Planck-scale correction cannot be ignored.

This provides an opportunity for a pure-kinematics test: if a 10~TeV photon
collides with a photon of 0.03~eV and produces
an electron-positron pair the case $n=1$, $\eta \sim -1$
for the PKV0 test theory is ruled out.
A 10~TeV photon and a 0.03~eV photon can produce an
electron-positron pair according to ordinary special-relativistic
kinematics (and its associated requirement $E_{th} = m_e^2 /\epsilon$),
but they cannot produce an
electron-positron pair according to PKV0 kinematics with $n=1$
and $\eta \sim -1$.

For positive $\eta$ the situation is somewhat different.
While negative $\eta$ increases the energy requirement for electron-positron
pair production, positive $\eta$ decreases
the energy requirement for electron-positron
pair production.
In some cases where one would expect
electron-positron
pair production to be forbidden the PKV0 test theory with positive $\eta$
would instead allow it. But once a process is allowed there is no
guarantee that it will actually occur, not without some information
on the description of dynamics (that allows us to evaluate cross sections).
As in the case of photon decay, one must conclude that
a pure-kinematics framework
can be falsified when it predicts that a process cannot occur
(if instead the process is seen) but in principle it cannot be falsified when
it predicts that a process is allowed. Of course, here too one should gradually
develop balanced criteria taking into account the remarks I offered in Subsec.~3.3.1
concerning the plausibility (or lack thereof) of conspiracies between modifications
of kinematics and modifications of the strengths of interaction.

Concerning the level of sensitivity that we can expect to achieve also in
this case one can robustly claim that Planck-scale sensitivity is within
our reach. This, as anticipated above, is best seen considering
the ``TeV photons''  emitted by
some blazars, for which (as they travel toward our Earth detectors)
the photons of the
infrared diffuse extragalactic background
are potential targets for electron-positron pair production.
In estimating the sensitivity achievable with this type of
analyses it is necessary to take into account the fact that,
besides the form of the threshold condition, there are at least
three other factors that play a role in establishing the level of
absorption of TeV photons  emitted by a given Blazar:
our knowledge of the type of signal emitted by the Blazar (at the source),
the distance of the blazar,
and most importantly the density of
the infrared diffuse extragalactic background.

The availability of observations of the relevant type has increased
very significantly over these past few years.
For example for the blazar ``Markarian~501'' (at a redshift of $z=0.034$)
and the Blazar ``H1426+428'' (at a redshift of $z=0.129$)
robust observations up to the 20-TeV range have been
reported~\cite{ahaMKcut501,h1426},
and for the blazar ``Markarian~421''  (at a redshift of $z=0.031$)
observations of photons of energy up to 45 TeV has been reported~\cite{tev45},
although a more robust signal is seen once again up
to  the 20-TeV range~\cite{krennMKcut,ahaMKcut}.

The key obstruction for
translating these observations into an estimate of the
effectiveness of pair-production absorption
comes from the fact that measurements of the
density of the infrared diffuse extragalactic background
are very difficult,
and as a result our experimental information
on this density is still affected by
large uncertainties~\cite{dirbe,voelk,berezin,firbREVIEW}.

The observations do show convincingly that some absorption is
occurring~\cite{ahaMKcut501,h1426,tev45,krennMKcut,ahaMKcut}.
I should in particular stress the fact that
the analysis of the combined X-ray/TeV-gamma-ray spectrum
for the Markarian 421 blazar, as discussed in particular in
Ref.~\cite{steckXRAY}, provides rather compelling evidence.
The X-ray part of the spectrum allows to predict the
TeV-gamma-ray part of the spectrum
in a way that is rather insensitive on our
poor knowledge of the source. This in turn allows us
to establish in a source-independent way that some absorption
is occurring.

For the associated quantum-spacetime-phenomenology analysis
the fact that some absorption
is occurring does not allow to infer much: the analysis
will become more and  more effective as the quantitative characterization
of the effectiveness of absorption becomes more and more precise
(as measured by the amount of deviation from the
level of absorption expected within a classical-spacetime analysis
that would still be compatible with the observations).
And we are not yet ready to make any definite
statement about this absorption levels. This is not only
a result of our rather poor knowledge
of the infrared diffuse extragalactic background, but it is also due
to the status of the observations, which still presents us with some
apparent puzzles. For example, it is not yet fully understood
why, as observed by some authors~\cite{ahaMKcut501,krennMKcut,ahaMKcut,voelk},
there is a difference between the absorption-induced cutoff energy found
in data
concerning Markarian~421, $E_{\mathrm{mk421}}^{\mathrm{cutoff}} \simeq 3.6$~TeV,
and the corresponding cutoff estimate obtained
from Markarian-501 data, $E_{\mathrm{mk501}}^{\mathrm{cutoff}} \simeq 6.2$~TeV.
And the observation of TeV $\gamma$-rays emitted by
the blazar H1426+428, which is significantly more distant
than Markarian 421 and  Markarian 501, does show a level of absorption
which is higher than the ones inferred for Markarian~421
and  Markarian~501, but (at least assuming a certain description~\cite{h1426}
of the infrared diffuse extragalactic background) the H1426+428
TeV luminosity ``seems to
exceed the level anticipated from the current models of TeV blazars
by far''~\cite{h1426}.

Clearly the situation requires further clarification,
but it seems reasonable to expect that within a few years
we should fully establish facts such as ``$\gamma$-rays with energies
up to 20 TeV are absorbed by the infrared diffuse
extragalactic background''\footnote{While
some observers understandably argue that the
the residual grey areas that I discussed impose us to still be extremely prudent,
even at the present time one could legitimately describe as robust~\cite{newjourn}
the observational evidence indicating that some $\gamma$-rays with energies
up to 20 TeV are absorbed by the infrared diffuse
extragalactic background.
And some authors (see, e.g., Ref.~\cite{steckIRSTRONG}) actually see
in the presently-available data
an even sharper level of agreement with the classical-spacetime picture,
which would translate in having already at the present time achieved
Planck-scale sensitivity.}.
This would imply that at least some photons with energy
smaller than $\sim$~200~meV
can create an electron-positron pair
in collisions with a 20~TeV $\gamma$-ray. In turn this would imply for
the PKV0 test theory, with $n=1$, that necessarily $\eta \geq - 50$
(i.e., either $\eta$ is positive or $\eta$ is negative with
absolute value smaller than 50).
This means that this strategy of analysis will soon take us
robustly at sensitivities that are less than
a factor of a 100 away from Planck-scale sensitivities,
and it is natural to expect that further refinements of these measurements
will eventually take us at Planck-scale sensitivity and beyond.

The line of reasoning needed to establish whether this Planck-scale
sensitivity could apply to pure-kinematics frameworks is somewhat
subtle. One could simplistically state that when we see a process
which is forbidden by a certain set of laws of kinematics then those laws
are falsified. However, in principle this statement is correct only when we have full
knowledge of the process, including a full determination of the
momenta of the incoming particles.
In the case of the absorption of  multi-TeV gamma rays from blazars
it is natural to assume that this absorption
be due to interactions with infrared photons, but we are not in a position
to exclude that
the absorption be due to higher-energy background photons.
We should therefore contemplate the possibility that
the PKV0 kinematics be implemented within a framework
in which the description of dynamics is such to introduce
a large-enough modification of cross sections to allow absorption
of multi-TeV blazar gamma rays by background photons of energy
higher than  200~meV. As mentioned above repeatedly, I advocate a balanced
perspective for this sort of issues, which should not extend  all the way to assuming
wild conspiracies centered on very large changes in cross sections, even when
testing a pure-kinematics framework.
But, as long as a consensus on criteria for such a balanced approach is not established,
it is difficult to attribute a quantitative confidence level
to experimental bounds on a pure-kinematics
framework through mere observation of some absorption
of multi-TeV blazar gamma rays.

This concerns are of course not applicable to test theories which
do provide a description of dynamics, such as the FTV0 test theory,
with its effective-field-theory setup.
However, for the FTV0 test theory
one must take into account the fact
that the modification of the dispersion relation carries
opposite sign for the two polarizations of the photon
and might have an helicity dependence in the case of electrons and positrons.
So also in the case of the FTV0 test theory,
as long as observations only provide evidence of
some absorption of TeV gamma rays (without much to say about
the level of agreement with the amount of absorption expected in the
classical-spacetime picture), and are therefore consistent
with the hypothesis that only one of the polarizations of the photon
is being absorbed,
only rather weak limits can be established.

\subsubsection{Threshold anomalies and modified energy-momentum conservation}
\label{DSR-analysis}

Also for the derivation of threshold anomalies
combining a modification of the law of energy-momentum conservation
with the modification of the dispersion relation can lead
to results that are very different from the case where only the modifications
of the dispersion relations are assumed.
This is a feature which I already stressed in the case
of the analysis of photon stability.
In order to establish it also for
threshold anomalies
let me consider again an example of ``DSR-inspired'' modified law
of energy-momentum conservation.
So I assume
that the
modification of the law of energy-momentum
conservation for the case
of  $\gamma \gamma \rightarrow e^+ e^-$
takes the form
\begin{eqnarray}
E + \epsilon - \frac{\eta}{E_p} \vec{P} {\cdot} \vec{p} \simeq E_+
+ E_- -  \frac{\eta}{E_p} \vec{p}_+ {\cdot} \vec{p}_- ~,
 ~~~~~~~~~~~~~
\label{consgammagamma}\\
 \vec{P} +  \vec{p}
 +  \frac{\eta}{E_p} E \vec{p}
 + \frac{\eta}{E_p} \epsilon \vec{P}  \simeq \vec{p}_+ + \vec{p}_-
 +  \frac{\eta}{E_p} E_+ \vec{p}_-
 +  \frac{\eta}{E_p} E_- \vec{p}_+ ~
\label{consgammagamma2}
\end{eqnarray}
where I denoted with $\vec{P}$ the momentum of the photon of energy $E$
and I denoted with  $\vec{p}$ the momentum of the photon
of energy $\epsilon$.

Using these (\ref{consgammagamma}),(\ref{consgammagamma2}) and the ``$n=1$''
dispersion relation
 one obtains
(keeping only terms that are meaningful
for $\epsilon \ll m_e \ll E_{th} \ll E_p$)
\begin{equation}
E_{th} \simeq \frac{m_e^2}{\epsilon }
~,
\label{thrTREbis}
\end{equation}
i.e., one ends up
with the same result as in the special-relativistic case.

This shows very emphatically that modifications of the
law of energy-momentum conservation can compensate
for the effects on threshold derivation produced by modified
dispersion relations.
The cancellation should typically be only partial
but in cases when the two modifications are ``matched exactly''
there is no left-over effect. The fact that a DSR-inspired modification
of the law of conservation of energy-momentum produces
this exact matching admits a tentative interpretation which the
interested reader can find in Refs.~\cite{gacdsr,gacdsr2010}.


\subsection{Photopion production threshold anomalies and the
  cosmic-ray spectrum}
\label{photopion-production}

In the preceding Subsection~\ref{pair-production} I discussed the implications of
possible Planck-scale effects for the
process $\gamma \gamma \rightarrow e^+ e^-$, but of course this is not the only process
in which Planck-scale effects can be important. In particular, there has been strong
interest~\cite{kifune,ita,gactp,aus,tedOLDgood,gacpion,orfeupion,alfaroUHECR,nguhecr}
in the analysis
of the ``photopion production'' process, $p \gamma \rightarrow p \pi$.
As I already stressed in Section~\ref{simple-example}, interest in
the photopion-production process originates from its role
in our description of the high-energy portion of the
cosmic-ray spectrum.
The ``GZK cutoff'' feature of that spectrum is linked directly
to the value of the minimum (threshold) energy
required for cosmic-ray protons to produce pions in
collisions with CMBR photons~\cite{greisenGZK,zatsepinGZK}
(also see, e.g., Refs.~\cite{mooreUHECR,domokosUHECR}).
The argument suggesting that Planck-scale modifications of the dispersion relation
may affect significantly the estimate of this threshold energy
is of course completely analogous to the one discussed in the preceding
Section~\ref{pair-production} for $\gamma \gamma \rightarrow e^+ e^-$.
However, the derivation is somewhat more tedious:
in the case of $\gamma \gamma \rightarrow e^+ e^-$
the calculations are simplified
by the fact that both outgoing particles have mass $m_e$
and both incoming particles are massless,
whereas for the threshold conditions for the
photopion-production process one needs to
handle the kinematics for a head-on
collision between a soft photon of energy $\epsilon$
and a high-energy particle of mass $m_p$ and momentum $\vec{k}_p$
producing  two (outgoing) particles with masses $m_p$,$m_\pi$ and
momenta $\vec{k}_{p}'$,$\vec{k}_\pi$.
The threshold can then be conveniently~\cite{gactp} characterized as a relationship
describing the minimum value, denoted by $k_{p,th}$, that the
spatial momentum of the incoming particle
of mass $m_p$ must have in order for the process to be allowed
for given value $\epsilon$ of the photon energy:
\begin{equation}
k_{p,th} \simeq {(m_p + m_\pi)^2 - m_p^2 \over 4 \epsilon}
+ \eta {k_{p,th}^{2+n} \over 4 \epsilon E_p^n} \left(
{m_p^{1+n} + m_\pi^{1+n} \over (m_p + m_\pi)^{1+n}} -1 \right)
~
\label{lithresh2}
\end{equation}
(dropping terms that are further suppressed
by the smallness of $E_{p}^{-1}$ and/or the smallness of $\epsilon$
or $m_{p,\pi}$).

Notice that whereas in discussing the pair-production threshold relevant
for observations of TeV gamma rays I had immediately specialized~(\ref{displeadbisNEW})
to the case $n=1$, here I am contemplating values of $n$ that are even greater
than 1. One could of course admit $n>1$ also for the
pair-production threshold analysis, but it would be a mere academic exercise, since
it is easy to verify that in that case Planck-scale sensitivity is within reach only
for $n$ not significantly greater than 1.
Instead (as I briefly stressed already in Section~\ref{simple-example})
the role of the photopion-production threshold in cosmic-ray analysis is such
that even for the case of values of $n$ as high as 2 (i.e., even
for the case of effects
suppressed quadratically by the Planck scale)
Planck-scale sensitivity is not unrealistic.
In fact, using for $m_p$ and $m_\pi$ the values of the masses
of the proton and the pion and for $\epsilon$ a typical CMBR-photon energy
one finds that for negative $\eta$ of order 1 (effects introduced at the Planck scale)
the shift of the threshold codified in~(\ref{lithresh2}) is gigantic
for $n=1$ and still observably large~\cite{ita,gactp} for $n=2$.

For negative $\eta$ the Planck-scale correction shifts the
photopion-production threshold to higher values
with respect to the standard classical-spacetime prediction,
which estimates the photopion-production threshold scale to be of
about $5 \cdot 10^{19}$~eV.
Assuming\footnote{It used to be natural to expect~\cite{berezin}
that indeed the highest energy cosmic rays are protons.
This is however changing rather rapidly in light of recent dedicated
studies using Auger data~\cite{augerPRL2010,gaisserUHECR2011,stanevUHECR2011},
which favour  a significant contribution from heavy nuclei.
The implications for the Lorentz symmetry analysis of the differences
between protons and heavy nuclei, while significant in the detail (see, e.g.,
Ref.~\cite{maccioneUHECR2011}), are not as large as one might naively
expect. This is due to the fact that
 it just happens to be the case that the photodisintegration
  threshold is reached when the energy of
  typical heavy nuclei, say Fe, is $\sim 5 \cdot 10^{19}$~eV, i.e.,
  just about the value of the photopion-production threshold
  expected for cosmic-ray protons.}
that the observed cosmic rays of highest energies are protons,
when the spectrum reaches the photopion-production threshold
one should first encounter a pileup of cosmic rays with energies
just in the neighborhood of the threshold scale,
and then above the threshold
the spectrum should be severely depleted.
The pileup results from the fact that
protons with above-threshold energy tend to
loose energy through photopion production and slow down until
their energy is comparable to the threshold energy.
The depletion above the threshold is the counterpart of this pileup
(protons emitted at the source with energy above the threshold
tend to reach us, if they come to us from far enough, with energy
 comparable to the threshold energy).

The availability in this cosmic-ray context of
Planck-scale sensitivities for values of $n$ all the way up to $n=2$
was already fully established by the year 2000~\cite{ita,gactp}.
The debate then quickly focused on establishing what exactly the observations
were telling us about the photopion-production threshold.
The fact that the AGASA cosmic-ray observatory was reporting~\cite{agasa}
 evidence of a behaviour of the spectrum that was of the type expected in this Planck-scale
picture generated a lot of interest.
However, more recent cosmic-ray observations, most notably the ones reported
by the Pierre Auger observatory~\cite{augerdata1,augerPLB2010},
appear to show no evidence of unexpected behaviour. There is
even some evidence~\cite{augersources}
(see however the updated Ref.~\cite{augersourcesNEW})
suggesting that to
the highest-energy observed cosmic rays one can associate some relatively
nearby sources, and that all this is occurring at scales
that could fit within the standard picture of the photopion-production threshold,
without Planck scale effects.

These results reported by the Pierre Auger Observatory
are already somewhat beyond
the ``preliminary'' status, and  we should soon have
at our disposal very robust cosmic-ray
data, which should be easily converted into actual
experimental bounds on the parameters
of Planck-scale test theories.

Among the key ingredients which are still missing I should
assign priority to the mentioned issue of correlation of cosmic-ray
observations with
the large scale distribution of matter in the nearby universe
and the issue of the composition of cosmics rays (protons versus
heavy nuclei).
The rapidly-evolving~\cite{augersources,augersourcesNEW}
picture of correlations
with matter in the nearby universe focuses on
cosmic-ray events with energy $\geq 5.7 \cdot 10^{19}\mathrm{\ eV}$,
while the growing evidence
of a significant heavy-nuclei component at high energies
is limited so far at energies $\leq 4 \cdot 10^{19}\mathrm{\ eV}$.
And this state of affairs,  as notably stressed for example in
Ref.~\cite{gaisserUHECR2011},
limits our insight on several issues relevant for the understanding
of the origin of cosmic rays and the related issues for tests
of Lorentz symmetry, since it leaves open several options for the
nature and distance of the sources above and below,
say,  $5 \cdot 10^{19}\mathrm{\ eV}$.

Postponing then more definite claims on the situation on
the experimental side, let me stress however that
there is indeed a lot at stake in these studies for
 the hypothesis of quantum-spacetime-induced
Planck-scale departures from Lorentz symmetry.
Even for pure-kinematics test theories this type of data analysis
is rather strongly relevant. For example, the kinematics of the PKV0 test theory
forbids (for negative $\eta$ of order 1 and $n \le 2$)
photopion production when the incoming proton energy is in the neighborhood
of $5 \cdot 10^{19}$~eV and the incoming photon has typical CMBR energies.
 For reasons that I already stressed (for other contexts)
 previously in this review, in order to establish
 a robust experimental limit on pure-kinematics scenarios using the role of the
 photopion-production threshold
 in the cosmic-ray spectrum it would be necessary to also exclude that other
 background photons
 (not necessarily CMBR photons) be
 responsible for the observed cutoff\footnote{We are here dealing again
with the limitations that pure-kinematics particle-reaction analyses suffer when the properties of the incoming particles are not fully under control.
The pure kinematics of the PKV0
test theory definitely forbids (for negative $\eta$ of order $1$ and $n \le 2$)
pion production resulting from collisions between a $5 \cdot 10^{19}$~eV proton
and a CMBR photon. But it allows
pion production resulting from collisions between a $5 \cdot 10^{19}$~eV proton
and more energetic photons, and in order to exclude that possibility one
ends up formulating assumptions
about dynamics (low density of relevant photons may be compensated
by unexpected increase in cross section).}.
It appears likely that such a level of understanding
of the cosmic-ray spectrum will be achieved in the not-so-distant future.

For the FTV0 test theory, since it goes beyond pure kinematics,
one is not subject to similar concerns\cite{maccione:2009ju}. However, the fact that it admits the
possibility of different effects for the two helicities of the incoming proton,
complicates and renders less sharp
this type of cosmic-ray analyses. It does however lead to intriguing hypothesis:
 for example, exploiting the possibility
of helicity dependence of the Planck scale effect for protons one can rather naturally
end up with a scenario that predicts a pileup/cutoff structure somewhat similar
to the one of the standard classical-spacetime analysis, but softer as a result
of the fact that only roughly half of the protons would be allowed to loose
energy by photopion production.

Of course, also for the photopion-production threshold one finds
exactly the same mechanism, which I discussed in some detail for the
pair-production threshold, of possible compensation between the
effects produced by modified dispersion relations and the effects
produced by modified laws of energy-momentum conservation.
So the analysis of frameworks where both the dispersion relation
and the energy-momentum conservation law are modified,
as typical in DSR scenarios~\cite{gacdsr2010},
should take into account that added element of complexity.


\subsection{Pion non-decay threshold and cosmic-ray showers}
\label{pion-non-decay}

Also relevant for the analysis of cosmic-ray observations is another
aspect of the possible implications of
quantum-spacetime-motivated
Planck-scale departures from Lorentz
symmetry: the possibility of a suppression of pion decay at ultrahigh energies.
While in some
cases departures from Lorentz symmetry allow the decay
of otherwise stable particles (as in the case of $\gamma \rightarrow e^+ e^-$,
discussed above, for appropriate choice of values of parameters),
it is indeed also possible for
departures from Lorentz symmetry to either introduce a threshold value of
the energy of the particle above which a certain decay channel for that particle
is totally forbidden~\cite{colgla,dedenko},
or introduce some sort of suppression of the decay probability which increases
with energy and becomes particularly effective above a certain
threshold value of
the energy of the decaying particle~\cite{gacpion,orfeupion,siglpion}.
This may be relevant~\cite{dedenko,gacpion} for the description of the air showers
produced by cosmic rays, whose structure depends rather sensitively
on certain decay probabilities, particularly the one for
the decay $\pi \rightarrow \gamma \gamma$.

The possibility of suppression at ultrahigh energies of the
decay $\pi \rightarrow \gamma \gamma$ has been considered
from the quantum-gravity-phenomenology perspective primarily
adopting PKV0-type frameworks~\cite{gacpion,orfeupion}.
Using the kinematics of the PKV0 test theory
one easily arrives~\cite{gacpion} at the following
relationship between the opening angle $\phi$ of the directions of the momenta
of the outgoing photons,
the energy of the pion ($E_\pi$) and the energies ($E$ and $E'=E_\pi-E$)
 of the outgoing photons:
\begin{eqnarray}
\cos(\phi)    \! = \!& {2 E E' - m_\pi^2
+ 3 \eta E_\pi E E'/E_p
\over
2 E E' + \eta E_\pi E E'/E_p} ~.
\label{pithresh}
\end{eqnarray}
This relation shows that, for positive $\eta$,
 at high energies the phase space available to the decay
is anomalously reduced:
for given value of $E_\pi$ certain values of $E$
that would normally be accessible to the decay are no longer
accessible (they would require $\cos \theta > 1$).
This anomaly starts to be noticeable at pion energies of
order $(m_\pi^2/L_p)^{1/3} \sim 10^{15}$~eV, but only
very gradually (at first only a small portion of the available
phase space is excluded).

This is rather intriguing
since there is a report~\cite{dedenko} of experimental evidence of anomalies
for the structure of the air showers produced by cosmic rays,
particularly their
longitudinal development. And it has been argued in Ref.~\cite{dedenko}
that these unexpected features of the
longitudinal development of air showers could be explained in terms
of a severely reduced decay probability for pions of energies
of $10^{15}$~eV and higher.
This is still to be considered a very preliminary observation,
not only because of the need to acquire data of better quality on
the development of air showers, but also because of the role~\cite{gacpion}
that our limited control of nonperturbative QCD has in setting
our expectations for what air-shower development should look like
without new physics.

It is becoming rather ``urgent'' to reassess this issue in light
of recent data on cosmic rays and and cosmic-ray shower development.
Such an exercise has not been made for a few years now,
and for the mentioned Auger data, with the associated debate on
the composition of cosmic rays, the analysis of shower development
(and therefore of the hypothesis of some suppression of pion decay)
is acquiring increasing
significance~\cite{stanevUHECR2011,augershowers,aloisioshowers,wilkshowers}.

As for the other cases in which I discussed effects of modifications
of the dispersion relation for kinematics of particle reactions,
also for this pion-decay argument scenarios hosting both a modified
dispersion relation and modifications of the law of conservation
of energy-momentum, as typical in DSR scenarios,
can host~\cite{gacdsr2010}
a compensation of the correction terms.


\subsection{Vacuum Cerenkov and other anomalous processes}

The quantum-spacetime-phenomenology analyses I have reviewed so far
have played a particularly significant role in the rapid growth of the
field of quantum-spacetime phenomenology over the last decade.
This is particularly true for
the analyses of the pair-production threshold for gamma rays
and
of the photopion-production
threshold for cosmic rays,  in which the data relevant
for the Planck-scale
effect under study  can be perceived as
providing some encouragement for new physics.
One can indeed legitimately argue~\cite{aus,piranIRnew} that the observed level of absorption of TeV
gamma rays is low enough to justify speculations about ``new physics''
(even though, as mentioned, there are ``conventional-physics
descriptions'' of the relevant data).
The opportunities for Planck scale physics to play a role in the neighborhood
of the GZK  scale of the cosmic-ray spectrum are becoming slimmer,
as stressed in Section~\ref{photopion-production},
but still it has been an important sign of maturity for quantum-spacetime
phenomenology
to play its part in the debate that for a while was generated
by the preliminary and tentative indications
of an anomaly around the ``GZK cutoff''.
And it is interesting how the hypothesis a pion-stability
threshold, another Planck-scale-motivated hypothesis, also plays a role
in the assessment of the present status of studies of ultra-high-energy
cosmic rays.

I am giving a disproportioned attention to
the particle-interaction analyses described in
Sections~\ref{pair-production}, \ref{photopion-production}, \ref{pion-non-decay}
because they are the most discussed and clearest evidence
in support of the claim that
quantum-spacetime Planck-scale phenomenology
does have the ability to discover its target new physics, so much so that
some (however tentative) ``experimental puzzles'' have been considered
and are being considered from the quantum-spacetime perspective.

But  it is of course
important to also consider the implications of
quantum-spacetime-inspired Planck-scale departures from
Lorentz symmetry, and particularly Planck-scale modifications of the
dispersion relation, for all possible particle-physics processes. And
a very valuable type of particle-physics processes to be considered
are the ones that are forbidden in a standard special-relativistic
setup but could be allowed in presence of Planck-scale departures from
Lorentz symmetry. These processes could be called ``anomalous
processes'', and in the analysis of some of them one does find
opportunities for Planck-scale sensitivity, as already here discussed
for the case of the process $\gamma \rightarrow e^- e^+ $, in
Section~\ref{photon-stability}.

For a comprehensive list (and more detailed discussion) of other analyses
of anomalous processes,
which are relevant for the whole subject of the study of possible departures from Lorentz
symmetry (within or without quantum spacetime),
readers can rely on Refs.~\cite{mattinLRR,tedAP} and references therein.

I will here just briefly mention one more significant example of
anomalous process that is relevant from a quantum-spacetime-phenomenology
perspective: the ``vacuum Cerenkov''
process, $e^- \rightarrow e^- \gamma $,
which in certain scenarios~\cite{mattinLRR,tedAP,altschuCHERE}
with broken Lorentz symmetry is allowed above a threshold value
of electron energy. This is analyzed in close analogy with
the discussion a reviewed above for
the process $\gamma \rightarrow e^- e^+ $
(which is another example of anomalous particle interaction),
which
I discussed in Section~\ref{photon-stability}.

Since we have no evidence at present of vacuum-Cerenkov processes
the relevant analyses are of the type that sets limits on the parameters
of some test theories.
Clearly this observational evidence against vacuum-Cerenkov processes
is also relevant for pure-kinematics test teories, but in ways that it is difficult
to quantify, because of the dependence on the strength of the interactions (an aspect
of dynamics). So here too one should contemplate the implications
of these findings from the perspective of the remarks I offered in Subsec.~3.3.1
concerning the plausibility (or lack thereof) of conspiracies between modifications
of kinematics and modifications of the strengths of interaction.

Of course, within
the FTV0 test theory one can rigorously analyze
the  vacuum-Cerenkov process, and there actually,
if one arranges for opposite-sign dispersion-relation correction terms
for the two helicities of the electron, one can in principle have
helicity-changing $e^- \rightarrow e^- \gamma $ at any energy (no threshold),
but estimates performed~\cite{mattinLRR,tedAP}  within the FTV0 test theory show that
the rate is extremely small at low energies.

Above the threshold for
helicity-preserving $e^- \rightarrow e^- \gamma $
the FTV0 rates are substantial, and this in particular
would allow an analysis with Planck-scale sensitivity that relies
on observations of 50-TeV gamma rays from the Crab nebula.
The argument is based on several assumptions (but all apparently robust)
and its effectiveness is somewhat limited by the combination of parameters
allowed by FTV0 setup and by the fact that for these 50-TeV gamma rays
we observe from the Crab nebula we can only reasonably guess a part of the
properties of the emitting particles.
According to the most commonly adopted model the relevant gamma rays
are emitted by the Crab nebula as a result of inverse Compton processes,
and from this one infers~\cite{mattinLRR,tedAP,altschu} that for electrons of
energies up to 50 TeV
the vacuum Cerenkov process is still ineffective, which in turn allows one
to exclude certain corresponding regions of the FTV0 parameter space.


\subsection{In-vacuo dispersion for photons}\label{invacuosubsec}

Analyses of thresholds for particle-physics processes, discussed in the previous
subsections, played a particularly important role in the development of
quantum-spacetime
phenomenology over the last decade, also because the relevant studies were
already at Planck-scale sensivity. In June 2008, with the launch of the Fermi (/GLAST)
space telescope~\cite{glast1,glast2,glast3,unoSCIENCE,fermiGRB090510,fermiPHYSREPORT}
we  gained access to Planck-scale effects also
for in-vacuo dispersion.
These studies deserve particular interest
because they have broad applicability to
quantum-spacetime test theories
of the fate
of Lorentz/Poincar\'e symmetry at the Planck scale.
 In the previous subsections I stressed
how the analyses of thresholds for particle-physics processes
provided information that is rather strongly model dependent, and dependent on
the specific choices of parameters within a given model. The type of insight gained
through in-vacuo-dispersion studies is instead significantly more robust.

A wavelength dependence of the speed of photons is
obtained~\cite{grbgac,shore} from a
modified dispersion relation, if one assumes the velocity to be
still described by $v = dE/dp$. In particular, from the dispersion relation
of the PKV0 test theory one obtains
(at ``intermediate energies'', $m < E \ll E_p$)
a velocity law of the form
\begin{equation}
v \simeq 1 - \frac{m^2}{2 E^2} +  \eta \frac{n+1}{2}
\frac{E^n}{E_p^n} ~.
\label{velLIVbis}
\end{equation}
Arguments and semi-heuristic derivations in support of this
type of speed law for massless particles have been
reported\footnote{For the related subject of the description
of light propagation in models of emergent spacetime, see, e.g.,
Ref.~\cite{fotiniPRD} and references therein.}
both in the spacetime-noncommutativity
literature (see, e.g., Refs.~\cite{gacmaj,piotrKAPPAspeed})
and in the loop-quantum-gravity
literature (see, e.g., Refs.~\cite{gampul,urrutia,thiemLS}).

On the basis of the speed law (\ref{velLIVbis}) one would find that
two simultaneously-emitted photons should reach the detector at
different times if they carry different energy. And this
time-of-arrival-difference effect can be
significant~\cite{grbgac,schaefer,piranKARP,wagnerGRB,falconeGRB}
in the analysis of
short-duration gamma-ray bursts that reach us from cosmological
distances. For a gamma-ray burst it is not uncommon\footnote{Up to 1997
the
distances from the gamma-ray bursters to the Earth were not
established experimentally.
Starting with the 1997 result of Ref.~\cite{bepposax}
we are now able to establish, through
a suitable analysis of the
gamma-ray-burst ``afterglow'',
the distance
between the gamma-ray bursters and the Earth for a significant portion of
all detected bursts. Sources at a distance of $\sim 10^{10}$ light years ($\sim 10^{17}$~s) are
not uncommon.} that the time
travelled before reaching our Earth detectors be of order $T \sim
10^{17}$~s. Microbursts within a burst can have very short
duration, as short as $10^{-3}$~s, and this
should suggest that the photons that compose such a microburst are all
emitted at the same time, up to an uncertainty of $10^{-3}$~s.
Some of the photons in these bursts have energies that extend
even above~\cite{unoSCIENCE} 10~GeV, and for two photons with energy
difference of order $\Delta E \sim 10$~GeV a $\Delta E/E_p$
speed difference over a time of travel of $10^{17}$~s would lead~\cite{unoEdue}
to a difference in times of arrival
of order $\Delta t \sim \eta T \Delta \frac{E}{E_p} \sim \eta \cdot 1$~s which
is not negligible\footnote{Of course there are ordinary-physics effects
that could be relevant for these analyses, such as ordinary electromagnetic dispersion,
but it is easy to show~\cite{grbgac,bombelli}
that already at energies of, say, a few GeV these ordinary-physics
effects would be negligible with respect to the candidate quantum-gravity effect
here considered.} with respect to the
typical variability time scales one expects for the astrophysics
of gamma-ray bursts.
Indeed,
it is rather clear~\cite{unoEdue,granotFERMIERA}
that the studies of gamma-ray bursts
conducted by the Fermi telescope
provide us access to testing Planck-scale effects, in the linear-modification (``$n=1$'')
scenario.

Of course these tests do not actually use (\ref{velLIVbis}) since
for redshifts of 1 and higher, spacetime curvature/expansion is a very
tangible effect.
And this introduces nonnegligible complications.
Most results in quantum-spacetime research hinting at modifications
of the dispersion relation, and possible associated energy/momentum dependence
of the speed of massless particles, where derived working essentially
in the flat-spacetime/Minkowski limit: it is obvious that analogous effects would
also be present when spacetime expansion is switched on, but it is not obvious
how formulas should be generalized to that case.
In particular, the formula (\ref{velLIVbis}) is essentially unique for ultrarelativistic
particles in the flat-spacetime limit: we are only interested in leading-order formulas
and the difference between  $(E/E_p)^n$ and, say, $p^2E^{n-2}/E_p^n$
is of course negligible for ultrarelativistic particles (with $p^2 \gg m^2$).
How spacetime expansion renders these considerations more subtle
 is visible already in the
case of de Sitter expansion. Adopting conformal coordinates in de Sitter spacetime,
with metric $ds^2=dt^2-a^2(t) \, dx^2$ (and $a(t)=e^{Ht}$)
we have for ultrarelativistic particles (with $p^2 \gg m^2$) the velocity formula
\begin{equation}
v \simeq  a^{-1}(t)- \frac{m^2}{2 p^2} a(t)  ~,
\label{velSTANDARDds}
\end{equation}
so already in the undeformed case the coordinate velocity (from which physical
time delays will be derived) depends not only on momentum but also
on the scale factor $a(t)$. It is then not obvious how one should describe
leading-order Planck-scale corrections to this, going like some power of momentum.
It is natural to make the {\it ansatz}
\begin{equation}
v \simeq  a^{-1}(t)- \frac{m^2}{2 p^2} a(t) + \eta \frac{n+1}{2}
\frac{p^n}{E_p^n}  a^k(t) ~,
\label{velDEFds}
\end{equation}
with however the integer $k$ being at this point one more phenomenological parameter
to be determined experimentally.
Arguments on which value of the integere $k$ could be most ``natural''
were reported in Refs.~\cite{ellisREDSHIuno,piranRedShFirst,piranRedSh,ellisREDSHIdue},
ultimately leading to a consensus~\cite{piranRedSh,ellisREDSHIdue} converging
on describing $k = -n$ as the most natural choice. I shall not dwell much on this: let
me just confirm that I would also give priority to the case $k = -n$, but doing this
in such a way not to by-pass the obvious fact that the value of $k$ would have
to be determined experimentally (and Nature might well have chosen a value
for $k$ different from $-n$).

Assuming that indeed $k = -n$ one would expect
for simultaneously emitted massless particles in a Universe parametrized by the cosmological
parameters $\Omega_m$,$\Omega_\Lambda$, $H_0$ (evaluated today)
a momentum-dependent difference in times of arrival at a telescope
given by
\begin{equation}
\Delta t \simeq   \eta \frac{n+1}{2H_0}
\frac{p^n}{E_p^n} \int_0^z dz' \frac{(1+z')^n}{\sqrt{\Omega_m (1+z')^3 + \Omega_\Lambda}} ~,
\label{timedelayDS}
\end{equation}
where $p$ is the momentum of the particle when detected at the telescope.

Actually Planck-scale sensitivity to in-vacuo disperson
can also be provided by observations of
TeV flares from certain active galactic nuclei, at redshifts much smaller
than 1 (cases such that spacetime expansion is not much tangible).
In particular,
studies of TeV flares from
Mk501 and PKS2155--304 performed by
 the MAGIC~\cite{magicGENERAL} and HESS~\cite{hessgeneral} observatories
have established~\cite{emnrevASTRO,magic,ellisUNO,hessPRL2008,hess2011,hessREVIEW}
bounds on the scale of dispersion, for the linear-effects (``$n=1$'')
scenario, at about $1/10$ of the Planck scale.

But the present best constraints on quantum-spacetime-induced
in-vacuo dispersion are derived from observations
of gamma-ray bursts reported by
the Fermi telescope.
There are so far four Fermi-detected gamma-ray bursts which are
particularly significant for the hypothesis of in-vacuo dispersion:
GRB090816C~\cite{unoSCIENCE}
GRB090510~\cite{fermiGRB090510}
GRB090902B~\cite{grb090902B}
GRB090926A~\cite{grb090926A}.
The data for each one of these bursts has the strength of constraining
the scale of in-vacuo dispersion,
for the linear-effects (``$n=1$'')
scenario, at better than $1/10$ of the Planck scale.
In particular, GRB090510 was a truly phenomenal short burst~\cite{fermiGRB090510}
and the structure of its observation allows us to conservatively establish
that
the scale of in-vacuo dispersion,
for the linear-effects (``$n=1$'')
scenario, is higher than $1.2$ times the Planck scale.

The simplest way to do such analyses is to take one high-energy
photon observed from the burst and take as reference its delay $\Delta t$
with respect
to the burst trigger: if one could exclude conspiracies such that the
specific photon was emitted before the trigger (we cannot really exclude it, but
we would consider that as very unlikely, at least with present knowledge)
evidently $\Delta t$ would have to be bigger than any delay caused by
the quantum-spacetime effects.
This in turn allows us, for the case of
GRB090510, to establish
the limit at $1.2$ times the Planck scale~\cite{fermiGRB090510}.
And interestingly even more sophisticated techniques
of analysis, using not a single photon but the whole structure
of the high-energy observation of GRB090510,
also encourage the adoption of a limit at
$1.2$ times the Planck scale~\cite{fermiGRB090510}.
It has also been noticed~\cite{nemiroff090510} that if one takes at face value
the presence of
high-energy photon bunches observed for GRB090510,
as evidence that these photons were emitted nearly simultaneously
at the source and they are being detected nearly simultaneously,
then the bound inferred could be even two orders of magnitude above
the Planck scale~\cite{nemiroff090510}.

I feel that at least
the limit at $1.2$ times the Planck scale is reasonably safe/conservative.
But it is obvious that here we would feel more comfortable with a wider
collection of gamma-ray bursts usable for our analyses.
This would allow us to balance, using high statistics,
the challenges for such studies of in-vacuo dispersion
that (as for other types of
studies based on observations in astrophysics I discussed earlier)
originate from the fact that we only have tentative models of the source of the signal.
In particular,
 the engine mechanisms causing the bursts of gamma rays also introduces
 correlations at the source between
the energy of the emitted photons and the time of their emission.
This was in part expected by some astrophysicists~\cite{piranKARP},
and Fermi data allows one to infer it at levels even beyond
expectations~\cite{unoSCIENCE,fermiGRB090510,meszarosUNO,grbMistery,guettaSHORTstrange,ghiselliniDUE}.
On a single observation of gamma-ray-burst event such at-the-source correlations
are in principle indistinguishable from the effect we expect from in-vacuo dispersion,
which indeed is a correlation between times of arrival and energies of the photons.
And another challenge I should mention originates from
the necessity of understanding at least partly
the ``precursors'' of a gamma-ray burst, another
feature that was already expected and to some extent known~\cite{lazzatiPRECURSOR},
but recently came to be known as a more significant effect than
expected~\cite{fermiGRB090510,troja090510}.

So we will reach a satisfactory ``comfort level'' with our bounds on
in-vacuo dispersion only with ``high statistics'', a relatively large collection~\cite{unoEdue}
 of gamma-ray bursts usable for our analyses.
High statistics always helps but in this case
it will also provide a qualitatively new handle for the data analysis:
a relatively large collection of high-energy gamma-ray bursts,
inevitably distributed over different values of redshift,
would help our analyses also because comparison of bursts at
different redshifts
can be exploited to  achieve
results that are essentially free from uncertainties originating from our lack of knowledge
of the sources.
This is due to the fact that
 the structure of in-vacuo dispersion is such that the effect
should grow in predictable manner with redshift,
whereas we can exclude that the exact same dependence on redshift
(if any) could characterize
the correlations at the source between
the energy of the emitted photons and the time of their emission.

In this respect we might be experiencing a case of tremendous bad luck:
as mentioned we really  still only have four gamma-ray bursts to work with,
GRB090816C~\cite{unoSCIENCE}
GRB090510~\cite{fermiGRB090510}
GRB090902B~\cite{grb090902B}
GRB090926A~\cite{grb090926A},
but on the basis of how Fermi observations had been going for the first 13 months
of operation we were led to hope that by this time (end of 2012), after
50 months of operation of Fermi, we might have had as many as 15
such bursts and perhaps 4 or 5 bursts of outstanding interest for
in-vacuo dispersion, comparable to GRB090510.
Indeed all these four bursts we keep using from the Fermi data set
were observed during the first 13 months of operation (in particular GRB090510
was observed during the 10th month of operation)
and we got from Fermi really nothing else of any use for us over the
last 37 months.
If our luck turns around we should be able to claim for quantum-spacetime
phenomenology a first small but tangible success: ruling out at least the
specific hypothesis of Planck-scale in-vacuo dispersion, at least specifically
for the case of linear-effects (``$n=1$'').

This said about the opportunities and challenges facing this
phenomenology of in-vacuo dispersion, let me, in closing this subsection,
offer a few additional remarks on the broader picture.
From a broader quantum-spacetime-phenomenology perspective
it is noteworthy that, while in the analyses discussed in the previous subsections
the amplifier of the Planck-scale effect was provided by a large boost,
in this in-vacuo-dispersion case the amplification is due primarily
to the long propagation times, which essentially render the analysis
sensitive to the accumulation~\cite{polonpap}
of very many minute Planck-scale effects.
For propagation times that are realistic in controlled Earth experiments,
in which one perhaps could manage to study the propagation of photons of, say, TeV energies,
over distances of, say, $10^6$~m,
the in-vacuo dispersion would still induce, even for $n=1$,
only time delays of order $\sim 10^{-18}$~s.

In-vacuo-dispersion analyses of gamma-ray bursts are also extremely popular
within the quantum-spacetime-phenomenology community because of
the very limited number of assumptions on which they rely. One comes
very close to having a direct test of a Planck-scale modification of the dispersion
relation.
In comparing the PKV0 and the FTV0 test theories, one could exploit the fact that
whereas for the PKV0 test theory the
Planck-scale-induced time-of-arrival difference would affect a multi-photon microburst
by producing a difference in the ``average arrival time''
 of the signal
in different energy channels, within the FTV0 test theory, for an ideally unpolarized signal,
 one would
expect a dependence of the time-spread of a microburst that grows
with energy, but no effect for
the average arrival time
in different energy channels.
This originates of course from the polarization dependence imposed
by the structure of the FTV0 test theory: for low-energy channels
the whole effect will anyway be small, but in the highest-energy channels
the fact that the two polarizations travel at different speed will manifest itself
as spreading in time of the signal, without any net average-time-of-arrival effect
for an ideally unpolarized signal.
Since there is evidence
 that at least some gamma-ray bursts
are somewhat far from being ideally unpolarized
(see evidence of polarization reported, e.g.,
in Refs.~\cite{integralGRBPOLAR,toma1GRBPOLAR,toma2GRBPOLAR}),
one could also
exploit a powerful correlation: within the FTV0 test theory one expects to find
some bursts with sizeable energy-dependent average-time-of-arrival
differences between energy channels
(for bursts with some predominant polarization),
 and some bursts (the ones with no net polarization) with much less
average-time-of-arrival differences between energy channels but a sizeable
difference in time spreading in the different channels.
Polarization-sensitive observations of gamma-ray bursts would of course
allow to look directly for the polarization dependence predicted by the FTV0 test theory.

Clearly these in-vacuo
dispersion studies using gamma rays in the GeV--TeV range
provide us at present with the cleanest opportunity
to look for Planck-scale modifications of the dispersion relation.
Unfortunately, while they do provide us comfortably with
Planck-scale sensitivity to linear ($n=1$) modifications of the
dispersion relation, they are unable to probe significantly the case of quadratic
($n=2$) modifications.

And, while, as stressed, these studies apply
to a wide range of quantum-spacetime scenarios
with modified dispersion relations, mostly as a result of their
insensitivity to the whole issue about
description of dynamical aspects of a quantum-spacetime theory,
 one should be aware of the fact that it might be inappropriate
to characterize these studies as tests that must necessarily apply
to all quantum-spacetime
pictures with modified dispersion relations. Most notably,
the assumption of obtaining the velocity law from the dispersion relation
through the formula $v = dE/dp$ may or may not be valid in a given
quantum-spacetime picture.
Validity of the formula $v = dE/dp$ essentially requires that
the theory is still ``Hamiltonian'', at least in the sense that the velocity along the $x$ axis
is obtained from the commutator with a Hamiltonian ($v_x \sim [x,H]$),
and that the Heisenberg commutator preserves its
standard form ($[x,p_x]\sim \hbar$ so that $x \sim \partial / \partial p_x$).
Especially this second point
is rather significant since
heuristic arguments of the type also used to motivate
 modified dispersion relations
suggest~\cite{ahluDB,kappaHP,stringsHP,mignemiHP,galanDSRhp,meljanacHP}
that the Heisenberg commutator might have to be
 modified in the quantum-spacetime realm.


\subsection{Quadratic anomalous in-vacuo dispersion for neutrinos}

Observations of gamma rays in the GeV--TeV range could provide us a very
sharp picture of Planck-scale-induced dispersion if it happens to
be a  linear ($n=1$) effect, but, as stressed above, one would need
observations of similar quality for photons of significantly higher energies
in order to gain access to scenarios with quadratic ($n=2$) effects
of Planck-scale-induced dispersion.
The prospect of observing photons with energies up to $10^{18}$~eV
at ground observatories~\cite{rissehomola,unoEdue}
is very exciting, and should be pursued very forcefully~\cite{unoEdue},
but it represents an opportunity whose viability still remains to be fully
established.
And in any case we expect photons of such high energies to be absorbed
rather efficiently by background soft photons (e.g., CMBR photons)
so that we could not observe them from very distant sources.

One possibility that could be considered~\cite{aemn1}
is the one of 1987a-type supernovae;
however such supernovae
are typically seen at distances not greater than some $10^5$ light years.
And the fact that neutrinos from 1987a-type supernovae can be definitely observed
up to energies of at least tens of TeV's is not enough to
compensate for the smallness of the distances (as compared to typical gamma-ray-burst
distances).
As a result using  1987a-type supernovae one might have serious difficulties~\cite{aemn1}
even to achieve Planck-scale sensitivity
for linear ($n=1$)
modifications of the dispersion relation,
and going beyond linear order  clearly is not possible.

The most advanced plans for in-vacuo-dispersion
studies with sensitivity up to quadratic ($n=2$) Planck-scale
modifications of the dispersion relation actually
exploit~\cite{ellisgrqc9911055,kingNeutri,gacneutrinos,piranNeutriNat}
(also see, for a similar argument within a somewhat different framework,
Ref.~\cite{bertoNeutri})
once again the extraordinary
properties of gamma-ray bursters, but their neutrino emissions rather
than their production of photons.
Indeed, according to
current models~\cite{grbNEUTRINOnew,waxneutri},
gamma-ray bursters should also emit a substantial amount of
high-energy neutrinos.
Some neutrino observatories should soon observe neutrinos with energies
between $10^{14}$ and $10^{19}$~eV, and one could either (as it appears
to be more feasible~\cite{piranNeutriNat}) compare
the times of arrival of these neutrinos emitted by
gamma-ray bursters to the corresponding times of arrival of
low-energy photons or compare the times of arrivals of different-energy neutrinos
(which however might require larger statistics than it seems natural to expect).

In assessing the significance of these foreseeable studies of neutrino propagation
within different test theories, one should again take into account issues
revolving around the possibility of
anomalous reactions.
In particular, in spite of the weakness of their interactions with other particles,
within an effective-field-theory setup neutrinos can be affected by
Cherenkov-like processes at levels which are experimentally significant~\cite{cohegla},
though not if the scale of modification of the dispersion relation is as high
as the Planck scale.
The recent overall analysis of modified dispersion for neutrinos
in quantum field theory given in Ref.~\cite{libeneutri}
shows that for the linear ($n=1$) case we are presently able to establish constraints
at levels of about $10^{-2}$
times the Planck scale (and of course even further from the Planck scale for the quadratic case, $n=2$).


\subsection{Implications for neutrino oscillations}

It is well established~\cite{colgla,brust,emnEscape,gamboneutri,winsta,joyneutri}
that flavour-dependent
 modifications to the energy-momentum dispersion relations for neutrinos
 may lead to neutrino oscillations even if neutrinos are massless.
This point is not directly relevant for the three test theories I have chosen
to use as frameworks of reference for this review. The PKV0 test theory
adopts universality of the modification of the dispersion
relation, and also the FTV0 test theory describes flavour-independent effects (its
effects are ``nonuniversal'' only in relation to polarization/helicity).
Still I should mention this possibility both because clearly flavour-dependent
effects may well attract gradually more interest from quantum-spacetime
phenomenologists
(some valuable analyses have
already been produced; see, e.g.,  Refs.~\cite{mattinLRR,tedAP} and references therein),
and because even for researchers focusing on flavour-independent effects
it is of course important to be familiar with constraints that may be set on
flavour-dependent scenarios (those constraints, in a certain sense, provide motivation
for the adoption of flavour independence).

Most studies of neutrino oscillations induced by violations of Lorentz symmetry were actually
not motivated by quantum-gravity/quantum-spacetime
research (they were part of the general Lorentz-symmetry-test
research area) and assumed that the flavour-dependent violations would take the form of
a flavour-dependent speed-of-light scale~\cite{colgla}, which essentially corresponds to
the adoption of a dispersion relation of the type~(\ref{displeadbisNEW}), but with $n=0$,
and flavour-dependent values of $\eta$.
A few studies have considered the case\footnote{Also noteworthy is the analysis
reported in Ref.~\cite{carcorNEUTRI}, which argues that neutrino
oscillations may play a role also for other aspects of quantum-spacetime phenomenology,
in addition
to their use in relation to flavour-dependent Planck-scale
modifications of the dispersion relation.} $n=1$  with flavour-dependent $\eta$,
which is instead mainly of interest from
a quantum-spacetime perspective\footnote{This is in part due to
the fact that ``naive quantum gravity'' is not a renormalizable theory, and as a result
the restriction to power-counting renormalizable correction terms (which is standard outside
quantum-gravity research) is expected not to be necessarily applicable to quantum-gravity
research.}, and found~\cite{brust,emnEscape,winsta} that for $n=1$
from~(\ref{displeadbisNEW}) one naturally ends up with oscillations lengths that depend
quadratically on the inverse of the energies of the particles ($L \sim E^{-2}$),
whereas in the case $n=0$ (flavour-dependent speed-of-light scale)
such a strong dependence on the inverse of the energies is not possible~\cite{brust}.
In principle, this opens an opportunity for the discovery of manifestations of the
flavour-dependent $n=1$
case through studies of neutrino oscillations~\cite{brust,winsta};
however, at present there is no evidence of a role for these effects in neutrino
oscillations and therefore the relevant data analyses produce bounds~\cite{brust,winsta}
on flavour dependence of the dispersion relation.

In a part of the next section I shall comment again on neutrino oscillations,
but in relation to the possible role of quantum-spacetime-induced decoherence
(rather than Lorentz-symmetry violations).


\subsection{Synchrotron radiation and the Crab Nebula}

Another opportunity to set limits on test theories with Planck-scale modified
dispersion relations is provided by the study of the
implications of modified dispersion relations for synchrotron
radiation~\cite{jaconature,newjourn,tedsteck,stefaCrab,repEllisCrab,urruSYNCH,altschuSYNCH}.
An important point for these analyses~\cite{jaconature,tedsteck,stefaCrab}
is the observation
that in the conventional (Lorentz-invariant) description of synchrotron
radiation one can estimate the characteristic energy $E_c$ of
the radiation through a semi-heuristic derivation~\cite{jackson}
leading to the formula
\begin{equation}
E_c \simeq {1 \over
R {\cdot} \delta {\cdot} [v_\gamma - v_e]}
~,
\label{omegacjack}
\end{equation}
where $v_e$ is the speed of the electron,
$v_\gamma$ is the speed
of the photon, $\delta$ is the angle of outgoing radiation,
and $R$ is the radius of curvature of
the trajectory of the electron.

Assuming that the only Planck-scale modification in this formula
should come from the velocity law (described using $v=dE/dp$
in terms of the modified dispersion relation),
one finds that in some instances the characteristic energy of
synchrotron
radiation may be significantly modified by the presence of
Planck-scale modifications of the dispersion relation.
This originates from the fact that, for example,
according to~(\ref{velLIVbis}), for $n=1$
and $\eta < 0$,
an electron
cannot have a speed that exceeds the
value $v^{max}_{e} \simeq 1 - (3/2) (|\eta | m_e/E_p)^{2/3}$,
whereas in special relativity $v_{e}$ can take values arbitrarily close to 1.

As an opportunity to test such a modification of the
value of the synchrotron-radiation characteristic energy one
can attempt to use data~\cite{jaconature}
on photons emitted by the Crab nebula.
This must be done with caution since
the observational information on synchrotron radiation being emitted
by the Crab nebula is rather indirect: some of the photons we observe
from the Crab nebula are attributed to sychrotron processes, but only on the basis
of a (rather successful) model, and the value of the
relevant magnetic fields is also not directly measured.
But the level of Planck-scale sensitivity that could be within the
reach of this type of analysis is truly impressive:
assuming that indeed the observational situation has been properly
interpreted, and relying on the mentioned assumption that
the only modification to be taken into account is the
one of the velocity law,
one could~\cite{jaconature,stefaCrab} set limits on the parameter $\eta$
of the PKV0 test theory that go several orders of magnitude beyond  $|\eta| \sim 1$,
for negative $\eta$ and $n=1$, and even for quadratic ($n=2$) Planck-scale modifications
the analysis would fall ``just short'' of reaching Planck-scale sensitivity (``only'' a
few orders of magnitude away from $|\eta| \sim 1$ sensitivity for $n=2$).

However, the assumptions of this type of analysis, particularly the assumption
that nothing changes but the velocity law, cannot even be investigated within
pure-kinematics test theories, such as the PKV0 test theory.
Synchrotron radiation is due to the acceleration
of the relevant charged particles and therefore implicit
in the derivation of the formula~(\ref{omegacjack})
is a subtle role for dynamics~\cite{newjourn}.
From a quantum-field-theory perspective the process of
synchrotron-radiation emission
can be described in terms
of Compton scattering of the electrons
with the virtual photons of the magnetic field, and its analysis
is therefore rather sensitive even to details of the description
of dynamics in a given theory.
Indeed, essentially
this synchrotron-radiation phenomenology has focused on the FTV0 test theory
and its generalizations, so that one can rely on the
familiar formalism of quantum field theory.
Making reasonably prudent assumptions
on the correct model of the source one can establish~\cite{stefaCrab}
valuable (subPlanckian!) experimental bounds on the parameters of the FTV0 test theory.


\subsection{Birefringence and observations of polarized radio
  galaxies}

As I stressed already a few times earlier in this review,
the FTV0 test theory, as a result of a rigidity of the adopted effective-field-theory
framework, necessarily predicts birefringence, by assigning different speeds to different
photon polarizations. Birefringence is a pure-kinematics effect, so it can be also included
in straightforward generalizations of the PKV0 test theory, if one assigns different
dispersion relation to
different
photon polarizations and then assumes that the speed is obtained from the dispersion
relation via the standard $v = dE/dp$ relation.

I have already discussed some ways in which birefringence may affect
other tests of dispersion-inducing (energy-dependent)
modifications of the dispersion relation, as in the example
of searches of time-of-arrival/energy correlations for observations
of gamma-ray bursts.
The applications I already discussed use the fact that for large enough
travel times birefringence essentially splits a group of simultaneously-emitted
photons with roughly the same energy and without characteristic polarization
 into two temporally and spatially separated
groups of photons, with different circular polarization (one group being delayed with respect
to the other as a result of the polarization-dependent speed of propagation).

Another feature that can be exploited is the fact that even for travel times that
are somewhat shorter than the ones achieving a separation into two groups of photons
the same type of birefringence can already effectively erase~\cite{gleiser1,gleiser2} any
linear polarization
that might have been there to begin with, when the signal was emitted.
This observation can be used in turn to argue that
for given magnitude of the birefringence effects and given values of the distance
from the source
it should be impossible to observe linearly polarized light, since the polarization
should have been erased along the way.

Using observations of polarized light
from distant radio galaxies~\cite{mattinLRR,gleiser1,gleiser2,jackbire,jackbireB,maBIREF}
one can achieve comfortably
Planck-scale sensitivity
(for ``$n=1$'' linear modifications of the dispersion relation)
 to birefringence effects following this strategy.
In particular,
the analysis reported in Ref.~\cite{gleiser1,gleiser2}
leads to a limit of $|\eta_\gamma| < 2 \cdot 10^{-4}$
on the parameter $\eta_\gamma$ of the FTV0 test theory.
And more recent studies of this type allowed to establish
even more stringent bounds (see Refs.~\cite{mattinLRR,liberati2009}
and references therein).

Interestingly,
 even for this strategy based on the effect of removal of linear polarization,
gamma-ray bursts could in principle provide formidable opportunities.
And there was a report~\cite{polarGRB}
of observation of polarized MeV gamma rays in the prompt
emission of the gamma-ray burst GRB021206,
which would have allowed to establish very powerful bounds on energy-dependent
birefringence.
 However, the report of
Ref.~\cite{polarGRB}
has been challenged (see, e.g., Ref.~\cite{nopolarGRB,nopolarGRBbis}).
Still, experimental studies of polarization
for gamma-ray bursts continue to be a very active area of research
(see, e.g.,
Refs.~\cite{integralGRBPOLAR,toma1GRBPOLAR,toma2GRBPOLAR}),
and it is likely that this will gradually become the main avenue
for constraining quantum-spacetime-induced birefringence.


\subsection{Testing modified dispersion relations in the lab}\label{inthelab}

Over this past decade
there has been growing awareness of the fact that data analyses
with good sensitivity to effects introduced genuinely at the Planck scale
are not impossible, as once thought. It is at this point rather well
known, even outside the quantum-gravity/quantum-spacetime
community, that Planck-scale sensitivity
is achieved in certain (however rare) astrophysics studies.
It would of course be very very valuable if we could establish
the availability of analogous tests in controlled laboratory setups, but this
is evidently more difficult, and opportunities are rare and of limited reach.
Still I feel it is important to keep this goal as a top priority, so in this
subsection I mention a couple of illustrative examples, which can
at least show that laboratory tests are possible.
Considering these objectives it makes sense to focus here again
on quantum-spacetime-motivated
Planck-scale modifications of the dispersion relation, so that the
estimates of sensitivity levels achievable in a controlled laboratory
setup can be compared to the corresponding studies
in astrophysics.

One possibility is to use
 laser-light interferometry
to look for in-vacuo-dispersion effects.
In Ref.~\cite{gaclaem} two examples of interferometric setups
were discussed in some detail, with the common feature of making
use of a frequency doubler, so that part of the beam would be
for a portion of its journey through the interferometer at double
the reference frequency of the laser beam feeding the interferometer.
The setups must of course be such that the interference pattern is
sensitive to the fact that, as a result of in-vacuo dispersion,
there is a nonlinear relation between the phase advancement of a beam
at frequency $\omega$ and a beam at frequency $2 \omega$.
For my purposes here it suffices to discuss briefly one such interferometric
setup. Specifically let me give a brief description of a setup in which the
frequency (or energy) is the parameter characterizing the splitting
of the photon state, so the splitting is in energy space (rather than
the more familiar splitting in configuration space, in which two parts
of the beam actually follow geometrically different paths).
The frequency doubling could be
accomplished using a ``second harmonic generator''~\cite{shgref}
so that if a wave reaches the frequency doubler with frequency $\omega$ then,
after passing through the frequency doubler, the outgoing wave
in general consists of two components, one at
 frequency $\omega$ and the other at frequency $2 \omega$.

If two such frequency doublers are placed along the path of the beam
at the end one has a beam with several components, two of which have
frequency $2 \omega$:  the
transmission of the component which left the first frequency doubler
as a $2 \omega$ wave, and another component which
is the result of frequency doubling of that part of the beam which went
through the first frequency doubler without change in the frequency.
Therefore, the final $2 \omega$ beam represents an
interferometer in energy space.

As shown in detail in Ref.~\cite{gaclaem} the intensity of
this $2 \omega$ beam takes a form of the type
\begin{equation}
I^{(2\omega)} = I_a + I_b \cos \left( \alpha + (k^\prime - 2 k) L \right) ~,
\end{equation}
where
 $L$ is the distance between the two frequency doublers,
 $I_a$ and $I_b$ are $L$-independent (they depend on the amplitude of the original
 wave and the effectiveness of the frequency doublers~\cite{gaclaem}),
 the phase $\alpha$ is also $L$-independent and is obtained combining
 several contributions to the phase (both a contribution
 from the propagation of the wave and a contribution
 introduced by the frequency doublers~\cite{gaclaem}),
 $k$ is the wave number corresponding to the frequency $\omega$ through the
 dispersion relation,
 and $k^\prime$ is the wave number corresponding to the frequency $2 \omega$ through the
dispersion relation (since the dispersion relation is Planck-scale modified one expects
departures from the special-relativistic result $k^\prime = 2k$).

Since the intensity only depends on the distance $L$ between the frequency doublers
through the Planck-scale correction
to the phase, $(k^\prime - 2 k) L$, by exploiting a setup that allows to vary $L$ one should
rather easily disentangle the Planck-scale effect. And one finds~\cite{gaclaem} that
the accuracy achievable with modern interferometers
is sufficient to achieve Planck-scale sensitivity (e.g., sensitivity to $|\eta|\sim 1$
in the PKV0 test theory with $n=1$).
It is of course rather optimistic to assume that the accuracy achieved in standard interferometers
would also be achievable with this peculiar setup, particularly since it would require
the optics aspects of the setup (such as lenses) to work with that high accuracy
simultaneously with two beams of different wavelength. Moreover, it would require
some very smart technique to vary the distance between the frequency doublers without
interfering with the effectiveness of the
optics aspects of the setup.
So in practice we would not be presently capable of using such setups
to set Planck-scale-sensitive limits on in-vacuo dispersion, but the fact that
the residual obstructions are of rather mundane technological nature encourages us to
think that in the not-so-distant future tests of Planck-scale
in-vacuo dispersion in controlled laboratory experiments will be possible.

Besides in-vacuo dispersion another aspect of the physics of Planck-scale
modified dispersion relations that we should soon be able to test in controlled
laboratory experiments is the one concerning anomalous thresholds, at least
in the case of the $\gamma \gamma \rightarrow e^+ e^-$ process which I already considered
from an astrophysics perspective in Section~\ref{pair-production}.
It is in fact not so far from our present technical capabilities
to set up collisions
between, say, 10~TeV photons and, say, 0.03~eV photons,
thereby reproducing essentially the situation of the analysis of blazars
that I discussed in Section~\ref{pair-production}.
And notice that with respect to the analysis of observations of blazars
such controlled laboratory studies would give much more powerful indications.
In particular, for the analysis  of observations
of blazars that I discussed in Section~\ref{pair-production} a key limitation
on our ability to translate the data into experimental bounds on parameters
of a pure-kinematics framework was due to the fact that (even assuming we are indeed
seeing absorption of multiTeV photons) the astrophysics context does not
allow us to firmly establish whether the absorption is indeed due to
the infrared component of the intergalactic background radiation (as expected) or instead
is due to a higher-energy component of the background (in which case the absorption would instead
be compatible with some corresponding Planck-scale pictures).
If collisions between 10~TeV photons and 0.03~eV photons in the lab
do produce pairs, since we would in that case have total control on the properties
of the particles in the in state of the process,
then we will have firm pure-kinematics bounds on the parameters of certain corresponding
Planck scale test theories (such as the PKV0 test theory).

These laboratory studies of Planck-scale-modified dispersion relations
could of course be adapted also to the FTV0 test theory, by simply
introducing some handles on the polarization of the photons
that are placed under observation
(also see Refs.~\cite{gharibyan2003,gharibyan2012}),
with sensitivity not far from Planck-scale sentivity in controlled laboratory experiments.


\subsection{On test theories without energy-dependent modifications of dispersion relations}

Readers for which this review is the first introduction to the world of quantum-spacetime
phenomenology might be surprised that
this long section,
with an ambitious title announcing related tests of Lorentz symmetry
was so heavily biased toward probing the form of the energy-momentum
dispersion relation.
Other aspects of the implications of Lorentz (and Poincar\'e) symmetry
did intervene, such as the law of energy-momentum conservation and its deformations
(and the form of the interaction vertices and their deformations), and are in part
probed through the data analyses that I reviewed, but the
feature that clearly is at center stage is the structure of
the dispersion relation.
The reason for this is rather simple: researchers that recognize themselves
as ``quantum-spacetime phenomenologists'' will consider a certain data analysis
as part of the field if that analysis concerns an effect
that can be robustly linked to quantum properties of spacetime
(rather than, for example, some classical-field background)
and if the analysis exposes  the
availability of Planck-scale sensitivities, in the sense I described above.
At least according to the results obtained so far,
the aspect of Lorentz/Poincar\'e symmetry which is most robustly
challenged by the idea of a quantum spacetime is the form
of the dispersion relation,
and this is also an aspect of Lorentz/Poincar\'e symmetry
for which the last decade of work on this phenomenology
robustly exposed opportunities for Planck-scale sensitivities.

For the type
of modifications of the dispersion relation that I considered in this section
we have at present rather robust evidence of their applicability in
certain noncommutative pictures of spacetime, where the noncommutativity is
very clearly introduced at the Planck scale. And
several independent (although all semi-heuristic) arguments suggest
that the same general type of modified dispersion relations
should apply to the ``Minkowski limit'' of Loop Quantum Gravity, a framework
where a certain type of discretization of spacetime structure is introduced
genuinely at the Planck scale.
Unfortunately, these two frameworks
are so complex that one does not manage to analyze
spacetime symmetries much beyond building a ``case'' (and not a waterproof case)
for modified dispersion relations.

Of course a broader range of
Lorentz-symmetry tests could be valuable for quantum-spacetime research,
but without the support of a derivation
 it is very hard to
argue that the relevant effects are being probed with sensitivities that
are significant from a quantum-spacetime/Planck-scale perspective.
Think for example of a framework,
such as the one adopted in Ref.~\cite{colgla}, in which the form
of the dispersion relation is modified, but not in energy-dependent way:
one still has dispersion relations
of the type $E^2 = c_\#^2 p^2 +  m_\#^2$, but with a different value of
the velocity scale $c_\#$ for different particles.
This is not necessarily a picture beyond the realm of possibilities one could
consider from a quantum-spacetime perspective, but
there is no known quantum-spacetime picture that has provided direct
support for it. And
 it is also essentially impossible to
estimate what accuracy must be achieved in measurements
of, say, $c_{\mathrm{proton}}-c_{\mathrm{electron}}$, in order to reach Planck-scale
sensitivity. Some authors  qualify
as ``Planckian magnitude'' of this type of effects
the case in which the dimensionless parameter
has value of the order of the ratio of the mass of the particles involved
in the process versus the Planck scale
(as
in  $c_{\mathrm{proton}}-c_{\mathrm{electron}} \sim (m_{\mathrm{proton}} \pm
m_{\mathrm{electron}})/E_p$)
but this arbitrary criterion  clearly does not amount to establishing
genuine Planck-scale sensitivity, at least as long as we do not have
a derivation
starting with spacetime quantization at the Planck scale that actually finds
such magnitudes of this sort of effects.

Still it is true that the general structure of the quantum-gravity problem
and the structure of some of the quantum spacetimes
which are being considered for
the Minkowski limit of quantum gravity might host
a rather wide range of departures
from classical Lorentz symmetry. Correspondingly a broad range
of Lorentz-symmetry tests could be considered of potential interest.

I shall not review here this broader Lorentz-symmetry-tests literature, since it is not
specific to quantum-spacetime research
(these are tests that could be done and in large part
were done even before the development of research on Lorentz
symmetries from within
the quantum-spacetime literature)
and it has already been reviewed very effectively
in Ref.~\cite{mattinLRR}.
Let me just stress that for these broad searches of departures from Lorentz
symmetry one of course needs test theories with many parameters. Formalisms
that are well suited for a systematic programme of such searches are already
at a rather advanced stage of
development~\cite{repCOLLAkoste1,repCOLLAkoste2,sme1,sme2,blumnew,laemSME1,laemSME2}
(also see Ref.~\cite{toulouse}),
and in particular the ``standard-model-extension'' framework~\cite{repCOLLAkoste1,repCOLLAkoste2,sme1,sme2}
has reached a high level of adoption
of preference for theorists and experimentalists as the language in which
to characterize the results of systematic multi-parameter Lorentz-symmetry-test
data analyses.
The ``Standard Model Extension''
was originally conceived~\cite{sme1} as a generalization of the Standard Model of particle-physics interactions
restricted to power-counting-renormalizable correction terms, and as such it was
of limited interest for the bulk of the quantum-spacetime/quantum-gravity community:
since  quantum gravity is not a (perturbatively) renormalizable theory
many quantum-spacetime researchers
would be unimpressed with Lorentz-symmetry tests
restricted to powercounting-renormalizable correction terms.
However, over these last few years~\cite{blumnew}
 most theorists involved in studies of the ``Standard Model Extension''
have started to add correction terms that are not powercounting
renormalizable.\footnote{A warning to readers: whereas originally the
denomination ``Standard Model Extension'' was universally used
to describe a framework implementing the restriction
to powercounting-renormalizable correction terms,
recently (see, e.g., Ref.~\cite{blumnew}) some theorists
describe as ``Standard Model Extension'' the generalization
that includes correction terms that are not powercounting
renormalizable, while they describe as a ``Minimal Standard Model Extension''
the case with the original restriction to powercounting-renormalizable correction terms.
Still, even as I write this review, many authors
(in particular the near totality of experimentalists involved in such studies)
continue to adopt the original
description of the ``Standard Model Extension'',
restricted to powercounting-renormalizable correction terms,
and this may create
some confusion (for example experimentalists reporting results on
the ``Standard Model Extension'' are actually, according to the terminology
now used by some theorists, describing experimental limits on
the ``Minimal Standard Model Extension'').}
 A good entry point for the literature on limits on the parameters
of the ``Standard Model Extension''
is provided by Refs.~\cite{mattinLRR,blumnew,kosteDATATABLES}.

From a quantum-gravity-phenomenology perspective it is useful to contemplate
the differences between alternative strategies
for setting up a ``completely general'' systematic investigation
of possible violations of Lorentz symmetry.
In particular,
it has been stressed (see, e.g., Refs.~\cite{laemSME1,laemSME2})
that violations of Lorentz symmetry
can be introduced directly
at the level of the dynamical equations,
without assuming (as done in the Standard Model Extension)
the availability
of a Lagrangian generating the dynamical equations.
This is of course more general than the Lagrangian approach: for example,
the generalized Maxwell equation discussed in Ref.~\cite{laemSME1,laemSME2}
predicts effects that go beyond the Standard Model Extension.
And charge conservation, which automatically comes out from the Lagrangian
approach, can be violated in models generalizing the field
equations~\cite{laemSME1,laemSME2}.
The comparison of the
Standard-Model-Extension approach and of the approach based on generalizations introduced
directly at the level of the dynamical equations illustrates how different ``philosophies''
lead to different strategies for setting up a ``completely general'' systematic investigation
of possible departures from Lorentz symmetry.
By removing the assumption of the availability of a Lagrangian the second approach
is ``more general''. Still no ``general approach'' can be absolutely general: in
principle one could always consider removing an extra layer of assumptions.
Of course, as the topics I have reviewed in this section illustrate,
from a quantum-spacetime-phenomenology perspective it is not necessarily appropriate to
seek the most general parametrizations. On the contrary we would like to single out some
particularly promising candidate quantum-spacetime effects (as in the case of modified
dispersion relations) and focus our efforts accordingly.

\newpage


\section{Other Areas of UV Quantum-Spacetime
  Phenomenology}
\label{other-areas}

Tests of Lorentz symmetry, and particularly of the form of the dispersion relation,
probably make up
something of the order of the half of the whole
quantum-spacetime-phenomenology literature.
The other half is spread over a few other, evidently less developed,
 research lines.
Nonetheless for some of these other research lines
the literature has reached some nonnegligible maturity, and even those
which are at preliminary stages of development
could be precious potential opportunities for
quantum-spacetime research.

Evidently the most challenging part of this review work concerns
these other components of quantum-spacetime-phenomenology
research, since it is harder to summarize and organize intelligibly
the results and scopes of research programs which are still
in early stages of development.
But it is also the part of this review that could be most valuable, since
there are already some
works~\cite{polonpap,tedAP,subirMDRreview,hosseLee}
attempting to summarize and review, although more concisely than done
here in
the previous section,  the results obtained
by the quantum-spacetime-phenomenology of Planck-scale modified
dispersion relations.

In reporting on the work done in these other quantum-spacetime-phenomenology
reseach lines I shall use as one of the guiding
concepts the one of assessing whether a given research programme
concerns UV (ultraviolet) quantum-spacetime effects or IR (infrared)
quantum-spacetime effects.
The typical situation for a UV quantum-spacetime effect is
that it takes the form of correction terms that grow with
the energy of the particles, and whose significance is therefore
increasingly high as the energy of the particles increases.
For any given standard-physics (no-quantum-spacetime) prediction $A_0$
this will take the general form
\begin{equation}
A_0 \longrightarrow A_0 \left( 1 + \eta_\# \frac{E^n}{E_p^n} \right) \qquad\qquad
(\mathrm{with~context/theory~specific~numerical~factor}\, \eta_\# )
\label{contextspecific}
\end{equation}

This is the type of quantum-spacetime effects that traditionally
one expects to be inevitably produced by any form of spacetime
quantization, and is the focus of this section.
The possibility of ``IR quantum-spacetime effects'', effects
that are due to Planck-scale spacetime quantization but
somehow are significant in some deep-infrared regime,
came to the attention of the community only rather recently,
emerging mainly from work
on ``IR/UV mixing in quantum spacetime'', and I shall focus on
it in the next section.

\subsection{Preliminary remarks on fuzziness}

In this review, as natural for phenomenology,
 I am primarily looking at quantum-spacetime effects from the perspective
of the type of pre-quantum-spacetime laws that they affect (so we have ``departures from
classical spacetime symmetries'', ``violations of the quantum-mechanical coherence'',
and so on).
And our experimental opportunities are such that the main focus
is on how spacetime quantization could affect particle propagation
(and, for a restricted sample of phenomenological opportunities, interactions
among particles).
For this section on ``other UV quantum-spacetime effects''
a significant role (noticeable, more or less explicitly, in several subsections)
will be played by the idea that quantum-spacetime effects may introduce
an additional irreducible contribution to the fuzziness of the worldlines of particles.

This should be contrasted to the content of the previous
Section~\ref{symmetry-tests} ,
which focused mainly on phenomenological proposals
involving mechanisms for systematic
departures from the currently-adopted laws of propagation
of (and interaction among) particles.
In most cases such systematic effects
amount to departures from the
predictions of Lorentz symmetry
(such as a systematic dependence
of the velocity of a massless particle on its energy,
which would produce a systematic
difference between the arrival times of high-energy and low-energy
photons that are simultaneously emitted).

If it ends up being the case that the correct quantum-spacetime picture does
not provide us any such systematic effects, then we will be left with
non-systematic effects, i.e. ``fuzziness''~\cite{iucaapap}.
In looking for such effects we can be guided by the intuition that
spacetime quantization might act as an environment inducing
apparently random fluctuations in certain observables. For example, by
distance fuzziness one does not describe an effect that would
systematically gives rise to larger (or smaller) distance-measurement
results, but rather one describes a sort of new uncertainty principle
for distance measurements.

This distinction between systematic and nonsystematic effects can be
easily characterized
for any given
observable ${\hat X}$
for which the pre-quantum-spacetime theoretical prediction can be
described in terms of a ``prediction'' $X$ and, possibly, a fundamental
(ordinarily quantum mechanical) ``uncertainty'' $\delta X$. The effects of
spacetime quantization
in general could lead~\cite{iucaapap} to a new
prediction $X'$ and a new uncertainty $\delta X'$.
One would attribute to quantum spacetime
the effects
\begin{equation}
(\Delta X)_{\mathrm{QG}} \equiv X' - X
\label{deltasyste}
\end{equation}
and
\begin{equation}
(\delta X)_{\mathrm{QG}} \equiv \delta X' - \delta X~.
\label{deltanonsyste}
\end{equation}
One can speak of purely systematic
quantum-spacetime effect when $(\Delta X)_{\mathrm{QG}} \neq 0$
and $(\delta X)_{\mathrm{QG}} = 0$,
while the opposite case, $(\Delta X)_{\mathrm{QG}} = 0$ and $(\delta X)_{\mathrm{QG}} > 0$,
can be qualified as
purely non-systematic. It is perhaps likely that for many observables
 both types of quantum-spacetime effect be present
simultaneously, but of course it is natural that at least the first stages
of development of a quantum-spacetime phenomenology on an observable  ${\hat X}$
be focused  on one or the other special case ($(\delta X)_{\mathrm{QG}} = 0$ or $(\Delta X)_{\mathrm{QG}} = 0$).
Clearly the discusions of effects given in Section~\ref{symmetry-tests} were all
with $(\delta X)_{\mathrm{QG}} = 0$, while for most of the proposals
discussed in this section the main focus
will be on the effects characterized by $(\Delta X)_{\mathrm{QG}} = 0$.


\subsection{Spacetime foam, distance fuzziness and interferometric
  noise}
\label{foam-fuzziness-noise}
The scenarios for spacetime fuzziness that are most studied from
a quantum-spacetime perspective are intuitively linked to the notion
of ``spacetime foam'', championed by Wheeler and studied extensively
in the quantum-gravity literature, more or less directly, for several decades
(see, e.g., Refs.~\cite{wheelerFORSEFOAM,deserFOAM1957,colemanFOAM,
hawkingFOAM,carlipFOAM,garayFOAM,woodardfoam}).
From a modern perspective here one is attempting to characterize the physics
of matter particles as effectively occurring in an ``environment'' of short-distance quantum-gravitational degrees of freedom. And one may expect that for propagating particles with wavelength much larger than the Planck length, when it may be appropriate to integrate out these short-distance
quantum-gravitational degrees of freedom, the main residual effect of short-distance
gravity would indeed be an additional contribution to the fuzziness of worldlines.

While in full-fledged quantum-spacetime theories,
such a Loop Quantum Gravity, such analyses are still beyond our reach,
one can find partial encouragement for this intuition in recent
progress for the understanding of
 quantum gravity in 3D (2+1-dimensional) spacetime.
 Studies such as the ones reported in
 Refs.~\cite{kodadsr,jurekkodadsr,meusSchr3DQG,carlip3DQGlrr,oritigirelli3DQGhopf,freidellivine3DQGhopf,noui3DQGhopf,schr3dQG2012}
 establish that for 3D quantum gravity (exploiting the much lower complexity than for the 4D case) we are able to perform the task needed for studies of spacetime foam: we can actually integrate out gravity, reabsorbing its effects into novel properties for a gravity-free propagation of particles. And foaminess is formalized in the fact that this procedure integrating out gravity
 leaves us with a theory of free particles in a noncommutative spacetime,
 Refs.~\cite{kodadsr,jurekkodadsr,oritigirelli3DQGhopf,freidellivine3DQGhopf},
 specifically a spacetime with ``Lie-algebra noncommutativity''
\begin{equation}
[x_\alpha,x_\beta] = i \kappa^\gamma_{\alpha \beta} x_\gamma ~.
\label{liealgNCST}
\end{equation}
(in particular the choice of $\kappa^\gamma_{\alpha \beta}$ as the Levi-Civita tensor
is the one suggested by the direct derivation given in Ref.~ \cite{schr3dQG2012}).
In other words, upon integrating out the gravitational degrees of freedom, the quantum dynamics of matter fields coupled to 3D gravity is effectively
described~\cite{freidellivine3DQGhopf},
by matter fields in a noncommutative spacetime, a fuzzy/foamy spacetime.

While the only direct/deductive derivations of such results are for 3D quantum gravity,
it is natural to take that as a starting point for the study of real 4D quantum gravity,
whereas analogous results are still unavailable.
And a sizable literature has been devoted to the search of
possible experimental manifestations of ``spacetime foam''.
Several subsections of this section concern related phenomenological proposals.
I start in this subsection with spacetime-foam test
 theories whose structure renders them well suited for interferometric tests.

\subsubsection{Spacetime foam as interferometric noise}
The first challenge for a phenomenology investigating the possibility of spacetime
foam originates in the fact that Wheeler's spacetime foam intuition,
 while carrying strong conceptual appeal, cannot on its own be used for phenomenology,
 since it is not characterized in terms of observable properties.
 The phenomenology then is based on test theories inspired by the spacetime-foam
 intuition.

A physical/operative
definition of at least one aspect of spacetime foam
is given in Refs.~\cite{gacgwi,gacgwiB,bignapap,nggwi} and is well suited for
a phenomenology based on interferometry\footnote{Interestingly,
this simple scheme for modeling spacetime-foam effects
also provides the basis for the proposal put forward in
Refs.~\cite{ahluwaliaFOAMandASYMMETRY,ahluTALKwithMYNAPA1}
of a mechanism that could be responsible
for the cosmological matter-antimatter asymmetry.}. According to this
definition the fuzziness/foaminess of a spacetime is
established\cite{gacgwi,gacgwiB,bignapap,nggwi}
on the basis of an analysis of strain noise\footnote{Since modern interferometers were
planned to look for classical gravity waves (gravity waves are
their sought ``signal''), it is reasonable to denominate
as ``noise'' all test-mass-distance fluctuations that are not due to
gravity waves. I adopt this terminology which reflects
the original objectives of modern interferometers, even though
this terminology is somewhat awkward for the type of
quantum-spacetime-phenomenology studies I am
discussing, in which interferometers would be used for searches of
quantum-gravity-induced distance fluctuations (and therefore in
these studies quantum-gravity-induced distance fluctuations would
play the role of ``signal'').}
in interferometers set up in that spacetime.
In achieving their remarkable
accuracy modern interferometers must deal with several
classical-physics strain noise sources (e.g., thermal and
seismic effects induce fluctuations in the relative positions of
the test masses). And importantly strain noise sources associated
with effects due to ordinary quantum mechanics are also significant
for modern interferometers (the combined minimization of
\textit{photon shot noise} and \textit{radiation pressure noise} leads
to a noise source which originates from ordinary quantum
mechanics~\cite{saulson}). One can give
an operative definition~\cite{gacgwi,bignapap} of fuzzy/foamy
spacetime in terms of a corresponding
 additional source of strain noise. A
theory in which the concept of distance is fundamentally fuzzy in
this operative sense would be such that the read-out of an
interferometer would still be noisy (because of quantum-spacetime
effects) even in the idealized limit in which all
classical-physics and ordinary-quantum-mechanics noise sources are
completely eliminated/subtracted.

\subsubsection{A crude estimate for laser-light interferometers}
Before even facing the task of developing test theories for spacetime foaminess
in interferometry it is best to first check whether there is any
chance of using realistic interferometric setups to uncover effects
as small as expected if introduced at the Planck scale.
A first encouraging indication comes from identifying the presence
of a huge amplifier in modern interferometers:
a well-known quality of these modern
interferometers
is their ability to detect gravity waves
of amplitude $\sim 10^{-18}$~m
by carefully monitoring distances of order $\sim 10^4$~m,
and this should provide opportunities for
 an ``amplifier'' which is of order $10^{22}$.

This also means that our modern interferometers have outstanding control
over noise sources, which is ideal for the task here at hand, involving
scenarios for how quantum-spacetime effects may contribute an
additional source of noise in such interferometers.
Clearly the noise we could conceivably
see emerging from spacetime quantization
should be modeled in terms of some random vibrations.
Evidently random vibrations are particularly difficult to characterize.
For example there is in general no spendable notion of ``amplitude'' of random vibrations.
The most fruitful way to characterize them, also for the purposes of comparing
their ``intensity'' to other non-random sources of vibration which might affect
the same system, is by using the power spectral density.
Let me then introduce some notation, which will also prove useful when
I move on to discuss crude models of quantum-spacetime-induced noise.
For this I simple-mindedly consider the readout of an interferometer
as $h(t)$, given by the position $x(t)$ of a mirror divided by a reference
length scale $L$ ($h(t) = x(t)/L$), and adjust the reference frame so that
on average $x(t)$ vanishes, $\mu_x =0$.
Given some rules for fluctuations of this  readout
one can indeed be interested in its power spectral density $\Sigma(\omega)$,
in principle computable via~\cite{saulson}
\begin{equation}
\Sigma(\omega) =  \int_{-\infty}^{\infty} \, d\tau \, \mu_{[h(t)h(t+\tau)]} \, e^{-2\pi i \omega \tau}
\label{psd}
\end{equation}
where $\mu_{[h(t)h(t+\tau)]}$ depends only on $\tau$ and is the
value expected
on average for $h(t)h(t+\tau)$ in presence of the vibration/fluctuation process
of interest in the analysis.

Having characterized the noise source in terms of its
power spectral density we can then easily get out some primary characteristics,
such as its root mean square deviation $\sigma_h$, which for cases
of zero-mean noise, such as the one I am considering, will be given by the expectation
of $h^2$. This can be expressed in terms of the power spectral density
as follows~\cite{saulson}
\begin{equation}
\sigma^2_h = \mu_{h^2} = \int_{-\infty}^{\infty} \, d\omega \, \Sigma (\omega)~.
\label{sigmaFROMrhoIDEAL}
\end{equation}
In experimental practice, for a frequency-band limited signal ($f_{max}$)
and a finite observation time ($T_{obs}$), this relation will take the shape
\begin{equation}
\sigma^2_h \simeq  \int_{1/T_{obs}}^{f_{max}} \, d\omega \, \Sigma (\omega) ~.
\label{sigmaFROMrho}
\end{equation}

In modern interferometers such as LIGO~\cite{ligoSCIENCE,ligoREVIEW2009}
and VIRGO~\cite{virgo1997,virgo2008}
the power spectral density of the noise
is controlled at a level of  $\Sigma (\omega) \sim 10^{-44}\mathrm{\ Hz}^{-1}$
at observation frequencies $\omega$ of
about $100~Hz$,
and in turn this (also considering the length of the arms of these modern
interferometers) implies~\cite{ligoSCIENCE,ligoREVIEW2009,virgo1997,virgo2008}
that for a gravity wave
with $100~Hz$ frequency the detection threshold is indeed
 around $ 10^{-18}$~m.

The challenge here for quantum-spacetime phenomenologists is to
characterize the relevant quantum-spacetime effects
 in terms of a novel contribution $\Sigma^{[QG]}(\omega)$
to the power spectral density of the noise.
If at some point experimentalists will manage to bring the total noise $\Sigma(\omega)$,
for some range of observation frequencies $\omega$, below the level  predicted by
a certain quantum-spacetime test theory then that test theory will be ruled out.

Is there any hope for a reasonable quantum-spacetime
test theory to predict noise
at a level comparable to the ones that are within the reach of modern
interferometry? Well, this is the type of question that one can only
properly address in the context of  models,
but it may be valuable to first use dimensional
analysis, assuming the most optimistic behavior of the quantum-spacetime effects,
and check if the relevant order of magnitude is at all providing any
encouragement for the painful (if at all doable) analysis
of the relevant issues in quantum-spacetime models.

To get what is likely to be the most optimistic (and certainly the simplest,
but not necessarily the most realistic)
Planck-scale estimate of the effect let us assume that quantum-spacetime noise
is ``white noise'', $\Sigma^{[QG]}(\omega) =\Sigma_0$ (frequency independent), so that
it is fully specified
by a single dimensionful number setting the level of this white noise.
And since $\Sigma$
carries units of $\mathrm{\ Hz}^{-1}$ one easily notices~\cite{gacgwiB} a
tempting simple {\it naive}
estimate in terms of the Planck length and the
speed-of-light\footnote{While most formulas
in this review adopt $\hbar=c=1$ conventions,
I do make some exceptions (with explicit $\hbar$ or $c$)
when I believe it can help to characterize the conceptual ingredients of the formula.}
scale: $\Sigma_0 \sim L_p/c$, which, since $L_p/c \sim 10^{-44}\mathrm{\ Hz}^{-1}$,
encouragingly happens to be just at the
mentioned level of sensitivity of LIGO-VIRGO-type
interferometers.
This provides some initial encouragement for a phenomenology based on
interferometric noise, though of course only within the limitations of
a very crude and naive estimate.

\subsubsection{A simple-minded mechanism for noise in laser-light
  interferometers}
My next task is going beyond assuming for simplicity that the quantum-spacetime
noise be white and beyond adopting
a naive dimensional-analysis estimate of what could constitute a Planck-scale
level of such a noise.
The ultimate objective here would be
to analyze an
interferometer in the framework of a compelling quantum-spacetime
theory, but this is beyond
our capabilities at present. We can however start things off by
identifying some semi-heuristic pictures
(the basis for a test theory) with
effects introduced genuinely at the Planck scale that turn out to produce strain
noise at the level accessible with modern interferometers.

Having in mind this objective let us take as starting point for a first naive
picture of spacetime fuzziness the popular arguments
suggesting that the Planck scale should also set some absolute
limitation on the measurability of distances.
And let us (optimistically) assume that this translates in
the fact that any experiment in which a distance $L$ plays
a key role (meaning that one is either measuring $L$ itself
or the observable quantity under study depends strongly on $L$)
is affected by a mean square deviation $\sigma_L^2$.

It turns out to be useful~\cite{gacgwi,bignapap}
to consider this $\sigma_L^2$ as a possible stepping stone toward the strain-noise
power-spectrum estimate.
And a particulary striking picture arises by assuming
that the
distances $L$ between the test masses of an interferometer be
affected by Planck-length
fluctuations of random-walk type
occurring at a rate of one per Planck time ($\sim 10^{-44}$~s),
so that~\cite{gacgwi,bignapap}
\begin{equation}
\sigma_L^2 \simeq L_p T  \simeq L_p L
~~~~~~~~~~~~~~~~~~~~~~
[\textrm{random~walk~case}]
\label{randomwalkie}
\end{equation}
where $T$ is the time
scale over which the experiment monitors the distance $L$,
and I am assuming the use of ultrarelativistic particles ($T \simeq L$).

It is perhaps noteworthy that  $\sigma_L^2 \simeq L_p T$ can be motivated
independently (without having in mind the idea of such effective spacetime fluctuations)
on the basis of some aspects of the quantum-gravity problem~\cite{gacmpla}.
And the study of certain quantum-spacetime pictures which have been of interest
to the quantum-gravity community, such as the $\kappa$-Minkowski noncommutative
spacetime of Eq.~(\ref{kappaminkB}), provide some support for this
random-walk picture: from $[x_j,t]=i L_P x_j$ one could guess
roughly a law of the form $\sigma_x^2 \sim \delta x \, \delta t \sim L_P x$.

Some arguments inspired by the ``holography paradigm''
for quantum gravity~\cite{nggwi,ngSelectedTopics,ngFOAM2011}
suggest even weaker effects, characterized by
\begin{equation}
\sigma_L^2 \simeq L_p^{4/3} L^{2/3}
~~~~~~~~~~~~~~~~~~~~~~
[\textrm{holography-inspired~case}]
\label{holofuzzy}
\end{equation}
And interestingly also this {\it ansatz}  $\sigma_L^2 \simeq L_p^{4/3} L^{2/3}$ had
been independently proposed in the quantum-gravity literature on the basis of
a perspective on the quantum-gravity problem
(see Ref.~\cite{ng1994,karo,diosi}), which originally involved in no way spacetime fuzziness.

Probably the most conservative (and pessimistic) expectation
for spacetime fuzziness one can find in the quantum-spacetime literature
is the one omitting any opportunity for amplification by the involvement
of a long observation time
(see, e.g., parts of Refs.~\cite{garay,sabineMINLENreview})
\begin{equation}
\sigma_L^2 \simeq L_p^{2}
~~~~~~~~~~~~~~~~~~~~~~[\textrm{weakest~case}]
\label{weakfuzzy}
\end{equation}

The random-walk case is  the most typical textbook study case for
random noise. Its power spectral density goes like $\omega^{-2}$,
so one should have
\begin{equation}
\Sigma_L^{[QG;rw]} \simeq  \frac{L_P}{\omega^2} ~
\label{rhorw}
\end{equation}
which gives
\begin{equation}
\sigma^2_h \sim \frac{\sigma^2_L}{L^2}
 \simeq \frac{1}{L^2} \int_{1/T_{obs}}^{f_{max}} \, d\omega \, \frac{L_P}{\omega^2}
\simeq \frac{L_P T_{obs}}{L^2} ~
\label{sigmarw}
\end{equation}
(so, for $L \sim T_{obs}$ one indeed finds  $\sigma^2_L \sim L_P L$).

Analogously one can associate
to the ``holographic noise'' of Eq.~(\ref{holofuzzy})
a power spectral density going like $\omega^{-5/3}$,
so one should have
\begin{equation}
\Sigma_L^{[QG;holo]} \simeq  \frac{L_P^{4/3}}{\omega^{5/3}} ~
\label{rhoholo}
\end{equation}
which indeed gives
$\sigma^2_L \simeq L_p^{4/3} T_{obs}^{2/3}    \simeq L_p^{4/3} L^{2/3}   $.

And finally for the $\sigma_L^2 \simeq L_p^{2}$
case of Eq.~(\ref{weakfuzzy})
a rough but valuable approximate description of
the power spectral density goes like $\omega^{-1}$,
so one should have
\begin{equation}
\Sigma_L^{[QG;weak]} \simeq  \frac{L_P^{2}}{\omega} ~
\label{rhoweak}
\end{equation}
which indeed gives
$\sigma^2_L \simeq L_p^{2}$.

It is then tempting
to obtain from these estimates
of the quantum-spacetime-induced distance uncertainty
an estimate for the quantum-spacetime-induced strain noise,
by simply dividing by the square of the length of the arms of
the interferometer, $\Sigma^{[QG]}=\Sigma_L^{[QG]}/L^2$.
This would be the way to proceed if we were converting distance noise
into strain noise, but really here we are obtaining a rough estimate of strain
noise from an estimate of distance uncertainty, and I shall therefore
proceed in some sense {\it sub judice}
(see in particular my comments below concerning the large number of photons
collectively used for producing the accurate measurements of a modern interferometer).
Assuming that indeed $\Sigma^{[QG]}=\Sigma_L^{[QG]}/L^2$,
and taking as reference value an observation frequency of $\omega \sim 100$~Hz,
one would get for the three cases I discussed the following estimates of strain noise
at $100$~Hz, for arm lengths of a few kilometers:
\begin{equation}
\Sigma^{[QG;weak]} \sim  10^{-78}Hz^{-1}    ~,~~~
\Sigma^{[QG;holo]} \sim  10^{-52}Hz^{-1}    ~,~~~
\Sigma^{[QG;rw]} \sim  10^{-38}Hz^{-1}  ~.
\label{rhoestimates}
\end{equation}
These estimates are rather naive but it is nonetheless interesting to compare
them to the levels of noise control achieved experimentally.
As mentioned, around 100 Hz both LIGO and VIRGO
achieve noise control at the
level of strain noise of $\Sigma \sim  10^{-44}Hz^{-1}$,
so estimates like
$\Sigma^{[QG;weak]}$ and $\Sigma^{[QG;holo]}$
would be safe, but the estimate $\Sigma^{[QG;rw]}$
must be excluded: the estimate $\Sigma^{[QG;rw]}$ would assign
more noise of quantum-spacetime
origin than the total noise that LIGO and VIRGO managed to control
(which of course would include the hypothetical quantum-spacetime-induced noise).
In spite of the crudeness of the derivations I discussed so far, this does give
a rather worthy input for those who fancy the random-walk picture,
as I shall stress in the next subsubsection.

Before I get to that issue let me stress
that there is a possible source of confusion for terminology (and content)
in the literature.
In the quantum-gravity literature
there has been some discussion for several years
  of ``holography-inspired noise''
in the sense of my Eq.~(\ref{rhoholo})
and of Refs.~\cite{nggwi,ngSelectedTopics,ngFOAM2011}.
More recently a different mechanism for quantum-spacetime-induced noise,
also labeled as ``holography inspired'',
was proposed in a series of papers by Hogan~\cite{hoganNOISE,hoganGEO600,hogan2012}.
There is no relation between the two ``holography-inspired'' proposals
for quantum-spacetime-induced interferometric noise.
I do not think it is particularly important at the present
time to establish which one (if any) of the two proposals
is more directly inspired by holography.
I must instead stress that the holographic noise
of Refs.~\cite{nggwi,ngSelectedTopics,ngFOAM2011}
is a rather mature proposal, centered on  Eq.~(\ref{rhoholo}) and meaningful
at least as a quantum-spacetime test theory in the sense I just described.
Instead it is probably fair to describe the alternative version of holographic noise
more recently proposed in Ref.~\cite{hoganNOISE,hoganGEO600,hogan2012}
as a young proposal still looking for some maturity: it does not amount
to any however wild variant of the description of interferometric noise
I summarized here above, and actually it is claimed~\cite{hogan2012}
to be immune not only to the sort of interferometric noise I discussed in this subsection
 but also
to all other effects that have been typically associated with spacetime
quantization in the literature.
It would be a quantum-spacetime picture whose effects ``can only be detected in an experiment that coherently compares transverse positions
over an extended spacetime volume to extremely high precision,
and with high time resolution or bandwidth''~\cite{hogan2012} .
Evidently some work is still needed on
the conceptual aspects (as a rigorous theory of spacetime quantization)
and on the phenomenological aspects (as a computably predictive and broadly applicable
test theory of spacetime quantization)
of this proposal. Only time
will tell if this present lack of maturity is due to intrinsic unsurmountable
limitations of the proposal or is simply
a result of the fact that the proposal was made only rather recently
(so there was not much time for this maturity to be reached).
I should note that at some point,
in spite of its lack of maturity, this proposal
started to attract some pronounced interest
in relation to reports by the GEO600 interferometer~\cite{geo600general}
of unexplained excess noise~\cite{geo600noise}:
it had been claimed~\cite{hoganGEO600} that Hogan's version
of holographic noise could match exactly the anomaly that was
being reported by GEO600.
It appears however that experimenters at GEO600 have achieved
recently a better understanding of their noise sources,
and no unexplained contribution is at this point reported
(this is at least implicit in Ref.~\cite{geo600nonoise} and
is highlighted at www.aei.mpg.de/hannover-en/05-research/GEO600).
The brief season
of the ``GEO600 anomaly'' (at some point known among specialists
as the ``mystery noise'') is over.

\subsubsection{Insight already gained and ways to go beyond it}
At the present time the ``state of the art'' of
phenomenologically-spendable descriptions of Planck-scale-induced
strain noise does not go much beyond the simple-minded estimates I just described
in relation to equations (\ref{rhorw}),(\ref{rhoholo}),(\ref{rhoweak}).
But some lessons were nonetheless learned, as it usually happens even with
the most humble phenomenology. And these lessons do point toward some
directions worthy of exploration in the future.
In this subsubsection I highlight some of these lessons and possible
future developments.

Among the few steps of simple derivation which I described in the previous
subsubsection
evidently much scrutiny should be particularly directed toward the
assumption $\Sigma^{[QG]}=\Sigma_L^{[QG]}/L^2$:
I got to motivate some candidate forms for a  $\Sigma_L^{[QG]}$
using essentially the sort of arguments that usually allow us to establish
uncertainty principles for single particles, such as the ones taking as starting
point a postulated noncommutativity of single-particle coordinates;
however, the strain noise  $\Sigma^{[QG]}$ relevant for
our interferometers is not at all a single-particle feature.
Let me use the example of random-walk fuzziness for illustrating
how the relationship between single-particle quantum-spacetime arguments
and interferometric strain noise could be more subtle than assumed
in $\Sigma^{[QG]}=\Sigma_L^{[QG]}/L^2$.
For this I shall follow
 Ref.~\cite{iucaapap}
(a similar thesis was also reported in Ref.~\cite{ngFOAM2011}).
I specialize the more general idea of random-walk quantum-spacetime fuzziness
in the sense of assuming that each single photon in an interferometer
experiences a random-walk path:
 a random Planck-length fluctuation per Planck-time would
affect the path of each photon of the beam.
This would imply in particular that as a photon goes from one mirror of the interferometer
to the other, over a distance $L$,
it reaches its destination with an uncertainty corresponding
to  $\sigma_L^2 \sim L_P T \sim L_P L$.
However, the interferometer (and this is key to its outstanding sensitivity)
does not depend on determining the position of each single photon
in the beam: on the contrary the key observable is the {\emph{average}}
position of the photons composing the beam, which may be viewed as the
putative ``position of the mirror'' (when such a beam reaches the mirror).
If $L$ now is viewed as the distance between  positions of mirrors defined
in this way,
rather than as the distance of propagation of an individual photon,
then evidently the result is an estimate $\sigma_L^2 \sim L_P T \sim L_P L/N_\gamma$,
where $N_\gamma$ is the (very large!) number of photons contributing
to each such determination of the ``position of the mirror''.

While the noise levels produced by a random-walk {\it ansatz}
assuming $\Sigma^{[QG]}=\Sigma_L^{[QG]}/L^2$ are, as stressed
in the previous subsubsection, already ruled out by the achievements of LIGO and VIRGO,
this single-particle picture of a random-walk scenario,
which evidently leads us to assume
\begin{equation}
\Sigma^{[QG]}=\frac{\Sigma_L^{[QG]}}{N_\gamma L^2}
\label{ngammarescaling}
\end{equation}
 is still safely compatible with the noise results of
 LIGO and VIRGO, thanks to the large $N_\gamma$ suppression.

 This observation of course is not specific to the random-walk scenario.
 A similar  $N_\gamma$ suppression could be naturally expected
 for the holographic noise scenario of Eq.~(\ref{rhoholo}).
 As discussed in the previous subsubsection that
 holographic noise scenario would be safe from LIGO/VIRGO bounds
 even without the $N_\gamma$ suppression. (In some sense
 that holographic noise scenario would
 turn into unpleasantly ``too safe from LIGO/VIRGO'', i.e.
 probably beyond
 the reach of any foreseeable interferometric experiment, if it were
 to take into account the plausible $N_\gamma$ suppression).

 Concerning the scenario for weak quantum-spacetime-induced fuzziness,
 the one of Eq.~(\ref{rhoweak}), contemplating the possibility of
 an $N_\gamma$ suppression is of mere academic interest:
 those noise levels are so low, even without the possible
 additional  $N_\gamma$ suppression, that we should exclude their
 testability for the foreseeable future.

 But for random-walk noise and  for the holographic-noise
 scenario of Eq.~(\ref{rhoholo}) this issue of a
 possible $N_\gamma$ suppression needs to be investigated
 and understood. This is probably not for the LIGO/VIRGO season:
 LIGO and VIRGO have not found any excess noise so far, and at
 this point it is unlikely they will ever find it.
 But a completely new drawing board for phenomenology would materialize
 with the advent of LISA~\cite{lisa}: LISA will operate at lower
 observational frequencies $\omega$ than LIGO/VIRGO-type interferometers,
 which is important form the quantum-spacetime perspective
 since both
 random-walk noise, as described by Eq.~(\ref{rhorw}),
 and the holographic-noise
 scenario of Eq.~(\ref{rhoholo}) predict effects that increase
 at lower observational frequencies\footnote{The fact that,
 according to some of these test theories,
 quantum-spacetime-induced noise
becomes increasingly significant as the characteristic frequency
of observation is lowered,
also opens the way to possible studies~\cite{laemgwi}
using cryogenic resonators,
which are
rigid optical interferometers with good sensitivities down to frequencies
of about $10^{-6}$~Hz.}. The outlook of such LISA quantum-spacetime-noise
studies may then well depend
 on issues such as the  possible $N_\gamma$ suppression.

 I should also stress that the analysis of
 these opportunities for quantum-spacetime phenomenology
from experiments  operating at low
 observational frequencies $\omega$, is perhaps
 the most significant and most robust conceptual achievement
 of the sort of phenomenology of spacetime foam that I am discussing
 in this subsection.
When these pictures were first proposed it was seen by many as a total surprise that
one could contemplate Planck-scale effects at frequencies of observation of only 100~Hz.
The naive argument goes something
like ``Planck-scale-induced noise must be studied at Planck frequency'',
and the Planck frequency is $E_p/\hbar \sim 10^{43}$~Hz.
However in analyzing actual pictures of quantum-spacetime fuzziness, even the
simple-minded ones described above,
one becomes familiar with
well-known facts establishing
(and we should expect this lesson to apply even to more sophisticated
picture of quantum-spacetime-induced fuzziness)
that discrete fluctuation mechanisms
tend to produce very significant effect at low
 observational frequencies $\omega$, with
typical behaviors of the type $\omega^{-|\alpha|}$,
 even when their charateristic
 time scale is ultrashort.

\subsubsection{Distance fuzzyness for atom interferometers}

Since the phenomenology of the implications of spacetime foam for interferometry
is at an early stage of development, at the present time it may be
premature to enter into detailed discussions of what type of
interferometry might be best suited for uncovering
quantum-spacetime/Planck-scale effects.
Accordingly, in this Section~\ref{foam-fuzziness-noise} I focused
by default on the simplest case of interferometric studies, the one
using a laser-light beam.
However, in recent times atom interferometry has reached equally astonishing
levels of sensitivity and for several interferometric measurements
it is presently the best choice.
Laser-light interferometry is still preferred for certain
well-established techniques of interferometric studies of spacetime observables,
as in the case of searches of gravity waves, and the observations
I reported above for the phenomenology of strain noise induced
by quantum-spacetime effects appear to be closely linked to the issues
encountered in the search of gravity waves.
However, it seems plausible that soon there will be
some atom-interferometry setups that are competitive
for gravity-wave searches (see, e.g., Refs.~\cite{tinovetrano, dimopGWI}).
This in turn might imply that
searches of quantum-spacetime-induced strain noise could rely
on atom interferometry.

The alternative between light and matter interferometry might prove
valuable at later more mature stages of this phenomenology.
It is likely  that different test theories will give different
indications  in this respect, so that
atom interferometry might provide the tightest constraints on
some spacetime-foam test theories whereas
laser-light interferometry might provide the best constraints
on other spacetime-foam test theories.
Of course a key aspect of the description of Planck-scale effects
for atom interferometry to be addressed by
the test theories (and hopefully, some day, by some fully-developed
quantum-spacetime/quantum-gravity theories)
is the role played by the mass of the atoms.
With respect to laser-light interferometry the case of atom interferometry
challenges us with at least two more variables to be controlled at the
theory level, which are the mass of the atoms and their compositeness.
How do these two aspects of atom interferometry interface with
the quantum-spacetime features that are here of interest?
Do they effectively turn out to introduce suppressions of the relevant effects
or on the contrary could be exploited for seeing the effects?
For none of the quantum-spacetime models that are presently studied
we have reached a level of understanding of physical implications robust enough
for us to answer confidently these questions. And perhaps we should
also worry about (or exploit) another feature which is in principle
tunable  in atom interferometry, which is  the velocity
of the particles in the beam.


\subsection{Fuzziness for waves propagating
  over cosmological distances}

Interferometric studies of spacetime-foam are another rare example
of tests of quantum-spacetime effects that can be conducted in
a controlled laboratory setup (also see Subsection~\ref{inthelab}).
But also for this type of studies astrophysics may turn into the
most powerful arena. Indeed the studies I discussed in the previous section,
which started toward the end of the 1990s, inspired a few years later
some follow-up studies from the astrophysics side.
As it should be expected, the main opportunities come from
observations of waves that have propagated over very large distances,
thereby possibly accumulating a significant collective effect of the fuzziness
encountered along the way to our detectors.

\subsubsection{Time spreading of signals}\label{fuzzytimes}

An implication of distance fuzziness that one should naturally consider
for waves propagating over large distances
is the possibility of ``time spreading'' of the signal:
if at the source the signal only lasted a certain very short time,
but the photons that compose the signal travel a large distance $L$,
affected by uncertainty $\sigma_L^2$, before reaching
our detectors the observed spread of times of arrival might carry little
trace of the original time spread at the source and be instead a manifestation
of the quantum-gravity-induced $\sigma_L$. If the distance $L$
is affected by a quantum-spacetime uncertainty $\sigma_L$
then different photons
composing the signal will effectively travel distances that are not all exactly
given by $L$ but actually differ from $L$ and from each other
up to an amount $\sigma_L$.

Again it is of particular interest to test laws
of the type  discussed in the previous section,
but it appears that these effects
would be unobservably small
even in the case that provides the strongest effects,
which is the
random-walk {\it ansatz} $\sigma_L^2 \simeq  L_p  L \simeq  L_p T$
(assuming ultrarelativistic particles, for which $L$ is at least roughly
equal to the time duration $T$ of the journey).
To see this let me consider
once again gamma-ray bursts, which as mentioned often travel
for times of the order of $10^{17}$~s before reaching our
Earth detectors and are sometimes characterized by time structures
(microbursts within the burst) that have  duration
as short as $10^{-3}$~s. Values of $\sigma_L^2$ as small
as $\sigma_L^2 \sim c^2 10^{-8}\mathrm{\ s}^2$ could be noticeable in
the analysis of such bursts.
However, the estimate $\sigma_L^2 \simeq c L_p  T$
only provides $\sigma_L^2 \sim c^2 10^{-27}\mathrm{\ s}^2$
and is therefore much beyond our reach.

I shall comment in the later subsection \ref{causalsetphen} on
an alternative formulation of phenomenology of
quantum-spacetime-induced
worldline fuzziness,
the one in Ref.~\cite{scargleCAUSALSETS}
inspired by the causal-set approach (the approach on which
subsection \ref{causalsetphen} focuses).

\subsubsection{Fuzziness from nonsystematic symmetry-modification effects}\label{nonsyste}
As a possible alternative way to model spacetime fuzziness
there has been some interest~\cite{nguhecr,uncpa1,grilloNONSYST}
in the possibility that there might be  effects
resembling the ones here
discussed in Section~\ref{symmetry-tests},
which are systematic deviations from the predictions
of Poincar\'e symmetry, but are ``nonsystematic'' in the sense discussed
at the beginning of this section.
The possibility of fuzziness of particles worldlines governed  by $E/E_p$,
mentioned in the previous subsubsection, is an example
of such nonsystematic violations of Poincar\'e symmetry.

This speculations are not on firm ground on the theory side, in the sense
that there is not much in support of this among available results
on actual analysis of formalizations of spacetime quantization.
But it is legitimate to expect that this might be due to our limited abilities
in mastering these complex formalisms. After all, as suggested
for example in Ref.\cite{nguhecr}, if spacetime geometry is fuzzy
then it may be inevitable for the operative procedures that give sense
to the notion of energy and momentum of a particle to also be fuzzy.

This sort of picture could have tangible observational consequences.
For example, it can inspire,
as suggested in Refs.~\cite{unoEdue,iucaapap,uncpa1},
scenarios such that spacetime fuzziness effectively produces an uncertainty in the
velocity of particles of order $E/E_p$.
This would give rise to magnitude of these nonsystematic effects comparable to
the one discussed in subsection \ref{invacuosubsec} for the corresponding
systematic effects.
After a journey of $\sim 10^{17}$~s the acquired fuzziness of arrival times
could be within the reach~\cite{unoEdue}  of
suitably arranged  gamma-ray-burst studies.
There is however no significant effort to report here on establishing bounds
following this strategy.

There are instead some studies of this phenomenological picture~\cite{nguhecr,uncpa1,grilloNONSYST}
which take as starting
point the possibility, here discussed in
Sections~\ref{pair-production} and \ref{photopion-production}, of
 modifications of the dispersion relation leading to
 modifications of the threshold requirements for certain
particle-production processes, such as the case of two incoming
photons producing an outgoing electron-positron
pair. Refs.~\cite{nguhecr,uncpa1,grilloNONSYST}  considered the
possibility of a non-systematic quantum-spacetime-induced deformation of
the dispersion relation, specifically the case in which the classical
relation $E^2 = p^2 + m^2$ still holds on average, but  for a given
particle with large momentum $\vec{p}$, energy would be somewhere in
the range
\begin{equation}
|\vec{p}| + {m^2 \over 2 |\vec{p}|}
- { |\eta| \over 2} {\vec{p}^2  \over E_{p}}
\le E \le
|\vec{p}| + {m^2 \over 2 |\vec{p}|}
+ {|\eta| \over 2} {\vec{p}^2  \over E_{p}}
~,
\label{dispnons}
\end{equation}
with some (possibly Gaussian) probability distribution.
A quantum-spacetime theory with this feature should
be characterized by a fundamental
value of $\eta$, but each given particle would
satisfy a dispersion relation of the type
\begin{equation}
E \simeq  |\vec{p}| + {m^2 \over 2 |\vec{p}|}
+ {{\tilde {\eta}} \over 2} {\vec{p}^2  \over E_{p}}
~,
\label{disptilde}
\end{equation}
with $-|\eta| \le {\tilde {\eta}} \le |\eta|$.

In analyses such as the one discussed in
Section~\ref{pair-production} (for observations of gamma rays from
blazars) one would then consider electron-positron pair production
in a head-on photon-photon collision assuming that  one of the photons is very hard
while the other one is very soft.
To leading order,
for the soft photon only the energy $\epsilon$
is significant (for an already small $\epsilon$ the
actual value of ${\tilde {\eta}}$ will not matter in leading order).
So the soft photon
can, in leading order, be treated as satisfying a classical
dispersion relation.
In a quantum-spacetime theory predicting
such non-systematic effects, the hard photon would be characterized both
by its energy $E$
and its value of ${\tilde {\eta}}$. In order to establish whether
a collision between two such photons can produce an electron-positron
pair, one should establish whether, for some admissible values
of ${\tilde {\eta}}_+$ and ${\tilde {\eta}}_-$ (the values
of ${\tilde {\eta}}$ pertaining to the outgoing positron and the electron
respectively), the conditions for energy-momentum conservation
can be satisfied.
The process will be allowed if
\begin{equation}
E \ge  {m^2 \over \epsilon}
- {{\tilde {\eta}} \over 4} {E^3  \over \epsilon E_{p}}
+ {{\tilde {\eta}}_+ + {\tilde {\eta}}_- \over 16} {E^3  \over \epsilon E_{p}}
~.
\label{thr1prelim}
\end{equation}
Since ${\tilde {\eta}}$, ${\tilde {\eta}}_+$ and ${\tilde {\eta}}_-$
are bound to the range between $-|\eta|$ and $|\eta|$,
the process is only allowed, independently of the
value of ${\tilde {\eta}}$, if the condition
\begin{equation}
E \ge  {m^2 \over \epsilon}
- {3|{\eta}| \over 8} {E^3  \over \epsilon E_{p}}
\label{thr1}
\end{equation}
is satisfied.
This condition defines the actual threshold in the non-systematic-effect
scenario. Clearly in this sense the threshold
is inevitably decreased by the non-systematic effect.
However, there is only a tiny chance that a given photon
would have ${\tilde {\eta}} = |{\eta}|$, since this is
the limiting case of the range allowed by the nonsystematic effect,
and unless ${\tilde {\eta}} = |{\eta}|$ the process will still not be allowed
even if
\begin{equation}
E \simeq  {m^2 \over \epsilon}
- {3|{\eta}| \over 8} {E^3  \over \epsilon E_{p}} ~.
\label{thr1b}
\end{equation}
Moreover, even assuming ${\tilde {\eta}} = |{\eta}|$,
the energy value described by~(\ref{thr1b}) will only be sufficient
to create an electron positron pair with
${\tilde {\eta}}_+ = - |{\eta}|$ and ${\tilde {\eta}}_- = -|{\eta}|$,
which again are isolated points at the extremes of the relevant
probability distributions.
Therefore the process becomes possible at the energy level
described by~(\ref{thr1b}) but it remains extremely
unlikely, strongly suppressed by the small probability
that the values of ${\tilde {\eta}}$, ${\tilde {\eta}}_+$
and ${\tilde {\eta}}_-$ would satisfy the kinematical requirements.

With reasoning of this type, one can easily develop
an intuition for the dependence on the energy $E$,
for fixed value of $\epsilon$ (and treating
${\tilde {\eta}}$, ${\tilde {\eta}}_+$
and ${\tilde {\eta}}_-$ as totally unknown), of the likelihood
that the pair-production process can occur:
(i)~when~(\ref{thr1}) is not satisfied the process is not allowed;
(ii)~as the value of $E$ is increased above the
value described by~(\ref{thr1b}), pair production becomes
less and less suppressed by the relevant probability
distributions for
${\tilde {\eta}}$, ${\tilde {\eta}}_+$
and ${\tilde {\eta}}_-$,
but some suppression remains up to the value of $E$
that satisfies
\begin{equation}
E \simeq  {m^2 \over \epsilon}
+ {3|{\eta}| \over 8} {E^3  \over \epsilon E_{p}} ~;
\label{thr1c}
\end{equation}
(iii)~finally for energies $E$ higher than the one described by
(\ref{thr1c}), the process is kinematically allowed for all
values of ${\tilde {\eta}}$, ${\tilde {\eta}}_+$
and ${\tilde {\eta}}_-$, and therefore the likelihood
of the process is just the same as in the classical-spacetime theory.

This describes a single photon-photon
collision taking into account the nonsystematic effects.
One should next consider that
for a hard photon travelling toward our Earth detectors from a distant
astrophysical source
there are many opportunities to collide with soft photons
with energy suitable for pair production to occur (the mean
free path is much shorter than the distance between the source
and the Earth). Thus one expects~\cite{uncpa1,grilloNONSYST}
 that even a small probability of producing an electron-positron
pair in a single collision would be sufficient
to lead to the disappearance of the hard photon before
reaching our detectors. The probability is small in a
single collision with a soft background photon, but
the fact that there are, during the long journey,
many such pair-production opportunities
renders it likely that in one of the many collisions
the hard photon would indeed disappear into an electron-positron
pair. For this specific scheme of non-systematic effects
it appears therefore that a characteristic prediction
is that the detection of such hard photons from distant
astrophysical sources should start being significantly suppressed
already at the energy level
described by~(\ref{thr1b}), which is \emph{below} the threshold corresponding to the
classical-spacetime kinematics.

It is interesting~\cite{iucaapap,uncpa1,unoEdue} to contemplate in this case the possibility
that systematic and nonsystematic effects may both be present. It is not unnatural
to arrange the framework in such a way that the systematic effects tend to
give higher values of the threshold energy, but then the nonsystematic effects would
allow (with however small probability) configurations below threshold to produce
the electron-positron pair. And for very large propagation distances (very
many ``target soft photons''
available) the nonsystematic effect can essentially erase~\cite{uncpa1} the systematic effect
(no noticeable upward shift of the threshold).

I here illustrated the implications of nonsystematic effects within a given scenario
and specifically for the case of observations of gamma rays from
blazars. But of course one can implement the non-systematic effects
in some alternative ways and the study of the observational implications
can consider other contexts. In this respect I should bring to the attention
of my readers the studies of non-systematic effects for ultra-high-energy cosmic
rays reported in Refs.~\cite{grilloNONSYST,weilerUHECRns,mattiUHECRns}.

Combinations of systematic
and nonsystematic effects can also be relevant~\cite{iucaapap,unoEdue}
for studies of the correlations between times of arrival and energy of simultaneously-emitted particles.
For that type of studies
both the systematic and the nonsystematic effects could leave an observable
trace~\cite{unoEdue} in the data, codified in the mean arrival time and the standard deviation
of arrival times found in different energy channels.

\subsubsection{Blurring images of distant sources}

The two examples of studies in astrophysics
of quantum-spacetime-induced distance fuzziness
I discussed in subsubsections \ref{fuzzytimes} and \ref{nonsyste}
have only been moderately popular. I have left as last item of this
subsection the most intensely studied opportunity
in astrophysics
for quantum-spacetime-induced distance fuzziness.
These are studies essentially looking for effects blurring the images
of distant sources.

It is interesting that these studies
were started by Ref.~\cite{lieu} which cleverly combined
some aspects of Refs.~\cite{gacgwi,gacgwiB,bignapap,nggwi},
providing the main concepts for the proposal here summarized
in subsection \ref{foam-fuzziness-noise},
and some aspects of Ref.~\cite{grbgac},
providing the main concepts for the proposal
here summarized in subsection \ref{invacuosubsec}.
Ref.~\cite{lieu} was interested in the same phenomenology of distance
fuzziness introduced and analyzed for controlled interferometers in
 Refs.~\cite{gacgwi,gacgwiB,bignapap,nggwi},
 but looked for opportunities to perform analogous tests using the whole Universe
 as laboratory, in the sense first introduced in Ref.~\cite{grbgac}.

Gradually over the last decade this became a rather active research area,
as illustrated by the studies reported in Refs.~\cite{lieu,jackNEWastroint,coule,ragazzoniALALIEU,gallouALALIEU2003,jackALALIEUprl,chenALALIEU2006,maziaALALIEU2006,fuzzyseen,maziaALALIEU2007,maziaALALIEU2009,tamburiniALALIEU2011,jackALALIEU2011,maziaALALIEU2012}.

The phenomenological idea is powerfully simple:
effects of quantum-spacetime-induced spacetime fuzziness
had been shown~\cite{gacgwi,gacgwiB,bignapap,nggwi}
to be potentially relevant for LIGO/VIRGO-like (and LISA-like)
intereferometers exploiting not only
the distance-monitoring accuracy of those interferometers, but also the fact
that such accuracy on distance monitoring is achieved for rather large terrestrial
distances. Essentially the Universe gives much larger distances
for us to monitor~\cite{grbgac}, and although we can monitor them with accuracy
inferior to the one of a LIGO/VIRGO-like
intereferometer,
 on balance the astrophysics route may be advantageous
also for studies of
quantum-spacetime-induced spacetime fuzziness~\cite{lieu}.

As for Refs.~\cite{gacgwi,gacgwiB,bignapap,nggwi},
here reviewed in subsection \ref{foam-fuzziness-noise},
the core  intuition here is that
 the quantum-spacetime contribution
  to the fuzziness of a particle's worldline
might grow with propagation distance.
And collecting
the scenarios here
summarized in equations (\ref{rhorw}),(\ref{rhoholo}),(\ref{rhoweak})
one arrives at a one-parameter family of
phenomenological {\it ansatzae} for the characterization of this dependence
of fuzziness on distance
\begin{equation}
\sigma^2_L \simeq  L_P^{2\alpha} L^{2-2\alpha} ~.
\label{alphafamily}
\end{equation}
with $1/2 \leq \alpha \leq 1$.

An assumption shared by most explorations~\cite{lieu,jackNEWastroint,coule,ragazzoniALALIEU,gallouALALIEU2003,jackALALIEUprl,chenALALIEU2006,maziaALALIEU2006,fuzzyseen,maziaALALIEU2007,maziaALALIEU2009,tamburiniALALIEU2011,jackALALIEU2011,maziaALALIEU2012} of this phenomenological
avenue is that from (\ref{alphafamily}) there would also follow
an associated uncertainty in the specification of momenta
\begin{equation}
\delta E \sim L_P^\alpha ~ E^{1+\alpha}~,~~~
\delta p \sim L_P^\alpha ~ p^{1+\alpha}~.
\label{fuzzyMOMENTUM}
\end{equation}
I must stress that this (however plausible)
deduction of heuristic arguments has not been
confirmed in any explicit model of spacetime quantization.
And it plays a crucial role in most astrophysics tests of
distance fuzziness: from (\ref{fuzzyMOMENTUM})
it is easy to see~\cite{lieu} that it follows that (assuming a classical-wave description
still is admissible when such effects are nonnegligible)
there should be a mismatch between uncertainty in the group velocity and uncertainty
in the phase velocity of a classical wave, and this in turn proves to
be a very powerful tool for the phenomenology. During a propagation time $T=L/v_g$
($v_g$ being the group velocity) the phase of a wave advances by
$\Delta \phi = 2 \pi (v_p/v_g)(L/\lambda)$ (where $v_p$ is the phase velocity
and $\lambda$ is the wavelength).
There are two schools of intuition concerning how quantitatively spacetime
fuzziness should scrumble the phase of a wave.
According to Ref.~\cite{lieu} and followers the effect should go like
\begin{equation}
\delta \phi \simeq \frac{E^\alpha}{E_p^\alpha} \frac{L}{\lambda}
= \frac{E^{1+\alpha}}{E_p^\alpha} L
\label{phaselieu}
\end{equation}
whereas according to Ref.~\cite{jackNEWastroint,jackALALIEUprl} and followers the effect should grow more slowly with the distance of propagation, going like
\begin{equation}
\delta \phi \simeq \frac{E}{E_p^\alpha} L^{1-\alpha}
\label{phasejack}
\end{equation}

As first observed in Ref.~\cite{fuzzyseen},
the alternative formulas (\ref{phaselieu}) and (\ref{phasejack})
should anyway be improved to account for redshift.
For the case of Eq.~(\ref{phasejack})
Ref.~\cite{fuzzyseen} proposes the following
\begin{equation}
\delta \phi
\sim  \frac{E}{E_p^\alpha}
\frac{1-\alpha}{H_0 q_0} \int_0^\infty dz \, (1+z) L^{-\alpha}
\left( 1 - \frac{1-q_0}{\sqrt{1+2 z q_0}} \right)
\label{phasejackred}
\end{equation}
where $q_0$ is the decelaration parameter, $q_0=\Omega_0/2 - \Lambda/(3 H_0^2)$
and $L$ is the luminosity distance, $L=[zq_0+(1-q_0)(1-\sqrt{1+2zq_0})]/(H_0q_0^2)$
($\Lambda$, $H_0$ and $\Omega_0$ here denote, as usual,
respectively the cosmological constant, the Hubble parameter and the matter fraction).

Evidently this phenomenology still has a bit too many quantitative details subjectable
to further scrutiny and a bit too many alternative scenarios.
This is the result of the fact that works on
actual formalizations of spacetime quantization,
while encouraging the general intuitive picture,
have been unable to provide detailed guidance.
And the heuristic arguments based on these preliminary studies
have been unable to narrow the range of possibilities.
But pursuing this path further appears to be an exciting opportunity
for quantum-spacetime phenomenology, and we should therefore persevere.
In particular, based on the (however alternative) estimates
given by (\ref{phaselieu}), (\ref{phasejack}) and (\ref{phasejackred})
several authors (see, e.g., Refs.~\cite
{jackALALIEUprl,fuzzyseen,tamburiniALALIEU2011})
have concluded that a phenomenology based on blurring of the images
of distant sources can provide Planck-scale sensitivity for a rather
broad range of possible phenomenological test theories and
for values of $\alpha$ significantly greater than $1/2$,
possibly~\cite{tamburiniALALIEU2011}
going all the way up to values of $\alpha$ close to $1$.
In Ref.~\cite{fuzzyseen} one even finds a preliminary data analysis
suggesting that for observations of quasars
there might be a trend towards lower observed Strehl ratios
with increasing redshift, which would provide encouragement
for hope of discovering quantum-spacetime-induced image blurring.

The main opportunities appear to be provided indeed by observations
of distant quasars~\cite{jackALALIEUprl,fuzzyseen,tamburiniALALIEU2011},
whose dimensions are small and are rather abundantly observed
at high redshift.


\subsection{Planck-scale modifications of CPT symmetry and
  neutral-meson studies}

Investigations of spacetime symmetries and distance fuzziness are evidently
relevant for some of the core features of the idea of spacetime quantization.
My next task concerns CPT-symmetry tests, and the possibility that
indirectly some scenarios for the quantization of spacetime
might affect CPT symmetry.

A complication, but also an opportunity, for quantum-spacetime-motivated
tests of CPT symmetries comes from the fact that CPT symmetry
should be and is tested even independently of the quantum-spacetime motivation.
From this perspective the situation is somewhat analogous to
what I discussed earlier in this review concerning quantum-spacetime-motivated
tests of Lorentz symmetry. The quantum spacetime literature can provide
special motivation for probing CPT symmetry in certain specific ways,
but there is already plenty of motivation, even without
 quantum-spacetime research, for testing CPT symmetry as broadly
 as possible~\cite{maianiCPTdafne,okunCPT,fidecaroCPT}.

 Also in this case the Standard Model Extension provides a much appreciated and widely
 adopted formalization, finding a good balance between
 the desire of searching for violations of CPT symmetry (and/or,
 as mentioned, violations of Lorentz symmetry)
 within the confines of quantum field theory
 but allowing for both effects that have
 been discussed from the quantum-spacetime perspective and effects for which
 so far there is no quantum-spacetime motivation.
 I shall here focus on the hypothesis of quantum-spacetime-induced
 and Planck-scale-magnitude CPT violation effects, so I shall not
 review the broad subject of CPT violation within the Standard Model Extension,
for which readers can anyway find valuable reviews and
perspectives in
Refs.~\cite{kostcpt,bertocpt,repCOLLAkoste1,kosteCPTprl,kosteJackiw,kosteLorentzCPT2001,kosteDATATABLES}
(also see parts of Ref.~\cite{mattinLRR}).

Another issue that should always be kept in mind in relation to CPT symmetry
is the fact that it can be derived as a theorem for local quantum field theories with Lorentz invariance.
In approaches based on local field theory there it is then natural to perform
combined studies of CPT and Lorentz symmetry\footnote{The interplay
 between violations of Lorentz symmetry and violations of CPT symmetry,
 which is part of the objectives being pursued within the context provided
 by the Standard Model Extension, is also the subject of a lively debate, as can be seen for
 example from Refs.~\cite{greenbergONcpt,chaichianONcpt}.}.
 The notion of spacetime quantization at the Planck scale involves however some aspects
 of nonlocality (at least the notion of points that coincide with accuracy better than
 the Planck length is typical abandoned) and in most quantum-spacetime studies
 of the fate of CPT symmetry the expectation is that these aspects of non-locality
 might be primarily responsible for the conjectured violations of CPT symmetry.

I shall not attempt to summarize here the results on violations
of CPT symmetry arising from spacetime quantization
not introduced at the Planck scale  (but rather at some
much lower scale), for which however readers can find valuable
starting points to the related literature in
Refs.~\cite{chaichianCPTncft,gaumeCPTncft,pospelovCPTncft,jabbariCPTncft,balaCPTncft}
and references therein.

Consistently with the scopes of this review work I shall focus here
exclusively on scenarios for violations of CPT symmetry based
on nonclassicality (``quantization'') of spacetime
introduced at the Planck scale.
As a result of some technical challenges, which  I already mentioned
 in Subsection~\ref{cptgeneral}, this literature can only rely on preliminary
 theory results, but does suggest convincingly
  that Planck-scale sensitivity to quantum-spacetime-induced violations
 of CPT symmetry  is within our reach.

\subsubsection{Broken-CPT effects from Liouville strings}
In the case of test of CPT symmetry it is easier for me to start from
discussing the availability of Planck-scale sensitivity, postponing briefly
some comments on test theories based on the idea of spacetime quantization.

There is a sizable literature establishing that
 CPT symmetry can be tested with Planck-scale sensitivity
 in the neutral-kaon and the neutral-B systems
(see, e.g., Refs.~\cite{ehns,elmn,huetpesk,floreacpt}).
It turns out that in
these neutral-meson systems there are plenty of opportunities for
Planck-scale departures from CPT
symmetry to be amplified. In particular, the
neutral-kaon system hosts the peculiarly small mass difference
between long-lived and the short-lived
kaons $|M_L - M_S|/M_{L,S} \sim 7 {\cdot} 10^{-15}$
and other small numbers naturally show up in the analysis of the system,
such as the ratio $|\Gamma_L - \Gamma_S|/M_{L,S} \sim 1.4 \cdot 10^{-14}$.
And for certain types of departures
from CPT symmetry the inverse of one of these small
numbers amplifies the small CPT-violation
effect~\cite{ehns,elmn,huetpesk,floreacpt}.
In particular, this mechanism turns out to provide sufficient
amplification for
Planck-scale effects inducing
a difference of order $M_{K^0}^2/E_p$
 between the terms
on the diagonal of the
$K^0$, ${\bar{K^0}}$ mass matrix (exact classical CPT symmetry
would of course require the terms on the diagonal to be identical).
It should be noticed that  $M_{K^0}/E_p \sim 10^{-19}$,
which is not overwhelmingly smaller than the
mentioned $|M_L - M_S|/M_{L,S} \sim 7 {\cdot} 10^{-15}$.

A much studied quantum-spacetime description of violations
of CPT symmetry is centered on the mentioned Liouville-strings
approach~\cite{emn,elmn}, particularly with its
description of spacetime foam and its non-classical description
of time, involving a non-trivial role for the Liouville field~\cite{emnchaos}.
This model is in particular the reference for the
analysis of Planck-scale limits on quantum-spacetime-induced
CPT violation reported by the CPLEAR collaboration on
the basis of studies of  neutral kaons~\cite{cplear}
(also see the related results reported using neutral-kaon
data gathered at the particle-physics
laboratory in Frascati~\cite{cplear,kloe1,didomALL}).
Interestingly the Liouville-string
model hosts both departures from CPT symmetry and decoherence,
and I find most convenient to discuss it in the later part of this section which
is devoted to decoherence studies.

Let me here rather
highlight a more recent development which is in part inspired by
these Liouville-string studies.
It was recently observed (primarily in Refs.~\cite{nickBern1,nickBern2})
that quantum-spacetime
scenarios with violations of CPT symmetry might also require
some corresponding modifications of the recipe for obtaining multiparticle states
from singleparticle states for identical particles.
This may in particular apply to the neutral-kaon $K_0-\bar{K_0}$ system,
since standard CPT transformations take $K_0$ into $\bar{K_0}$ but violations
of CPT symmetry are likely to also induce a modification of the link
between $K_0$ and $\bar{K_0}$.

Refs.~\cite{nickBern1,nickBern2} proposed
a phenomenology inspired by this argument and based on the
following parametrization of the state $|i>$ initially produced by a $\phi$-meson decay:
\begin{eqnarray}\label{nickbern}
|i> \propto \big( (|K_S(p),K_L(-p)>-|K_L(p),K_S(-p)>) + \omega
(|K_S(p),K_S(-p)>-|K_L(p),K_L(-p)>) \big)
\end{eqnarray}
where the complex parameter $\omega$ essentially characterizes the level of
contamination of the state $|i>$ by the (otherwise unexpected)
C-even component $|K_S(p),K_S(-p)>-|K_L(p),K_L(-p)>$.

Stringent constraints on $\omega$ can be placed by performing measurements
of the chain of processes $\phi \rightarrow K K \rightarrow X Y$,
in which first the $\phi$ meson decays into a pair of neutral kaons and then one of the kaons
decays at time $t_1$ into a final state $X$ while the other kaon
decays at time $t_2$ into a final state $Y$.
By following this strategy the KLOE experiment~\cite{kloe1,kloe2}
at DA$\Phi$NE
is setting~\cite{didomOMEGA,didomALL}
experimental limits on $\omega$ at the level $10^{-3}$
($|Re(\omega) | < 10^{-3}$, $|Im(\omega)| <10^{-3}$).

It is not easy at present to establish robustly
what level of sensitivity to $\omega$ could really
amount to Planck-scale sensitivity,
but it is noteworthy that there are
semi-quantitative/semi-heuristic estimates based on a
certain  intuition
for spacetime foam
suggesting~\cite{nickBern1,nickBern2,nickCPTshort}
that sensitivities in the neighborhood
of $\omega \sim 10^{-3}$, $\omega \sim 10^{-4}$ could already be significant.

\subsubsection{Departures from classical CPT symmetry from spacetime
noncommutativity at the Planck scale}

Another formalism for spacetime quantization at the Planck scale
where violations of CPT symmetry have been discussed to some extent
is the one of the
so-called ``$\kappa$-Minkowski
spacetime noncommutativity''~\cite{majrue,kpoinap,gacmaj}.
A first hint that this might be appropriate comes from the
fact that the $\kappa$-Minkowski formalism is one of those
providing support for the possibility
of modifications of the dispersion relation
of the form $m^2 \simeq E^2 - \vec{p}^2 + {\lambda  E  } \vec{p}^2/2$,
with $\lambda$ of the order of the Planck length.
It may be relevant for the relation between particles and antiparticles
(for which CPT symmetry is a crucial player)
that for the values of $E$ allowed by the dispersion relation for given $|\vec{p}|$
one does not recover the ordinary result (with its traditional
two solutions of equal magnitude and opposite sign); instead,
one finds that the two solutions $E_+$, $E_-$ are given by
\begin{equation}
E_{\pm} \simeq - {\lambda \over 2 } \vec{p}^2 \pm
\sqrt{m^2+  \vec{p}^2} ~.
\label{sols}
\end{equation}
The fact that the solutions  $E_+$ and $E_-$
are not exactly opposite may suggest that
one should make room
for a mismatch  $\delta M$ of the terms
on the diagonal of the $K^0, {\bar{K^0}}$ mass matrix,
of order
\begin{equation}
|\delta M| \sim {| E_{+}| - | E_{-}| \over | E_{+}| + | E_{-}|} 2 M
 \simeq {\lambda } {\vec{p}^2 M \over
\sqrt{ M^2+  \vec{p}^2}} ~.
\label{stimadm}
\end{equation}
The most significant feature of this description of $\delta M$
 is its momentum
dependence, and, for given $|\lambda|$, $|\delta M|$
is an increasing function of $|\vec{p}|$,
quadratic in the non-relativistic limit
and linear in the ultra-relativistic limit.
Therefore among experiments achieving
comparable $\delta M$ sensitivity the ones studying more energetic kaons
are going to lead to more stringent bounds on $\lambda$.

Considering that, as mentioned, neutral-kaon experiments
at $\Phi$ factories are now sensitive at the
level $\delta M \sim 10^{-18}$~GeV,
one infers a sensitivity to this type of candidate quantum gravity
effect that, for kaons of momenta of about 110~MeV
(at the $\phi$ resonance), corresponds to sensitivity
to values of $|\lambda|$ around $10^{-32}$~m, i.e., not far
(just 3 orders of magnitude away) from the Planck scale.
Because of the premium on high momenta of this scenario
better limits could be set using experiments with high-momentum kaons
of the type of Fermilab's E731~\cite{fermilabE731a,fermilabE731b}.
And studies with neutral B mesons of relatively high momenta
could also be valuable from this perspective.

We are however at a very early stage of understanding
of the fate of CPT symmetry in these spacetimes with quantization
at the Planck scale. Specifically for the case of $\kappa$-Minkowski
spacetime, analyses such as the one in
Ref.~\cite{gacmaj} suggest that CPT symmetry is deformed rather
than broken/lost. Indeed in $\kappa$-Minkowski
the anomalies one can presently preliminarily see for CPT symmetry
are all linked to the peculiarity of $P$-parity transformations.
It appears that in $\kappa$-Minkowski $P$-parity transformations for momenta
should not take a momentum $\vec{p}$ into $-\vec{p}$, but
rather $\vec{p} \rightarrow \ominus \vec{p}$,
where  $\ominus \vec{p}$ denotes the so-called antipode
operation: $\ominus \vec{p} \equiv -\vec{p}e^{-\lambda p_0}$
(where $\lambda$ denotes again
the $\kappa$-Minkowski noncommutativity length scale).


\subsection{Decoherence studies with kaons and atoms}

\subsubsection{Spacetime foam as decoherence effects and the
  ``$\alpha$, $\beta$, $\gamma$ test theory''}

As stressed earlier in this review the idea of ``spacetime foam'' appears to appeal
to everyone involved in quantum-spacetime research, but this is in  part due to the fact
that this idea is not really well defined, not by the
qualitative intuitive picture proposed by Wheeler.
Of course, in order to set up a phenomenology for effects induced by this spacetime
foam it is instead necessary to provide for it physical/experimentally-meaningful
characterizations. I already discussed one possible such characterization, given in terms
of distance fuzziness and associated strain noise for interferometry.
Another attempt to characterize physically spacetime foam can be found in Refs.~\cite{elmn,emn}
(other valuable perspectives on this subject can be found in Refs.~\cite{floreacpt,garayDECO}),
focusing on the possibility that the rich dynamical properties of spacetime foam
might act as a decoherence-inducing environment.

The main focus of Refs.~\cite{elmn,emn} has been the neutral-kaon system, whose
remarkably delicate balance of scales provides opportunities not only for very sensitive
tests of CPT symmetry but also for very sensitive tests of decoherence.
Refs.~\cite{elmn,emn} essentially propose a test theory,
based on the mentioned Loiuville-strings idea,
 for spacetime-foam-induced
decoherence in the
neutral-kaon system. This test theory
adopts the formalism of density matrices
and is centered on
the following evolution equation for the neutral-kaon reduced density matrix $\rho$:
\begin{eqnarray}
 \partial_t \rho = i [\rho,H] + \delta \! H \, \rho
\label{ehnseq}
\end{eqnarray}
where $H$ is an ordinary-quantum-mechanics Hamiltonian and $\delta \! H_{mn}$
(with indices $m,n$ running from 1 to 4: $\{m,n\} \in \{1,2,3,4\}$)
is the spacetime-foam-induced
decoherence matrix, taken to be such
that $\delta \! H_{1n}=\delta \! H_{2n}=\delta \! H_{n1}=\delta \! H_{n2}=0$,
while $\delta \! H_{34}=\delta \! H_{43}=-2\beta$, $\delta \! H_{33}=-2\alpha$,
and $\delta \! H_{44}=-2\gamma$. Therefore the test theory is fully specified
upon fixing $H$ and giving some definite values to the parameters $\alpha$, $\beta$, $\gamma$.

It should be stressed that this test theory necessarily violates CPT symmetry
whenever $\delta \! H \neq 0$. Additional CPT violating features may be introduced
in the ordinary-quantum-mechanics Hamiltonian $H$, by allowing for differences in masses and/or
differences in widths between particles and antiparticles.
Therefore this test theory is an example of framework that could be used in
a phenomenology looking simultaneously both for departures from CPT symmetry of types
admissible within ordinary quantum mechanics and for departures from CPT symmetry
that require going beyond quantum mechanics (by allowing for decoherence).
And it is noteworthy that the two types of CPT violation
(within and beyond quantum mechanics) can be distinguished experimentally.

Concerning more directly decoherence various characterizations
of the effects of this test theory have been provided, and in particular
a valuable description of how significant the decoherence effects are (depending on
the values given to  $\alpha$, $\beta$, $\gamma$) is found looking at how the rate of kaon decay
into a pair of pions, $R_{2 \pi}$, evolves as a function
of time. This time evolution will in general take the form
\begin{eqnarray}
R_{2 \pi}(t)  = C_S e^{-\Gamma_S t} + C_L e^{-\Gamma_L t}
 + 2 C_I e^{-(\Gamma_L +\Gamma_S) t/2} \cos[(m_L-m_S) t - \phi]
 ~, \label{r2pi}
\end{eqnarray}
where the indices $S,L,I$ stand respectively for short-lived, long-lived, interference,
and the combination $\zeta \equiv 1 - C_I/\sqrt{C_S C_L}$ provides a good phenomenological
characterization of the amount of decoherence induced in the system~\cite{nickCPTshort}.

Using data gathered by the CPLEAR experiment~\cite{cplear} one can set bounds
on $\alpha , \beta , \gamma$ at the levels $\alpha \sim 10^{-17}$~GeV,  $\beta \sim 10^{-19}$~GeV,
and $\gamma \sim 10^{-21}$~GeV. A comparable limit on $\gamma$ has been placed by DA$\Phi$NE's
KLOE experiment, and in that case the analysis was based~\cite{nickCPTshort,kloe1,didomALL}
on entangled kaon states.

I should stress that this is clearly a quantum-spacetime picture (at least in as much
as it models spacetime foam) and the objective of the associated research program
is of introducing quantum/foamy properties of spacetime at the Planck scale,
but it is at present still unclear
which levels of sensitivity to $\alpha$, $\beta$, $\gamma$
would correspond to foaminess of spacetime at the Planck scale.
We are still unable to perform a
derivation starting  from foaminess at the Planck scale and
deriving corresponding values for $\alpha$, $\beta$, $\gamma$.
It is nonetheless encouraging that the
present experimental limits on these (dimensionful) parameters
are in a neighborhood of the Planck-scale-inspired
quantification $m_K/E_p \sim 10^{-19}$
(but it should be noticed that as much ``Planck-scale inspiration''
should be attributed, for example,
to the scale $m_K^2/E_p^2 \sim 10^{-38}$).

\subsubsection{Other descriptions of foam-induced decoherence for
  matter interferometry}

Another attempt to characterize spacetime foam as a decoherence-inducing medium
was developed by Percival and collaborators (see, e.g., Refs.~\cite{perci1,perci0,perci2}).
Also this approach assumes that ordinary quantum systems should all be treated
as open systems because of neglecting the degrees of freedom of the spacetime foam,
but, rather than a formalization using density matrices,
Refs.~\cite{perci1,perci0,perci2} adopt a formalism in which an open quantum
system is represented by a pure state diffusing in Hilbert space.
The dynamics of such states is formulated in terms of ``Primary state diffusion'',
an alternative to quantum theory with only one free parameter,  a time scale $\tau_0$ which one
can set to be the planck time $L_p/c$.

One way to charaterize  $\tau_0$ is through a formula for the proper time interval
for a timelike segment, which is given by~\cite{perci2}
\begin{eqnarray}
\Delta s \simeq |\Delta \xi (x)|^2 + \Delta \xi (x) \sqrt{\tau_0}~, \label{dsperci}
\end{eqnarray}
where $\Delta \xi (x)$ are point-dependent fluctuations induced by
the foaminess/quantization
of spacetime which are modelled within the proposed theory.

A key characteristic of this picture would be~\cite{perci2} a suppression
of the interference pattern for interferometers using beams of massive particles
(and such that of course an original beam is first split and then reunited to seek
an interference pattern). And the suppression increases with the mass of the particles,
so it could be more easily tested with atom interferometers (rather neutron interferometers).
Unfortunately a realistic analysis of an interferometer in the relevant primary-state-diffusion
formalism is much beyond the level of answers one is (at least presently) able to extract
from the primary-state-diffusion setup.
Ref.~\cite{perci2}
considered resorting to some
simple-minded simplifications, including
the assumption that the Hamiltonian be given by the mass together with projectors onto
the wave packets in the arms of the interferometer, neglecting the kinetic-energy
terms. Within such simplifications one does find that values of $\tau_0$ at or even a few
 orders of magnitude
below the Planck time would leave an observably large trace in modern atom interferometers.
These simplifications however amount to a model of the interferometer
which is much too crude (as acknowledged by the authors themselves~\cite{perci2})
and this does not allow us to explore meaningfully the possibility of genuine
Planck-scale sensitivities being achieved by this strategy.
One should already notice that by taking $\tau_0$ as the Planck
time it is not obvious that the effects are being introduced genuinely at the Planck scale,
since the nature of the effects is characterized not only by $\tau_0$ but also
by other aspects of the framework, such as the description of the fluctuations.
Moreover, even if all other aspects of the picture were understood,
indeed
the crudity of the model used for matter interferometers would still
not allow us to investigate the Planck-scale-sensitivity issue.

Recently, Ref.~\cite{tinoadded} and Ref.~\cite{bobcharles}
presented somewhat different pictures of
quantum-gravity-induced decoherence for atom interferometers. Several aspects of
the Percival setup are maintained but different intuitions are applied in some aspects of the
analysis. For example, Ref.~\cite{bobcharles} removes part of the assumptions
adopted by Percival and  collaborators,
 particularly in relation to the description of the ``quantum fluctuations''
of the metric,
 and proposes an estimate of the amount of
suppression of the interference pattern\footnote{In Ref.~\cite{bobcharles}
the main focus was again on atom interferometry, but following a closely related
approach the very recent Ref.~\cite{wangbinghamBOSEINSTEIN} made a proposal
for Bose-Einstein condensates.}
which is perhaps more intriguing from a phenomenology
perspective, since it would suggest that the effect is just beyond present sensitivities
(but within the reach of sensitivities achievable by atom interferometers in the not-so-distant
future).
For these recent proposals one is still (for reasons analogous
to the ones I just discussed for the Percival approach) unable to meaningfully explore
the issue of ``genuine Planck-scale sensitivity'', but they may represent a step in
the direction of a more detailed description of spacetime foam, if intended as
fluctuations of the metric.


\subsection{Decoherence and neutrino oscillations}
\label{decoherence-and-neutrinos}

The observations briefly discussed in the previous subsection that are relevant
for the study of manifestations of foam-induced decoherence in some laboratory
experiments (neutral-meson studies, atom interferometers) can be very naturally
applied also to neutrino astrophysics, as discussed in
the recent Ref.~\cite{nick08010872} and references therein
(also see
Refs.~\cite{floreaniniNEUTRI,ahluDECOneutri,winstaDECOneutri,cagoDECOneutri}).
Also in the neutrino context it is natural to attempt to develop test theories
codifying the intuition that spacetime foam may act as an environment,
so that neutrino observations
would have to be analyzed considering the relevant neutrino system as an open system.
And the evolution of the neutrino density matrix could be described
(in the same sense of
the description in Eq.~(\ref{ehnseq}) for neutral-meson systems)
by an evolution equation of the type
\begin{eqnarray}
 \partial_t \rho = i [\rho,H] + \delta \! H \, \rho
~. \label{ehnsNEUTRINO}
\end{eqnarray}
It is argued in Ref.~\cite{nick08010872} that such a formalization of the effects
of spacetime foam should generate a contribution to the mass difference between different netrinos,
and could give rise to neutrino oscillations constituting a ``gravitational MSW effect''.

As an alternative to the setup of Eq.~(\ref{ehnsNEUTRINO})
one could consider~\cite{nick08010872,nickHEPPH0606048}
the possibility of random (Gaussian)
fluctuations
of the background space-time metric over which the neutrinos propagates.
For the random metric one can take~\cite{nick08010872,nickHEPPH0606048}
a formalization of the type
\begin{eqnarray}
g^{\mu \nu}
=\left(
\begin{array}{c c}
- (a_1 +1)^2+a_2^2     -a_3 (a_1 +1)+a_2(a_4 +1)\\
-a_3 (a_1 +1)+a_2(a_4 +1)     -a_3^2 + (a_4 +1)^2
\end{array}
\right)
\label{nickrandomG}
\end{eqnarray}
and enforce~\cite{nick08010872,nickHEPPH0606048} for the random gauissian variables $a_i$
a parametrization based on parameters $\sigma_i$ (one per each $a_i$)
such that $<a_i>=0$ and $<a_i a_j>=\delta_{ij} \sigma_i$.
These fluctuations of the metric are found~\cite{nick08010872,nickHEPPH0606048} to
induce decoherence even when the neutrinos
are assumed to evolve according to a standard Hamiltonian setup,
\begin{eqnarray}
 \partial_t \rho = i [\rho,H] ~.
\label{ehnsSTANDARD}
\end{eqnarray}
But the decoherence effects generated in this framework with standard Hamiltonian
evolution in a nonstandard (randomly-fluctuating) metric, are significantly different
from the ones generated with the nonstandard evolution equation~(\ref{ehnsNEUTRINO})
in a standard classical metric. In particular,
in both cases one obtains neutrino-transition probabilities with decoherence-induced
exponential damping factors in
front of the oscillatory terms, but in the framework with evolution equation~(\ref{ehnsNEUTRINO})
the scaling with the oscillation length (time) is naturally linear~\cite{nick08010872,nickHEPPH0606048},
whereas adopting
standard Hamiltonian evolution in a fluctuating metric it is natural~\cite{nick08010872,nickHEPPH0606048}
to have quadratic scaling with the oscillation length (time).

The evidence growing for ordinary-physics neutrino oscillations,
which one expects to be much more significant than the foam-induced ones,
provides of course a formidable challenge for the phenomenology based on these
test theories for foam-induced decoherence in the neutrino sector.
Some preliminary ideas on how to overcome this difficulty
are described in Ref.~\cite{nick08010872}.
From the strict quantum-spacetime-phenomenology perspective of requiring to establish
that the relevant measurements could be sensitive to  effects
introduced genuinely at the Planck scale, these neutrino-decoherence test theories
must face challenges that I already discussed for a few other test theories:
there is at present no rigorous/constructive
derivation of the values of the parameters of these test theories
from a description (be it a full quantum-spacetime theory or simply a toy model)
of effects introduced genuinely at the Planck scale, so one can only express these parameters
in terms of the Planck scale using some dimensional-analysis arguments.


\subsection{Planck-scale violations of the Pauli Exclusion Principle}\label{vipsection}
A case for Planck-scale sensitivity was recently made~\cite{balaCHIprl,balaCHIjhep}
also for the hypothesis of possible violations of the Pauli Exclusion Principle.
This has still not been metabolized by an appreciably wide
quantum-gravity community,
but it certainly deserves to be highlighted briefly in this review, since the
chances for gradually gaining a strong impact on quantum-spacetime phenomenology
are rather high.

As observed already a few times in this review,
the spin-statistics theorem assumes a classical spacetime with ordinary locality.
And it is therefore legitimate to speculate that small departures from the implications
of the spin-statistics theorem may arise in a quantum spacetime.
Some earlier suggestions that this might be the case can be found, e.g., in
Refs.~\cite{balaSPINstat,chakPAULI,michdarioSTAT}, but the setup then was not such that
one could see an emerging case for Planck-scale sensitivity.

The recent studies reported in Refs.~\cite{balaCHIprl,balaCHIjhep}
investigated this issues assuming the specific form of spacetime noncommutativity
given by
\begin{equation}
[x_i,x_j]=0~,~~~[x_0,x_i]=i \chi \epsilon_{kij}n_k x_j~,
\label{chibala}
\end{equation}
where $n_k$ are the components of a fixed spatial unit vector and the deformation
length scale $\chi$ can be interestingly taken to be of the order of the Planck length.

It is actually rather easy to show that this form of noncommutativity imposes
a corresponding modification of the ``flip operator'', i.e. the operator
that is used for symmetrization (anti-symmetrization) purposes
in the commutative-spacetime case.
In turn this gives rise to a deformed description of bosons and fermions.
And the end result is that certain  transitions that would be Pauli-forbidden
in a commutative spacetime are actually allowed, although of course they have a small
rate (suppressed by the smallness of $\chi$).

Computing these rates on the basis of (\ref{chibala}) is at present only possible
relying on an uncomfortable number of simplifying assumptions~\cite{balaCHIprl,balaCHIjhep},
but the outcome is nonetheless intriguing since it suggests that
sensitivity to values of $\chi$ of order the Planck length is within reach.
This exploits the high sensitivity toward possible violations
of the Pauli Exclusion Principle at ongoing experiments, such as Borexino~\cite{borexino}
and VIP~\cite{vip}.


\subsection{Phenomenology inspired by causal sets}\label{causalsetphen}
As mentioned most of the quantum-spacetime phenomenology
of this past decade has been inspired
by results on spacetime noncommutativity and/or Loop Quantum Gravity.
But several other approaches are getting closer to inspiring
phenomenological programmes.
I share the view of many quantum-spacetime phenomenologists
who are looking
at the approach based on
causal dynamical
triangulations~\cite{ambjorn1993,lollLRR,ambjornlollPRL,ambjornlollPRD,lollDESKTOP,ambjornloll2010}
as a maturing opportunity
for inspiring the phenomenology work.
And first indications are coming from the ``asymptotic safety
approach''~\cite{weinbergSAFETY,reuterSAFETY,percacciSAFETY,reuter0702051,reuterMinLength},
on which I shall comment in relation to a tangible proposal
for phenomenology later in this review.
And most definitely in recent years we have seen a blossoming
phenomenology emerging from the
the causal set program.

Since I just discussed, in the previous subsection, the concept
of non-systematic effects, I find it appropriate to place here an aside
on this recent phenomenology inspired by
the causal set program.
Indeed, because of the perspective that guides that research program,
most (if not all) new effects predicted
within the causal set program
will be of non-systematic type.

Causal sets are a discretization of spacetime that allows
the symmetries of general relativity to be preserved in the continuum
approximation~\cite{sorkinPRLcausalsets,sorkinRideoutCONTINUUM,hensonREVIEW}.
And causal sets can be used to construct simple
 models suitable for exploring
 possible manifestations of fuzziness of quantum spacetime.
Moreover, the causal set proposal has recently been combined with the loop representation
to formulate ``causal spin foams''~\cite{fotinileeCAUSAL},
thereby establishing a link to an already mature source of inspiration
for quantum-spacetime phenomenology.

Clearly some of the manifestations one must expect from
a causal-set setup
fall within the class of phenomena here already briefly
described in Subsection~\ref{fuzzytimes}:
at a coarse-grained level of analysis a causal-set background
should introduce an intrinsic limitation to the accuracy of lengths
and durations.
Several recent works were aimed at formalizing and modeling these
aspects of fuzziness for propagation~\cite{johnstonDISCRETE,johnstonCAUSALSETS,causalsetsLIEU}.
The preliminary indications that are emerging appear to suggest
that, if discreteness is indeed introduced at the Planck scale,
the effects are very soft (hard to detect).
Nonetheless we do have already a few examples
of studies aiming for tangible predictions
to be compared to actual data: for example,
Ref.~\cite{scargleCAUSALSETS} reports a causal-set-inspired analysis
of possible fuzziness of arrival times (the sort of effects I already discussed
in Sec.~\ref{fuzzytimes}), relevant for studies conducted at gamma-ray telescopes.

An intriguing effect
of random fluctuations in photon
polarization can also be motivated by the causal-set
framework~\cite{dowkerCausalphen}. The presently-available
models of this causal-set-induced effect
are to be viewed as very crude/preliminary,
particularly
since the present understanding of the framework
is still not at the point of providing a definite model of photons propagating
on a causal set background (from which then one could derive the
polarization-fluctuation feature).
Still this appears a very promising direction, especially since
experimental information on CMB polarization
is improving quickly and will keep improving in the coming years.

Presently the most tangible phenomenological plans inspired
by the causal-set framework revolve around
an effect~\cite{dowkerSorkinMPLA2004,dowkerSorkinPRD2009}
of Lorentz invariant diffusion in the 4-momentum of massive
particles. This is
an Ornstein-Uhlenbeck process,
a diffusion process on the mass-shell which results in a
stochastic evolution in spacetime.
An intuitive picture for this mechanism was given
in Ref.~\cite{dowkerSorkinMPLA2004}, by considering
a classical particle, of mass $m$ propagating on a random spacetime lattice.
The particle would then be constrained to move from point to point,
but the discretization is such that in order to ``reach the next point''
(remaining on the lattice)
the particle must `swerve' slightly, also adapting to the swerving its velocity $v$
(also see Ref.~\cite{philpoCAUSAL} for a comparison of possible variants
of the description of particle propagation in causal-set theory).
The change in velocity also amounts to the particle jumping to
 a different point on its mass shell.
 The net result of this swerving is
 that~\cite{dowkerSorkinMPLA2004,kaloperMATTINGLY,mattinglyCausalSets}
  a collection of particles initially with an energy-momentum distribution $\rho(p)$
  will diffuse in momentum space along their mass shell according to the
  equation
\begin{equation}
\frac{\partial \rho}{\partial \tau} = {\cal K} \nabla^2_{\cal P} \rho
  - \frac{1}{m} p^\mu \partial_\mu \rho~,
\label{rhopswerving}
  \end{equation}
  where~\cite{dowkerSorkinMPLA2004,kaloperMATTINGLY,mattinglyCausalSets}
   $ {\cal K}$ is the diffusion constant, $\nabla^2_{\cal P}$
  is the Laplacian in momentum space on the mass shell of the particle, $\tau$
  is the proper time, and $\partial_\mu$
  is an ordinary spacetime derivative.

  The tightest limit on $\cal K$ is ${\cal K} < 10^{-61}GeV^3$,
  and was obtained~\cite{kaloperMATTINGLY}
from limits on the amount of relic-neutrino contribution to hot dark matter.
This follows from the observation that energy on the mass shell is bound
from below by the mass, so that particles close to rest, when swerving,
essentially can only increase their energy.

Interestingly these  $\cal K$-governed effects can also be
relevant~\cite{mattinglyCausalSets}
for some of the phenomenology I already discussed in this review,
concerning the threshold requirements for certain particle-physics processes.
Essentially the expected implications of swerving for these threshold analyses
is similar to what I already discussed in the previous
Subsection~\ref{nonsyste}: in any given opportunity of interaction between, say,
a hard proton and a soft photon the swerving can effectively raise or lower
(from the perspective of the asymptotically incoming states)
the threshold requirements for, say, pion production.
It appears however that the bound ${\cal K} < 10^{-61}GeV^3$,
if applicable also to protons\footnote{It is not implausible that
  different particles would be characterized by different values of $\cal K$.
  This in particular could reflect the expectation that more pointlike
  particles, such as neutrinos, should be more sensitive to the spacetime-lattice
  structure with respect to composite particles (such as protons and, even more evidently,
  atoms).
  It appears however natural to assume~\cite{mattinglyCausalSets},
  at least in these first explorations of causal-set phenomenology,
  that different particles would have values of  $\cal K$
  that are not too far apart.}, brings the magnitude of such effects
  safely beyond the reach~\cite{mattinglyCausalSets}
   of ongoing cosmic-ray studies.

\subsection{Tests of the Equivalence Principle}

\subsubsection{Aside on tests of the Equivalence Principle
in the semiclassical-gravity limit}
I am focusing in this review on tests motivated by  (and on effects modeled
within) proposals of spacetime quantization at the Planck scale,
but concerning tests of the Equivalence Principle inspired
by quantum-spacetime models there is some merit into making
a small digression on tests of the Equivalence Principle
in the semiclassical limit of quantum gravity (where, by construction,
no quantum-spacetime effects could be seen).
This will allow me to set up
compellingly the issue of testing the Equivalence Principle
from a general quantum-gravity perspective and specifically
from the perspective of spacetime quantization at the Planck scale.

As I already discussed briefly in Section~\ref{intro}, there is
a long
tradition of phenomenological studies,
 concerning the semiclassical-gravity limit,
  based on a ``gravity version'' of the Schr\"{o}dinger equation
of the form
\begin{eqnarray}
\left[ - \left( {1 \over 2 \, M_I} \right) \vec{\nabla}^2
+ M_G \, \phi(\vec{r}) \right] \psi(t,\vec{r}) = i  \,
{\partial \, \psi(t,\vec{r}) \over \partial t} ~,
\label{epextra}
\end{eqnarray}
describing the dynamics of matter (with wave
function $\psi(t,\vec{r})$, inertial mass $M_I$ and gravitational mass $M_G$)
in an external gravitational potential $\phi(\vec{r})$.
Some of the most noteworthy results obtained within this framework
are the interferometric studies of the type first set up
by Colella, Overhauser and Werner~\cite{cow}, which established
that the Earth's gravitational field is strong enough to affect
the evolution of the wave
function $\psi$ in observably-large manner, and
the more recent evidence~\cite{natureNeutron}  that
ultracold neutrons falling towards a horizontal mirror
do form gravitational quantum bound states.

Of relevance here is the fact that
some of the issues that have been most extensively considered
by researchers involved in
these studies concern the Equivalence Principle.
This is signaled
by the adoption of separate notation for inertial and
gravitational mass in Eq.~(\ref{epextra}).
In principle the gravitational mass $M_G$
 governs the accrual of
gravity-induced phases, while the inertial mass $M_I$
 intervenes in determining
the ratio between wave vector and velocity vector in the Galilean
limit ($p \ll m$).
And even for  $M_G = M_I$ the mass does not factor out of the free-fall
evolution of the quantum state (but for $M_G = M_I$
one at least recovers~\cite{onofrioviola} a complete identification between
the effects of gravitation and the effects of acceleration).

Besides neutrons these studies can also be performed
with atoms~\cite{dharamEP}.
And interestingly, one can also
perform rather similar analyses in studying
neutrino oscillations, finding (see, e.g.,
Refs.~\cite{gasperiniEP,halperinLeung,camachoEP} and references therein),
that gravity may induce neutrino oscillations
if different neutrino flavors are coupled differently
to the gravitational field,
thereby violating the
Equivalence Principle.


\subsubsection{On the Equivalence Principle in quantum spacetime}

Evidently
searches of possible violations of the Equivalence Principle in the
semiclassical-gravity limit
of quantum gravity have significant intrinsic interest.
And some of these tests of the Equivalence Principle in
semiclassical-gravity limit
find also explicit motivation in approaches to the study of the full
quantum-gravity problem: most notably
the string-theory-inspired studies reported in
 Refs.~\cite{stringsEP,repDamourPOLY1,repDamourPOLY2,damour,damourEP2,repDamourVARYING},
and references
 therein,
 predict violations of the Equivalence Principle in the
semiclassical-gravity limit.

Returning to the main subject of this review, I should stress that
the idea of spacetime quantization at the Planck scale provides
a particularly crisp motivation for testing the Equivalence Principle.
The simplest way to see this comes from observing the role
of absolute and ideally sharp locality in the role that the Equivalence
Principle plays in classical gravity, in contrast to the large class
of qualitative very severe (though tiny) anomalies for locality
that the various known scenarios for spacetime quantization
(starting with spacetime noncommutativity for example)
provide.
Unfortunately often our present level of
mastery of the relevant formalisms
falls short of allowing us to investigate the fate of the Equivalence Principle.
I shall therefore just here briefly describe one illustrative example of promising attempt
to model how spacetime foam could affect the Equivalence Principle.
This is the objective of recent studies, reported in Ref.~\cite{clausEP}
and references therein,
in which spacetime foam is modeled in terms
 of small fluctuations of the metric on a given background metric\footnote{Similar
types of ``fuzzy metrics'' were considered
in Refs.~\cite{fordCMB,yuALAford,polarskiALAford} but I postpone their discussion
to the section devoted to quantum-spacetime cosmology.}.
The analysis of Ref.~\cite{clausEP}, which also involves
an averaging procedure over a finite space-time scale,
 ends up motivating the study of a modified
Schroedinger equation of the form
\begin{eqnarray}
\left[ - \left( {1 \over 2 \, m} \right) (\delta^{kl}+{\tilde{ \alpha}}^{kl})
  \partial_k \partial_l
- m \phi(\vec{r}) \right] \psi(t,\vec{r}) = i  \,
\partial_t \, \psi(t,\vec{r})
\label{epextraCL}
\end{eqnarray}
where the tensor ${\tilde{ \alpha}}^{kl}$ is a characterization of the spacetime
foaminess, and it is natural to consider the tensor ${\tilde{ m}}^{kl}$,
\begin{eqnarray}
({\tilde{ m}}^{kl})^{-1} \equiv {1 \over  m} (\delta^{kl}+{\tilde{ \alpha}}^{kl})
\label{epextraCL2}
\end{eqnarray}
as an anomalous inertial mass tensor which depends on the type
of particle and on the fluctuation scenario.
The particle-dependent rescaling  of the inertial mass
provides a candidate key manifestation of foam-induced violations of the Equivalence
Principle to be sought experimentally, in ways that are once again exemplified
by the COW experiments.

This very recent proposal illustrates a type of path that could be followed
to introduce violations of the Equivalence Principle originating genuinely
from spacetime quantization at
the Planck scale: one might find a way to describe spacetime foamyness in
terms of effects of genuinely Planckian size, and then elaborate the
implications of this spacetime foaminess for the Equivalence
Principle.
The formalization adopted in Ref.~\cite{clausEP}
is still too crude to allow such an explicit link between the Planck-scale
picture of spacetime foam and the nature and magnitude of the effects,
but provides a  significant step toward that direction.

\newpage


\section{Infrared Quantum-Spacetime Phenomenology}
\label{uvir}

Work on Planck-scale quantum-spacetime phenomenology is
a rather recent development, with a significant effort
taking place only over little more than a decade.
But one could already make a distinction
between ``traditional'' and ``novel'' quantum-spacetime phenomenology approaches.
The proposals I have reviewed in the previous two sections
cover the scopes of the ``traditional'' approach, considering UV
(ultraviolet) effects
that could be relevant for observations in astrophysics and/or in controlled-laboratory
experiments.
I devote this and the next section to the ``novel'' ideas that
Planck-scale quantization of spacetime could have valuable phenomenological
implications also in some IR (infrared) regimes
and/or that the tests could rely on cosmology.

Considering that these ``novel'' areas of quantum-spacetime phenomenology
are in a preliminarily exploratory phase I will adopt in some sense
lower standard in the selection of topics, meaning that I will mentioned
even some proposals that have not fully established a link
to a definite scheme of spacetime quantization and/or have not
fully established the availability of sensitivities that could be
compellingly linked with the introduction of spacetime quantization
at the Planck scale. I will rather rely on
an (inevitably subjective)
assessment of whether
the relevant proposals
provide valuable first steps in the direction of establishing
in a not-so-distant future
robust Planck-scale quantum-spacetime phenomenology.

\subsection{IR quantum-spacetime effects and UV/IR mixing}

In the long~\cite{stachelearly,crHISTO}, and so far inconclusive, search of
quantum gravity and quantum spacetime
the main strategy
was inspired by the discovery paradigm
of the 20th century, the ``microscope paradigm''
with discovery potential measured in terms of
the shortness of the distance scales probed.
But recent research has raised the possibility
that by quantizing spacetime at the Planck scale
one might have not only
some new phenomena in a
far-UV  regime, but also some new phenomena
 in a ``dual'' IR
regime. Actually,
as compellingly stressed in Ref.~\cite{cohenUVIR},
 our present understanding
of black-hole thermodynamics, and particularly
the scaling $S \propto R^2$ of the entropy
of a black hole of radius $R$,
suggests that such effects of ``UV/IR  mixing''
may be inevitable.
It is on the basis of apparently robust hypotheses concerning
the behavior of quantum gravity in the UV (Planckian) regime that
we arrive at this quadratic dependence, which is surprising with respect
to what one might expect in particular in quantum field theory,
where cubic scaling ($S \propto R^3$) naturally arises.
But this feature originating from the UV sector clearly should have its most profound
implications in the large-distance/IR regime since the difference
between quadratic dependence on the radius and cubic dependence on
the radius becomes more and more significant as the radius is
increased\footnote{For example, for Planck-length radii one might even imagine
to satisfy the Bekenstein-Hawking bound $S \leq R^2$
in a theory where
actually entropy scales with a cubic law of the type $S = \alpha R^3$,
if the proportionality parameter $\alpha$ is such
that $S = \alpha R^3  \leq R^2$. But evidently for any such $\alpha$ there
are values of the radius large enough that instead
the  Bekenstein-Hawking bound would be violated.}.

Another argument in favor of UV/IR mixing is found considering
a popular intuition for  quantum spacetime, which
relies on the introduction of an uncertainty principle for spacetime
itself (in addition to the Heisenberg one, which acts in phase space).
The link with UV/IR mixing can be seen already by simply
considering a principle of the form $\delta x \delta y \geq \lambda_*^2$
for spatial coordinates,
with $\lambda_*$ plausibly of the order of the Planck length.
This type of uncertainty relations would evidently imply that small uncertainty
in $x$ should require large uncertainty in $y$, and this suggests a link
between probing short distance scales (small $\delta x$) and probing
large distance scales (large $\delta y$).

For this last point we have more than general arguments:
 computations in a noncommutative spacetime compatible
with this sort of uncertainty relations,
the ``canonical spacetime'', with
noncommutativity of  coordinates governed
by  $[x_\mu ,x_\nu]=i \theta_{\mu \nu}$,
have found explicit manifestations of UV/IR mixing.
This is particularly evident when analyzing mass renormalization
within the most popular formalization of quantum field theories
in such canonical noncommutative spacetimes.
At one loop one finds terms in mass renormalization
of the form~\cite{dougnekr,szaboREVIEW,susskind}
(for a ``$\Phi^4$ scalar field theory'')
\begin{equation}
\Delta^{\rm renorm}_{m^2}=\frac1{32}\,\frac{g^2\Lambda_{\rm eff}^2}{\pi^2}-\frac1{32}\,
\frac{g^2m^2}{\pi^2}\,\log \frac{\Lambda_{\rm eff}^2}{m^2}+\mathcal{O}(g^4) \ ,
\label{massren}\end{equation}
where $\Lambda_{\rm eff}$ is a peculiar cutoff which can be expressed in
terms of a standard ultraviolet cutoff $\Lambda$,
the ``noncommutativity matrix''  $\theta_{\mu \nu}$ and the momentum $q_\mu$
of the particle as follows
\begin{equation}
\Lambda_{\rm eff}^2=\frac1{\Lambda^{-2}+q_\rho (\theta^2)^{\rho\sigma} q_\sigma} \ .
\label{Lambdaeff}\end{equation}
Removing the cutoff $\Lambda$ ($\Lambda\to\infty$)
one is left with $\Lambda_{\rm eff}^2=1/[{q_\rho (\theta^2)^{\rho\sigma} q_\sigma} ]$,
so that
\begin{equation}
\Delta^{\rm renorm}_{m^2} \sim \frac{g^2}{q_\rho (\theta^2)^{\rho\sigma} q_\sigma}
+ g^2m^2 \,\log [m^2  q_\rho (\theta^2)^{\rho\sigma} q_\sigma ] \, .
\label{massrenIR}\end{equation}
These power-law and logarithmic IR features are the result of the UV implications
of noncommutativity, which manifest themselves in
 a rearrangement of the renormalization
 procedure~\cite{dougnekr,szaboREVIEW,susskind}.
 In general the presence of such sharp features in the IR may be of some concern,
 since they have not (yet) been observed.
 And these concerns are of course more serious in the cases where
 these features are sharpest.
 It should however be noticed that different choices of the
 matrix $\theta_{\mu \nu}$ produce very different types of IR behavior,
 and it is well established
  that in presence (at least in the UV sector)
  of supersymmetry only the logarithmic IR features
  survive (the power-law corrections are removed by one of
  the standard supersymmetry-induced cancellation mechanisms).
  The least virulent IR scenario is obtained by assuming the presence of
  UV supersymmetry
and  choosing a ``light-like'' noncommutativity
matrix~\cite{lightconeNCFT1,lightconeASCHIERI}
($\theta_{\mu\nu} \theta^{\mu\nu} = \epsilon_{\mu\nu\rho\sigma} \theta^{\mu\nu} \theta^{\rho\sigma}=0$), so that the main IR feature is a modifications of the on-shell
relation of the form
\begin{eqnarray}
m^2 \simeq E^2 - p^2 + \chi_\theta \, m^2 \log \left( \frac{E+{\vec{p}} \cdot {\hat{u}}_\theta}{m} \right)
~.
 \label{softIR}
\end{eqnarray}
The  unit vector ${\hat{u}}_\theta$ describes a preferential
direction~\cite{lightconeNCFT1}
determined by the matrix $\theta_{\mu \nu}$,
while
the dimensionless parameter $\chi_\theta$ also allows
for an expected~\cite{susskind} dependence of the magnitude
of the effect on the specific particle under study: since the IR feature is
found in the renormalization procedure, and this in turn has an obvious
dependence on the interactions of a given field with other fields in
the theory, the coefficient of the logarithmic IR correction has
different value for different fields.

Also note that in the IR regime (small $p$) one can rewrite (\ref{softIR})
as follows
\begin{eqnarray}
m^2 \simeq
E^2 - p^2 + \chi \, m \, {\vec{p}} \cdot  {\hat{u}}_\theta
~,
 \label{softIRlinear}
\end{eqnarray}
so that the effect ultimately amounts to a correction that is linear in
momentum. Clearly this is a scenario where the IR implications
of UV/IR mixing are particularly soft.

Interestingly canonical noncommutativity is not the only quantum-spacetime
proposal which can motivate the study of
 UV/IR mixing.
This is suggested by the
perspective on
the semi-classical limit of Loop Quantum Gravity
that provided motivation for
the quantum-spacetime model of Refs.~\cite{urrutia,urrutiaPRD},
which also inspired the models considered
in Refs.~\cite{repCortesTRITIUM,josePRD,gacPRL2009}.
In this Loop-Quantum-Gravity-inspired scenario
one also finds~\cite{urrutia,urrutiaPRD},
modifications of the dispersion relation which are linear in momentum
in the IR regime,
and this has provided motivation for a phenomenology
based \footnote{I shall here
set aside,
like Refs.\cite{josePRD,gacPRL2009},
 the possibility of an helicity dependence~\cite{urrutiaPRD} of the effects.}
 on the IR dispersion relation~\cite{repCortesTRITIUM,josePRD,gacPRL2009}
\begin{eqnarray}
m^2 \simeq
E^2 - p^2 + \chi_{\hat p} \, m \, p
~,
 \label{softIRurrutia}
\end{eqnarray}
where $\chi_{\hat p}$ is a phenomenological parameter\footnote{There is no
consistent adoption of conventions for notation
in Refs.~\cite{urrutia,urrutiaPRD,repCortesTRITIUM,josePRD,gacPRL2009}.
I am here adopting a notation which appears to render
more transparent the connection between arguments for UV/IR mixing
from Loop Quantum Gravity and arguments
for UV/IR mixing
from spacetime noncommutativity.}
analogous to $\chi_\theta$.

\subsection{A simple model with soft UV/IR mixing and precision Lamb-shift
measurements}
 The long-wavelength behaviour of the two scenarios
 for ``soft  UV/IR mixing'' here
 summarized in Eq.~(\ref{softIRlinear}) and Eq.~(\ref{softIRurrutia})
 evidently differ only because of the fact that invariance under spatial rotations
 (lost in (\ref{softIRlinear})) is preserved by the scenario
 here described in Eq.~(\ref{softIRurrutia}).
 One could therefore
  consider simultaneously the two scenarios,
by observing
that the characterization of (\ref{softIRlinear}) in terms of $\chi_\theta$
and ${\hat{u}}_{\theta}$ is applicable
to the scenario of (\ref{softIRurrutia})
by replacing ${\hat{u}}_{\theta}$ with ${\hat p} \equiv {\vec{p}}/p $
and replacing $\chi_\theta$ with  $\chi_{\hat p}$.
However,
in light of the limited scopes of my review of results
on ``soft  UV/IR mixing'', I shall be satisfied with a
simplified description, assuming space-rotation invariance
and limiting my focus to the effects of dispersion relations of
the form
\begin{eqnarray}
m^2 \simeq
E^2 - p^2 + \xi \, \frac{m^2}{E_p} \, p
~,
 \label{mysoft}
\end{eqnarray}
where I also introduced a change of definition of the dimensionless coefficient,
rescaling it in a way that might be relevant for connecting the infrared effects
with the Planck scale ($\chi \rightarrow \xi m/E_p$, which however
provides no loss of generality if $\xi$
is allowed to be particle dependent).

The phenomenology of models such as this requires a complete
change of strategy with respect to the phenomenology of quantum-spacetime
UV effects which I discussed in previous parts of this review.
Whereas the typical search of those UV effects relied on low-precision
high-energy data, for the type of IR effects which I am now considering
the best options come from high-precision low-energy data.
A first example of this was given in Ref.~\cite{josePRD}, most notably
with a (however brief) discussion of how a dispersion relation of type
(\ref{mysoft}) could be relevant for Lamb-shift measurements.
Indeed, assuming (\ref{mysoft}) holds for the electron
then one should have a modification of the energy levels of the hydrogen atom.
And in light of the high precision of certain Lamb shift measurements
(which Ref.~\cite{josePRD} assesses as being better than one part in $10^{5}$,
also see, e.g., Refs.~\cite{lambMeasKINOSHITA,lambMeasBIRABEN})
one can use this observation to place valuable limits on parameters such
as $\xi$ (and $\chi$)
for the electron.

\subsection{Soft UV/IR mixing and atom-recoil experiments}
Evidently the {\it ansatz} (\ref{mysoft}) is such that if particles of
different mass had the same value\footnote{As stressed already
a few times in this review, this is not necessarily a natural expectation.
In particular, several arguments
appear to suggest that composite particles may be less sensitive to
quantum-spacetime effects than ``fundamental'' particles. Still, of course
it is interesting to study this issue experimentally, and a way to do that is to
look for opportunities for the ``universality assumption'' to break down.}
of $\xi$ then the effect would be seen more easily
 for heavier (more massive) types of particles.

I find particularly striking the case of measurements of the recoil
of Caesium (and Rubidium) atoms.
For Caesium one would assume, following (\ref{mysoft}), that
\begin{eqnarray}
m^2 \simeq E^2 - p^2 + \xi_{Cs} m^2 \frac{p}{M_P}~,
 \label{softIRcoldatom}
\end{eqnarray}
where $\xi_{Cs}$ is  the $\xi$ parameter for the case of Caesium atoms.

The measurement strategy we proposed in Ref.~\cite{gacPRL2009}
for testing (\ref{softIRcoldatom}) with atoms
is applicable to measurements
of the ``recoil frequency'' of atoms with experimental setups
involving one or more ``two-photon Raman
transitions''~\cite{Wicht02}.
The strategy of the analysis is best described by setting
initially aside the possibility of Planck-scale effects,
and looking at
the recoil of an atom in a two-photon Raman transition
from the perspective adopted
in Ref.~\cite{Wicht02},
which provides a convenient starting point for the Planck-scale generalization
which is here of interest.
One can impart momentum to an atom through a process involving absorption
of a photon of frequency $\nu$ and (stimulated)
emission, in the opposite direction,
of a photon of frequency $\nu'$. The frequency $\nu$ is computed taking into
account a resonance frequency $\nu_*$
of the atom and the momentum the atom acquires,  recoiling upon absorption
of the photon: $ \nu \simeq  \nu_* + ( h \nu_* + p)^2/(2 m) - p^2/(2m)$,
where $m$ is the mass of the atom (e.g. $m_{Cs} \simeq 124\mathrm{\ GeV}$ for Caesium),
and $p$ is its initial momentum.
The emission of the photon of frequency $\nu'$ must be such to deexcite the atom
and impart to it additional
momentum: $\nu' + (2 h \nu_* + p)^2/(2 m) \simeq  \nu_* + (h \nu_*+p)^2/(2 m)$.
Through this analysis one establishes that by measuring $\Delta \nu \equiv \nu - \nu'$,
in cases  where $\nu_*$ and $p$ can be accurately determined, one
actually measures $h/m$ for the atoms:
\begin{eqnarray}
     \frac{\Delta \nu}{ 2 \nu_* (\nu_* +p/h)} = \frac{h}{m} ~. \label{deltaomeNOEP}
\end{eqnarray}
This result has been confirmed experimentally with remarkable accuracy.
A powerful way to illustrate this success is provided by comparing
the results of atom-recoil
measurements of $\Delta \nu/[\nu_* (\nu_* +p/h)]$ and of measurements~\cite{gab08}
of $\alpha^2$, the square
of the fine structure constant. $\alpha^2$ can be expressed in terms of the mass $m$
of any given particle~\cite{Wicht02} through the Rydberg constant, $R_\infty$,
 and the mass of the electron, $m_{{e}}$, in the following
 way~\cite{Wicht02}: $ \alpha^2 = 2 R_\infty \frac{m}{m_{{e}}} \frac{h}{m}$.
Therefore according to Eq.~(\ref{deltaomeNOEP}) one should have
\begin{equation}
\frac{\Delta \nu}{ 2 \nu_* (\nu_* +p/h)} =
\frac{\alpha^2}{2 R_\infty}
\frac{m_e}{m_u} \frac{ m_u}{m} ~, \label{alphaJ2}
\end{equation}
where $m_u$ is the
atomic mass unit and $m$ is the mass of the atoms used in
measuring $\Delta \nu/[\nu_* (\nu_* +p/h)]$. The outcomes of atom-recoil measurements,
such as the ones with Caesium reported in Ref.~\cite{Wicht02},
are consistent with Eq.~(\ref{alphaJ2})
with the accuracy
of a few parts in $10^9$.
The fact that Eq.~(\ref{deltaomeNOEP}) has been verified to such a high degree
of accuracy proves to be very
valuable, since it turns out~\cite{gacPRL2009}
that modifications of the dispersion relation of type (\ref{softIRcoldatom})
 require a modification
of  (\ref{deltaomeNOEP}).
Following Ref.~\cite{gacPRL2009} one easily finds
\begin{eqnarray}
\Delta \nu \!  & \simeq & \! \frac{ 2 \nu_* (h \nu_* +p)}{m} +  \xi_{Cs} \frac{m}{M_P} \nu_*
~,
\label{DeltaOmegaLeading}
\end{eqnarray}
and in turn in place of Eq.~(\ref{alphaJ2})
one has
\begin{equation}
\frac{\Delta \nu}{ 2 \nu_* (\nu_* \! + \! p/h)} \!\! \left[ \! 1
\! - \xi_{Cs} \! \left( \! \frac{ m}{2 M_P} \! \right) \!\! \left( \! \frac{m}{h \nu_* +p} \! \right)
 \! \right] \!\! = \!\!
\frac{\alpha^2}{2 R_\infty}
\frac{m_e}{m_u} \frac{ m_u}{m} ~.
\nonumber
\end{equation}
This equation has been arranged
so that on the left-hand side it is easy to recognize that
the small quantum-spacetime effect
 in this specific context receives a sizable ``amplification'' by the large
 hierarchy of energy scales $m/(h \nu_* +p)$, which
in typical experiments of the type here of interest can
be~\cite{Wicht02}
of order $\sim 10^{9}$.

This turns out to be just enough to provide the desired ``Planck-scale sensitivity'':
one easily finds that combining the measurements on Caesium
reported in Ref.~\cite{Wicht02} and
the  determination
of $\alpha^2$ reported in Ref.~\cite{gab08},
one can establish~\cite{gacPRL2009} that $\xi_{Cs} = - 1.8 \pm 2.1$.

And it is interesting that besides tests of IR modifications of the dispersion relation
these atom-recoil studies can also be used to investigate possible IR modifications
of the law of conservation of momentum. An example of such an analysis
is given in Ref.~\cite{michJUREatom}.

\subsection{Opportunities for Bose-Einstein condensates}
The use of atoms in quantum-spacetime phenomenology immediately confronts us with
issues that are presently beyond the reach of available theoretical results.
A legitimate expectation is that quantum-spacetime effects for atoms could be weaker
than for the particles that compose atoms, as a result of the sort of ``average-out effects''
which is often expected in the quantum-spacetime literature. This would have to be modeled
by introducing an extra suppression factor (a sort of ``compositeness factor'') in addition
to the Planck-scale suppression which is standard in quantum-spacetime phenomenology.
Analyses not making room for such an additional suppression might overestimate
the Planck-scale-sensitivity reach of the relevant experiments.
On the other hand we are at present not sure whether or not such compositeness
suppression factors are truly needed, or at least if they are needed in all contexts
and in all quantum-spacetime models. For example, it is not unreasonable to imagine
that in appropriate quantum-spacetime models, when we achieve the ability
to analyze them in detail, we might find that as long as a particle is to be handled
as a quantum state (far from its classical limit) then it might be irrelevant for the magnitude
of quantum-spacetime effects whether the particle is composite or ``fundamental''.

This issue of compositeness will surely gradually take an important role in quantum-spacetime
research, but at present it is at a very preliminary stage of investigation, and I shall therefore
here set it aside. I do note however that if particles composed of a very large number of
particles experience Planck-scale effects unsuppressed by their compositeness, then not
only atoms but also (and perhaps more powerfully) Bose-Einstein condensates could
prove to be a very valuable opportunity for quantum-spacetime phenomenology.

And it is noteworthy  that in the recent quantum-spacetime-phenomenology
literature there has ben already a surge of interest in the possibilities offered
by Bose-Einstein condensates, as seen for example in
Refs.~\cite{wangbinghamBOSEINSTEIN,camachoBE2011,brisceseBEepl,briscesePLB2012}.
In particular,
Refs.~\cite{brisceseBEepl,briscesePLB2012} study Bose-Einstein condensates
adopting a perspective on soft UV/IR mixing
that is closely related to the one I discussed for atoms in the previous subsection.

\subsection{Soft UV/IR mixing and the end point of tritium beta decay}
Perhaps the most tempting opportunity for the phenomenology of UV/IR mixing
comes from studies of the low-energy beta decay spectrum of tritium,
$^3\!H \rightarrow ~^3\!\!\,H\!e + e^- + {\bar{\nu_e}}$,
which
have produced so far some rather puzzling results~\cite{repWeinheimer,repLobashev}.
It is well understood (see, e.g. Refs.~\cite{repMagueijoTRITIUM,repCohenTRITIUM})
that these puzzles could be addressed by introducing deformed rules of kinematics.
And it is intriguing that studies conducted
near the endpoint of tritium beta decay are the only known way to investigate
rather accurately the properties of neutrinos in a non-relativistic (non-ultrarelativistic)
regime, where their momenta could be comparable to their (tiny) masses.
So it would seem to be a very natural opportunity for advocating
UV/IR mixing as possible explanation. The evidence available so far is
however
not very encouraging for the hope of attributing the magnitude of the reported
 anomalies to IR effects induced by the Planck scale. Still it is noteworthy that
 specifically the simple model for soft UV/IR mixing which I described in the previous
 subsections has just the right structure for producing the sort of anomalies
 which is being reported, as was first stressed in Ref.~\cite{repCortesTRITIUM}.

The main point of Ref.~\cite{repCortesTRITIUM} is centered on the properties
of the function $K(E)$ conventionally used to characterize the
Kurie plot of tritium beta decay:
\begin{equation}
K(E) = \Big[ \int dp_\nu \, p_\nu^2 \, \delta(Q-E-E_\nu) \Big]^{1/2}
\label{cortestritiA}
\end{equation}
where $Q$ is the difference between initial and final masses of the
process, $Q \simeq M_{^3\!H} - M_{^3\!\!\,H\!e} -m_e$
(and therefore $Q$ is the
 sum of the neutrino energy, $E_\nu$,
and the kinetic energy of the electron , $E$).

Using standard dispersion relations evidently one finds
\begin{equation}
K(E) = \Big[ (Q-E)  \sqrt{(Q-E)^2-m_\nu^2)} \Big]^{1/2}
\label{cortestritiB}
\end{equation}
which however does not fit well the available data near the
endpoint~\cite{repWeinheimer,repLobashev}.
It was observed in
 Ref.~\cite{repCortesTRITIUM}
 that using instead a modified dispersion relation
of type (\ref{mysoft}), for negative $\xi$,
one obtains better agreement,
but this requires for $ \xi m_\nu^2 /E_p$ a value of a few eV.
In turn this implies a value\footnote{I am describing the analysis
of Ref.~\cite{repCortesTRITIUM} using the notation I adopted throughout
this section. Instead of $\xi$ Ref.~\cite{repCortesTRITIUM} used a parameter $\lambda$
linked to $\xi$ by $\lambda = - \xi m_\nu^2 /E_p$.} of  $\xi$ which
is extremely large with respect to the natural
quantum-spacetime estimate $\xi \sim 1$, and as a result
the case for a quantum-spacetime interpretation is rather weak
at present. Still this exciting experimental situation deserves
to be further pursued: perhaps we are modeling soft UV/IR mixing
correctly but we have developed the wrong intuition about the role
the Planck scale should play, or perhaps one should look at alternative
ways to  model UV/IR mixing.

\subsection{Non-Keplerian rotation curves from quantum-gravity effects}
In addition to precision measurements on particles of peculiarly low momentum,
another very clear opportunity for UV-IR mixing is provided by data on
the behavior of gravity on very large distance scales.
And in that context speculating about new-physics phenomena
is fully justified by
the observed non-Keplerian features of the rotation curves of galaxies or
clusters~\cite{nokepler}.
These non-Keplerian features
are usually interpreted as motivation for introducing dark matter
(or other non-quantum-gravity new physics,
such as MOND~\cite{mond}), but, in light of the recent awareness
of the possibility of UV/IR mixing, it is legitimate to speculate
that they may be at least in part due to quantum-spacetime effects.

The perspective one might adopt in trying to profit from
this opportunity is similar to when one works within
 standard quantum field theories and
derives an ``effective potential'' (usually obtained through calculation
of loop contributions) that corrects the tree-level classical potential.

And interestingly the type of modifications of dispersion relations that have
been motivated by quantum-spacetime research do automatically suggest
that the Newtonian potential should receive some corresponding corrections.
In fact, the Newtonian potential is produced by a static point source when
the field that mediates the
force described by the potential has energy-momentum space (inverse)
propagator $G^{-1}(E,p)=E^2-p^2$.
In general, if the field that mediates the force has different propagator,
$G_{def}^{-1}(E,p)$,
the Newtonian potential produced at the spatial point $\vec r$ by a point-like mass $M$, located
 at the origin,
is replaced by the potential obtained by computing~\cite{hellingCOULOMBscreen}
\begin{equation}
V(\vec r) = L_p^2 M \int \frac{d^3p}{2 \pi^2}~G_{def}(0,\vec p) ~e^{i \vec p \cdot \vec r} ~,
\label{PotenzialeDaPropagatore}
\end{equation}
i.e. the potential is the spatial Fourier transform
of the propagator evaluated at $E = 0$.

A more articulated argument for modifications of the Newton potential
at large distances from a quantum-spacetime perspective
has been put forward as part of the mentioned research programme
on ``asymptotic safety''.
This is done in Ref.~\cite{reuter0702051}, which indeed adopts as working
assumption the availability of a
quantum field theory of gravity whose underlying degrees of freedom are those of the
spacetime metric,  defined nonperturbatively as a fundamental, ``asymptotically-safe'' theory.
Obtaining definite predictions for
the rotation curves of galaxies or clusters within this formalism is presently well beyond our
technical capabilities. However, preliminary studies of the renormalization-group
behavior
provide encouragement for a certain level of analogy between this theory and non-Abelian Yang--Mills
theories, and, relying in part on this analogy,
Ref.~\cite{reuter0702051} argued that one could obtain
non-Keplerian features from renormalization.

\subsection{Aside on gravitational quantum well}
Another opportunity for studies of UV/IR mixing is provided by measurements
performed on neutron quantum states in the gravity field of the Earth,
such as the striking ones reported in Refs.~\cite{natureNeutron,repNesviz}.
I have nothing to report on this that would fit the main focus of this review,
concerning Planck-scale quantum pictures of spacetime, but it seemed worth
mentioning this nonetheless, especially in light of the fact that this class of
low-energy studies (candidates for the investigation of UV/IR mixing)
have already been analyzed from the perspective of some quantum spacetimes,
even though so far all such studies have introduced spacetime quantization
at scales that are very far from the Planck scale (much lower energy scales,
much greater
distance scales).

Since I am already here diverting from the main theme of the review
I shall be satisfied confining the discussion of quantum-spacetime studies
of the gravitational quantum well to the particularly interesting points made in
Refs.~\cite{repBertolami,repBanerjee,repSaha,repBrau}.
The studies in Refs.~\cite{repBertolami,repBanerjee,repSaha} all assumed ``canonical noncommutativity'' of spacetime
coordinates:
\begin{equation}
[x_j , x_k] = i \theta_{j k}~,~~~[x_j , x_0] = i \theta_{j 0}
\label{canonicalNCST}
\end{equation}
where I separated the space/space noncommutativity ($\theta_{j k} \neq 0$)
from the space/time noncommutativity ($\theta_{j 0} \neq 0$).

And Refs.~\cite{repBertolami,repBanerjee,repSaha} agree on the fact
that pure space/space noncommutativity ($\theta_{j 0} = 0$)
has no significant implications
for the gravitational quantum well. However,
Ref.~\cite{repSaha}  notices that
with space/time noncommutativity ($\theta_{j 0} \neq 0$)
there are tangible consequences for the gravitational quantum well
so that in turn one can use the measurement results
of Refs.~\cite{natureNeutron,repNesviz} to put bounds on
space/time noncommutativity\footnote{Though not relevant for my review,
it is interesting to note that, besides constraining certain quantum pictures of
spacetime, the measurement results reported in Refs.~\cite{natureNeutron,repNesviz}
can also be used to set an upper limit for an additional short-range fundamental
force~\cite{repNesviz}.},
although only at the level $\theta_{j 0} < 10^{-9}m^2 $
(whereas interest from the Planck-scale-quantum-spacetime side would
focus in some neighborhood of $\theta_{j 0}  \sim 10^{-70} m^2 $).

Refs.~\cite{repBertolami,repBanerjee}  make the choice of
combining space/space noncommutativity with a noncommutativity
of momentum space:
\begin{equation}
[p_j , p_k] = i \psi_{j k}
\label{pIpJnoncomm}
\end{equation}
It then turns out that this
noncommutativity
of momentum space does affect tangibly the analysis of
the gravitational quantum well. So that in turn one can
use the measurement results
of Refs.~\cite{natureNeutron,repNesviz} to place bounds
at the level $\psi_{j k} < 10^{-6}eV^2 $.

Ref.~\cite{repBrau}
 is an example of analysis of the
gravitational quantum well not from the viewpoint of spacetime noncommutativity,
but rather from the viewpoint of the
 scheme of spacetime quantization introduced
in Refs.~\cite{stringsHP,kempf}, which is centered on
a modification of the Heisenberg principle
\begin{equation}
[x_j,p_k] = i \delta_{j k}  (1 + \beta p^2)~.
\label{xJpKnoncomm}
\end{equation}
The parameter $\beta$ does turn out~\cite{repBrau} to affect
the analysis of the gravitational quantum well, and using the
 measurement results
of Refs.~\cite{natureNeutron,repNesviz} one can place bounds
at the level $\beta < 10^{-18}m^2 $.

\newpage


\section{Quantum-Spacetime Cosmology}
\label{QGC}
In the previous sections
I discussed several opportunities for investigating
candidate quantum-spacetime effects through
observations in astrophysics and, occasionally,
some controlled laboratory setups.
It is likely however that gradually cosmology will acquire more and more weight
in the search of manifestations of quantum-spacetime effects.
In the earliest stages of evolution of the Universe
the typical energies of particles were much higher than the ones we can presently
achieve, and high-energy particles are of course the ideal probes
for the short-distance structure of spacetime.
Over these past few years several studies that could be viewed as preparing the ground
for this use of cosmology have been presented in the literature.
Most of these proposals do not have yet the structure and robustness necessary
for actual phenomenological analyses, such as the ones setting bounds on the
parameters of a given quantum-spacetime picture.
But the overall picture emerging from these studies confirms
the expectation that cosmology has the potential of being a key player in
quantum-spacetime phenomenology.

For a combination of reasons I shall review this recent literature in
even more sketchy way than for other parts of this review.
This reflects the fact that I view this area as still at a very early stage
of development: we are probably just starting to learn what could be the observable manifestations
of quantization of spacetime in cosmology. Even in cases when the quantum-spacetime side
is reasonably well understood, the study of the implications for cosmology, when focused
on possible observably-large manifestations, is still in its infancy.
Moreover, most of the works done so far in this area do not even invoke
a definite role for spacetime quantization, which is the main focus of this review,
but rather find inspiration in generic features of the quantum-gravity problem,
or rely on string theory (which, as stressed, is a fully legitimate quantum-gravity
candidate, but is one such candidate that, as presently understood, would
rather lead us to assume that quantum properties of spacetime
are absent/negligible).

In light of these considerations the list of proposals and ideas that composes
this section is not representative of the list of scenarios being considered in
quantum-spacetime cosmology. It serves mainly the purposes of offering
some illustrative examples of how one might go about proposing
a quantum-spacetime-cosmology scenario and giving some strength to
my opening remarks foreseeing a great future for quantum-spacetime cosmology.

In the last subsection of this section I also, even more briefly, mention some
examples of quantum-gravity-cosmology proposals, which in their present formulation
do not invoke a role for quantization of spacetime (but could inspire future reformulations
centered on a quantum-spacetime perspective).


\subsection{Probing the trans-Planckian problem with modified
  dispersion relations}
In the long run, one of the most significant opportunities
for quantum-spacetime phenomenology could be
an aspect of quantum-spacetime cosmology: the trans-Planckian problem.
Inflation works in such a way that
some of the scales that are presently of
cosmological interest should have been trans-Planckian scales at the beginning of inflation,
and therefore cannot be handled
satisfactorily without (the correct)
quantum gravity~\cite{brandeREVIEWinfla,brandePRD2001,danielssonINITIAL}.
In extrapolating the evolution of cosmological perturbations according to linear theory to very early times, we are implicitly making the assumption that the theory remains perturbative to arbitrarily high energies. And it is easy to see that the expected
new physics at the Planck scale could affect our predictions. For example, if there was
a sharp Planck-scale cutoff in the theory, then, if inflation lasts many e-folding, the modes which represent fluctuations on galactic scales today
would not be present~\cite{brandeREVIEWinfla}
in the theory since their wavelength would have been
smaller than the cutoff length at the beginning of inflation.

While in the long run this might get very exciting,
I feel we are at present only at a very early stage of exploration of the
potentialities of this opportunity for quantum-spacetime phenomenology.
But there is growing awareness of this opportunity and the related
literature starts to grow large (see, e.g.,
Refs.~\cite{brandePRD2001,danielssonINITIAL,niemeyerCUTOFF,brandCUTOFF, kempf2001,chung0011241, mersini0101210, shiu0104102,
  shiu0411217, danielsson0606474, kempf0609123, burgess07082865,hannestadCOSMOcutoff,starobCOSMO,parentaniMDR,brande2012},
and references therein).
Many of these studies~\cite{brandCUTOFF, chung0011241, mersini0101210, shiu0104102,
  shiu0411217, danielsson0606474, kempf0609123,hannestadCOSMOcutoff}
   have probed the possibility that a short-distance
cutoff
might leave a trace in cosmology measurements such as the ones conducted on the cosmic microwave
background.

Among the scenarios that have been so far considered
in relation to the trans-Planckian problem
the ones that are more directly linked with the study of spacetime quantization
are those involving Planck-scale modifications of the dispersion
relation (see, e.g., Refs.~\cite{parentaniMDR,cosmodsr,machadoALAbrande}).
For example, one may consider the possibility~\cite{cosmodsr}
of dispersion
relations with a trans-Planckian branch where energy increases with decreasing
momenta, such as
\begin{eqnarray}
\omega^2 \simeq k^2 - \alpha_4 k^4 + \alpha_6 k^6
\label{joaosteph}
\end{eqnarray}
for appropriate choices of the parameters $\alpha_4$ and $\alpha_6$.
A radiation dominated universe
with particles governed by such modified dispersion relations
ends up being characterized~\cite{cosmodsr}
by negative radiation pressure and remarkably
may be governed by an inflationary equation of state, even without introducing
 an inflaton field.

These results (and those of Refs.~\cite{livcosmos1,livcosmos2,livcosmos3,livcosmos4})
 establish a connection with previous attempts
of replacing inflation by scenarios accommodating departures from Lorentz symmetry,
such as the scenario with a time-varying speed of light,
introduced in works by Moffat~\cite{moffatVSL} and in works
 by  Albrecht and Magueijo~\cite{magueVSL1}  (also see Ref.~\cite{barrowVSL}).
By postulating an appropriate time variation of the speed of light one
can affect causality in a way that is somewhat analogous to inflation:
very distant
regions of the Universe, which could have never been in causal contact with a time-independent
speed of light, could have been in causal contact at very early times if at those
early times the speed of light was much higher than at the present time.
As argued in the recent review given in Ref.~\cite{magueVSLrev},
this alternative to inflation
is rather severely constrained but still to be considered a viable alternative
to inflation.


\subsection{Randomly-fluctuating metrics and the cosmic microwave
  background}
Also relevant for cosmology are the mentioned studies suggesting
that spacetime quantization could effectively produce spacetime fuzziness/foam
amenable to description in terms of
a fluctuating spacetime metric~\cite{fordCMB,yuALAford,polarskiALAford}
(also see Refs.~\cite{fordBLURRING2004,fordSPECTRALeBLURRING}).
In particular one can consider~\cite{fordCMB}
fluctuating spacetime metrics amounting to a fluctuating lightcone.
Such  fluctuations of the lightcone have implications for
the arrival times of signals from distant sources that would result
in a broadening of the spectra.
It was observed in
Ref.~\cite{fordCMB}
that starting with a thermal spectrum one would end up
with a slightly different spectrum.
This can be summarized in a simple phenomenological formula
for spectrum distorsion~\cite{fordCMB}
\begin{equation}
F(\omega) = F_0(\omega) [1+f(\omega)]
\label{fordspectrum}
\end{equation}
where $F_0(\omega)$ is the spectrum expected without the light-cone fluctuation effects
and $f(\omega)$ encodes the corrections due to the lightcone fluctuations.
Ref.~\cite{fordCMB} provides arguments in support of the possibility
that the corrections due to lightcone fluctuations could get very large at large
frequencies, with $f(\omega)$ growing like $\omega^4$.
As we achieve better and better accuracy in the measurement of possible
high-frequency departures from
a thermal spectrum
for the cosmic microwave background we should then find evidence
of such effects~\cite{fordCMB} .

Ref.~\cite{polarskiALAford}
also explored the possibility that  lightcone fluctuations might have observable implications
for the gravitational wave background. The gravitational wave background
is always emitted much before the cosmic microwave background,
but it was found~\cite{polarskiALAford} that
the flat nature of the gravitational-wave-background spectrum is such that the effects
of lightcone fluctuations are negligible.

\subsection{Loop quantum cosmology}
An area of quantum-spacetime cosmology which is not
directly linked to the tools and scenarios considered in other areas of
quantum-spacetime phenomenology is the so-called
Loop Quantum Cosmology~\cite{bojoFIRST,bojoFIRSTb,bojoLRR,ashtekarLQC} .
This is a framework for implementing several effects seen to arise
for the quantum spacetime of loop quantum gravity in a cosmological setting.
The most popular formulations of loop quantum cosmology
are defined on ``minisuperspace'',
where one quantizes homogeneous spacetimes using the
methods of loop quantum gravity.
And one finds that the characteristic discreteness of
the loop-quantum-gravity spacetime quantization
change the dynamics of expanding universe models. These changes are
 particularly significant at high densities,
 giving rise to mechanisms avoiding classical singularities.

And one can consider the novel quantum-spacetime effects also at
later stages of the Universe expansion, when densities are lower
and the corrections  can be treated perturbatively in a gauge-invariant way.
This can be done in particular for linear perturbations around spatially
flat Friedmann--Robertson--Walker models, and the results are found to
be primarily characterized in terms
of ``inverse-volume corrections'', due to the fact that the quantized densitized triad has a discrete spectrum, with the value zero contained in the spectrum.
Essentially one finds that the loop-quantum-gravity quantum-spacetime effects
can be effectively described
in terms of a novel repulsive force~\cite{ashtekarLQC}.
This repulsion can compete with the standard gravitational attraction,
and can even become the dominant contribution, thereby evading the
singularity, when the curvature is strong.

In Refs.~\cite{bojoLRR,ashtekarLQC,bojoPRL2011},  and references therein,
readers can find a list of possible signatures of Loop Quantum Cosmology.
In my opinion, at present, such tests of predictions of Loop Quantum Cosmology
may tell us more about the choice of setup for incorporating the quantum-spacetime
effects, rather than providing actual information on the quantum structure of spacetime.
But as this novel approach keeps maturing it may well turn into a key resource for
probing experimentally the quantum structure of spacetime.


\subsection{Cosmology with running spectral dimensions}
As mentioned earlier in this review,
several formalisms relevant for the study of the quantum-gravity
problem have recently been shown to host the mechanism
of running spectral dimensions.
The spectral dimension of a spacetime is
essentially defined~\cite{visserRUNNINGsd} by considering a fictitious diffusion process,
with the spectral dimension given in terms
of the average return probability for given (fictitious) diffusion time.
When the number of spectral dimensions matches the number of
Hausdorff dimensions of a spacetime the return probability depends
on diffusion time
in a characteristic way that is indeed found in all models for large diffusion times.
But at short diffusion times one finds in several studies of interest for quantum-gravity
and quantum-spacetime research that the average return probability has properties
signaling a number of spectral dimensions smaller than the number of Hausdorff dimensions.

First results of this type were found in studies done within the framework of
causal dynamical triangulations\cite{lollFIRSTdimreduction},
naturally working with four Hausdorff dimensions and finding that the behaviour at
small diffusion times signaled two spectral dimensions.
Two spectral dimensions for small diffusion times was then also found
in studies inspired by asymptotic safety~\cite{safetydimreduction1,safetydimreduction2}
and studies based on Horava-Liftschitz gravity~\cite{hldimreduction}.
A somewhat different situation is found in studies inspired by
spacetime noncommutativity~\cite{dariodimreduction,michedimreduction1,michedimreduction2}
and by spin foams~\cite{foamdimreduction}
but still giving fewer than four spectral dimensions for small diffusion times
(also see Refs.~\cite{visserRUNNINGsd,carlipRUNNINGsd}).

These ``running spectral dimensions" could have very significant implications for cosmology, as suggested at least intuitively by the definition of spectral dimensions based on the dependence of the return probability on the diffusion time.   At present these potentialities are still  largely unexplored, but one possibility has been debated in Refs.~\cite{runningSDexp1,runningSDexp2,runningSDexp3}. Ref.~\cite{runningSDexp1}, citing as motivation some of the quantum-gravity studies exhibiting running spectral dimensions, proposed that such studies should motivate the search of indirect evidence of absence of gravitational degrees of freedom in the early Universe. However, Ref.~\cite{runningSDexp2}
more prudently observed that gravitational degrees of freedom are indeed absent in spacetimes with three or less Hausdorff dimensions, but may well be present in spacetimes with four Hausdorff dimensions but with three or less spectral dimensions.


\subsection{Some other quantum-gravity-cosmology proposals}
So far in this section, consistently with the main goals of this review,
I focused on cosmology proposals which are based on
(or at least are directly linkable to) some theories of quantum spacetime.
In this last subsection of the section I mention just a few
illustrative examples of ideas and proposals
which are still based on the quantum-gravity problem but without invoking
any definite quantum properties of spacetime.
It is not unlikely that future exploitations of the associated phenomenological
opportunities would involve spacetime quantization.

\subsubsection{Quantum-gravity-induced vector fields}
An active area of quantum-gravity cosmology focuses
on Lorentz-violating vector fields.
These studies do not have as reference scenarios for spacetime
quantization, but they are being linked generically to  the opportunities
that the quantum-gravity
problem provides for
the emergence of Lorentz-violating vector fields.
Several possible implications are being considered, including
the possibility that in presence
of such Lorentz-violating vector fields
the Universe might
experience a slower rate of expansion for a given matter content
(see, e.g., Ref.~\cite{limCarroll}).

It is also emerging that the implications of such
Lorentz-violating vector fields
would be rather significant for the cosmic microwave background~\cite{limMP}.
In particular, as stressed in other parts of this review,
the presence of Lorentz-violating vector fields
is often associated with energy-dependent birefringence.
And the cosmic microwave background,
since its radiation originates from
the surface of last scattering,
which is the most distant source of light,
can be a very powerful opportunity
to test anomalous features for the propagation
of photons.
Several techniques of data analysis have been developed  which are
capable of constraining birefringence of photon propagation
using cosmic-microwave-background data
(see, e.g.,
Refs.\cite{kamionkowskiPOLARIZATION,durrer2008,Kostelecky2007,giulialucaPROCSUPP}
and references therein).

\subsubsection{A semiclassical Wheeler--DeWitt-based description of the
  early Universe}
Cosmology is also an arena where
some interest is attracted by
studies of the semiclassical limit of quantum gravity.
These do not invoke a quantum-spacetime picture and may even not
rely on any given quantum-gravity proposal. They are rather viewed in
analogy~\cite{kieferSEMI}
with the semiclassical limits of other quantum theories:
one can consider, for example, the
corrections to the classical Maxwell action described by Heisenberg and Euler
(in a pre-QED era)
in terms of quantum fluctuations of
electrons and positrons, which can of course be now rederived~\cite{kieferSEMI}
from QED by integrating out
the fermions and expanding in powers of $\hbar$.

These studies of the semiclassical limit of quantum gravity
are often centered around the Wheeler--DeWitt equation.
While for the development of a full quantum-gravity theory the
Wheeler--DeWitt equation has proven to be extremely ``cumbersome'',
the fact that it is rather intuitively formulated is of course convenient for setting
up a semiclassical approximation
(see, e.g., Refs.~\cite{kieferSING,otherSEMI,montaniSEMI}).
A result that can be rather readily analyzed from a phenomenology perspective
is the one providing~\cite{kieferSING} correction terms for the Schroedinger equation,
obtained through a formal expansion of the Wheeler--DeWitt equation with respect to powers
of the Planck mass. Unsurprisingly the relevant correction terms are far too small to matter
in laboratory experiments~\cite{kieferSING}. However, it is plausible that such a procedure
could give rise to observably large effects in the description of the early stages of
evolution of the Universe. In particular, the semiclassical approximation set up
in Ref.~\cite{kieferSING} could be used rather straightforwardly to describe
corrections to the Schroedinger equation for higher multipoles on a Friedman background.


\subsubsection{No-singularity cosmology from string theory}
An interesting string-theory-inspired area of cosmology research
revolves around a scenario for singularity avoidance
linked to the availability
of duality tranformations, which allow to set up a suitable ``pre-Big-Bang''
scenario~\cite{veneCOSMOnew,veneCOSMOold1,veneCOSMOold2}.
In this scenario the
Universe starts inflating from an initial state characterized by very small curvature
and weak interactions. The small-curvature initial state is gravitationally unstable
and would naturally evolve~\cite{veneCOSMOnew,veneCOSMOold1,veneCOSMOold2}
into states with
higher curvature, until string-size (roughly Planck-scale-size) effects
are strong enough to induce a ``bounce'' into a decreasing-curvature regime. Instead
of a conventional hot big bang one would
have~\cite{veneCOSMOnew,veneCOSMOold1,veneCOSMOold2}
a ``hot big bounce'' in
which in particular the heating mechanism is provided by the quantum production
of particles in the pre-bounce phase characterized by high curvature and strong interactions.

For this string-inspired pre-big-bang scenario several possible
observational consequences have been
discussed~\cite{veneCOSMOnew,veneCOSMOold1,veneCOSMOold2},
including the one of a stochastic background of gravity waves due to a background
of gravitons from the pre-big-bang phase.
It appears to be plausible~\cite{veneCOSMOnew,veneCOSMOold1,veneCOSMOold2}
that the magnitude of the associated effects
might be within the range of
sensitivities of modern gravity-wave interferometers.

\newpage


\section{Quantum-Spacetime Phenomenology Beyond the Standard Setup}
\label{beyond-standard}
Most of the ideas for phenomenology I here reviewed are set up
following a common strategy. They reflect the expectation that the
characteristic scale of quantum-spacetime effects should be within a few
orders of magnitude of the Planck scale, and that it should be
possible (for studies conducted at scales much below the Planck scale)
to analyze quantum-spacetime effects using an expansion in  powers of
the Planck length. All this is inspired by
analogous strategies which have been very fruitful in other areas
of physics: many arguments indicate that the Planck scale is the scale
where the current theories break down, and usually the breakdown scale
is also the scale that governs the magnitude of the effects of the new
needed theory.
In the case of quantum-spacetime research the expectation of perturbative
effects suppressed by a large scale finds further motivation in
at least two observations:\\
$\star$ The effects we expect from spacetime quantization are rather striking,
qualitatively virulent departures from the structure of our current theories.
The fact that no trace of such ``easily noticeable''
effects has ever been seen surely provides further encouragement for
the expectation of perturbative effects suppressed by an ultralarge scale.\\
$\star$ I would list the evidence in favor of grandunification as an
even more significant source
of additional encouragement for
the expectation of perturbative effects suppressed by an ultralarge scale.
If that evidence is taken at face value (as, I would argue, we should, at least
as natural working assumption) it suggests  that particle physics
works well on its own up to a scale of about $10^{-3}$ the Planck scale.
If quantum-spacetime effects where non-perturbative in ways affecting grandunification
or if the scale of spacetime quantization was much lower than the Planck scale
it would then be hard to explain the (preliminary) success of the grandunification idea.

In light of this, surely quantum-spacetime
phenomenologists should continue to focus most of their efforts on
applications of the standard strategy,
assuming perturbative effects suppressed by a scale in some neighborhood
of the Planck scale. However, other scenarios and opportunities
should not be completely overlooked.
We are clearly presently unable to exclude that
the correct quantum  picture of spacetime
might turn out  to be unsuitable to the standard strategy
of quantum-spacetime phenomenology.
As a way to give some substance to this assessment,
in this section I briefly discuss examples of mechanisms that could
render ineffective the standard strategy of quantum-spacetime phenomenology.


\subsection{A totally different setup with Large Extra Dimensions}
Can the scale characteristic of quantum-spacetime effects be much lower
than the Planck scale? We surely know at least one mechanism by which the
quantum-gravity scale can be much lower than the Planck scale, and therefore
quantum-gravity models with spacetime quantization affected by this mechanism
would describe quantum-spacetime effects at a relatively low scale.
I am of course thinking of the popular scenarios with large extra dimensions.

Through these scenarios one can achieve a sizeable reduction
of the quantum-gravity scale
with the introduction of $D$ extra space dimensions~\cite{anto1,anto2,anto3,addlarge}
of finite size $R_*$.
Then the fundamental length scale $L_D$ characteristic of quantum gravity
in the 3+D+1-dimensional spacetime
can be much bigger than the Planck length.
The smallness of the Planck length
can emerge as the result of the fact that, as deduced from applying Gauss law in
the 3+D+1-dimensional context,
 the strength
of gravitation at distance scales larger than the size $R_*$ of the extra dimensions
in the ordinary (infinite-size) 3+1-dimensional spacetime
would be proportional to the square-root of the inverse of
the volume of the external compactified space multiplied
by an appropriate power of $L_D$.

These scenarios need to be tuned rather carefully in order to get a
phenomenologically viable picture. Essentially the only truly appealing
 possibility is the one of 2 extra dimensions
 of relatively ``large'' size, somewhere below millimeter
 size~\cite{addlarge} (perhaps $10^{-4}$ or $10^{-5}$ meters).
There might be other extra dimensions of smaller (possibly Planckian) size,
but for the desired phenomenology one needs two and only two extra dimensions
of relatively large size; otherwise one finds effects that either violate
known experimental facts or are too small to ever be tested.
But with these (however contrived) choices one does end up with a phenomenologically exciting scenario in which the fundamental length scale of
 quantum gravity $L_D$ is somewhere in the neighborhood
 of  the $(TeV)^{-1}$ length scale, and therefore
  within the reach
 of  particle-physics  experiments
 (see, e.g., Refs.~\cite{giudicelarge,giddingslarge,photolarge,appelquistlarge}).
Moreover, there are phenomenologically-relevant implications for the
behavior of (classical) gravity at submillimeter distances~\cite{addlarge,ledEXPERIMENT}.


\subsection{The example of hard UV/IR mixing}

The large-extra-dimension scenario is an example of inapplicability
of the standard setup of quantum-spacetime phenomenology due to the
fact that, within that scenario, the characteristic scale of quantum gravity
is not the Planck scale. There are also scenarios in which one may still
assume that quantum-spacetime effects are fundamentally introduced
at the Planck scale, but the
standard setup of quantum-spacetime phenomenology
is inapplicable because the most characteristic effects are not describable
in terms of an expansion in powers of the Planck length.

I have partly already discussed this possibility in the portion of this
review which was devoted to soft UV/IR mixing.
However, in that context one could basically fall back on roughly the
standard strategy of quantum-spacetime phenomenology, by looking
for an infrared scale playing the role of characteristic scale of the
infrared manifestations of the quantum properties of spacetime.
Let me here just stress that an even more pervasive revision of the
standard strategy of quantum-spacetime phenomenology would be required
in the case of hard UV/IR mixing, which might take the form of
correction terms with the behavior of inverse powers of momentum.
With hard UV/IR mixing one should expect that in certain contexts
the departures from known physical laws should be dramatic.
The most efficacious  tests of this hypothesis might not take the
shape of searches of small corrections to standard predictions
in ordinary contexts, but rather be based on the identification of
those peculiar contexts where the implications of UV/IR mixing are large.


\subsection{The possible challenge of not-so-subleading higher-order
  terms}

Some challenges for the standard setup of quantum-spacetime phenomenology may also be present
when the effects are genuinely introduced at the Planck scale and there is nothing peculiar
about the infrared sector.
In particular, just because this standard setup is based on a (truncated) expansion in powers
of the Planck length, it can happen that the formally sub-leading terms (higher powers of the Planck length),
which are usually neglected in leading-order analyses, are actually not really negligible.
The fact that experiments suitable for quantum-spacetime
phenomenology must host, as I stressed in several points of this
review, some ultralarge ordinary-physics dimensionless ``amplifiers''
could play a role in these concerns:
if some mechanism is allowing the tiny leading-order Planck-length
correction to be observably large it would not be so surprising to find
that the same (or some other) amplifier is also such that
some ``formally subleading'' Planck-length
corrections, neglected in the analysis, are significant.

And another possible source of concern can originate from the fact that
some of the contexts of interest for quantum-spacetime phenomenology are characterized
by several length scales:  expansions in powers of the Planck length
actually are of course
expansions in powers of some dimensionless quantity obtained dividing the Planck length
by a characteristic length scale of the physical context of interest, and
some ``pathologies''
may be encountered if there are several
candidate length scales for the expansion.

While I feel that these issues for the power expansion should not be ignored,
it is partly reassuring that the only explicit examples we seem to be
able to come up with
are rather contrived.
For example, in order to illustrate
the issues connected with the many length scales available
in certain contexts of interest for quantum-spacetime phenomenology,
I cannot mention anything more appealing than the following \textit{ad hoc}
formulation of
a deformation of the speed-energy relation applicable in the ``relativistic regime'' ($E\gg m$):
\begin{equation}
v \simeq 1 - \frac{m^2}{2 E^2} + \eta  L_p E \left(
\tanh \left( {L_p^2 E^6 \over m^4 } \right) -1 \right)
~.
\label{velLIVbSTRANGE}
\end{equation}
At low (but still ``relativistic'') energies this would fit within a picture that
has been much studied from
the quantum-spacetime-phenomenology perspective, the one of the
speed-energy relation $v \simeq 1 - m^2/(2 E^2) - \eta L_p E$. But whereas in the
relevant literature it is assumed that the
term $\eta L_p E$, if present, should always be the leading correction,
up to particle energies of the order of the Planck scale, $E \sim 1/L_p$,
from~(\ref{velLIVbSTRANGE}) one would find that the correction term is already
no longer leading at particle energies of the order of  $E \sim (m^2/L_p)^{1/3}$,
i.e., well below the Planck scale.
Of course here the point is that~(\ref{velLIVbSTRANGE})
is characterized by two distinct small quantities suitable for
the ``expansion in powers of $L_p$'':
the quantity $L_p E$ and the quantity $L_p E/m^2$.

\newpage


\section{Closing Remarks}
\label{closing-remarks}

Clearly the most significant development of these first few years
of quantum-spacetime phenomenology
has been our ability to uncover some experimental/observational contexts in which,
through appropriate
data analyses, we could gain access to effects introduced
genuinely at the Planck scale.
The compellingness of such instances of
genuine Planck-scale sensitivity, which was here most simply and
clearly illustrated in Section~\ref{simple-example},
should be contrasted to the more frequent
case of ``dimensional-analysis Planck scale sensitivities'', which typically
involve a description of a plausible quantum-spacetime
effect in terms of a dimensionless parameter,
estimated arbitrarily as a ratio of the Planck length and some characteristic length scale
of the problem.

Looking at the results I have summarized in this review
different readers, depending on how stringent are their criteria for genuine Planck-scale
sensitivity, will only recognize one, two or anyway very few examples.
Not much, but much better than expected even just 15 years ago.
And, as I also stressed, we do have, at this point, a rather encouraging
list of contexts in which, while the availability of genuine Planck-scale sensitivity has still
not been fully  established, it appears that  sensitivity to
effects introduced genuinely
at the Planck scale could be achieved in a not-so-distant future.

The fact that the development of this phenomenology is proving beneficial
for the study of the idea of spacetime quantization is perhaps best testified
by the fact that it is already managing to
truly affect the directions taken by more formal work on spacetime quantization,
especially in the areas of loop quantum gravity and spacetime noncommutativity.
Theorists in these areas follow the developments on the
phenomenology side and do their best (the technical challenges they are
facing are very severe) to derive results that can be exploited for the
opportunities in phenomenology that are being established.
In turn of course the phenomenology takes notice of the developments
on the theory side, finding in them new input for enlarging the list
of candidate quantum-spacetime effects that one could attempt
to investigate experimentally.

The goal of testing/falsifying rigorous theories of spacetime quantization
appears to be still beyond
our present reach. But while most of the work in quantum-spacetime phenomenology so far
has relied on simple-minded test theories describing candidate quantum-spacetime effects,
I see first indications of a
phase of further maturation of this phenomenology, in which we will
actually test/falsify at least the most virulent rigorous formalizations
of quantum spacetime.
Planck-scale theories formulated in noncommutative versions of Minkowski
spacetime are the example where we are presently closer to this goal.

The (however limited) information presently available to us appears to
provide a clear invitation
to continue to focus most of our efforts in the search of effects describable
in terms of a (low-energy) expansion in powers of the Planck length,
though other opportunities clearly should not be overlooked.
Concerning the type of data on which quantum-spacetime phenomenology can rely,
I have attempted to maintain throughout this review some visible separations between
different  proposals on the basis of whether they concern astrophysics,
cosmology or controlled laboratory experiments.
It is very clear that astrophysics has provided so far the
most fruitful arena, but cosmology has the greatest potential reach
(although for the most part this potential
has not yet materialized) .
The role played so far in quantum-spacetime phenomenology by
controlled laboratory experiments is rather marginal, but it would
be important for the
future development  of quantum-spacetime phenomenology
to find more opportunities for controlled laboratory experiments.


\newpage

\begin{thebibliography}{100}

\bibitem{burgessLRR}
C. Burgess,
\newblock Living Rev.Rel. 7 (2004) 5, gr-qc/0311082.

\bibitem{hanwillen}
T. Han and S. Willenbrock,
\newblock Phys.Lett. B616 (2005) 215, hep-ph/0404182.

\bibitem{natureNeutron}
V.V. Nesvizhevsky et~al.,
\newblock Nature 415 (2002) 297.

\bibitem{stachelearly}
J. Stachel,
\newblock Early history of quantum gravity,
\newblock Black Holes, Gravitational Radiation and the Universe, edited by B.
  Iyer and B. Bhawal, Kluwer Academic Publisher, Netherlands, 1999.

\bibitem{polonpap}
G. Amelino-Camelia,
\newblock Lect. Notes Phys. 541 (2000) 1, gr-qc/9910089.

\bibitem{percacciHeisenberg}
R. Percacci and G.P. Vacca,
\newblock Class.Quant.Grav. 27 (2010) 245026, arXiv:1008.3621.

\bibitem{dvaliHeisenberg}
G. Dvali, S. Folkerts and C. Germani,
\newblock Phys.Rev. D84 (2011) 024039, arXiv:1006.0984.

\bibitem{donoghue1993}
J. Donoghue,
\newblock Phys.Rev.Lett. 72 (1994) 2996, gr-qc/9310024.

\bibitem{bellucci1996}
A.A. Akhundov, S. Bellucci and A. Shiekh,
\newblock Phys.Lett. B395 (1997) 16, gr-qc/9611018.

\bibitem{kirilin2002}
I. Khriplovich and G. Kirilin,
\newblock J.Exp.Theor.Phys. 95 (2002) 981, gr-qc/0207118.

\bibitem{donoghue2003}
N. Bjerrum-Bohr, J. Donoghue and B.R. Holstein,
\newblock Phys. Rev. D 67 (2003) 084033, hep-th/0211072.

\bibitem{kirilin2004}
I. Khriplovich and G. Kirilin,
\newblock J.Exp.Theor.Phys. 98 (2004) 1063, gr-qc/0402018.

\bibitem{woodard2010}
S. Park and R. Woodard,
\newblock Class. Quantum Grav. 27 (2010) 245008, arXiv:1007.2662.

\bibitem{mead}
C.A. Mead,
\newblock Phys. Rev. 135 (1964) B849.

\bibitem{venekonmen1}
G. Veneziano,
\newblock Europhys. Lett. 2 (1986) 199.

\bibitem{venekonmen2}
D. Gross and P. Mende,
\newblock Nucl. Phys. B303 (1988) 407.

\bibitem{venekonmen3}
D. Amati, M. Ciafaloni and G. Veneziano,
\newblock Phys. Lett. B216 (1989) 41.

\bibitem{venekonmen4}
K. Konishi, G. Paffuti and P. Provero,
\newblock Phys. Lett. B234 (1990) 276.

\bibitem{padma}
T. Padmanabhan,
\newblock Class. Quant. Grav. 4 (1987) L107.

\bibitem{dopliPLB}
S. Doplicher, K. Fredenhagen and J.E. Roberts,
\newblock Phys. Lett. B331 (1994) 39.

\bibitem{ahlu1994}
D.V. Ahluwalia,
\newblock Phys. Lett. B339 (1994) 301, gr-qc/9308007.

\bibitem{ng1994}
Y. Ng and H. van Dam,
\newblock Mod. Phys. Lett. A9 (1994) 335.

\bibitem{gacmpla}
G. Amelino-Camelia,
\newblock Mod. Phys. Lett. A9 (1994) 3415, gr-qc/9603014.

\bibitem{garay}
L. Garay,
\newblock Int. J. Mod. Phys. A10 (1995) 145, gr-qc/9403008.

\bibitem{casadio}
F. Scardigli and R. Casadio,
\newblock Class. Quant. Grav. 20 (2003) 3915, hep-th/0307174.

\bibitem{chandralimit1}
S. Chandrasekhar,
\newblock Astrophys. J. 74 (1931) 81.

\bibitem{chandralimit2}
S. Chandrasekhar,
\newblock Mon. Not. Roy. Astron. Soc. 95 (1935) 207.

\bibitem{cow}
R. Colella, A. Overhauser and S. Werner,
\newblock Phys. Rev. Lett. 34 (1975) 1472.

\bibitem{sakurai}
J. Sakurai,
\newblock Modern Quantum Mechanics, rev. ed. (Addison-Wesley, Reading, MA,
  1994).

\bibitem{gasperiniEP}
M. Gasperini,
\newblock Phys. Rev. D 38 (1988) 2635.

\bibitem{dharamEP}
G. Adunas, E. Rodriguez-Milla and D. Ahluwalia,
\newblock Gen. Relativ. Gravit. 33 (2001) 183, gr-qc/0006022.

\bibitem{cowEPviol}
K. Littrel, B. Allman and S. Werner,
\newblock Phys. Rev. A 56 (1997) 1767.

\bibitem{anan1}
J. Anandan,
\newblock Phys. Lett.  (1984) 280.

\bibitem{anan2}
J. Anandan,
\newblock Class. Quantum Grav. 1 (1984) L51.

\bibitem{meadPHEN}
C.A. Mead,
\newblock Phys.Rev. 143 (1966) 990.

\bibitem{ehns}
J. Ellis et~al.,
\newblock Nucl. Phys. B241 (1984) 381.

\bibitem{elmn}
J. Ellis et~al.,
\newblock Phys. Rev. D 53 (1996) 3846, hep-ph/9505340.

\bibitem{huetpesk}
P. Huet and M. Peskin,
\newblock Nucl. Phys. B 434 (1995) 3.

\bibitem{floreacpt}
F. Benatti and R. Floreanini,
\newblock Nucl. Phys. B 488 (1997) 335.

\bibitem{cplear}
CPLEAR, R. Adler et~al.,
\newblock Phys. Lett. B364 (1995) 239, hep-ex/9511001.

\bibitem{perci1}
I. Percival,
\newblock Proc. Roy. Soc. A451 (1995) 503.

\bibitem{perci0}
I. Percival,
\newblock Physics World 10 (1997) 43.

\bibitem{perci2}
I. Percival and W. Strunz,
\newblock (1996), quant-ph/9607011.

\bibitem{kostesamuel}
V. Kosteleck{\'{y}} and S. Samuel,
\newblock Phys. Rev. D 39 (1989) 683.

\bibitem{kostcpt}
V. Kosteleck{\'{y}} and R. Potting,
\newblock Phys. Rev. D 51 (1995) 3923, hep-ph/9501341.

\bibitem{grbgac}
G. Amelino-Camelia et~al.,
\newblock Nature 393 (1998) 763, astro-ph/9712103.

\bibitem{gampul}
R. Gambini and J. Pullin,
\newblock Phys. Rev. D 59 (1999), gr-qc/9809038.

\bibitem{schaefer}
B. Schaefer,
\newblock Phys. Rev. Lett. 82 (1999) 4964, astro-ph/9810479.

\bibitem{gacgwi}
G. Amelino-Camelia,
\newblock Nature 398 (1999) 216, gr-qc/9808029.

\bibitem{gacgwiB}
G. Amelino-Camelia,
\newblock Nature 410 (2001) 1065, gr-qc/0104086.

\bibitem{bignapap}
G. Amelino-Camelia,
\newblock Phys. Rev. D 62 (2000) 024015, gr-qc/9903080.

\bibitem{nggwi}
Y. Ng and H. van Dam,
\newblock Found. Phys. 30 (2000) 795, gr-qc/9906003.

\bibitem{kifune}
T. Kifune,
\newblock Astrophys. J. 518 (1999) L21, astro-ph/9904164.

\bibitem{ita}
R. Aloisio et~al.,
\newblock Phys. Rev. D 62 (2000) 053010, astro-ph/0001258.

\bibitem{gactp}
G. Amelino-Camelia and T. Piran,
\newblock Phys. Rev. D 64 (2001) 036005, astro-ph/0008107.

\bibitem{aus}
R.J. Protheroe and H. Meyer,
\newblock Phys. Lett. B493 (2000) 1, astro-ph/0005349.

\bibitem{kluz}
W. Kluzniak,
\newblock Astroparticle Physics 11 (1999) 117.

\bibitem{billetal}
S.D. Biller et~al.,
\newblock Phys. Rev. Lett. 83 (1999) 2108, gr-qc/9810044.

\bibitem{crLIVING}
C. Rovelli,
\newblock Living Rev. Rel. 1 (1998) 1, gr-qc/9710008.

\bibitem{ashtelewandREVIEW}
A. Ashtekar and J. Lewandowski,
\newblock Class. Quantum Grav. 21 (2004) R53, gr-qc/0404018.

\bibitem{leeLQGrev}
L. Smolin,
\newblock (2003), hep-th/0303185.

\bibitem{thieREV}
T. Thiemann,
\newblock Lect. Notes Phys. 631 (2003) 41, gr-qc/0210094.

\bibitem{ashtekarLQGreview2012}
A. Ashtekar,
\newblock (2012), arXiv:1201.4598.

\bibitem{urrutia}
J. Alfaro, H. Morales-Tecotl and L. Urrutia,
\newblock Phys. Rev. Lett. 84 (2000) 2318, gr-qc/9909079.

\bibitem{thiemLS}
T. Thiemann,
\newblock (2001), gr-qc/0110034.

\bibitem{kodadsr}
G. Amelino-Camelia, L. Smolin and A. Starodubtsev,
\newblock Class. Quant. Grav. 21 (2004) 3095, hep-th/0306134.

\bibitem{bojoLQGphen}
M. Bojowald, H. Morales-Tecotl and H. Sahlmann,
\newblock Phys. Rev. D 71 (2005) 084012, gr-qc/0411101.

\bibitem{ashtereview}
A. Ashtekar,
\newblock (1999), gr-qc/9901023.

\bibitem{crHISTO}
C. Rovelli,
\newblock (2000), gr-qc/0006061.

\bibitem{leePW}
L. Smolin,
\newblock Physics World 12 (1999) 79.

\bibitem{carlip}
S. Carlip,
\newblock Rept. Prog. Phys. 64 (2001) 885, gr-qc/0108040.

\bibitem{thooftDiscret}
G. 't~Hooft,
\newblock Class. Quantum Grav. 13 (1996) 1023.

\bibitem{rigidfoam}
S. Bernadotte and F. Klinkhamer,
\newblock Phys. Rev. D 75 (2007) 024028, hep-ph/0610216.

\bibitem{suda1}
A. Perez and D. Sudarsky,
\newblock Phys. Rev. Lett. 91 (2003) 179101, gr-qc/0306113.

\bibitem{suda2}
J. Collins et~al.,
\newblock Phys. Rev. Lett. 93 (2004) 191301, gr-qc/0403053.

\bibitem{suda3}
D. Sudarsky, L. Urrutia and H. Vucetich,
\newblock Phys. Rev. Lett. 89 (2002) 231301, gr-qc/0204027.

\bibitem{tuning}
P.M. Crichigno and H. Vucetich,
\newblock Phys. Lett. B651 (2007) 313, hep-th/0607214.

\bibitem{calmethsuPRL}
X. Calmet, S.D.H. Hsu and D. Reeb,
\newblock Phys. Rev. Lett. 101 (2008) 171802, arXiv:0805.0145.

\bibitem{wilczekUNI}
S. Robinson and F. Wilczek,
\newblock Phys. Rev. Lett. 96 (2006) 231601, hep-th/0509050.

\bibitem{tomsNature}
D.J. Toms,
\newblock Nature 468 (2010) 56.

\bibitem{calmethsuPRD}
X. Calmet, S.D.H. Hsu and D. Reeb,
\newblock Phys. Rev. D 81 (2010) 035007, arXiv:0911.0415.

\bibitem{newjourn}
G. Amelino-Camelia,
\newblock New J. Phys. 6 (2004) 188, gr-qc/0212002.

\bibitem{mattinLRR}
D. Mattingly,
\newblock Living Rev. Rel. 8 (2005) 5, gr-qc/0502097.

\bibitem{giddings2011qgstrings}
S.B. Giddings,
\newblock (2011), arXiv:1105.6359.

\bibitem{anto1}
I. Antoniadis,
\newblock Phys. Lett. B246 (1990) 377.

\bibitem{anto2}
J.D. Lykken,
\newblock Phys. Rev. D 54 (1996) 3693, hep-th/9603133.

\bibitem{anto3}
E. Witten,
\newblock Nucl. Phys. B471 (1996) 135, hep-th/9602070.

\bibitem{addlarge}
N. Arkani-Hamed, S. Dimopoulos and G.R. Dvali,
\newblock Phys. Lett. B429 (1998) 263, hep-ph/9803315.

\bibitem{ledMARCH}
N. Arkani-Hamed, S. Dimopoulos and J. March-Russell,
\newblock Phys. Rev. D 63 (2001) 064020, hep-th/9809124.

\bibitem{rubakovLED}
V.A. Rubakov,
\newblock Phys. Usp. 44 (2001) 871, hep-ph/0104152.

\bibitem{damourEP2}
T. Damour, F. Piazza and G. Veneziano,
\newblock Phys. Rev. Lett. 89 (2002) 081601, gr-qc/0204094.

\bibitem{banerLANDAUgood}
R. Banerjee,
\newblock Mod. Phys. Lett. A17 (2002) 631, hep-th/0106280.

\bibitem{dougnekr}
M. Douglas and N. Nekrasov,
\newblock Rev. Mod. Phys. 73 (2001) 977, hep-th/0106048.

\bibitem{szaboREVIEW}
R.J. Szabo,
\newblock Phys. Rept. 378 (2003) 207, hep-th/0109162.

\bibitem{wittenPT}
E. Witten,
\newblock Phys. Today 49 (1996) 24.

\bibitem{carloleeAREA}
C. Rovelli and L. Smolin,
\newblock Nucl. Phys. B442 (1995) 593, gr-qc/9411005.

\bibitem{ashtAREA}
A. Ashtekar and J. Lewandowski,
\newblock Class. Quant. Grav. 14 (1997) A55, gr-qc/9602046.

\bibitem{rovellispeziale2003}
C. Rovelli and S. Speziale,
\newblock Phys. Rev. D 67 (2003) 064019, gr-qc/0205108.

\bibitem{rovellispeziale2011}
C. Rovelli and S. Speziale,
\newblock Phys. Rev. D 83 (2011) 104029, arXiv:1012.1739.

\bibitem{bojoFIRST}
M. Bojowald,
\newblock Phys. Rev. Lett. 86 (2001) 5227, gr-qc/0102069.

\bibitem{bojoFIRSTb}
A. Ashtekar, M. Bojowald and J. Lewandowski,
\newblock Adv. Theor. Math. Phys. 7 (2003) 233, gr-qc/0304074.

\bibitem{bojoLRR}
M. Bojowald,
\newblock Living Rev. Rel. 8 (2005) 11, gr-qc/0601085.

\bibitem{ashtekarLQC}
A. Ashtekar,
\newblock J.Phys.Conf.Ser. 360 (2012) 012001.

\bibitem{majrue}
S. Majid and H. Ruegg,
\newblock Phys. Lett. B334 (1994) 348, hep-th/9405107.

\bibitem{kpoinap}
J. Lukierski, H. Ruegg and W. Zakrzewski,
\newblock Ann. Phys. 243 (1995) 90.

\bibitem{wessNCYM}
J. Madore et~al.,
\newblock Eur. Phys. J. C16 (2000) 161, hep-th/0001203.

\bibitem{gacmaj}
G. Amelino-Camelia and S. Majid,
\newblock Int. J. Mod. Phys. A15 (2000) 4301, hep-th/9907110.

\bibitem{balaSPINstat}
A.P. Balachandran et~al.,
\newblock Int. J. Mod. Phys. A21 (2006) 3111, hep-th/0508002.

\bibitem{connes1995}
A. Connes,
\newblock J. Math. Phys. 36 (1995) 6194.

\bibitem{connesBOOK}
A. Connes,
\newblock {Noncommutative geometry} (Academic Press, 1994).

\bibitem{emn}
J. Ellis, N.E. Mavromatos and D. Nanopoulos,
\newblock Phys. Lett. B293 (1992) 37, hep-th/9207103.

\bibitem{emnreview}
J. Ellis, N.E. Mavromatos and D. Nanopoulos,
\newblock (1994), hep-th/9405196.

\bibitem{aemn1}
G. Amelino-Camelia et~al.,
\newblock Int. J. Mod. Phys. A12 (1997) 607, hep-th/9605211.

\bibitem{nickREVIEW2010}
N.E. Mavromatos,
\newblock Int. J. Mod. Phys. A25 (2010) 5409, arXiv:1010.5354.

\bibitem{sorkinPRLcausalsets}
L. Bombelli et~al.,
\newblock Phys. Rev. Lett. 59 (1987) 521.

\bibitem{sorkinRideoutCONTINUUM}
D.P. Rideout and R.D. Sorkin,
\newblock Phys. Rev. D 63 (2001) 104011, gr-qc/0003117.

\bibitem{ambjorn1993}
J. Ambj{\o}rn, J. Jurkiewicz and C.F. Kristjansen,
\newblock Nucl.Phys. B393 (1993) 601, hep-th/9208032.

\bibitem{lollLRR}
R. Loll,
\newblock Living Rev.Rel. 1 (1998) 13, gr-qc/9805049.

\bibitem{ambjornlollPRL}
J. Ambj{\o}rn, J. Jurkiewicz and R. Loll,
\newblock Phys. Rev. Lett. 93 (2004) 131301, hep-th/0404156.

\bibitem{ambjornlollPRD}
J. Ambj{\o}rn, J. Jurkiewicz and R. Loll,
\newblock Phys. Rev. D 72 (2005) 064014, hep-th/0505154.

\bibitem{lollDESKTOP}
R. Loll,
\newblock Class. Quant. Grav. 25 (2008) 114006, arXiv:0711.0273.

\bibitem{ambjornloll2010}
J. Ambj{\o}rn, J. Jurkiewicz and R. Loll,
\newblock (2010), arXiv:1004.0352.

\bibitem{lollFIRSTdimreduction}
J. Ambj{\o}rn, J. Jurkiewicz and R. Loll,
\newblock Phys.Rev.Lett. 95 (2005) 171301, hep-th/0505113.

\bibitem{runningSDexp1}
J.R. Mureika and D. Stojkovic,
\newblock Phys.Rev.Lett. 106 (2011) 101101, arXiv:1102.3434.

\bibitem{runningSDexp2}
T.P. Sotiriou, M. Visser and S. Weinfurtner,
\newblock Phys.Rev.Lett. 107 (2011) 169001, arXiv:1104.1223.

\bibitem{runningSDexp3}
J. Mureika and D. Stojkovic,
\newblock Phys.Rev.Lett. 107 (2011) 169002, arXiv:1109.3506.

\bibitem{weinbergSAFETY}
S. Weinberg,
\newblock (1996), hep-th/9702027.

\bibitem{reuterSAFETY}
M. Reuter,
\newblock Phys. Rev. D 57 (1998) 971, hep-th/9605030.

\bibitem{percacciSAFETY}
D. Dou and R. Percacci,
\newblock Class. Quant. Grav. 15 (1998) 3449, hep-th/9707239.

\bibitem{reuter0702051}
M. Reuter and H. Weyer,
\newblock Int. J. Mod. Phys. 15 (2006) 2011, hep-th/0702051.

\bibitem{reuterMinLength}
M. Reuter and J.M. Schwindt,
\newblock JHEP 01 (2006) 070, hep-th/0511021.

\bibitem{emergentLIBER}
C. Barcelo, M. Visser and S. Liberati,
\newblock Int.J.Mod.Phys. D10 (2001) 799, gr-qc/0106002.

\bibitem{emergentVOLOVI}
G. Volovik,
\newblock JETP Lett. 89 (2009) 525, arXiv:0904.4113.

\bibitem{emergentPADMA}
T. Padmanabhan,
\newblock AIP Conf.Proc. 939 (2007) 114, arXiv:0706.1654.

\bibitem{emergentNCST1}
H. Steinacker,
\newblock JHEP 0712 (2007) 049, arXiv:0708.2426.

\bibitem{emergentNCST2}
H.S. Yang,
\newblock Int.J.Mod.Phys. A24 (2009) 4473, hep-th/0611174.

\bibitem{emergentSINDO}
L. Sindoni,
\newblock SIGMA 8 (2012) 027, arXiv:1110.0686.

\bibitem{emergentHU}
B. Hu,
\newblock Int.J.Theor.Phys. 44 (2005) 1785, gr-qc/0503067.

\bibitem{thooftDISSIPA}
G. 't~Hooft,
\newblock Class.Quant.Grav. 16 (1999) 3263, gr-qc/9903084.

\bibitem{huDISSIPA}
B. Hu,
\newblock Int.J.Theor.Phys. 41 (2002) 2091, gr-qc/0204069.

\bibitem{gacdsr}
G. Amelino-Camelia,
\newblock Int. J. Mod. Phys. 11 (2002) 35, gr-qc/0012051.

\bibitem{gacdsrB}
G. Amelino-Camelia,
\newblock Phys. Lett. B510 (2001) 255, hep-th/0012238.

\bibitem{jurekkodadsr}
L. Freidel, J. Kowalski-Glikman and L. Smolin,
\newblock Phys. Rev. D 69 (2004) 044001, hep-th/0307085.

\bibitem{smolinDSR2008}
L. Smolin,
\newblock (2008), arXiv:0808.3765.

\bibitem{chaichan}
M. Chaichian et~al.,
\newblock Phys. Lett. B604 (2004) 98, hep-th/0408069.

\bibitem{wessfiore}
G. Fiore and J. Wess,
\newblock Phys. Rev. D 75 (2007) 105022, hep-th/0701078.

\bibitem{balaIRUV}
A.P. Balachandran, A. Pinzul and A.R. Queiroz,
\newblock Phys. Lett. B668 (2008) 241, arXiv:0804.3588.

\bibitem{hawkingCPT}
S.W. Hawking,
\newblock Phys. Rev. D 14 (1976) 2460.

\bibitem{pageCPT}
D.N. Page,
\newblock Phys. Rev. Lett. 44 (1980) 301.

\bibitem{waldCPT}
R. Wald,
\newblock Phys. Rev. D 21 (1980) 2742.

\bibitem{pagegaume1}
D.N. Page,
\newblock Gen. Rel. Grav. 14 (1982) 299.

\bibitem{pagegaume2}
L. Alvarez-Gaume and C. Gomez,
\newblock Commun. Math. Phys. 89 (1983) 235.

\bibitem{emnWcpt}
J. Ellis, N. Mavromatos and D. Nanopoulos,
\newblock Phys. Lett. B293 (1992) 142, hep-ph/9207268.

\bibitem{bertocpt}
O. Bertolami et~al.,
\newblock Phys. Lett. B395 (1997) 178.

\bibitem{ahlucpt}
D. Ahluwalia,
\newblock Mod. Phys. Lett. A13 (1998) 2249, hep-ph/9807267.

\bibitem{muracpt}
H. Murayama and T. Yanagida,
\newblock Phys. Lett. B520 (2001) 263.

\bibitem{klinkCPT}
F.R. Klinkhamer and C. Rupp,
\newblock Phys. Rev. D 70 (2004) 045020, hep-th/0312032.

\bibitem{emnchaos}
J.R. Ellis, N.E. Mavromatos and D.V. Nanopoulos,
\newblock Chaos Solitons Fractals 10 (1999) 345, hep-th/9805120.

\bibitem{hawkingDECO}
S.W. Hawking,
\newblock Commun. Math. Phys. 87 (1982) 395.

\bibitem{rubakovDECO}
G.V. Lavrelashvili, V.A. Rubakov and P.G. Tinyakov,
\newblock Nucl. Phys. B299 (1988) 757.

\bibitem{pullinDECOprl}
R. Gambini, R.A. Porto and J. Pullin,
\newblock Phys. Rev. Lett. 93 (2004) 240401.

\bibitem{kempf}
A. Kempf and G. Mangano,
\newblock Phys. Rev. D 55 (1997) 7909, hep-th/9612084.

\bibitem{ahluDB}
D. Ahluwalia,
\newblock Phys. Lett. A275 (2000) 31.

\bibitem{kappaHP}
A. Blaut et~al.,
\newblock Phys. Lett. B582 (2004) 82, hep-th/0312045.

\bibitem{diffcalc1}
A. Sitarz,
\newblock Phys. Lett. B 349 (1995) 42.

\bibitem{diffcalc2}
S. Majid and R. Oeckl,
\newblock Commun. Math. Phys. 205 (1999) 617.

\bibitem{gacdsr2010}
G. Amelino-Camelia,
\newblock Symmetry 2 (2010) 230, arXiv:1003.3942.

\bibitem{camachoEPdeco}
A. Camacho,
\newblock Int. J. Mod. Phys. 10 (2001) 767, gr-qc/0107028.

\bibitem{ellisNONUNI}
J. Ellis et~al.,
\newblock Nature 428 (2004) 386, astro-ph/0309144.

\bibitem{clausEP}
E. Goklu and C. L\"ammerzahl,
\newblock Class. Quant. Grav. 25 (2008) 105012, arXiv:0801.4553.

\bibitem{stringsEP}
T.R. Taylor and G. Veneziano,
\newblock Phys. Lett. B213 (1988) 450.

\bibitem{repDamourPOLY1}
T. Damour and A.M. Polyakov,
\newblock Nucl. Phys. B423 (1994) 532, hep-th/9401069.

\bibitem{repDamourPOLY2}
T. Damour and A.M. Polyakov,
\newblock Gen. Rel. Grav. 26 (1994) 1171, gr-qc/9411069.

\bibitem{damour}
T. Damour, F. Piazza and G. Veneziano,
\newblock Phys. Rev. D 66 (2002) 046007, hep-th/0205111.

\bibitem{repDamourVARYING}
T. Damour,
\newblock Astrophys. Space Sci. 283 (2003) 445, gr-qc/0210059.

\bibitem{repUzan}
J.P. Uzan,
\newblock Rev. Mod. Phys. 75 (2003) 403, hep-ph/0205340.

\bibitem{repMartins}
C.J.A.P. Martins et~al.,
\newblock Phys. Lett. B585 (2004) 29, astro-ph/0302295.

\bibitem{kostegrav}
A.V. Kostelecky and J.D. Tasson,
\newblock Phys.Rev. D83 (2011) 016013, arXiv:1006.4106.

\bibitem{volovik}
G.E. Volovik,
\newblock Found. Phys. 33 (2003) 349, gr-qc/0301043.

\bibitem{laugh}
R.B. Laughlin,
\newblock Int. J. Mod. Phys. A18 (2003) 831, gr-qc/0302028.

\bibitem{laughchap}
G. Chapline et~al.,
\newblock Int. J. Mod. Phys. A18 (2003) 3587, gr-qc/0012094.

\bibitem{jacoTHERM}
T. Jacobson,
\newblock Phys.Rev.Lett. 75 (1995) 1260, gr-qc/9504004.

\bibitem{verliTHERM}
E.P. Verlinde,
\newblock JHEP 1104 (2011) 029, arXiv:1001.0785.

\bibitem{padmaTHERM}
T. Padmanabhan,
\newblock Rept.Prog.Phys. 73 (2010) 046901, arXiv:0911.5004.

\bibitem{dsrREVijmp2002}
G. Amelino-Camelia,
\newblock Int. J. Mod. Phys. 11 (2002) 1643, gr-qc/0210063.

\bibitem{unruhNONLOCAL}
R. Schutzhold and W.G. Unruh,
\newblock JETP Lett. 78 (2003) 431, gr-qc/0308049.

\bibitem{dedeoNONLOCAL}
S. DeDeo and C. Prescod-Weinstein,
\newblock (2008), arXiv:0811.1999.

\bibitem{sabinePRL}
S. Hossenfelder,
\newblock Phys. Rev. Lett. 104 (2010) 140402, arXiv:1004.0418.

\bibitem{bob}
G. Amelino-Camelia et~al.,
\newblock Phys. Rev. Lett. 106 (2011) 071301, arXiv:1006.2126.

\bibitem{leeLIMITEDinertial}
L. Smolin,
\newblock (2010), arXiv:1007.0718.

\bibitem{michJUREoct2010}
M. Arzano and J. Kowalski-Glikman,
\newblock Class. Quantum Grav. 28 (2011) 105009, arXiv:1008.2962.

\bibitem{principleRL}
G. Amelino-Camelia et~al.,
\newblock Phys.Rev. D84 (2011) 084010, arXiv:1101.0931.

\bibitem{kowadsr}
J. Kowalski-Glikman,
\newblock Phys. Lett. A286 (2001) 391, hep-th/0102098.

\bibitem{gacroxkowa}
N.R. Bruno, G. Amelino-Camelia and J. Kowalski-Glikman,
\newblock Phys. Lett. B522 (2001) 133, hep-th/0107039.

\bibitem{leedsrPRL}
J. Magueijo and L. Smolin,
\newblock Phys. Rev. Lett. 88 (2002) 190403, hep-th/0112090.

\bibitem{leedsrPRD}
J. Magueijo and L. Smolin,
\newblock Phys. Rev. D 67 (2003) 044017, gr-qc/0207085.

\bibitem{jurekDSRnew}
J. Kowalski-Glikman and S. Nowak,
\newblock Int. J. Mod. Phys. 12 (2003) 299, hep-th/0204245.

\bibitem{rainbowDSR}
J. Magueijo and L. Smolin,
\newblock Class. Quant. Grav. 21 (2004) 1725, gr-qc/0305055.

\bibitem{repJUREK1}
J. Kowalski-Glikman and S. Nowak,
\newblock Phys. Lett. B539 (2002) 126, hep-th/0203040.

\bibitem{repJUREK2}
J. Kowalski-Glikman and S. Nowak,
\newblock Class. Quant. Grav. 20 (2003) 4799, hep-th/0304101.

\bibitem{repJUREK3}
J. Kowalski-Glikman,
\newblock Phys. Lett. B547 (2002) 291, hep-th/0207279.

\bibitem{repAHLUWA}
D. Ahluwalia,
\newblock (2002), gr-qc/0212128.

\bibitem{repDaszk}
M. Daszkiewicz et~al.,
\newblock Int. J. Mod. Phys. A20 (2005) 4925, hep-th/0410058.

\bibitem{repRembie}
J. Rembielinski and K.A. Smolinski,
\newblock Bull. Soc. Sci. Lett. Lodz 53 (2003) 57, hep-th/0207031.

\bibitem{hossePRD}
S. Hossenfelder,
\newblock Phys. Rev. D 75 (2007) 105005, hep-th/0702016.

\bibitem{judesvisser}
S. Judes and M. Visser,
\newblock Phys. Rev. D 68 (2003) 045001, gr-qc/0205067.

\bibitem{stefanoDSRinterpr}
S. Liberati, S. Sonego and M. Visser,
\newblock Phys. Rev. D 71 (2005) 045001, gr-qc/0410113.

\bibitem{dsrgzk}
G. Amelino-Camelia,
\newblock Int. J. Mod. Phys. 12 (2003) 1211, astro-ph/0209232.

\bibitem{susskind}
A. Matusis, L. Susskind and N. Toumbas,
\newblock JHEP 12 (2000) 002, hep-th/0002075.

\bibitem{lukieFT}
P. Kosinski, J. Lukierski and P. Maslanka,
\newblock Czech.J.Phys. 50 (2000) 1283, hep-th/0009120.

\bibitem{gacmich}
G. Amelino-Camelia and M. Arzano,
\newblock Phys. Rev. D 65 (2002) 084044, hep-th/0105120.

\bibitem{gampulDENSI}
R. Gambini, R. Porto and J. Pullin,
\newblock Class. Quant. Grav. 21 (2004) L51, gr-qc/0305098.

\bibitem{kosinski:2002gu}
P. Kosinski and P. Maslanka,
\newblock Phys. Rev. D 68 (2003) 067702, hep-th/0211057.

\bibitem{mignemi:2003ab}
S. Mignemi,
\newblock Phys. Lett. A316 (2003) 173, hep-th/0302065.

\bibitem{daszkiewicz:2003yr}
M. Daszkiewicz, K. Imilkowska and J. Kowalski-Glikman,
\newblock Phys. Lett. A323 (2004) 345, hep-th/0304027.

\bibitem{jurekREV}
J. Kowalski-Glikman,
\newblock Lect. Notes Phys. 669 (2005) 131, hep-th/0405273.

\bibitem{tedAP}
T. Jacobson, S. Liberati and D. Mattingly,
\newblock Annals Phys. 321 (2006) 150, astro-ph/0505267.

\bibitem{unoEdue}
G. Amelino-Camelia and L. Smolin,
\newblock Phys. Rev. D 80 (2009) 084017, arXiv:0906.3731.

\bibitem{stefaSIGLemn}
L. Maccione, S. Liberati and G. Sigl,
\newblock Phys. Rev. Lett. 105 (2010) 021101, arXiv:1003.5468.

\bibitem{kosteEM}
V.A. Kostelecky and M. Mewes,
\newblock Phys.Rev. D80 (2009) 015020, arXiv:0905.0031.

\bibitem{rob}
R. Myers and M. Pospelov,
\newblock Phys. Rev. Lett. 90 (2003) 211601, hep-ph/0301124.

\bibitem{gacBIREFgiulia}
G. Gubitosi et~al.,
\newblock JCAP 0908 (2009) 021, arXiv:0904.3201.

\bibitem{pospeSYMM}
S. Groot~Nibbelink and M. Pospelov,
\newblock Phys. Rev. Lett. 94 (2005) 081601, hep-ph/0404271.

\bibitem{pospeSYMM2}
P.A. Bolokhov, S.G. Nibbelink and M. Pospelov,
\newblock Phys. Rev. D 72 (2005) 015013, hep-ph/0505029.

\bibitem{pospeHL}
M. Pospelov and Y. Shang,
\newblock Phys.Rev. D85 (2012) 105001, arXiv:1010.5249.

\bibitem{anselmi1}
D. Anselmi and M. Halat,
\newblock Phys. Rev. D 76 (2007) 125011, arXiv:0707.2480.

\bibitem{anselmi2}
D. Anselmi,
\newblock Phys. Rev. D 79 (2009) 025017, arXiv:0808.3475.

\bibitem{horava}
P. Horava,
\newblock Phys. Rev. D 79 (2009), arXiv:0901.3775.

\bibitem{visser}
T.P. Sotiriou, M. Visser and S. Weinfurtner,
\newblock JHEP 10 (2009) 033, arXiv:0905.2798.

\bibitem{tedOLDgood}
T. Jacobson, S. Liberati and D. Mattingly,
\newblock Phys. Rev. D 66 (2002) 081302, hep-ph/0112207.

\bibitem{gacpion}
G. Amelino-Camelia,
\newblock Phys. Lett. B528 (2002) 181, gr-qc/0107086.

\bibitem{seth}
T. Konopka and S. Major,
\newblock New J. Phys. 4 (2002) 57, hep-ph/0201184.

\bibitem{orfeupion}
O. Bertolami,
\newblock Lect. Notes Phys. 633 (2003) 96, hep-ph/0301191.

\bibitem{maccione:2008iw}
L. Maccione and S. Liberati,
\newblock J. Cosmol. Astropart. Phys.  (2008), arXiv:0805.2548.

\bibitem{steckGlashow}
F. Stecker and S. Glashow,
\newblock Astropart. Phys. 16 (2001) 97, astro-ph/0102226.

\bibitem{lehnertTHRESH}
R. Lehnert,
\newblock Phys. Rev. D 68 (2003) 085003, gr-qc/0304013.

\bibitem{repLiberati}
T. Jacobson, S. Liberati and D. Mattingly,
\newblock Phys. Rev. D 67 (2003) 124011, hep-ph/0209264.

\bibitem{maREVIEW}
L. Shao and B.Q. Ma,
\newblock Mod.Phys.Lett. A25 (2010) 3251, arXiv:1007.2269.

\bibitem{ahaMKcut501}
HEGRA, F. Aharonian et~al.,
\newblock (2000), astro-ph/0011483.

\bibitem{h1426}
HEGRA, F. Aharonian et~al.,
\newblock (2002), astro-ph/0202072.

\bibitem{tev45}
K. Okumura et~al.,
\newblock Astrophys. J. 579 (2002) L9, astro-ph/0209487.

\bibitem{krennMKcut}
F. Krennrich et~al.,
\newblock Astrophys. J. 560 (2001) L45, astro-ph/0107113.

\bibitem{ahaMKcut}
HEGRA, F. Aharonian et~al.,
\newblock Astron. Astrophys. 393 (2002) 89, astro-ph/0205499.

\bibitem{dirbe}
D. Finkbeiner, M. Davis and D. Schlegel,
\newblock Astrophys. J. 544 (2000) 81, astro-ph/0004175.

\bibitem{voelk}
H. Volk,
\newblock (2002), astro-ph/0210297.

\bibitem{berezin}
V. Berezinsky,
\newblock Phys. Atom. Nucl. 66 (2003) 423, astro-ph/0107306.

\bibitem{firbREVIEW}
M.G. Hauser and E. Dwek,
\newblock Ann. Rev. Astron. Astrophys. 39 (2001) 249, astro-ph/0105539.

\bibitem{steckXRAY}
A. Konopelko et~al.,
\newblock Astrophys. J. 597 (2003) 851, astro-ph/0302049.

\bibitem{steckIRSTRONG}
F. Stecker,
\newblock Astropart. Phys. 20 (2003) 85, astro-ph/0308214.

\bibitem{alfaroUHECR}
J. Alfaro and G. Palma,
\newblock Phys. Rev. D 65 (2002) 103516, hep-th/0111176.

\bibitem{nguhecr}
Y. Ng et~al.,
\newblock Phys. Lett. B 507 (2001) 236.

\bibitem{greisenGZK}
K. Greisen,
\newblock Phys. Rev. Lett. 16 (1966) 748.

\bibitem{zatsepinGZK}
G.T. Zatsepin and V.A. Kuzmin,
\newblock JETP Lett. 4 (1966) 78.

\bibitem{mooreUHECR}
O. Gagnon and G.D. Moore,
\newblock Phys. Rev. D 70 (2004) 065002, hep-ph/0404196.

\bibitem{domokosUHECR}
S. Kovesi-Domokos and G. Domokos,
\newblock Nucl. Phys. Proc. Suppl. 151 (2006) 38, hep-ph/0501281.

\bibitem{augerPRL2010}
Pierre Auger Observatory, J. Abraham et~al.,
\newblock Phys. Rev. Lett. 104 (2010) 091101, arXiv:1002.0699.

\bibitem{gaisserUHECR2011}
T.K. Gaisser,
\newblock (2010), arXiv:1010.5996.

\bibitem{stanevUHECR2011}
T. Stanev,
\newblock (2010), arXiv:1011.1872.

\bibitem{maccioneUHECR2011}
A. Saveliev, L. Maccione and G. Sigl,
\newblock (2011), arXiv:1101.2903.

\bibitem{agasa}
M. Takeda et~al.,
\newblock Phys. Rev. Lett. 81 (1998) 1163.

\bibitem{augerdata1}
Auger, E. Parizot,
\newblock (2007), arXiv:0709.2500.

\bibitem{augerPLB2010}
Pierre Auger collaboration, J. Abraham et~al.,
\newblock Phys. Lett. B685 (2010) 239, arXiv:1002.1975.

\bibitem{augersources}
J. Abraham et~al.,
\newblock Science 318 (2007) 938.

\bibitem{augersourcesNEW}
Pierre Auger Observatory, P. Abreu et~al.,
\newblock Astropart. Phys. 34 (2010) 314, arXiv:1009.1855.

\bibitem{maccione:2009ju}
L. Maccione et~al.,
\newblock J. Cosmol. Astropart. Phys.  (2009), arXiv:0902.1756.

\bibitem{colgla}
S.R. Coleman and S.L. Glashow,
\newblock Phys. Rev. D 59 (1999) 116008, hep-ph/9812418.

\bibitem{dedenko}
E.E. Antonov et~al.,
\newblock JETP Lett. 73 (2001) 446.

\bibitem{siglpion}
M. Galaverni and G. Sigl,
\newblock Phys. Rev. Lett. 100 (2008) 021102, arXiv:0708.1737.

\bibitem{augershowers}
Pierre Auger collaboration, J. Abraham et~al.,
\newblock (2009), arXiv:0906.2319.

\bibitem{aloisioshowers}
R. Aloisio et~al.,
\newblock Phys. Rev. D 77 (2008) 025007, arXiv:0706.2834.

\bibitem{wilkshowers}
G. Wilk and Z. Wlodarczyk,
\newblock (2010), arXiv:1006.1781.

\bibitem{piranIRnew}
U. Jacob and T. Piran,
\newblock Phys. Rev. D 78 (2008), arXiv:0810.1318.

\bibitem{altschuCHERE}
B. Altschul,
\newblock Phys. Rev. Lett. 98 (2007) 041603, hep-th/0609030.

\bibitem{altschu}
B. Altschul,
\newblock Phys. Rev. Lett. 96 (2006) 201101, hep-ph/0603138.

\bibitem{glast1}
J.P. Norris et~al.,
\newblock (1999), astro-ph/9912136.

\bibitem{glast2}
A. de~Angelis,
\newblock (2000), astro-ph/0009271.

\bibitem{glast3}
N. Omodei,
\newblock AIP Conf. Proc. 836 (2006) 642, astro-ph/0603762.

\bibitem{unoSCIENCE}
A. Abdo et~al.,
\newblock Science 323 (2009) 1688.

\bibitem{fermiGRB090510}
Fermi LAT and Fermi GBM, A. Abdo et~al.,
\newblock Nature 462 (2009) 331.

\bibitem{fermiPHYSREPORT}
P.F. Michelson, W.B. Atwood and S. Ritz,
\newblock Rept. Prog. Phys. 73 (2010) 074901, arXiv:1011.0213.

\bibitem{shore}
G.M. Shore,
\newblock Contemp. Phys. 44 (2003) 503, gr-qc/0304059.

\bibitem{fotiniPRD}
A. Hamma et~al.,
\newblock Phys. Rev. D 81 (2010) 104032, arXiv:0911.5075.

\bibitem{piotrKAPPAspeed}
P. Czerhoniak,
\newblock Mod.Phys.Lett. A15 (2000) 1823, hep-th/0012066.

\bibitem{piranKARP}
T. Piran,
\newblock Lect. Notes Phys. 669 (2005) 351, astro-ph/0407462.

\bibitem{wagnerGRB}
R. Wagner,
\newblock AIP Conf. Proc. 1112 (2009) 187, arXiv:0901.2932.

\bibitem{falconeGRB}
A.D. Falcone et~al.,
\newblock (2008), arXiv:0810.0520.

\bibitem{bepposax}
E. Costa et~al.,
\newblock Nature  (1997), astro-ph/9706065.

\bibitem{bombelli}
L. Bombelli and O. Winkler,
\newblock Class. Quantum Grav. 21 (2004) L89, gr-qc/0403049.

\bibitem{granotFERMIERA}
J. Granot et~al.,
\newblock (2009), arXiv:0905.2206.

\bibitem{ellisREDSHIuno}
J. Ellis et~al.,
\newblock Astropart.Phys. 25 (2006) 402, astro-ph/0510172.

\bibitem{piranRedShFirst}
M. Rodriguez~Martinez and T. Piran,
\newblock JCAP 0604 (2006) 006, astro-ph/0601219.

\bibitem{piranRedSh}
U. Jacob and T. Piran,
\newblock JCAP 0801 (2008) 031, arXiv:0712.2170.

\bibitem{ellisREDSHIdue}
J. Ellis et~al.,
\newblock Astropart.Phys. 29 (2008) 158, astro-ph/0510172.

\bibitem{magicGENERAL}
MAGIC Collaboration, D. Ferenc,
\newblock Nuclear Instruments and Methods in Physics Research A 553 (2005) 274
  .

\bibitem{hessgeneral}
HESS Collaboration, J. Hinton,
\newblock New Astron.Rev. 48 (2004) 331, astro-ph/0403052.

\bibitem{emnrevASTRO}
J. Ellis et~al.,
\newblock Astrophys.J. 535 (2000) 139, astro-ph/9907340.

\bibitem{magic}
MAGIC, J. Albert et~al.,
\newblock Phys. Lett. B668 (2008) 253, arXiv:0708.2889.

\bibitem{ellisUNO}
J. Ellis, N. Mavromatos and D. Nanopoulos,
\newblock Phys. Lett. B 674 (2009) 83, arXiv:0901.4052.

\bibitem{hessPRL2008}
F. Aharonian et~al.,
\newblock Phys. Rev. Lett. 101 (2008) 170402, arXiv:0810.3475.

\bibitem{hess2011}
HESS, A. Abramowski et~al.,
\newblock Astropart. Phys. 34 (2011) 738, arXiv:1101.3650.

\bibitem{hessREVIEW}
J. Bolmont and A. Jacholkowska,
\newblock Adv. Space Res. 47 (2011) 380, arXiv:1007.4954.

\bibitem{grb090902B}
A. Abdo et~al.,
\newblock Astrophys.J. 706 (2009) L138, arXiv:0909.2470.

\bibitem{grb090926A}
J. Sacahui et~al.,
\newblock Astrophys.J. 755 (2012) 127, arXiv:1203.1577.

\bibitem{nemiroff090510}
R.J. Nemiroff, J. Holmes and R. Connolly,
\newblock Phys.Rev.Lett. 108 (2012) 231103, arXiv:1109.5191.

\bibitem{meszarosUNO}
K. Toma, X.F. Wu and P. Meszaros,
\newblock Astrophys.J. 707 (2009) 1404, arXiv:0905.1697.

\bibitem{grbMistery}
M. Lyutikov,
\newblock (2009), arXiv:0911.0349.

\bibitem{guettaSHORTstrange}
A. Corsi, D. Guetta and L. Piro,
\newblock (2009), arXiv:0905.1513.

\bibitem{ghiselliniDUE}
G. Ghirlanda, G. Ghisellini and L. Nava,
\newblock (2009), arXiv:0909.0016.

\bibitem{lazzatiPRECURSOR}
D. Lazzati,
\newblock Mon. Not. Roy. Astron. Soc. 357 (2005) 722, astro-ph/0411753.

\bibitem{troja090510}
E. Troja, S. Rosswog and N. Gehrels,
\newblock Astrophys. J. 723 (2010) 1711, arXiv:1009.1385.

\bibitem{integralGRBPOLAR}
P. Laurent et~al.,
\newblock Phys. Rev. D 83 (2011) 121301, arXiv:1106.1068.

\bibitem{toma1GRBPOLAR}
D. Yonetoku et~al.,
\newblock (2011), arXiv:1111.1779.

\bibitem{toma2GRBPOLAR}
K. Toma et~al.,
\newblock (2012), arXiv:1208.5288.

\bibitem{stringsHP}
A. Kempf, G. Mangano and R. Mann,
\newblock Phys. Rev. D 52 (1995) 1108, hep-th/9412167.

\bibitem{mignemiHP}
S. Mignemi,
\newblock Phys. Rev. D 68 (2003) 065029, gr-qc/0304029.

\bibitem{galanDSRhp}
P. Galan and G.A. Mena~Marugan,
\newblock Int.J.Mod.Phys. 16 (2007) 1133, gr-qc/0702027.

\bibitem{meljanacHP}
S. Meljanac and A. Samsarov,
\newblock Int.J.Mod.Phys. A26 (2011) 1439, arXiv:1007.3943.

\bibitem{rissehomola}
M. Risse and P. Homola,
\newblock Mod. Phys. Lett. A22 (2007) 749, astro-ph/0702632.

\bibitem{ellisgrqc9911055}
J. Ellis et~al.,
\newblock Gen. Rel. Grav. 32 (2000) 1777, gr-qc/9911055.

\bibitem{kingNeutri}
S. Choubey and S.F. King,
\newblock Phys. Rev. D 67 (2003) 073005, hep-ph/0207260.

\bibitem{gacneutrinos}
G. Amelino-Camelia,
\newblock Int. J. Mod. Phys. 12 (2003) 1633, gr-qc/0305057.

\bibitem{piranNeutriNat}
U. Jacob and T. Piran,
\newblock Nature Physics 3 (2007) 87.

\bibitem{bertoNeutri}
O. Bertolami and C.S. Carvalho,
\newblock Phys. Rev. D 61 (2000) 103002, gr-qc/9912117.

\bibitem{grbNEUTRINOnew}
P. Meszaros et~al.,
\newblock astro-ph/0305066  (2003), astro-ph/0305066.

\bibitem{waxneutri}
E. Waxman,
\newblock Phil. Trans. Roy. Soc. Lond. A365 (2007) 1323, astro-ph/0701170.

\bibitem{cohegla}
A.G. Cohen and S.L. Glashow,
\newblock Phys.Rev.Lett. 107 (2011) 181803, arXiv:1109.6562.

\bibitem{libeneutri}
L. Maccione, S. Liberati and D.M. Mattingly,
\newblock (2011), arXiv:1110.0783.

\bibitem{brust}
R. Brustein, D. Eichler and S. Foffa,
\newblock Phys. Rev. D 65 (2002) 105006, hep-ph/0106309.

\bibitem{emnEscape}
J. Ellis, N.E. Mavromatos and D. Nanopoulos,
\newblock Phys. Rev. D 65 (2002) 064007, astro-ph/0108295.

\bibitem{gamboneutri}
P. Arias et~al.,
\newblock Phys. Lett. B650 (2007) 401, hep-ph/0608007.

\bibitem{winsta}
D. Morgan et~al.,
\newblock (2007), arXiv:0705.1897.

\bibitem{joyneutri}
J. Christian,
\newblock Phys. Rev. D 71 (2005), gr-qc/0409077.

\bibitem{carcorNEUTRI}
J.M. Carmona, J.L. Cortes and J. Indurain,
\newblock (2007), arXiv:0709.2267.

\bibitem{jaconature}
T. Jacobson, S. Liberati and D. Mattingly,
\newblock Nature 424 (2003) 1019, astro-ph/0212190.

\bibitem{tedsteck}
T. Jacobson et~al.,
\newblock Phys. Rev. Lett. 93 (2004) 021101, astro-ph/0309681.

\bibitem{stefaCrab}
L. Maccione et~al.,
\newblock JCAP 0710 (2007) 013, arXiv:0707.2673.

\bibitem{repEllisCrab}
J. Ellis, N.E. Mavromatos and A.S. Sakharov,
\newblock Astropart. Phys. 20 (2004) 669, astro-ph/0308403.

\bibitem{urruSYNCH}
R. Montemayor and L.F. Urrutia,
\newblock Phys. Rev. D 72 (2005) 045018, hep-ph/0505135.

\bibitem{altschuSYNCH}
B. Altschul,
\newblock Phys. Rev. D 72 (2005) 085003, hep-th/0507258.

\bibitem{jackson}
J. Jackson,
\newblock Classical Electrodynamics, 3rd ed. (John Wiley \& Sons, New York,
  1999).

\bibitem{gleiser1}
R. Gleiser and C. Kozameh,
\newblock Phys. Rev. D 64 (2001) 083007, gr-qc/0102093.

\bibitem{gleiser2}
R. Gleiser, C. Kozameh and F. Parisi,
\newblock Class. Quant. Grav. 20 (2003) 4375, gr-qc/0304048.

\bibitem{jackbire}
S.M. Carroll, G.B. Field and R. Jackiw,
\newblock Phys. Rev. D 41 (1990) 1231.

\bibitem{jackbireB}
V. Kosteleck{\'{y}} and M. Mewes,
\newblock Phys. Rev. Lett. 87 (2001) 251304, hep-ph/0111026.

\bibitem{maBIREF}
L. Shao and B.Q. Ma,
\newblock Phys. Rev. D 83 (2011) 127702, arXiv:1104.4438.

\bibitem{liberati2009}
S. Liberati and L. Maccione,
\newblock Ann. Rev. Nucl. Part. Sci. 59 (2009) 245, arXiv:0906.0681.

\bibitem{polarGRB}
W. Coburn and S. Boggs,
\newblock Nature 423 (2003) 415.

\bibitem{nopolarGRB}
R. Rutledge and D. Fox,
\newblock Mon. Not. Roy. Astron. Soc. 350 (2004) 1272, astro-ph/0310385.

\bibitem{nopolarGRBbis}
S. Boggs and W. Coburn,
\newblock (2003), astro-ph/0310515.

\bibitem{gaclaem}
G. Amelino-Camelia and C. L\"ammerzahl,
\newblock Class. Quant. Grav. 21 (2004) 899, gr-qc/0306019.

\bibitem{shgref}
E. Sauter,
\newblock Nonlinear Optics (John Wiley \& Sons, New York, 1996).

\bibitem{gharibyan2003}
V. Gharibyan,
\newblock Phys.Lett. B611 (2005) 231, hep-ex/0303010.

\bibitem{gharibyan2012}
V. Gharibyan,
\newblock (2012), arXiv:1207.7297.

\bibitem{repCOLLAkoste1}
D. Colladay and V. Kosteleck{\'{y}},
\newblock Phys. Rev. D 55 (1997) 6760, hep-ph/9703464.

\bibitem{repCOLLAkoste2}
D. Colladay and V. Kosteleck{\'{y}},
\newblock Phys. Rev. D 58 (1998) 116002, hep-ph/9809521.

\bibitem{sme1}
V. Kosteleck{\'{y}} and C.D. Lane,
\newblock Phys. Rev. D 60 (1999) 116010, hep-ph/9908504.

\bibitem{sme2}
V. Kosteleck{\'{y}} and M. Mewes,
\newblock Phys. Rev. D 66 (2002) 056005, hep-ph/0205211.

\bibitem{blumnew}
R. Bluhm,
\newblock Lect. Notes Phys. 702 (2006) 191, hep-ph/0506054.

\bibitem{laemSME1}
C. L{\"{a}}mmerzahl,
\newblock Class. Quant. Grav. 15 (1998) 13.

\bibitem{laemSME2}
C. L{\"{a}}mmerzahl, A. Macias and H. Mueller,
\newblock Phys. Rev. D 71 (2005) 025007, gr-qc/0501048.

\bibitem{toulouse}
P. Gaete and C. Wotzasek,
\newblock Phys. Rev. D 75 (2007), hep-ph/0607321.

\bibitem{kosteDATATABLES}
V. Kosteleck{\'{y}} and N. Russell,
\newblock Rev. Mod. Phys. 83 (2011) 11, arXiv:0801.0287.

\bibitem{subirMDRreview}
S. Sarkar,
\newblock Mod.Phys.Lett. A17 (2002) 1025, gr-qc/0204092.

\bibitem{hosseLee}
S. Hossenfelder and L. Smolin,
\newblock Phys.Canada 66 (2010) 99, arXiv:0911.2761.

\bibitem{iucaapap}
G. Amelino-Camelia,
\newblock Mod. Phys. Lett. A17 (2002) 899, gr-qc/0204051.

\bibitem{wheelerFORSEFOAM}
J. Wheeler,
\newblock Ann. Phys. 2 (1957) 604.

\bibitem{deserFOAM1957}
S. Deser,
\newblock Rev. Mod. Phys. 29 (1957) 417.

\bibitem{colemanFOAM}
S.R. Coleman,
\newblock Nucl.Phys. B307 (1988) 867.

\bibitem{hawkingFOAM}
S. Hawking,
\newblock Phys. Rev. D 53 (1996) 3099, hep-th/9510029.

\bibitem{carlipFOAM}
S. Carlip,
\newblock Phys.Rev.Lett. 79 (1997) 4071, gr-qc/9708026.

\bibitem{garayFOAM}
L.J. Garay,
\newblock Phys.Rev.Lett. 80 (1998) 2508, gr-qc/9801024.

\bibitem{woodardfoam}
R. Woodard,
\newblock Rept.Prog.Phys. 72 (2009) 126002, arXiv:0907.4238.

\bibitem{meusSchr3DQG}
C. Meusburger and B. Schroers,
\newblock Class. Quantum Grav. 20 (2003) 2193, gr-qc/0301108.

\bibitem{carlip3DQGlrr}
S. Carlip,
\newblock Living Rev. Rel. 8 (2005) 1, gr-qc/0409039.

\bibitem{oritigirelli3DQGhopf}
F. Girelli, E.R. Livine and D. Oriti,
\newblock Nucl.Phys. B708 (2005) 411, gr-qc/0406100.

\bibitem{freidellivine3DQGhopf}
L. Freidel and E.R. Livine,
\newblock Phys.Rev.Lett. 96 (2006) 221301, hep-th/0512113.

\bibitem{noui3DQGhopf}
E. Joung, J. Mourad and K. Noui,
\newblock J.Math.Phys. 50 (2009) 052503, arXiv:0806.4121.

\bibitem{schr3dQG2012}
P.K. Osei and B.J. Schroers,
\newblock J.Math.Phys. 53 (2012) 073510, arXiv:1109.4086.

\bibitem{ahluwaliaFOAMandASYMMETRY}
D.V. Ahluwalia and M. Kirchbach,
\newblock Int. J. Mod. Phys. 10 (2001) 811, astro-ph/0107246.

\bibitem{ahluTALKwithMYNAPA1}
D. Ahluwalia,
\newblock (2002), gr-qc/0202098.

\bibitem{saulson}
P. Saulson,
\newblock Fundamentals of Interferometric Gravitational Wave Detectors (World
  Scientific, Singapore; River Edge, NJ, 1994).

\bibitem{ligoSCIENCE}
A. Abramovici et~al.,
\newblock Science 256 (1992) 325.

\bibitem{ligoREVIEW2009}
LIGO Scientific Collaboration, B. Abbott et~al.,
\newblock Rept.Prog.Phys. 72 (2009) 076901, arXiv:0711.3041.

\bibitem{virgo1997}
B. Caron et~al.,
\newblock Class. Quantum Grav. 14 (1997) 1461.

\bibitem{virgo2008}
F. Acernese et~al.,
\newblock Class. Quantum Grav. 25 (2008) 184001.

\bibitem{ngSelectedTopics}
Y.J. Ng,
\newblock Mod. Phys. Lett. A18 (2003) 1073, gr-qc/0305019.

\bibitem{ngFOAM2011}
W.A. Christiansen et~al.,
\newblock Phys. Rev. D 83 (2011) 084003, arXiv:0912.0535.

\bibitem{karo}
F. Karolyhazy,
\newblock Nuovo Cim. A42 (1966) 390.

\bibitem{diosi}
L. Diosi and B. Lukacs,
\newblock Nuovo Cim. B108 (1993) 1419, gr-qc/9302028.

\bibitem{sabineMINLENreview}
S. Hossenfelder,
\newblock (2012), arXiv:1203.6191.

\bibitem{hoganNOISE}
C.J. Hogan,
\newblock Phys. Rev. D 77 (2008) 104031, arXiv:0712.3419.

\bibitem{hoganGEO600}
C. Hogan,
\newblock Phys. Rev. D 78 (2008), arXiv:0806.0665.

\bibitem{hogan2012}
C. Hogan,
\newblock (2012), arXiv:1208.3703.

\bibitem{geo600general}
B. Willke et~al.,
\newblock Class. Quantum Grav. 19 (2002) 1377.

\bibitem{geo600noise}
H. Luck et~al.,
\newblock J.Phys.Conf.Ser. 228 (2010) 012012, arXiv:1004.0339.

\bibitem{geo600nonoise}
M. Prijatelj et~al.,
\newblock Class. Quantum Grav. 29 (2012) 055009.

\bibitem{lisa}
G. Heinzel et~al.,
\newblock Class. Quantum Grav. 23 (2006) S119.

\bibitem{laemgwi}
S. Schiller et~al.,
\newblock Phys. Rev. D 69 (2004) 027504, gr-qc/0401103.

\bibitem{tinovetrano}
G. Tino and F. Vetrano,
\newblock Class. Quantum Grav. 24 (2007) 2167, gr-qc/0702118.

\bibitem{dimopGWI}
S. Dimopoulos et~al.,
\newblock Phys. Lett. B 678 (2009) 37, arXiv:0712.1250.

\bibitem{scargleCAUSALSETS}
J.D. Scargle and S.N. Simic,
\newblock (2009), arXiv:0912.3839.

\bibitem{uncpa1}
G. Amelino-Camelia, Y.J. Ng and H. Vanm~Dam,
\newblock Astropart. Phys. 19 (2003) 729, gr-qc/0204077.

\bibitem{grilloNONSYST}
R. Aloisio et~al.,
\newblock Astropart. Phys. 19 (2003) 127, astro-ph/0205271.

\bibitem{weilerUHECRns}
M. Jankiewicz et~al.,
\newblock Astropart.Phys. 21 (2004) 651, hep-ph/0312221.

\bibitem{mattiUHECRns}
S. Basu and D. Mattingly,
\newblock Class.Quant.Grav. 22 (2005) 3029, astro-ph/0501425.

\bibitem{lieu}
R. Lieu and L. Hillman,
\newblock Astrophys. J. 585 (2003) L77, astro-ph/0301184.

\bibitem{jackNEWastroint}
Y.J. Ng, H. van Dam and W.A. Christiansen,
\newblock Astrophys. J. 591 (2003) L87, astro-ph/0302372.

\bibitem{coule}
D. Coule,
\newblock Class. Quant. Grav. 20 (2003) 3107, astro-ph/0302333.

\bibitem{ragazzoniALALIEU}
R. Ragazzoni, M. Turatto and W. Gaessler,
\newblock Astrophys. J. 587 (2003) L1, astro-ph/0303043.

\bibitem{gallouALALIEU2003}
R. Le~Gallou,
\newblock Astropart.Phys. 20 (2004) 703, astro-ph/0304560.

\bibitem{jackALALIEUprl}
W. Christiansen, Y. Ng and H. van Dam,
\newblock Phys.Rev.Lett. 96 (2006) 051301, gr-qc/0508121.

\bibitem{chenALALIEU2006}
Y. Chen and L. Wen,
\newblock (2006), gr-qc/0605093.

\bibitem{maziaALALIEU2006}
M. Maziashvili,
\newblock (2006), hep-ph/0605146.

\bibitem{fuzzyseen}
E. Steinbring,
\newblock Astrophys.J. 655 (2007) 714, astro-ph/0610422.

\bibitem{maziaALALIEU2007}
M. Maziashvili,
\newblock Phys.Lett. B666 (2008) 364, arXiv:0708.1472.

\bibitem{maziaALALIEU2009}
M. Maziashvili,
\newblock Astropart.Phys. 31 (2009) 344, arXiv:0901.2405.

\bibitem{tamburiniALALIEU2011}
F. Tamburini et~al.,
\newblock Astron.Astrophys. 533 (2011) A71, arXiv:1108.6005.

\bibitem{jackALALIEU2011}
E.S. Perlman et~al.,
\newblock (2011), arXiv:1110.4986.

\bibitem{maziaALALIEU2012}
M. Maziashvili,
\newblock (2012), arXiv:1206.4388.

\bibitem{maianiCPTdafne}
L. Maiani,
\newblock Nuclear Physics A 623 (1997) 16 .

\bibitem{okunCPT}
L.B. Okun,
\newblock (1996), hep-ph/9612247.

\bibitem{fidecaroCPT}
M. Fidecaro and H.J. Gerber,
\newblock Rept. Prog. Phys. 69 (2006) 1713, hep-ph/0603075.

\bibitem{kosteCPTprl}
V. Kosteleck{\'{y}},
\newblock Phys. Rev. Lett. 80 (1998) 1818, hep-ph/9809572.

\bibitem{kosteJackiw}
R. Jackiw and V. Kosteleck{\'{y}},
\newblock Phys. Rev. Lett. 82 (1999) 3572, hep-ph/9901358.

\bibitem{kosteLorentzCPT2001}
V. Kosteleck{\'{y}} and R. Lehnert,
\newblock Phys. Rev. D 63 (2001) 065008, hep-th/0012060.

\bibitem{greenbergONcpt}
O.W. Greenberg,
\newblock Phys. Rev. Lett. 89 (2002) 231602, hep-ph/0201258.

\bibitem{chaichianONcpt}
M. Chaichian et~al.,
\newblock Phys. Lett. B699 (2011) 177, arXiv:1103.0168.

\bibitem{chaichianCPTncft}
M. Chaichian, K. Nishijima and A. Tureanu,
\newblock Phys. Lett. B568 (2003) 146, hep-th/0209008.

\bibitem{gaumeCPTncft}
L. Alvarez-Gaume and M.A. Vazquez-Mozo,
\newblock Nucl. Phys. B668 (2003) 293, hep-th/0305093.

\bibitem{pospelovCPTncft}
I. Mocioiu, M. Pospelov and R. Roiban,
\newblock Phys. Rev. D 65 (2002) 107702, hep-ph/0108136.

\bibitem{jabbariCPTncft}
M. Sheikh-Jabbari,
\newblock Phys.Rev.Lett. 84 (2000) 5265, hep-th/0001167.

\bibitem{balaCPTncft}
E. Akofor et~al.,
\newblock JHEP 08 (2007) 045, arXiv:0706.1259.

\bibitem{kloe1}
KLOE, R. Versaci,
\newblock (2007), hep-ex/0701008.

\bibitem{didomALL}
A.K.C. Di~Domenico,
\newblock Discoveries in Flavour Physics at $e^{+}e^{-}$ Colliders, edited by
  L. Benussi et~al., , Frascati Physics Series Vol. XLI, pp. 79--85, Frascati,
  2006, INFN.

\bibitem{nickBern1}
J. Bernabeu, N.E. Mavromatos and J. Papavassiliou,
\newblock Phys. Rev. Lett. 92 (2004) 131601, hep-ph/0310180.

\bibitem{nickBern2}
J. Bernabeu, N. Mavromatos and S. Sarkar,
\newblock Phys. Rev. D 74 (2006) 045014, hep-th/0606137.

\bibitem{kloe2}
KLOE, M. Testa,
\newblock (2008), arXiv:0805.1969.

\bibitem{didomOMEGA}
A. Di~Domenico,
\newblock arXiv:0904.1976.

\bibitem{nickCPTshort}
N.E. Mavromatos,
\newblock PoS KAON (2008) 041, arXiv:0707.3422.

\bibitem{fermilabE731a}
M. Woods et~al.,
\newblock Phys. Rev. Lett. 60 (1988) 1695.

\bibitem{fermilabE731b}
J.R. Patterson et~al.,
\newblock Phys. Rev. Lett. 64 (1990) 1491.

\bibitem{garayDECO}
L.J. Garay,
\newblock Int. J. Mod. Phys. A14 (1999) 4079, gr-qc/9911002.

\bibitem{tinoadded}
C. Simon and D. Jaksch,
\newblock Phys.Rev. A70 (2004) 052104, quant-ph/0406007.

\bibitem{bobcharles}
C.T. Wang, R. Bingham and J. Mendonca,
\newblock Class. Quant. Grav. 23 (2006) L59, gr-qc/0603112.

\bibitem{wangbinghamBOSEINSTEIN}
C.H.T. Wang, R. Bingham and J.T. Mendonca,
\newblock (2010), arXiv:1002.2962.

\bibitem{nick08010872}
N.E. Mavromatos et~al.,
\newblock Phys. Rev. D 77 (2008) 053014, arXiv:0801.0872.

\bibitem{floreaniniNEUTRI}
F. Benatti and R. Floreanini,
\newblock JHEP 02 (2000) 032, hep-ph/0002221.

\bibitem{ahluDECOneutri}
D.V. Ahluwalia,
\newblock Mod. Phys. Lett. A16 (2001) 917, hep-ph/0104316.

\bibitem{winstaDECOneutri}
D. Morgan et~al.,
\newblock Astropart. Phys. 25 (2006) 311, astro-ph/0412618.

\bibitem{cagoDECOneutri}
A.M. Gago et~al.,
\newblock (2002), hep-ph/0208166.

\bibitem{nickHEPPH0606048}
N.E. Mavromatos and S. Sarkar,
\newblock Phys. Rev. D 74 (2006) 036007, hep-ph/0606048.

\bibitem{balaCHIprl}
A. Balachandran, A. Joseph and P. Padmanabhan,
\newblock Phys.Rev.Lett. 105 (2010) 051601, arXiv:1003.2250.

\bibitem{balaCHIjhep}
A. Balachandran and P. Padmanabhan,
\newblock JHEP 1012 (2010) 001, arXiv:1006.1185.

\bibitem{chakPAULI}
B. Chakraborty et~al.,
\newblock J.Phys. A39 (2006) 9557, hep-th/0601121.

\bibitem{michdarioSTAT}
M. Arzano and D. Benedetti,
\newblock Int.J.Mod.Phys. A24 (2009) 4623, arXiv:0809.0889.

\bibitem{borexino}
Borexino Collaboration, G. Bellini et~al.,
\newblock Phys.Rev. C81 (2010) 034317, arXiv:0911.0548.

\bibitem{vip}
S. Bartalucci et~al.,
\newblock Phys.Lett. B641 (2006) 18.

\bibitem{hensonREVIEW}
J. Henson,
\newblock (2006), gr-qc/0601121.

\bibitem{fotinileeCAUSAL}
F. Markopoulou and L. Smolin,
\newblock Nucl. Phys. B508 (1997) 409, gr-qc/9702025.

\bibitem{johnstonDISCRETE}
S. Johnston,
\newblock Class. Quantum Grav. 25 (2008) 202001, arXiv:0806.3083.

\bibitem{johnstonCAUSALSETS}
S. Johnston,
\newblock Phys. Rev. Lett. 103 (2009) 180401, arXiv:0909.0944.

\bibitem{causalsetsLIEU}
F. Dowker, J. Henson and R. Sorkin,
\newblock Phys. Rev. D 82 (2010) 104048, arXiv:1009.3058.

\bibitem{dowkerCausalphen}
C.R. Contaldi, F. Dowker and L. Philpott,
\newblock Class. Quant. Grav. 27 (2010) 172001, arXiv:1001.4545.

\bibitem{dowkerSorkinMPLA2004}
F. Dowker, J. Henson and R.D. Sorkin,
\newblock Mod. Phys. Lett. A19 (2004) 1829, gr-qc/0311055.

\bibitem{dowkerSorkinPRD2009}
L. Philpott, F. Dowker and R.D. Sorkin,
\newblock Phys. Rev. D 79 (2009) 124047, arXiv:0810.5591.

\bibitem{philpoCAUSAL}
L. Philpott,
\newblock Class. Quantum Grav. 27 (2010) 042001, arXiv:0911.5595.

\bibitem{kaloperMATTINGLY}
N. Kaloper and D. Mattingly,
\newblock Phys. Rev. D 74 (2006) 106001, astro-ph/0607485.

\bibitem{mattinglyCausalSets}
D. Mattingly,
\newblock Phys. Rev. D 77 (2008) 125021, arXiv:0709.0539.

\bibitem{onofrioviola}
L. Viola and R. Onofrio,
\newblock Phys. Rev. D 55 (1997) 455, quant-ph/9612039.

\bibitem{halperinLeung}
A. Halprin, C.N. Leung and J.T. Pantaleone,
\newblock Phys. Rev. D 53 (1996) 5365, hep-ph/9512220.

\bibitem{camachoEP}
A. Camacho,
\newblock Mod. Phys. Lett. A14 (1999) 2545, gr-qc/9911112.

\bibitem{fordCMB}
R. Di~Stefano et~al.,
\newblock (2001), astro-ph/0107001.

\bibitem{yuALAford}
H.w. Yu and P.X. Wu,
\newblock Phys. Rev. D 68 (2003) 084019, gr-qc/0308065.

\bibitem{polarskiALAford}
D. Polarski and P. Roche,
\newblock Mod.Phys.Lett. A20 (2005) 499, gr-qc/0501021.

\bibitem{cohenUVIR}
A.G. Cohen, D.B. Kaplan and A.E. Nelson,
\newblock Phys. Rev. Lett. 82 (1999) 4971, hep-th/9803132.

\bibitem{lightconeNCFT1}
O. Aharony, J. Gomis and T. Mehen,
\newblock JHEP 09 (2000) 023, hep-th/0006236.

\bibitem{lightconeASCHIERI}
P. Aschieri,
\newblock Nucl. Phys. B617 (2001) 321, hep-th/0106281.

\bibitem{urrutiaPRD}
J. Alfaro, H.A. Morales-Tecotl and L.F. Urrutia,
\newblock Phys. Rev. D 66 (2002) 124006, hep-th/0208192.

\bibitem{repCortesTRITIUM}
J.M. Carmona and J.L. Cortes,
\newblock Phys. Lett. B494 (2000) 75, hep-ph/0007057.

\bibitem{josePRD}
J.M. Carmona and J.L. Cortes,
\newblock Phys. Rev. D 65 (2002) 025006, hep-th/0012028.

\bibitem{gacPRL2009}
G. Amelino-Camelia et~al.,
\newblock Phys. Rev. Lett. 103 (2009) 171302, arXiv:0911.1020.

\bibitem{lambMeasKINOSHITA}
T. Kinoshita,
\newblock Rep. Prog. Phys. 59 (1996) 1459.

\bibitem{lambMeasBIRABEN}
M. Weitz et~al.,
\newblock Phys. Rev. A 52 (1995) 2664.

\bibitem{Wicht02}
A. Wicht et~al.,
\newblock Phys. Script. T102 (2002) 82.

\bibitem{gab08}
D. Hanneke, S. Fogwell and G. Gabrielse,
\newblock Phys. Rev. Lett. 100 (2008) 120801, arXiv:0801.1134.

\bibitem{michJUREatom}
M. Arzano, J. Kowalski-Glikman and A. Walkus,
\newblock Europhys.Lett. 90 (2010) 30006, arXiv:0912.2712.

\bibitem{camachoBE2011}
J. Rivas, A. Camacho and E. Goeklue,
\newblock (2011), arXiv:1112.3303.

\bibitem{brisceseBEepl}
F. Briscese, M. Grether and M. de~Llano,
\newblock Europhys.Lett. 98 (2012) 60001, arXiv:1204.4670.

\bibitem{briscesePLB2012}
F. Briscese,
\newblock Physics Letters B 718 (2012) 214, arXiv:1206.1236.

\bibitem{repWeinheimer}
C. Weinheimer et~al.,
\newblock Phys. Lett. B460 (1999) 219.

\bibitem{repLobashev}
V.M. Lobashev et~al.,
\newblock Phys. Lett. B460 (1999) 227.

\bibitem{repMagueijoTRITIUM}
M. Blasone, J. Magueijo and P. Pires-Pacheco,
\newblock Europhys. Lett. 70 (2005) 600, hep-ph/0307205.

\bibitem{repCohenTRITIUM}
A. Cohen and S. Glashow,
\newblock (2006), hep-ph/0605036.

\bibitem{nokepler}
F. Combes et~al.,
\newblock Galaxies and CosmologyAstronomy and Astrophysics Library, 2nd ed.
  (Springer, Berlin; New York, 2002).

\bibitem{mond}
M. Milgrom,
\newblock Astrophys. J. 270 (1983) 371.

\bibitem{hellingCOULOMBscreen}
R.C. Helling and J. You,
\newblock JHEP 0806 (2008) 067, arXiv:0707.1885.

\bibitem{repNesviz}
V.V. Nesvizhevsky et~al.,
\newblock Eur. Phys. J. C40 (2005) 479, hep-ph/0502081.

\bibitem{repBertolami}
O. Bertolami et~al.,
\newblock Phys. Rev. D 72 (2005) 025010, hep-th/0505064.

\bibitem{repBanerjee}
R. Banerjee, B. Dutta~Roy and S. Samanta,
\newblock Phys. Rev. D 74 (2006) 045015, hep-th/0605277.

\bibitem{repSaha}
A. Saha,
\newblock Eur. Phys. J. C51 (2007) 199, hep-th/0609195.

\bibitem{repBrau}
F. Brau and F. Buisseret,
\newblock Phys. Rev. D 74 (2006) 036002, hep-th/0605183.

\bibitem{brandeREVIEWinfla}
R.H. Brandenberger,
\newblock (1999), hep-ph/9910410.

\bibitem{brandePRD2001}
J. Martin and R.H. Brandenberger,
\newblock Phys. Rev. D 63 (2001) 123501, hep-th/0005209.

\bibitem{danielssonINITIAL}
U. Danielsson,
\newblock Phys. Rev. D 66 (2002) 023511, hep-th/0203198.

\bibitem{niemeyerCUTOFF}
J.C. Niemeyer,
\newblock Phys. Rev. D 63 (2001) 123502, astro-ph/0005533.

\bibitem{brandCUTOFF}
R. Brandenberger and J. Martin,
\newblock Mod. Phys. Lett. A16 (2001) 999, astro-ph/0005432.

\bibitem{kempf2001}
A. Kempf,
\newblock Phys. Rev. D 63 (2001) 083514, astro-ph/0009209.

\bibitem{chung0011241}
C.S. Chu, B.R. Greene and G. Shiu,
\newblock Mod. Phys. Lett. A16 (2001) 2231, hep-th/0011241.

\bibitem{mersini0101210}
L. Mersini-Houghton, M. Bastero-Gil and P. Kanti,
\newblock Phys. Rev. D 64 (2001) 043508, hep-ph/0101210.

\bibitem{shiu0104102}
R. Easther et~al.,
\newblock Phys. Rev. D 64 (2001) 103502, hep-th/0104102.

\bibitem{shiu0411217}
B. Greene et~al.,
\newblock JCAP 0502 (2005) 001, hep-th/0411217.

\bibitem{danielsson0606474}
U.H. Danielsson,
\newblock (2006), astro-ph/0606474.

\bibitem{kempf0609123}
A. Kempf and L. Lorenz,
\newblock Phys. Rev. D 74 (2006) 103517, gr-qc/0609123.

\bibitem{burgess07082865}
C.P. Burgess,
\newblock PoS P2GC (2006) 008, arXiv:0708.2865.

\bibitem{hannestadCOSMOcutoff}
J. Hamann et~al.,
\newblock JCAP 0809 (2008) 015, arXiv:0807.4528.

\bibitem{starobCOSMO}
A.A. Starobinsky and I.I. Tkachev,
\newblock JETP Lett. 76 (2002) 235, astro-ph/0207572.

\bibitem{parentaniMDR}
J. Macher and R. Parentani,
\newblock Phys. Rev. D 78 (2008) 043522, arXiv:0804.1920.

\bibitem{brande2012}
R. Brandenberger,
\newblock (2012), arXiv:1204.6108.

\bibitem{cosmodsr}
S. Alexander, R. Brandenberger and J. Magueijo,
\newblock Phys. Rev. D 67 (2003) 081301, hep-th/0108190.

\bibitem{machadoALAbrande}
U. Machado and R. Opher,
\newblock Class. Quantum Grav. 29 (2012) 065003, arXiv:1102.4828.

\bibitem{livcosmos1}
J. Gamboa, J. Lopez-Sarrion and A. Polychronakos,
\newblock Phys. Lett. B 634 (2006) 471, hep-ph/0510113.

\bibitem{livcosmos2}
S. Kanno and J. Soda,
\newblock Phys. Rev. D 74 (2006), hep-th/0604192.

\bibitem{livcosmos3}
K. Nozari and B. Fazlpour,
\newblock Gen. Relativ. Gravit. 38 (2006) 1661, gr-qc/0601092.

\bibitem{livcosmos4}
K.j. Hamada et~al.,
\newblock (2007), arXiv:0705.3490.

\bibitem{moffatVSL}
J.W. Moffat,
\newblock Int. J. Mod. Phys. 2 (1993) 351, gr-qc/9211020.

\bibitem{magueVSL1}
A. Albrecht and J. Magueijo,
\newblock Phys. Rev. D 59 (1999) 043516, astro-ph/9811018.

\bibitem{barrowVSL}
J. Barrow,
\newblock Phys. Rev. D 59 (1999) 043515.

\bibitem{magueVSLrev}
J. Magueijo,
\newblock Rept. Prog. Phys. 66 (2003) 2025, astro-ph/0305457.

\bibitem{fordBLURRING2004}
J. Borgman and L. Ford,
\newblock Phys. Rev. D 70 (2004) 064032, gr-qc/0307043.

\bibitem{fordSPECTRALeBLURRING}
R. Thompson and L. Ford,
\newblock Phys. Rev. D 74 (2006) 024012, gr-qc/0601137.

\bibitem{bojoPRL2011}
M. Bojowald, G. Calcagni and S. Tsujikawa,
\newblock Phys.Rev.Lett. 107 (2011) 211302, arXiv:1101.5391.

\bibitem{visserRUNNINGsd}
T.P. Sotiriou, M. Visser and S. Weinfurtner,
\newblock Phys.Rev. D84 (2011) 104018, arXiv:1105.6098.

\bibitem{safetydimreduction1}
O. Lauscher and M. Reuter,
\newblock JHEP 0510 (2005) 050, hep-th/0508202.

\bibitem{safetydimreduction2}
M. Reuter and F. Saueressig,
\newblock JHEP 1112 (2011) 012, arXiv:1110.5224.

\bibitem{hldimreduction}
P. Horava,
\newblock Phys.Rev.Lett. 102 (2009) 161301, arXiv:0902.3657.

\bibitem{dariodimreduction}
D. Benedetti,
\newblock Phys.Rev.Lett. 102 (2009) 111303, arXiv:0811.1396.

\bibitem{michedimreduction1}
M. Arzano et~al.,
\newblock Phys.Rev. D84 (2011) 125002, arXiv:1107.5308.

\bibitem{michedimreduction2}
E. Alesci and M. Arzano,
\newblock Phys.Lett. B707 (2012) 272, arXiv:1108.1507.

\bibitem{foamdimreduction}
L. Modesto,
\newblock Class.Quant.Grav. 26 (2009) 242002, arXiv:0812.2214.

\bibitem{carlipRUNNINGsd}
S. Carlip, R. Mosna and J. Pitelli,
\newblock Phys.Rev.Lett. 107 (2011) 021303, arXiv:1103.5993.

\bibitem{limCarroll}
S.M. Carroll and E.A. Lim,
\newblock Phys. Rev. D 70 (2004) 123525, hep-th/0407149.

\bibitem{limMP}
E. Lim,
\newblock Phys. Rev. D 71 (2005) 063504, astro-ph/0407437.

\bibitem{kamionkowskiPOLARIZATION}
M. Kamionkowski,
\newblock Phys.Rev.Lett. 102 (2009) 111302, arXiv:0810.1286.

\bibitem{durrer2008}
T. Kahniashvili, R. Durrer and Y. Maravin,
\newblock Phys. Rev. D 78 (2008), arXiv:0807.2593.

\bibitem{Kostelecky2007}
V. Kosteleck{\'{y}} and M. Mewes,
\newblock Phys. Rev. Lett. 99 (2007), astro-ph/0702379.

\bibitem{giulialucaPROCSUPP}
G. Gubitosi and L. Pagano,
\newblock Nucl.Phys.Proc.Suppl. 194 (2009) 69.

\bibitem{kieferSEMI}
C. Kiefer,
\newblock (1993), gr-qc/9312015.

\bibitem{kieferSING}
C. Kiefer and T. Singh,
\newblock Phys. Rev. D 44 (1991) 1067.

\bibitem{otherSEMI}
R. Parentani,
\newblock Phys. Rev. D 56 (1997) 4618, gr-qc/9703008.

\bibitem{montaniSEMI}
S. Mercuri and G. Montani,
\newblock Int. J. Mod. Phys. 13 (2004) 165, gr-qc/0310077.

\bibitem{veneCOSMOnew}
M. Gasperini and G. Veneziano,
\newblock (2007), hep-th/0703055.

\bibitem{veneCOSMOold1}
R. Brustein et~al.,
\newblock Phys. Lett. B361 (1995) 45, hep-th/9507017.

\bibitem{veneCOSMOold2}
R. Brustein et~al.,
\newblock Phys. Rev. D 51 (1995) 6744, hep-th/9501066.

\bibitem{giudicelarge}
G. Giudice, R. Rattazzi and J. Wells,
\newblock Nucl. Phys. B544 (1999) 3, hep-ph/9811291.

\bibitem{giddingslarge}
S. Giddings and S. Thomas,
\newblock Phys. Rev. D 65 (2002) 056010, hep-ph/0106219.

\bibitem{photolarge}
K. Dienes, E. Dudas and T. Gherghetta,
\newblock Phys. Lett. B436 (1998) 55, hep-ph/9803466.

\bibitem{appelquistlarge}
T. Appelquist, H.C. Cheng and B. Dobrescu,
\newblock Phys. Rev. D 64 (2001) 035002, hep-ph/0012100.

\bibitem{ledEXPERIMENT}
C.D. Hoyle et~al.,
\newblock Phys. Rev. Lett. 86 (2001) 1418, hep-ph/0011014.

\end{thebibliography}
\end{document}